\documentclass[a4paper,10pt]{report}

\usepackage{tocloft}
\tocloftpagestyle{fancy}
\usepackage{latexsym}
\usepackage{hyphenat}
\usepackage[top=2.5cm, bottom=2.5cm, left=4.0cm, right=2.5cm]{geometry}
\usepackage{amsmath,amssymb}
\usepackage{color}
\usepackage{graphicx}
\usepackage[colorlinks=true, citecolor=blue, anchorcolor=black, linkcolor=blue, urlcolor=blue, breaklinks=true]{hyperref}
\usepackage{verbatim}
\usepackage{epic, eepic}
\usepackage{psfrag}
\usepackage{pstricks}
\usepackage{lscape}
\usepackage{float}
\usepackage{arev}

\usepackage[font=small,format=plain,labelfont=bf]{caption}
\usepackage{subfig}
\usepackage{fancyhdr}
\usepackage{overpic}
 \usepackage[raggedright]{titlesec}
 \newlength{\PictHOffset}
\newlength{\PictVOffset}
\usepackage{textpos}

\usepackage{epigraph}
\newcommand\prefixtext[1]{%
\ifvmode\else\\\@empty\fi
\noalign{%
\penalty0%
\vbox{\mathstrut}%
\penalty10000%
\vskip-\baselineskip
\penalty10000%
\vbox to 0pt{%
\normalbaselines
\ifdim\linewidth=\columnwidth
\else
\parshape\@ne
\@totalleftmargin\linewidth
\fi
\vss
\noindent#1\par}%
\penalty10000%
\vskip-\baselineskip}%
\penalty10000}

\clearpage{\pagestyle{empty}\cleardoublepage}

\makeatletter
\def\cleardoublepage{\clearpage\if@twoside \ifodd\c@page\else
\hbox{}
\vspace*{\fill}
\begin{center}

\end{center}
\vspace{\fill}
\thispagestyle{empty}
\newpage
\if@twocolumn\hbox{}\newpage\fi\fi\fi}
\makeatother

\usepackage{svn-multi}
\svnid{$Id: thesis_ebutler.tex 498 2011-03-31 12:00:27Z ebutler $}
\definecolor{yellow}{RGB}{0,0,0} 

\newcommand{\grad}{\nabla}
\newcommand{\Hbar}{\mathrm{\bar{H}}}
\newcommand{\pbar}{ \mathrm{\bar{p}} }
\newcommand{\Chi}{\mathrm{X}}

\newcommand{\jphysb}{J. Phys. B: At. Mol. Opt. Phys. }
\newcommand{\aipconf}{AIP Conf. Proc. }

\newcommand{\mtotex}[1]{\immediate\write18{m4 #1.m4 | gpic -t > #1.tex}}

\begin{document}

    \renewcommand{\chaptermark}[1]{%
    \markboth{#1}{}}

    \setcounter{page}{0}

       \noindent\hspace*{-\PictHOffset}%
   \raisebox{\PictVOffset}[0pt][0pt]{\makebox[0pt][l]{%
      \includegraphics[width=\paperwidth]{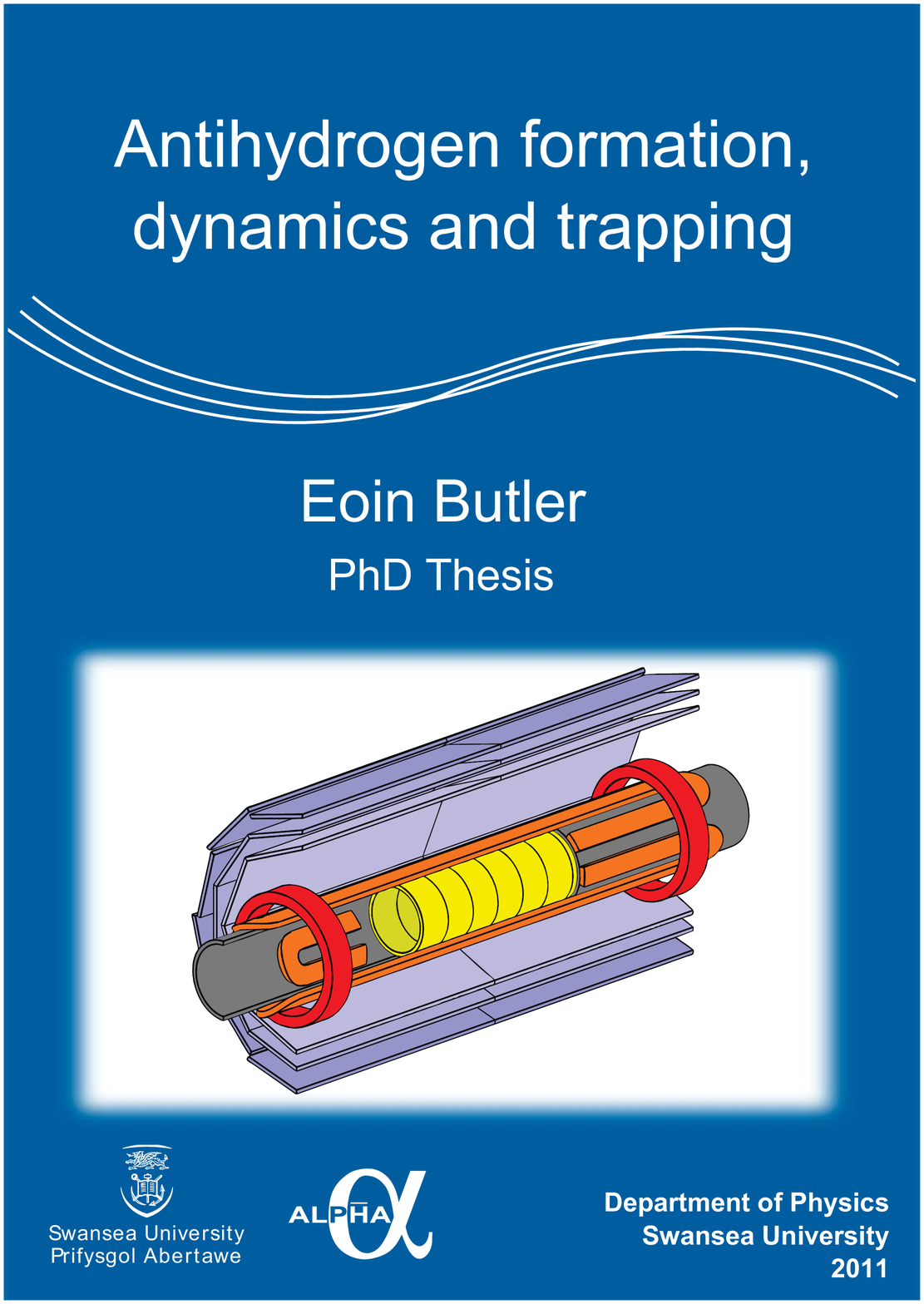}}}

\begin{titlepage}

\vspace*{4cm}

\begin{center}

{\huge \bfseries Antihydrogen formation, dynamics and trapping}\\[1.5cm]

{\Large PhD Thesis}\\[1.0cm]

{\Large Eoin Butler}\\[2.5cm]
{\large Department of Physics, Swansea University, Swansea, United~Kingdom\\ and\\ ALPHA Collaboration, CERN, Geneva, Switzerland}\\[1cm]

\begin{minipage}{0.3\textwidth}
\includegraphics[width=\textwidth]{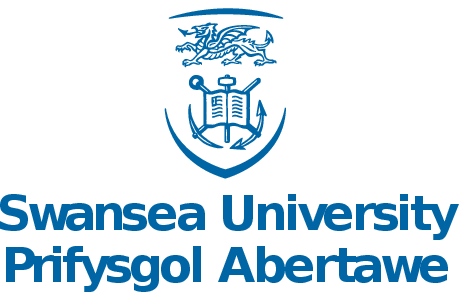}
\end{minipage}
\hspace{2cm}
\begin{minipage}{0.3\textwidth}
\includegraphics[width=\textwidth, clip=true, trim = 0 10cm 0 10cm]{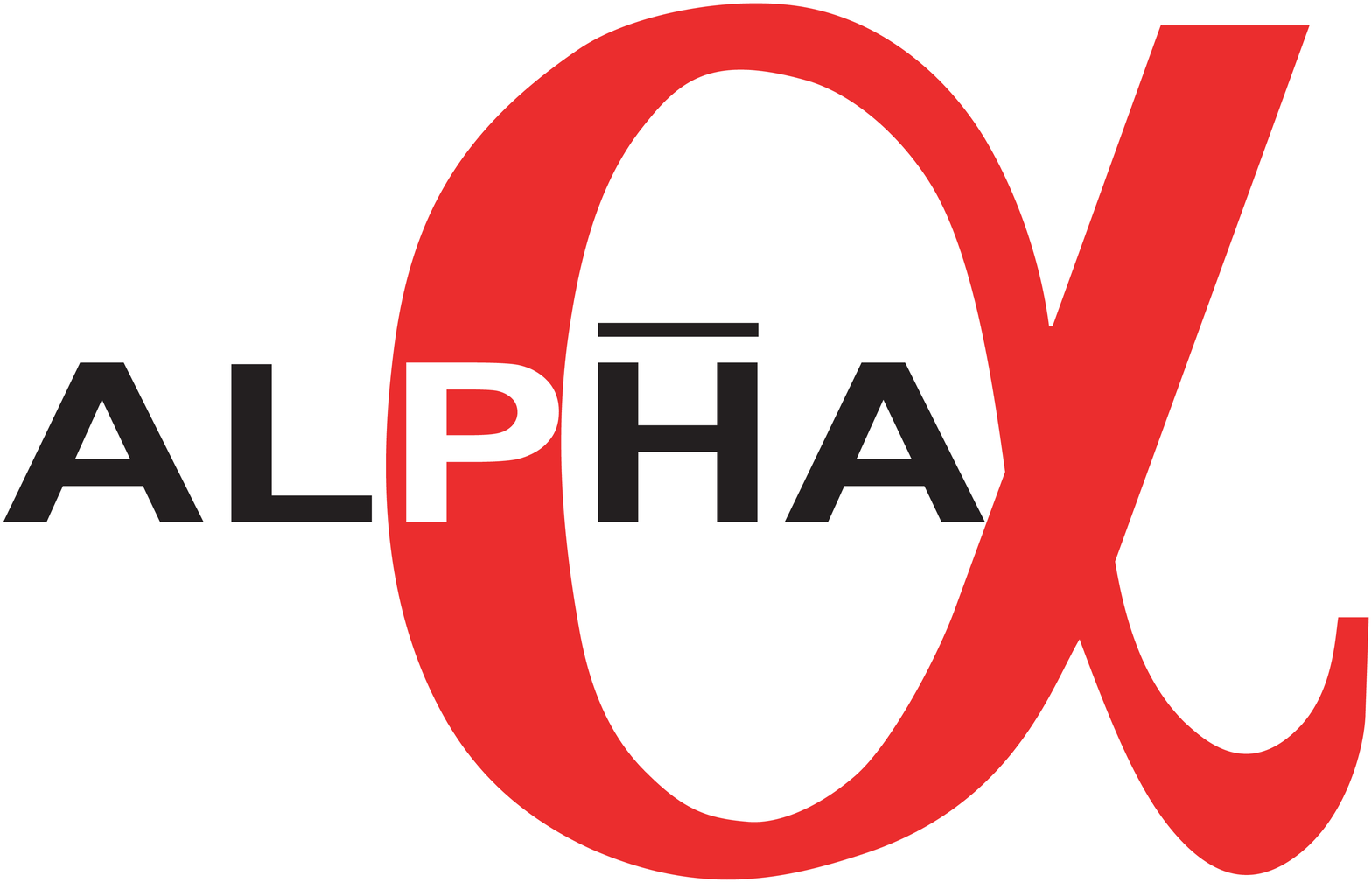}
\end{minipage}

\vspace{1cm}

{\large Submitted to Swansea University in fulfilment of the requirements for the Degree of Doctor of Philosophy}\\[0.5cm]

{\large 2011}

\end{center}

\end{titlepage}

    \renewcommand{\thepage}{\roman{page}}
    \cftsetpnumwidth{3em}

    \chapter*{Abstract}

\phantomsection
\addcontentsline{toc}{chapter}{Abstract}	

Antihydrogen, the simplest pure-antimatter atomic system, holds the promise of direct tests of matter-antimatter equivalence and CPT invariance, two of the outstanding unanswered questions in modern physics.
Antihydrogen is now routinely produced in charged-particle traps through the combination of plasmas of antiprotons and positrons, but the atoms escape and are destroyed in a minuscule fraction of a second.

The focus of this work is the production of a sample of cold antihydrogen atoms in a magnetic atom trap.
This poses an extreme challenge, because the state-of-the-art atom traps are only approximately 0.5~K deep for ground-state antihydrogen atoms, much shallower than the energies of particles stored in the plasmas.
This thesis will outline the main parts of the ALPHA experiment, with an overview of the important physical processes at work.
Antihydrogen production techniques will be described, and an analysis of the spatial annihilation distribution to give indications of the temperature and binding energy distribution of the atoms will be presented.
Finally, we describe the techniques needed to demonstrate confinement of antihydrogen atoms, apply them to a data taking run and present the results, making a definitive identification of trapped antihydrogen atoms.

    \chapter*{Declarations and Statements}

\phantomsection
\addcontentsline{toc}{chapter}{Declarations and Statements}	

\section*{Declaration}

This work has not been previously been accepted in substance for any degree and is not being concurrently submitted in candidature for any degree.

\vspace*{0.5cm}
\begin{tabular}{r c l}
	Signed: & \line(1,0){160} & (Candidate) \\
	& & \\
	Date : & \line(1,0){160} & ~ \\
\end{tabular}

\section*{Statement 1}

This thesis is the result of my own investigations, except where otherwise stated. Other sources are acknowledged by explicit references within the appended bibliography.

\vspace*{0.5cm}
\begin{tabular}{r c l}
	Signed: & \line(1,0){160} & (Candidate) \\
	& & \\
	Date : & \line(1,0){160} & ~ \\
\end{tabular}

\section*{Statement 2}

I hereby give consent for my thesis, if accepted, to be available for photocopying and for inter-library loan, and for the title and summary to be made available to outside organisations.

\vspace*{0.5cm}
\begin{tabular}{r c l}
	Signed: & \line(1,0){160} & (Candidate) \\
	& & \\
	Date : & \line(1,0){160} & ~ \\
\end{tabular}

    \phantomsection
    \addcontentsline{toc}{chapter}{Contents}
    \tableofcontents

    \chapter*{Acknowledgements}

\phantomsection
\addcontentsline{toc}{chapter}{Acknowledgements}	

I would like to express my gratitude to those who have kindly taken the time to share their knowledge and wisdom with me over the course of my studies. 
I could list every member of ALPHA here, but I must express my thanks in particular to Prof. Jeff Hangst, Prof. Joel Fajans, Prof. Francis Robicheaux, Dr. Paul Bowe, Dr. Makoto Fujiwara, and especially to my advisors Prof. Mike Charlton and Dr. Niels Madsen.
I would particularly like to thank them for creating an environment in which the work and opinions of a student are taken with the same value as those of an established scientist.

To my friends and colleagues of the ALPHA experiment, especially \nohyphens{Gorm Bruun Andresen, Tim Friesen, Will Bertsche, Andrew Humphries, Richard Hydomako,\linebreak[4] James Storey, Steve Chapman, Sarah Seif El Nasr-Storey, Daniel de Miranda Silveira, Dean Wilding, Crystal Bray, Mohammad Ashkezari, Chukman So, Marcelo Baquero-Ruiz and Andrea Gutierrez}, my heartfelt appreciation. 
There could be no better group of people to share the long shifts, the hard work, and the sweet victories with.

To all those others, too many to name, without whom this work would not have been possible.

I would like to acknowledge the financial support of the Leverhulme Trust, the University of Wales and the Engineering and Physical Sciences Research Council (EPSRC).
This work has made use of many free and open software projects, of which ROOT, GNU, Ubuntu and Latex are just a few. My thanks to those who gave freely of their time to make this possible.

    \chapter*{Foreword}

The work presented in this thesis took place at the ALPHA antihydrogen experiment, located in CERN, Geneva.
ALPHA is an experiment that requires a wide variety of skills and any results obtained are certainly only possible through the collaborative effort of the forty or so scientists working on ALPHA.

My personal contribution to ALPHA has ranged over the complete spectrum of disiplines, and it is difficult to point to any one area, task or result and say ``That's mine!''
In this thesis I have attempted to reflect, as closely as possible, my involvement in the ALPHA experiment, focussing on areas where I have had the most impact, while glossing over the details of others that are no less important.
As a graduate student permanently stationed at CERN, I have had the opportunity to be on the front line of the physics efforts over the past four experimental runs.
I have been closely involved with the experiments to trap and identify antihydrogen atoms, and took the role of the lead author on the recent publication \textit{Search for Trapped Antihydrogen}, as well as assisting in the analysis and editing for \textit{Trapped Antihydrogen}.
The analysis presented in chapter \ref{chp:vertex} is mostly my own work, but I have benefited from the advice and suggestions of my advisor, Dr. Niels Madsen, and from Prof. Francis Robicheaux.

Working on a physics experiment involves taking part in technical projects also, and I have been closely involved in the construction and maintenance of the ALPHA apparatus, in particular the assembly of the Penning-Malmberg trap, and I have benefited immensely from the opportunity to develop skills in these areas.

    \chapter*{Publications}

\phantomsection
\addcontentsline{toc}{chapter}{Publications}	

\chaptermark{Publications}

This is a list of the publications produced during the work reported in this thesis.

\section*{Peer-reviewed journals}

\nohyphens{
\begin{flushleft}

\begin{itemize}

\item{G.~B.~Andresen, M.~D.~Ashkezari, M.~Baquero-Ruiz, W.~Bertsche, P.~D.~Bowe, \textbf{E.~Butler}, C.~L.~Cesar, S.~Chapman, M.~Charlton, A.~Deller, S.~Eriksson, J.~Fajans,  T.~Friesen, M.~C.~Fujiwara, D.~R.~Gill, A.~Gutierrez, J.~S.~Hangst, W.~N.~Hardy, M.~E.~Hayden, A.~J.~Humphries, R.~Hydomako, S.~Jonsell, N.~Madsen, S.~Menary, P.~Nolan, A.~Olin, A.~Povilus, P.~Pusa, F.~Robicheaux, E.~Sarid, D.~M.~Silveira, C.~So, J.~W.~Storey, R.~I.~Thompson, D.~P.~van~der~Werf, J.~S.~Wurtele, and Y.~Yamazaki,  \textit{Centrifugal separation and equilibration dynamics in an electron-antiproton plasma}, in press at Phys.~Rev.~Lett (2011).}

\item{G.~B.~Andresen, M.~D.~Ashkezari, M.~Baquero-Ruiz, W.~Bertsche, P.~D.~Bowe, \textbf{E.~Butler}, P.~T.~Carpenter, C.~L.~Cesar, S.~Chapman, M.~Charlton, J.~Fajans, T.~Friesen, M.~C.~Fujiwara, D.~R.~Gill, J.~S.~Hangst, W.~N.~Hardy, M.~E.~Hayden, A.~J.~Humphries, R.~Hydomako, J.~L.~Hurt, S.~Jonsell, N.~Madsen, S.~Menary, P.~Nolan, K.~Olchanski, A.~Olin, A.~Povilus, P.~Pusa, F.~Robicheaux, E.~Sarid, D.~M.~Silveira, C.~So, J.~W.~Storey, R.~I.~Thompson, D.~P.~van~der~Werf, J.~S.~Wurtele, and Y.~Yamazaki, \textit{Autoresonant excitation of antiproton plasmas}, submitted to \href{http://dx.doi.org/doi:10.1103/PhysRevLett.106.025002}{Phys.~Rev.~Lett} \textbf{106} 025002 (2011).}

\item{G.~B.~Andresen, M.~D.~Ashkezari, M.~Baquero-Ruiz, W.~Bertsche, P.~D.~Bowe, \textbf{E.~Butler}, C.~L.~Cesar, S.~Chapman, M.~Charlton, A.~Deller, S.~Eriksson, J.~Fajans,  T.~Friesen, M.~C.~Fujiwara, D.~R.~Gill, A.~Gutierrez, J.~S.~Hangst, W.~N.~Hardy, M.~E.~Hayden, A.~J.~Humphries, R.~Hydomako, M.~J.~Jenkins, S.~Jonsell, L.~V.~J\o rgensen, L.~Kurchaninov, N.~Madsen, S.~Menary, P.~Nolan, K.~Olchanski, A.~Olin, A.~Povilus, P.~Pusa, F.~Robicheaux, E.~Sarid, S.~Seif~El~Nasr, D.M.~Silveira, C.~So, J.W.~Storey, R.~I.~Thompson, D.~P.~van~der~Werf, J.~S.~Wurtele, and Y.~Yamazaki , \textit{Trapped antihydrogen}, \href{http://dx.doi.org/doi:10.1038/nature09610}{Nature} \textbf{468} 673 (2010).}

\item{G.~B.~Andresen, M.~D.~Ashkezari, M.~Baquero-Ruiz, W.~Bertsche, P.~D.~Bowe, C.~C.~Bray, \textbf{E.~Butler}, C.~L.~Cesar, S.~Chapman, M.~Charlton, J.~Fajans, T.~Friesen, M.~C.~Fujiwara, D.~R.~Gill, J.~S.~Hangst, W.~N.~Hardy, R.~S.~Hayano, M.~E.~Hayden, A.~J.~Humphries, R.~Hydomako, S.~Jonsell, L.~V.~J\o rgensen, L.~Kurchaninov , R.~Lambo, N.~Madsen, S.~Menary, P.~Nolan, K.~Olchanski, A.~Olin, A.~Povilus, P.~Pusa, F.~Robicheaux, E.~Sarid, S.~Seif~El~Nasr, D.~M.~Silveira, C.~So, J.~W.~Storey, R.~I.~Thompson, D.~P.~van~der~Werf, D.~Wilding, J.~S.~Wurtele, and Y.~Yamazaki, \textit{Search for Trapped Antihydrogen},  \href{http://dx.doi.org/doi:10.1016/j.physletb.2010.11.004}{Phys. Lett. B} \textbf{695}, 95 (2011).}

\item{G.~B.~Andresen, M.D.~Ashkezari, M.~Baquero-Ruiz, W.~Bertsche,  P.~D.~Bowe, \textbf{E.~Butler}, C.~L.~Cesar, S.~Chapman, M.~Charlton, J.~Fajans, T.~Friesen, M.C.~Fujiwara,  D.~R.~Gill, J.~S.~Hangst, W.N.~Hardy, R.~S.~Hayano, M.~E.~Hayden, A.~Humphries, R.~Hydomako,  S.~Jonsell, L.~Kurchaninov, R.~Lambo, N.~Madsen, S.~Menary, P.~Nolan, K.~Olchanski, A.~Olin,  A.~Povilus, P.~Pusa,  F.~Robicheaux, E.~Sarid, D.M.~Silveira, C.~So, J.W.~Storey, R.I.~Thompson, D.P.~van~der~Werf,  D.~Wilding, J.~S.~Wurtele, and Y.~Yamazaki, \textit{Evaporative Cooling of Antiprotons to Cryogenic Temperatures}, \href{http://dx.doi.org/10.1103/PhysRevLett.105.013003}{Phys. Rev. Lett.}, \textbf{105} 013003 (2010).}

\item{G.~B.~Andresen, W.~Bertsche,  P.~D.~Bowe, C.~Bray, \textbf{E.~Butler}, C.~L.~Cesar, S.~Chapman, M.~Charlton, J.~Fajans, M.~C.~Fujiwara,  D.~R.~Gill, J.~S.~Hangst, W.~N.~Hardy, R.~S.~Hayano, M.~E.~Hayden, A.~J.~Humphries, R.~Hydomako,  L.~V.~J\o rgensen, S.~J.~Kerrigan, L.~Kurchaninov, R.~Lambo, N.~Madsen, P.~Nolan, K.~Olchanski, A.~Olin,  A.~Povilus, P.~Pusa,  F.~Robicheaux, E.~Sarid, S.~Seif~El~Nasr, D.~M.~Silveira, J.~W.~Storey, R.~I.~Thompson, D.~P.~van~der~Werf,  and Y.~Yamazaki, \textit{Antihydrogen formation dynamics in a multipolar neutral anti-atom trap}, \href{http://dx.doi.org/10.1016/j.physletb.2010.01.066}{Phys. Lett B}, \textbf{685} 141 (2010).}

\item{G.~B.~Andresen, W.~Bertsche,  P.~D.~Bowe, C.~C.~Bray, \textbf{E.~Butler}, C.~L.~Cesar, S.~Chapman, M.~Charlton, J.~Fajans, M.~C.~Fujiwara,  D.~R.~Gill, J.~S.~Hangst, W.~N.~Hardy, R.~S.~Hayano, M.~E.~Hayden, A.~J.~Humphries, R.~Hydomako,  L.~V.~J\o rgensen, S.~J.~Kerrigan, L.~Kurchaninov, R.~Lambo, N.~Madsen, P.~Nolan, K.~Olchanski, A.~Olin,  A.~P.~ Povilus, P.~Pusa,  E.~Sarid, S.~Seif~El~Nasr, D.~M.~Silveira, J.~W.~Storey, R.~I.~Thompson, D.~P.~van~der~Werf,  and Y.~Yamazaki, \textit{Anitproton, Positron, and Electron Imaging with a Microchannel Plate/Phosphor Detector}, \href{http://dx.doi.org/10.1063/1.3266967}{Rev. Sci. Inst.} \textbf{80}, 123701 (2009).}

\item{G.~B.~Andresen, W.~Bertsche,  C.~C.~Bray, \textbf{E.~Butler}, C.~L.~Cesar, S.~Chapman, M.~Charlton, J.~Fajans, M.~C.~Fujiwara,  D.~R.~Gill, W.~N.~Hardy, R.~S.~Hayano, M.~E.~Hayden, A.~J.~Humphries, R.~Hydomako,  L.~V.~J\o rgensen, S.~J.~Kerrigan, J.~Keller, L.~Kurchaninov, R.~Lambo, N.~Madsen, P.~Nolan, K.~Olchanski, A.~Olin,  A.~Povilus, P.~Pusa, F.~Robicheaux, E.~Sarid, S.~Seif~El~Nasr, D.~M.~Silveira, J.~W.~Storey, R.~I.~Thompson, D.~P.~van~der~Werf, J.~S.~Wurtele, and Y.~Yamazaki, \textit{Magnetic Multiple Induced Zero-Rotation Frequency Bounce-Resonant Loss in a Penning-Malmberg Trap Used For Antihydrogen Trapping}, \href{http://dx.doi.org/10.1063/1.3258840}{Phys. Plasmas}, \textbf{16} 100702 (2009).}

\item{G.~B.~Andresen, W.~Bertsche, P.~D.~Bowe, C.~C.~Bray, \textbf{E.~Butler}, C.~L.~Cesar, S.~Chapman, M.~Charlton, J.~Fajans, M.~C.~Fujiwara, R.~Funakoshi, D.~R.~Gill, J.~S.~Hangst, W.~N.~Hardy, R.~S.~Hayano, M.~E.~Hayden, R.~Hydomako, M.~J.~Jenkins, L.~V.~J\o rgensen, L.~Kurchaninov, R.~Lambo, N.~Madsen, P.~Nolan, K.~Olchanski, A.~Olin, A.~Povilus, P.~Pusa, F.~Robicheaux, E.~Sarid, S.~Seif~El~Nasr, D.~M.~Silveira, J.W.~Storey, R.I.~Thompson, D.P.~van~der~Werf, J.~S.~Wurtele, and Y.~Yamazaki, \textit{Compression of antiproton clouds for antihydrogen trapping}, \href{http://dx.doi.org/10.1103/PhysRevLett.100.203401}{Phys. Rev. Lett}, \textbf{100} 203401 (2008).}

\item{G.~B.~Andresen, W.~Bertsche, P.~D.~Bowe, C.~C.~Bray, \textbf{E.~Butler}, C.~L.~Cesar, S.~Chapman, M.~Charlton, J.~Fajans, M.C.~Fujiwara, R.~Funakoshi, D.R.~Gill, J.S.~Hangst, W.N.~Hardy, R.S.~Hayano, M.E.~Hayden, A.J.~Humphries, R.~Hydomako, M.J.~Jenkins, L.V.~J\o rgensen, L.~Kurchaninov, R.~Lambo, N.~Madsen, P.~Nolan, K.~Olchanski, A.~Olin, R.~D.~Page, A.~Povilus, P.~Pusa, F.~Robicheaux, E.~Sarid, S.~Seif~El~Nasr, D.M.~Silveira, J.W.~Storey, R.I.~Thompson, D.P.~van~der~Werf, J.~S.~Wurtele, and Y.~Yamazaki, \textit{A novel antiproton radial diagnostic based on octupole induced ballistic loss}, \href{http://dx.doi.org/10.1063/1.2899305}{Phys. Plasmas}, \textbf{15} 032107 (2008).}

\end{itemize}
\end{flushleft}
}
\section*{Peer-reviewed conference proceedings}

Only publications with the author of this thesis as the principal author have been included.
Six additional conference proceedings as a member of the ALPHA Collaboration are not listed.
\nohyphens{
\begin{flushleft}

\begin{itemize}
\item{\textbf{E.~Butler}, G.~B.~Andresen, M.~D.~Ashkezari, M.~Baquero-Ruiz, W.~Bertsche, P.~D.~Bowe, C.~C.~Bray, C.~L.~Cesar, S.~Chapman, M.~Charlton, J.~Fajans, T.~Friesen, M.~C.~Fujiwara, D.~R.~Gill, J.~S.~Hangst, W.~N.~Hardy, R.~S.~Hayano, M.~E.~Hayden, A.~J.~Humphries, R.~Hydomako, S.~Jonsell, L.~Kurchaninov, R.~Lambo, N.~Madsen, S.~Menary, P.~Nolan, K.~Olchanski, A.~Olin, A.~Povilus, P.~Pusa, F.~Robicheaux, E.~Sarid, D.~M.~Silveira, C.~So, J.~W.~Storey, R.~I.~Thompson, D.~P.~van~der~Werf, D.~Wilding, J.~S.~Wurtele, Y.~Yamazaki, \textit{Towards Antihydrogen Trapping and Spectroscopy at ALPHA}, accepted for publication in the proceedings of the conference on trapped charged particles and fundamental physics (TCP2010), Hyperfine Interactions (2010).}

\end{itemize}
\end{flushleft}

}

    \chapter*{Units and Notation}
\phantomsection
\addcontentsline{toc}{chapter}{Units and Notation}	

\chaptermark{Units and Notation}

\section*{Units}

All equations in this thesis use the \textit{Syst\`eme International} (SI) units \cite{SI}. 
In some cases, it has been more appropriate to quote a number using an SI prefix (\textit{e.g.} cm, ms), or using an SI-derived unit (\textit{e.g.} eV).
It has become common in the field of atom trapping to quote the energy of an atom in units of Kelvin (K). 
In all cases, the atom's energy should be understood to be Boltzmann's constant ($k_\mathrm{B}$) times the Kelvin quantity.
Units are written in an upright font, distinguishing them from variables, which are italicised.

\section*{Frequently used symbols}

Table \ref{tab:symbols} contains a list of the meanings of symbols commonly used in the mathematical expressions found in this thesis.
Scalar quantities are written italicised, while vector quantities are in boldface.
When a vector quantity (\textit{e.g} $\mathbf{B}$) is written as a scalar (\textit{e.g} $B$), it should be understood that the scalar refers to the vector's magnitude ($|\mathbf{B}|$).
Where a vector is written with a subscript corresponding to an axis (\textit{e.g} $B_z, E_r$), this quantity refers to the projection of that vector onto that axis.

\begin{table}
\begin{tabular} {l | l }
	$\mathbf{B}, B$			& magnetic field												\\
	$E$						& kinetic energy												\\
	$\mathcal{E}$			& atomic binding energy											\\
	$\mathbf{F}, F$			& electric field												\\
	$m$						& mass \textit{or} magnetic quantum number						\\
	$n$						& particle density \textit{or} principal quantum number			\\
	$N$						& number of particles											\\
	$\mathbf{p}, p$			& momentum														\\
	$t$						& time															\\
	$T$						& temperature													\\
	$U$						& potential energy												\\
	$\mathbf{v}, v$			& velocity														\\
	$\Gamma$				& rate (usually formation rate)									\\
	$\Phi$					& electric potential											\\
	$\omega$				& rotational frequency											\\
	$x, y, z$				& Cartesian coordinates											\\
	$r, \phi, z$			& cylindrical coordinates										
\end{tabular}
\caption{A list of frequently used symbols and their meanings.}
\label{tab:symbols}
\end{table}

\section*{Physical Constants}

Table \ref{tab:constants} contains a list of the physical constants used throughout this work, and their approximate values.
The exact values (and uncertainties) of the constants can be found in the CODATA database \cite{CODATA}.

\begin{table}
\begin{tabular} {l | l | l l}
	$c$						& speed of light											& 2.998 $\times$ $10^{8}$   & 	$\mathrm{m\,s^{-1}}$	\\
	$e$						& elementary charge											& 1.602 $\times$ $10^{-19}$ & 	$\mathrm{C}$	\\
	$k_\mathrm{B}$			& Boltzmann's constant										& 1.381 $\times$ $10^{-23}$ & 	$\mathrm{J\,K^{-1}}$	\\
	$h$						& Planck's constant											& 6.626 $\times$ $10^{-34}$ & 	$\mathrm{J\,s}$	\\
	$m_e$					& electron/positron mass									& 9.109 $\times$ $10^{-31}$ & 	$\mathrm{kg}$	\\
	$m_\pbar$				& proton/antiproton mass									& 1.673 $\times$ $10^{-27}$ & 	$\mathrm{kg}$	\\
	$\epsilon_0$			& permittivity of free space								& 8.854 $\times$ $10^{-12}$ & 	$\mathrm{F\,m^{-1}}$	\\
	$\mu_B$					& Bohr magneton												& 9.274 $\times$ $10^{-14}$ & 	$\mathrm{J\,T^{-1}}$	\\
	\end{tabular}
\caption{Definitions of the symbols and values of frequently used physical constants. }
\label{tab:constants}
\end{table}

    \clearpage

    \include{listfig}

    \renewcommand{\chaptermark}[1]{%
    \markboth{\chaptername
    \ \thechapter.\ #1}{}}

    \renewcommand{\thepage}{\arabic{page}}
    \setcounter{page}{1}
     \chapter{Antimatter}

\epigraph{If anybody says he can think about quantum physics without getting giddy, that only shows he has not understood the first thing about them.
}{Niels Bohr}

For every type of particle, there is a corresponding entity, known as its \textit{antiparticle}, with the same mass and lifetime, and the same magnitude, though opposite sign, of spin and electric charge.

For the electron, there is the positively charged positron, for the proton, the negatively charged antiproton and so forth through the entire particle zoo.
The symbols representing antiparticles are distinguished from those for particles by the addition of an overbar, or by changing the sign denoting the electric charge. So the proton ($\mathrm{p}$) becomes the antiproton ($\pbar$) and the electron ($\mathrm{e}^-$) becomes the positron ($\mathrm{e}^+$).
A small number of particles, most importantly the photon, are their own antiparticles.

The collision of a particle and its antiparticle can result in an annihilation -- the destruction of both and the conversion of their mass into energy through the famous equation
\begin{equation}
E = mc^2.
\label{eq:Emc2}
\end{equation}
Depending on the mass and kinetic energy of the particle-antiparticle pair, this energy appears as photons or showers of other, `daughter' particles. 
Annihilations conserve energy, momentum and electric charge, as well as baryon number, quark number, etc.

The quarks making up hadrons also have corresponding antiquarks - the anti-up for the up, the anti-down for the down. 
(Despite the naming convention, the anti-up quark and the down quark are distinct objects.)
Collisions of different hadrons can result in annihilation, even if the particles have different quark content, as quarks and anti-quarks of different `flavours' can combine to form mesons.
For example, the antiproton ($\mathrm{{\bar{u}\bar{u}\bar{d}}}$) can annihilate with the proton (uud) or the neutron (udd).

\section{History}

The first prediction of the existence of antimatter was made by Paul Dirac in 1931 when developing the relativistic extension of the Schr\"{o}dinger equation (now known as the Dirac equation) \cite{Dirac}.
When solving the equation for the electron, he found, as well as the expected solutions corresponding to an electron with spin-up and spin-down, two negative energy solutions.

Whether or not these solutions had a physical meaning was the subject of much debate at a time when quantum theory was in its infancy.
One possible interpretation was that they were simply protons, but this was quickly disregarded when it was pointed out that the negative-energy states had the same mass as the electron \cite{Oppenheimer}, and the proton has a mass almost 2,000 times that of the electron.
Dirac proposed the novel interpretation that the vacuum contained an infinite `sea' of negative energy states of the electron.
Electrons, being fermions, cannot occupy the same quantum state as another electron, so these states would `fill up'.
Thus, what we think of as empty vacuum is actually where all of the negative energy states are occupied, and none of the positive energy states.
If an electron was removed from a negative energy state, a hole would exist in the sea, and as Dirac wrote \cite{Dirac},
\begin{quote}
	``A hole, if there were one, would be a new kind of particle, unknown to experimental physics, having the same mass and opposite charge to an electron. We may call such a particle an anti-electron. Presumably the protons will have their own negative-energy states ... an unoccupied one appearing as an anti-proton.''
\end{quote}

\begin{figure}
\centering
\includegraphics[width = 0.5 \textwidth]{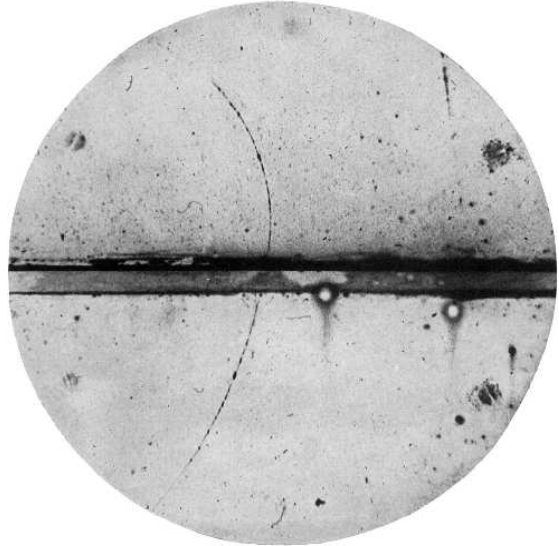}
\caption[A photograph of a positron track in a cloud chamber.]{A photograph of cloud chamber tracks showing a positron passing through a sheet of lead (from the bottom) and emerging with lower kinetic energy (from \cite{positron}).}
\label{fig:positron}
\end{figure}

In 1932, Carl Anderson reported the first observation of the anti-electron, today's positron \cite{positron}.
He was working using a cloud chamber, which makes the tracks of charged particles visible, and recorded photographs of cosmic rays. 
Particles with opposite signs of electric charge bend in opposite directions in a magnetic field, which allows electrons and positrons to be distinguished if their direction of motion is known.
Anderson distinguished downward-moving positrons from upward-going electrons by placing a sheet of lead in the cloud chamber.
Particles passing through the lead lost energy and moved on a trajectory with a correspondingly smaller radius of curvature, as shown in figure \ref{fig:positron}.

With the advent of high-energy particle colliders, a plethora of other particles and antiparticles were discovered in the mid $20^\mathrm{th}$ century.
These included the antiproton, discovered by Emilio Segr\`e and Owen Chamberlain in 1955 at the Bevatron accelerator.
They measured the mass of negatively-charged particles produced through collisions of protons with a copper target and identified a particle with the mass of the proton in the products \cite{antiproton}.

Shortly after the discovery of antimatter, scientists began to wonder if it was possible to have more complex antimatter structures.
The elementary particles of antimatter appeared to obey the same physical laws as those of matter, so it seemed plausible.
Were anti-atoms, anti-molecules, even anti-planets possible?.

The first anti-atoms, antihydrogen, were made at CERN \cite{CERN_Hbar} in 1995 and Fermilab \cite{Fermilab_Hbar} in 1998 through collisions of relativistic antiprotons with a nucleus.
In such interactions, some of the antiproton's energy could be converted to an electron-positron pair.
In a very small number of cases, the antiproton and the positron became bound together in an atom of antihydrogen.
They were detected by deflection of the charged antiproton beam; the antihydrogen atoms, being neutral, continued in a straight line to strike a detector.
These atoms, being produced at relativistic energies and living for a minuscule fraction of a second, were not amenable to study.

The first cold (by which it is meant non-relativistic) antihydrogen atoms were synthesised by the ATHENA experiment in 2002 at the Antiproton Decelerator at CERN \cite{ATHENA_Nature}.
Atoms with energies of order 1~eV were produced through the combination of trapped, charged clouds of antiprotons in a Penning-Malmberg trap.
Even though these particles were produced at energies several orders of magnitude lower than the first atoms, they still escaped the apparatus and annihilated in a few microseconds.
They were identified by detecting the spatially- and temporally-coincident annihilations of an antiproton and positron.

The most attractive route to study the properties of antihydrogen atoms is to confine them in an atom trap for at least a sizable fraction of a second.
The ALPHA collaboration was formed as a successor to ATHENA with the primary aim of confining the atoms in a magnetic trap. 
The work in this thesis describes the progress towards this next step in the history of antimatter.
As described in chapter \ref{chp:trapping}, in 2009, ALPHA observed the first signs of trapped antihydrogen \cite{SearchPaper}, followed by a definitive identification in 2010 \cite{ALPHA_Nature}.

\section{Motivations to study antimatter}

The study of antimatter systems, and in particular, their comparison to the equivalent matter systems, has the potential to enable direct, sensitive tests of a number of fundamental physical theories.

\subsection{Fundamental symmetries}

The theories that first postulated the existence of antimatter, as well as the more complete modern theories of quantum electrodynamics (QED) and quantum chromodynamics (QCD) predict that the matter and antimatter particles should obey the same physical laws.
This can be described in terms of fundamental symmetries, namely the charge conjuugation (C), parity (P) and time (T) symmetries.
A symmetry is said to be `good' with respect to a process or theory, if by applying it, the laws governing the process are unchanged.

Charge conjugation refers to changing the sign of all of the internal quantum numbers of the particles, including the electric charge.
Simplistically, charge conjugation can be thought of as replacing particles with their antiparticle equivalents, but as we shall shortly see, this is not always correct.
Parity conservation refers to changing the signs of the spatial dimensions,  equivalent to reflection in the origin.
Time reversal, as the name implies, is where time runs backwards.

At the beginning of quantum theory, it was thought that these symmetries were individually conserved in nature.
Newton's laws, for example, function equally well when any combination of the three symmetries are applied.
The same applies to Maxwell's equations and the early theories of quantum mechanics, including Schr\"{o}dinger's equation.

However, through experiments, it became clear that nature was not so obliging.
The helicity, or handedness, of a particle describes in which direction its spin points relative to its momentum. A particle with spin and momentum aligned has positive (right-handed) helicity; if they are anti-aligned, the particle has negative (left-handed) helicity.
Experiments led by Wu in 1956 measured the helicity of neutrinos and antineutrinos produced in beta-decays of nuclei and found that only left-handed neutrinos and right-handed anti-neutrinos are observed \cite{neutrino_CP}. 
This violates both the charge conjugation (C) and parity (P) symmetries, as charge conjugation would produce a left-handed antineutrino from a left-handed neutrino, and a parity transformation would produce a right-handed neutrino, neither of which are observed.
A massless neutrino, which would move at the speed of light, could not be overtaken to reverse the sign of its apparent momentum and thus its helicity, this implies that these symmetries are \textit{maximally} violated; the symmetric counterparts of the particles are \textit{never} observed.
This is complicated by fact that the neutrino is thought to have a non-zero rest-mass, which is needed to explain neutrino flavour oscillations \cite{NeutrinoFlavour}, -- anyway, neutrinos are very light and move at speeds very close to the speed of light, so overtaking them is not easily accomplished.

Even though C and P are not individually good symmetries, it was noticed that the combined operation CP still preserved the symmetry of the experiment, as it converted the left-handed neutrino into the right-handed anti-neutrino.
However, CP symmetry was in turn found to be violated in another experiment, using neutral kaons.
Neutral kaons exist in two states, known as the `K-short' ($\mathrm{K_S}$, with a lifetime of $\sim 9 \times 10^{-11}~\mathrm{s}$) and 'K-long' ($\mathrm{K_L}$, with a lifetime $\sim 5 \times 10^{-8}~\mathrm{s}$).
These states have different eigenvalues of CP symmetry; $\mathrm{K_S}$ has CP = +1, while $\mathrm{K_L}$ has CP = -1.
To conserve CP symmetry, when the kaon decays, it must be into a state with the same CP eigenvalue as the initial state. 
Thus, the $\mathrm{K_S}$ state should primarily decay to two pions (CP = +1) and the $\mathrm{K_L}$ to three pions (CP = -1).
However, in 1964, Cronin and Fitch observed a small fraction of decays of $\mathrm{K_L}$ into two pions \cite{CPviolation}, an instance of CP violation.
Other examples of CP-violating processes have since been discovered.

The CP-violating processes can be symmetrised by applying the third transformation, time or T. 
The resulting combined symmetry CPT, is conserved in all known processes.
In fact, the CPT theorem \cite{CPT_theorem}, \cite{CPT_theorem1}, \cite{CPT_theorem2} states that any quantum field theory of point-particles which is Lorentz invariant, and preserves causality will be symmetric under the combined transformation CPT.
The discovery of a CPT-violating process would force a major re-think of modern physics, possibly including the loss of one or more of the assumptions listed above.

It is now an active area of physics research to discover if CPT symmetry holds exactly for all processes, or if it too is violated in some processes.
The systems in which CPT violation is being sought include the oscillations of neutrinos and mesons, gravity (through its effects on atoms and the orbits of celestial bodies), and the spectroscopy of atoms and ions \cite{Kostelecky_dataTables}.
Of most relevance to the present work is the prediction that CPT violation may also manifest in differences in the energy spectra of atoms and their anti-atom counterparts \cite{Kostelecky_Hbar}. 
The study of hydrogen and antihydrogen is particularly attractive because of the very high precision ($\sim 2$ parts in in $10^{14}$ \cite{HydrogenPrecision}) to which the hydrogen spectrum is known.

\subsection{Matter-antimatter asymmetry}

While the fundamental CPT theorem is of great importance, a further question concerning antimatter is its apparent absence in the universe, also called `Baryon asymmetry'.
In the present understanding of the universe, matter and antimatter should have been created in equal quantities in the Big Bang, but as far as we can tell, the universe today is made up of only matter. 
Antimatter is seen only in the laboratory and in some exotic processes such as high-energy cosmic rays or radioactivity.
The Earth and its atmosphere are demonstrably made up entirely of matter, otherwise the matter and antimatter portions would annihilate each other.
Likewise, the fact that probes from Earth have landed on almost all of the planets and moons in the solar system has convincingly demonstrated that they too are made of matter.

We might think that antimatter is separated from matter in some other neighbourhood of the universe.
However, even intergalactic space is not completely empty, and we would expect to observe annihilations in the transition regions between pockets of matter and antimatter, and to be able to see antimatter cosmic rays.
Experiments \cite{AMS}, \cite{PAMELA_results} have not been able to detect antimatter concentrations in the observable universe.
It may be that the antimatter pockets have been placed beyond the horizon of the observable universe by cosmic inflation following the Big Bang.
In this case, light or particles from an antimatter pocket has simply not had enough time to reach us.

Another possible explanation is that matter and antimatter were not present in equal quantities shortly after the Big Bang, but that there was slightly more matter (perhaps one part in $10^9$). 
Almost all of the matter particles annihilated with all of the antimatter particles to produce energy, leaving the tiny fraction of matter from which we, the Earth and the rest of the physical universe is made.
This could be provided by either a bias towards matter in the initial conditions of the universe, or through a gradual evolution.

Matter-antimatter asymmetry is one of the greatest unsolved mysteries of modern science and gathers a great deal of attention.
The study of antihydrogen, to discover and investigate any differences between it and hydrogen, is one of the key efforts to understand this mystery.

\subsection{Antimatter and gravity}

The theory of general relativity incorporates the weak equivalence principle (WEP), which states that the gravitational acceleration of a body is independent of its composition.
This means, as we are familiar with, that objects of different masses fall with the same acceleration.
The WEP predicts the same for antimatter, but an experimental test has not been made.

A measurement of the gravitational acceleration of charged particles and anti-particles is complicated because of the much stronger coupling of the particles to any stray electric fields; the task is much simpler to perform with neutral antimatter -- antihydrogen.
The AEgIS experiment \cite{AEGIS}, under construction at the AD, plans the first measurement of the gravitational acceleration of antimatter by creating a horizontal beam of antihydrogen atoms and measuring their trajectories in the vertical direction using a series of gratings.
AEgIS plans to make a measurement of the gravitational acceleration $g$ of antimatter with around $\sim$1\% accuracy in a few years.

     \chapter{Apparatus}

\epigraph{Nothing is too wonderful to be true,\\if it be consistent with the laws of nature.}{Michael Faraday}

\begin{figure}[h]
\centering
\vspace{1cm}
\includegraphics[width=0.7\textwidth]{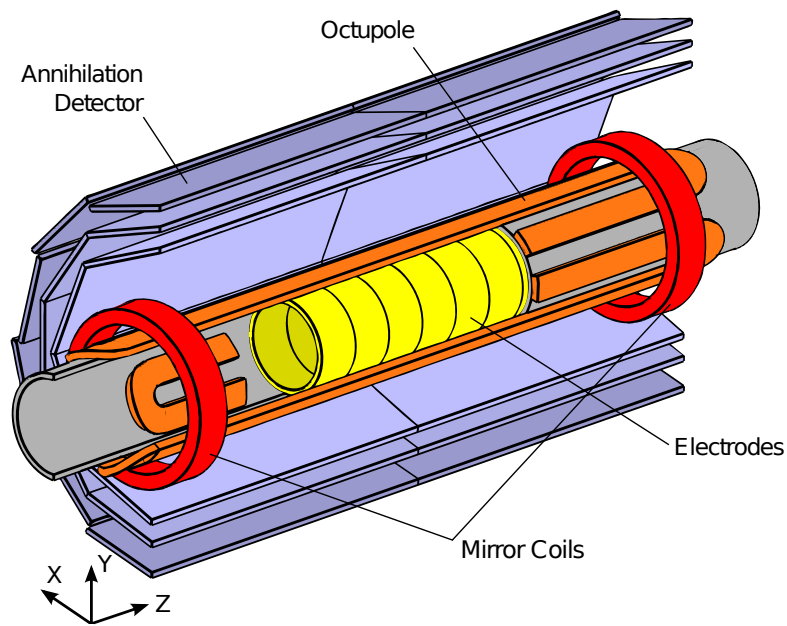}
\caption[A schematic view of the trapping section of the ALPHA apparatus]{A schematic view of the trapping section of the ALPHA apparatus, showing the Penning trap electrodes, the magnetic minimum trap and the annihilation vertex detector.}
\label{fig:Search_app}
\end{figure}

\begin{landscape}
\begin{figure}[p]
\centering
\includegraphics[width=1.5\textwidth]{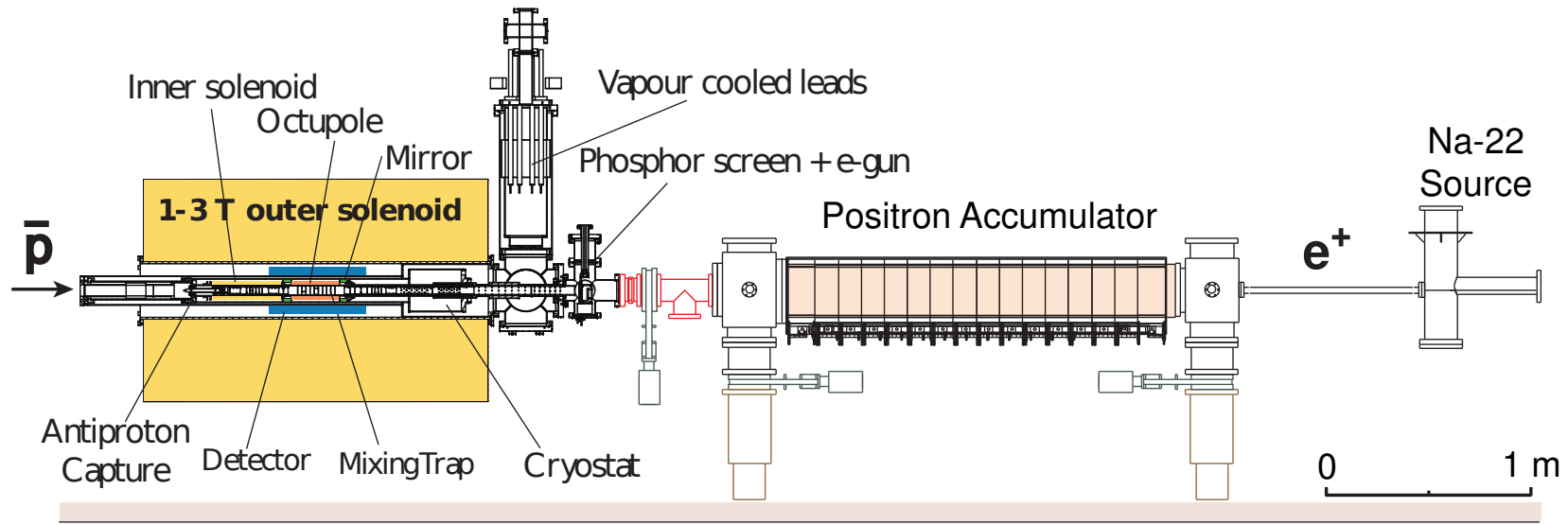}
\caption[A section through the ALPHA apparatus, also showing the positron accumulator]{A section through the ALPHA apparatus, also showing the positron accumulator (Image from the ALPHA Collaboration).}
\label{fig:ALPHA_section}
\end{figure}
\end{landscape}

\begin{figure}[p]
\centering
\includegraphics[width=\textwidth]{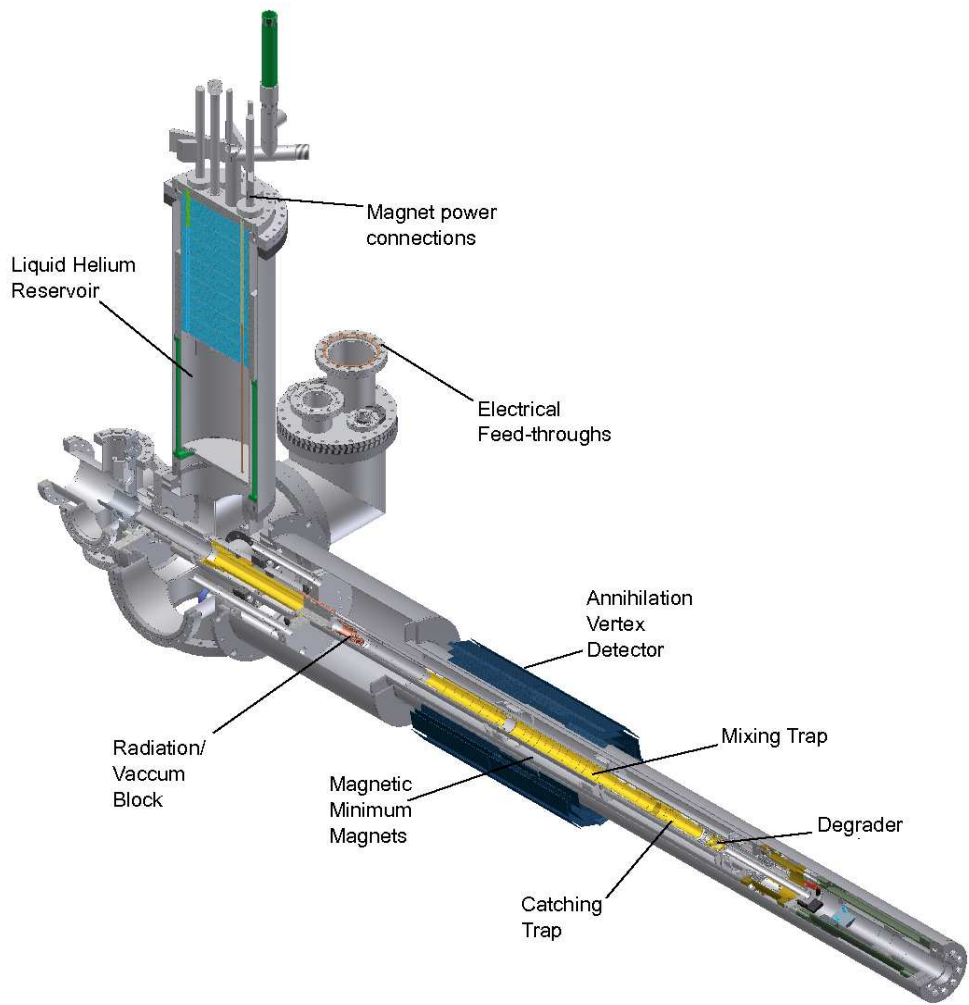}
\caption[A cut-away mechanical drawing of the internal parts of the ALPHA apparatus]{A cut-away mechanical drawing of the internal parts of the ALPHA apparatus. The horizontal portion is inserted into the external solenoid. The apparatus connects to the Antiproton Decelerator to the right and to the positron accumulator to the left.}
\label{fig:allofit}
\end{figure}

The ALPHA apparatus is designed to produce and confine the simplest neutral antimatter atomic system -- antihydrogen.
It is made up of a large number of subsystems -- particle sources, traps for charged and neutral particles, particle detectors, as well as control and data acquisition systems.
Figures \ref{fig:Search_app}, \ref{fig:ALPHA_section} and \ref{fig:allofit} show views of the ALPHA apparatus for reference in later discussions.

\clearpage

\clearpage

\section{Penning and Penning-Malmberg traps}

\subsection{Penning trap theory} \label{sec:penningTrap}

Penning traps are a widely-used class of charged particle traps. 
The prototypical Penning trap consists of a solenoidal magnetic field and a quadratic electric potential:
	\begin{subequations}\label{eq:penningFields}
		\begin{equation}\label{eq:penningFieldsB}
			\mathbf{B} = B_z\hat{z},
		\end{equation}
		\begin{equation}\label{eq:penningFieldsPhi}
			\phi = \frac{V_0}{2d^2}\left(z^2 - \frac{r^2}{2}\right),
		\end{equation}
	\end{subequations}
where $d$ is a characteristic trap dimension.
The electric potential is created by applying voltages to conductors whose edges lie along equipotential surfaces -- for equation \ref{eq:penningFieldsPhi}, the equipotentials are hyperboloids of rotation.

In this ideal Penning trap, the motion of a charged particle can be considered analytically.
A charged particle moving in the electric and magnetic fields is acted upon by the Lorentz Force, given by
	\begin{equation}\label{eq:lorentz}
		m\mathbf{\ddot{r}} = q\left(-\grad\phi + \mathbf{\dot{r}}\times\mathbf{B}\right).
	\end{equation}
The resulting motion can be considered as the combination of three independent oscillatory motions - an oscillation parallel to the magnetic field, at frequency $\omega_z$, a cyclotron motion transverse to the magnetic field at frequency $\omega_{c}'$ and a second rotation about the trap axis, due to the crossed electric and magnetic fields, and known as the magnetron motion, at frequency $\omega_m$.
A full mathematical treatment can be found in \cite{geonium}. 

	\begin{figure}[hbt]
	\centering
	\includegraphics[width=8cm]{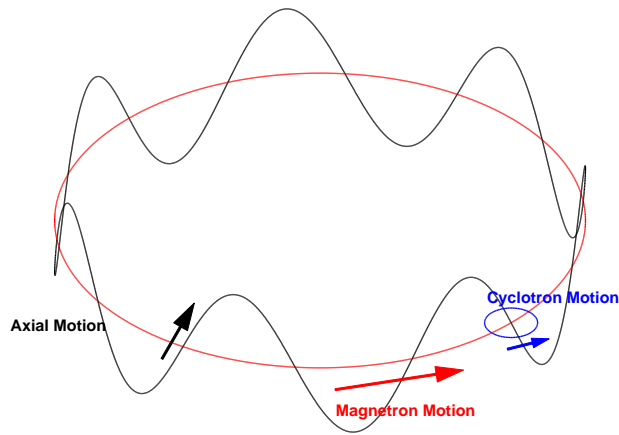}
	\caption{A schematic showing the three components of the motion of a particle in a Penning trap.}
	\label{fig:penningMotion}
	\end{figure}

$\omega_c'$ is closely related to the cyclotron frequency $\omega_c = \frac{eB}{m}$.
The frequencies are given by \cite{geonium}
\begin{eqnarray}
  \omega_c' &=& \omega_c - \omega_m \simeq \omega_c, \\ \label{eqn:omega_c}
  \omega_z &=& \sqrt{{2eV_0}/{m d^2}}, \\ \label{eqn:omega_z}
 \text{and}\hspace{1cm}\omega_m &=& \omega_z^2 / {2 \omega'_c}, \label{eqn:omega_m}
\end{eqnarray}

and obey the ordering
	\begin{equation}\label{eq:frequencyOrdering}
			\omega_c \gg \omega_z \gg \omega_m.
	\end{equation}
A schematic of the motion is shown in figure \ref{fig:penningMotion}, and typical frequencies for particles in the ALPHA experiment are shown in table \ref{table:oscillationFreq}.
It is notable that each of the types of motion have frequencies separated by several orders of magnitude.
The cyclotron motion has a high frequency and a small orbit, so it is often neglected when considering the motion of the particle; the transverse motion is considered to be due to the magnetron motion only.
	\begin{table}[H]
	\begin{center}
	\begin{tabular}{|l|l|l|l|} 
	\hline
	Particle & $\omega_c$ & $\omega_z$ & $\omega_m$ \\ \hline
	electron & 28 GHz & 17 MHz & 5.2 kHz\\ \hline
	positron & 28 GHz & 17 MHz & 5.2 kHz\\ \hline
	antiproton & 15.2 MHz & 400 kHz & 5.2 kHz\\ \hline
	\end{tabular}
	\caption{Typical frequencies of particles confined in the ALPHA trap at a magnetic field of 1~T.}
	\label{table:oscillationFreq}
	\end{center}
	\vspace{-12pt}
	\end{table}
	
While it is possible to manufacture hyperboloid electrodes with high precision to generate a pure quadratic potential, it is technically quite challenging.
In addition, the electrodes completely surround the trapping region and prevent access to the trap to load particles or perform measurements.
While a precise quadratic potential allows for an exact mathematical analysis of the particle motion to be performed, there are no special properties of a quadratic potential that improve the particle confinement.
One can replace the hyperboloid electrodes with stacked cylinders to form a Penning-Malmberg trap.
In this arrangement, many electrodes can be stacked together to form a very versatile trap.
The ends of the cylinders are open, which allows access for particles to be introduced or for diagnostic devices.
The ALPHA trap, described in section \ref{sec:ALPHAPenningTrap}, is an example of a Penning-Malmberg trap.
Particles stored in a Penning-Malmberg trap execute the same motions (though, in general, not with exactly the same frequencies) as particles in a Penning trap.

\subsection{Radiative energy loss}\label{sec:energyLoss}

A particle in a Penning-Malmberg trap is constantly accelerating, and therefore radiates electromagnetic energy, cooling to a limit imposed by the heating power from external sources (typically the thermal radiation from the surrounding apparatus).
The power radiated by an accelerating charge is given by the classical Larmor formula
\begin{equation} \label{eq:Larmor}
	\frac{\mathrm{d}E}{\mathrm{d}t} = - \frac{e^2 a^2}{6 \pi \epsilon_0 c^3},
\end{equation}
where $a$ is the acceleration.\newline
For each of the three components of the motion, this equation can be written as an exponential decay of the kinetic energy
\begin{equation}
\frac{\mathrm{d}E}{\mathrm{d}t} = -\frac{1}{\tau} E,
\end{equation}
where $\tau$ is given by the expressions in table \ref{table:coolingRates}.
The Penning traps at ALPHA are used to store electrons, positrons and antiprotons.
We can use the frequencies given in equations \ref{eqn:omega_c} -- \ref{eqn:omega_m}  and table \ref{table:oscillationFreq} to calculate the cooling times of each of the species in 1 T.
	\begin{table}
	\begin{center}
	\begin{tabular}{|l|c|c|c|} 
	\hline
	 & $\tau$ & $\tau_{e^-}$ & $\tau_{\pbar}$ \\ \hline
	cyclotron 	& $ \frac{3 \pi \epsilon_0 m c^3}{e^2 \omega_c^2}$		& 2.57 s & $1.6 \times 10^{10}$ s\\ \hline
	axial 		& $ \frac{6 \pi \epsilon_0 m c^3}{e^2 \omega_z^2}$		& $1.4 \times 10^7$ s & $4.6 \times 10^{13}$ s \\ \hline
	magnetron 	& $\frac{3 \pi \epsilon_0 m c^3}{e^2} \: \frac{\,^1/_2\omega_z^2 - \omega_m^2 }{\omega_m^4}$		& $1.6 \times 10^{22}$ s & $1.6 \times 10^{22}$ s\\ \hline
	\end{tabular}
	\caption{Characteristic cooling times of the motions of particles in a 1 T magnetic field.}
	\label{table:coolingRates}
	\end{center} 
	\vspace{-12pt}
	\end{table}

Antiprotons do not cool on a timescale relevant to the experiments -- other techniques are needed to reduce their energy, and these are described in chapter~\ref{chap:Cooling}.

For electrons and positrons, it is clear that energy is only effectively emitted from the cyclotron motion.
The axial motion is only cooled insofar as energy is transferred to the cyclotron motion through collisions.
In the limit that the rate of energy transfer is so high that the degrees of freedom parallel and perpendicular to the magnetic field remain approximately in equilibrium, the overall cooling time constant is $3/2$ times that given for the cyclotron motion in table \ref{table:coolingRates} \cite{JoelTemperatureAnis} -- 3.9~s for electrons in a 1~T field.
When the equilibration rate is low, the axial motion cools on the timescale of the time taken to transfer energy to the cyclotron motion.

We can rewrite the equation for the cyclotron cooling time in table \ref{table:coolingRates} using equation \ref{eqn:omega_c} as
\begin{equation}\label{eq:cyclotronCoolBField}
	\tau = \frac{3 \pi \epsilon_0 m^3 c^3}{e^4} \frac{1}{B^2}, 
\end{equation} 
displaying the important point that stronger magnetic fields result in faster cooling rates.
	
\subsection{The ALPHA trap} \label{sec:ALPHAPenningTrap}

\begin{figure}[p]
\centering
\includegraphics[width = 1.5\textwidth, angle = 90]{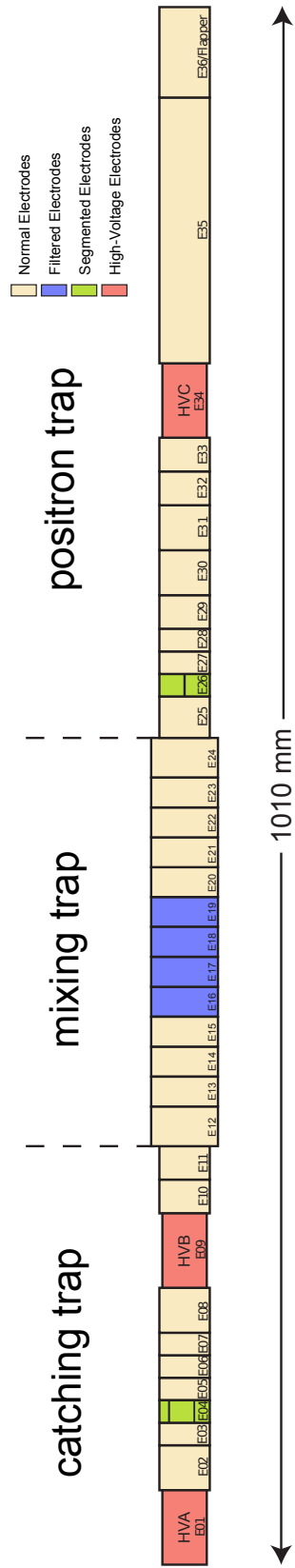}
\caption[The ALPHA Penning-Malmberg trap electrodes]{The ALPHA Penning-Malmberg trap electrodes. The trap is positioned in a 1~T solenoidal magnetic field directed along the trap axis. Several of the electrodes are filtered more heavily then the others (blue), are segmented to apply rotating-wall fields (green) or can sustain voltages up to several kV (red). }
\label{fig:ALPHAelectrodes}
\end{figure}

The ALPHA electrodes (figure \ref{fig:ALPHAelectrodes}) consist of thirty-five gold-plated aluminium cylinders, of varying radius and length. The electrodes can be individually biased by external amplifiers, and are electrically isolated from each other and the surrounding apparatus using synthetic ruby spheres and ceramic spacers.
The electrodes are placed in a 1 T solenoidal magnetic field directed along the axis of the cylinders, which is produced by an external superconducting magnet, seen in figure \ref{fig:ALPHA_section}.
There are a number of other magnets present which can be energised to change the magnetic field, but in normal operations, this `background field' is kept constant.

The trap is usually thought of as being composed of three distinct sections - the `catching trap', which includes two electrodes designed for use with high voltages, and is primarily used to capture, cool and accumulate antiprotons; the `positron trap', used to prepare positron plasmas; and the `mixing trap'.
The mixing trap, where antihydrogen production is carried out, lies between the other sections and is composed of thirteen electrodes constructed with large radius and very small thickness.
As will be discussed in section \ref{sec:magneticminimum}, this allows for the maximum possible trap depth for antihydrogen atoms.

Two electrodes - one in each of the catching and positron traps - are azimuthally segmented to allow for the application of phase-separated `rotating wall` fields (see section \ref{sec:rotatingWall}).

Each electrode is connected through copper coaxial cable (Caburn type Kap4) to a heatsunk circuit board, and from there to the outside world on stainless steel coaxial cable (Lakeshore type SS), chosen for its low thermal conductivity. 
The electrode excitations are applied by a 16-bit National Instruments PXI-6733 digital-to-analogue converter, with an output range of $\pm10~\mathrm{V}$, which is programmed before each experiment with the voltage and timing requirements (see section \ref{sec:Sequencer}).
The output of the NI PXI-6733 is amplified by a factor of 7.2 or 14, depending on the particular electrode, allowing control of the applied voltage to 2~mV or 4~mV within a range of $\pm72~\mathrm{V}$ or $\pm140~\mathrm{V}$ respectively.

It is important that the voltage applied to the electrodes be free of electronic noise.
Noise-induced fluctuations of the voltage applied to the electrodes can produce heating of stored particles through non-adiabatic changes to the potential \cite{NoiseMeasurements}, and has previously been used to artificially increase the temperature of a stored plasma in an antihydrogen experiment \cite{ATHENA_OnOff}.
Noise of sufficient amplitude that it reduces the depth of the confining well significantly can also cause loss of particles.

In an environment such as a particle accelerator, it is inevitable that some noise is transmitted to the amplifier chain from surrounding equipment.
The simplest solution is to place a low-pass filter at the last possible point before the electrodes.
The signals are filtered at the input to the vacuum chamber in multiple stages with an upper frequency cut-off (-3~dB point) of 50~kHz, with a selected number of channels undergoing additional filtering to approximately 1.5~kHz.
For experiments that require very fast signals, it is also possible to connect a high frequency signal through a high pass filter with a lower frequency cut-off at approximately 200 kHz.

While filtering is effective at removing most kinds of noise, it restricts the timescales on which manipulations can be performed, and there must be a trade-off against the need to access a particular frequency domain. 
Therefore it is also desirable to reduce the noise at source by careful design and choice of components, and to isolate and shield the electronics components from the remainder.
Assessment and reduction of the impact of electronic noise in ALPHA is an ongoing effort.

When designing experimental procedures, it is vital to know the potentials and fields experienced by the particles due to the voltages applied to the electrodes.
Generally, the total potential at any point inside the trap is a superposition of the contributions of an infinite number of charge elements residing on the electrode surfaces.
The most efficient way to produce maps of the potential is to calculate the contribution of each electrode to the potential at each point in the trap and combine these maps with appropriate scalings, proportional to the voltage applied to the electrode, to arrive at the total potential.
This was performed for the ALPHA trap using the OPERA finite element field solver produced by Vector Fields Ltd \cite{opera}.
The calculation is estimated to be accurate to 1 part in $10^6$ near the axis, and to 1 part in $10^5$ near the electrodes, where large electric fields exist.

\section{Magnetic trap} \label{sec:magneticTrap}

A Penning-Malmberg trap is ideal for confining the charged-particle ingredients of antihydrogen.
However, trapping depends on the particles having a non-zero electric charge, so a Penning trap cannot confine neutral particles such as antihydrogen atoms.
Trapping of neutral particles in static fields relies on acting on electric or magnetic dipole moments, and the gradients of the electric and magnetic fields in a Penning trap are much too small to hold most atoms.
Any antihydrogen atoms that are produced inside the trap will very quickly reach the trap walls and annihilate there. 
For instance, an atom with a kinetic energy of 1 K will have a velocity of $130~\mathrm{m\,s^{-1}}$, and will cross the trap volume in a few hundred microseconds. 
To make precision measurements with antihydrogen atoms, it is necessary to hold the atoms for a much longer time.

An antihydrogen atom has an intrinsic magnetic dipole moment $\boldsymbol{\mu}$ due to the spins of the antiproton and the positron, and the orbital motion of the positron.
The magnetic dipole moment due to the positron is the vector sum of the contributions from the orbital angular momentum $\mathbf{L}$ and the spin angular momentum $\mathbf{S}$,
	\begin{equation}\label{eq:muPos}
		\boldsymbol{\mu}_{e^+} = -g_L \mu_B \frac{\mathbf{L}}{\hbar} -g_S \mu_B \frac{\mathbf{S}}{\hbar},
	\end{equation}
where $g_L$ and $g_S$ are the gyromagetic ratios.
The magnetic dipole moment of the antiproton only has a contribution from the spin,
	\begin{equation}\label{eq:muPbar}
		\boldsymbol{\mu}_\pbar = -g_P \mu_N \frac{\mathbf{S_N}}{\hbar}.
	\end{equation}
The relative sizes of $\boldsymbol{\mu}_{e^+}$ and $\boldsymbol{\mu}_\pbar$ are dominated by the ratio of the positron and antiproton masses \cite{pdg}:
	\begin{equation}
		\frac{\mu_N}{\mu_B} = \frac{m_e}{m_\pbar} = 5.4 \times 10^{-4};
	\end{equation}
$g_L \mathbf{L}$, $g_S \mathbf{S}$, and $g_P \mathbf{S_N}$, are of the same order of magnitude.
In the ground state, the magnitude of $\mathbf{L}$ is 0, and the allowed values for projections of $\mathbf{S}$ onto the direction of $\mathbf{B}$ are $\pm \,^1/_2$, and we will take $g_S \simeq 2$, so 
	\begin{equation}\label{hbarMu}
		\mu_\Hbar = \pm \mu_B.
	\end{equation}

The energy of interaction of this dipole moment with a magnetic field is given by
	\begin{equation}
		U = -\boldsymbol{\mu_\Hbar}\cdot\mathbf{B}.
	\end{equation}
As the atoms moves through the magnetic field, the direction of $\boldsymbol{\mu}$ will stay aligned with the direction of $\mathbf{B}$ (a phenomenon known as `adiabatic following'), so in the ground state
	\begin{equation}\label{dipoleEnergy}
		U = \pm \mu_B B.
	\end{equation}
	
The existence of two solutions of opposite sign corresponds to two classes of particles.
Atoms with a positive magnetic dipole moment will be attracted to regions of high magnetic field strength, and are termed `high field seeking' atoms.
Conversely, atoms with a negative magnetic dipole moment will be attracted to regions with low magnetic field strength and are called `low field seeking' atoms.
Maxwell's equations forbid the presence of a three-dimensional static maximum of magnetic field in free space, but it is possible to construct a minimum.
Therefore, only low-field seeking atoms can be trapped using this method.

If the atom is in a state with larger $\mathbf{L}$, it can have a correspondingly larger magnetic dipole moment, and there will be $2(L+1)$ possible angular momentum states, half low-field seeking and half high-field seeking.

\subsection{Constructing a magnetic field minimum}
\label{sec:magneticminimum}

A magnetic field minimum can be constructed using either permanent magnets or current carrying wires. Practically, it is highly desirable that it is possible to switch off the trap, or to change its depth, so wires are generally preferred.

\begin{figure}[hbt]
\centering
\input{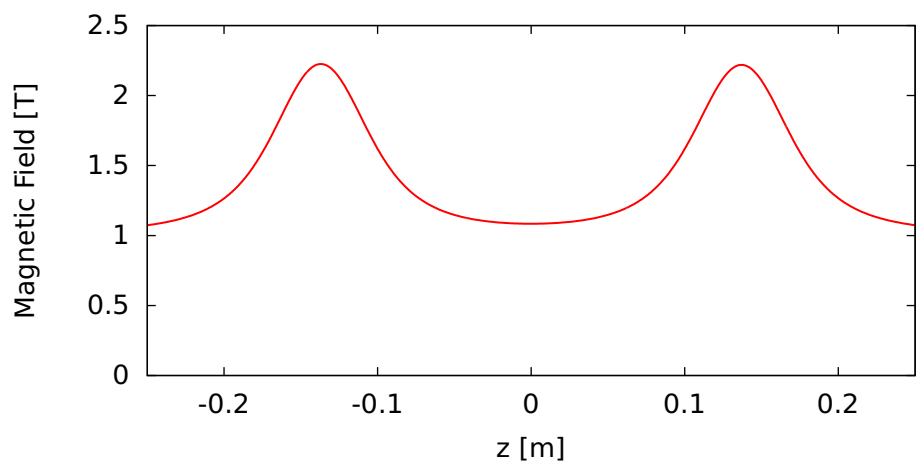}
\caption{The magnetic field strength along the axis of the ALPHA trap when the mirror coils are energised.}
\label{fig:magfieldaxial}
\end{figure}

Forming a magnetic field minimum in the axial direction (parallel to the Penning trap axis) is easily achieved by the addition of two additional cylindrical coils, one to either side of the trapping region.
These coils -- referred to as `mirror coils' increase the magnetic field locally producing a minimum of magnetic field between them (figure \ref{fig:magfieldaxial}).

\begin{figure}[hbt]
	\centering
	\input{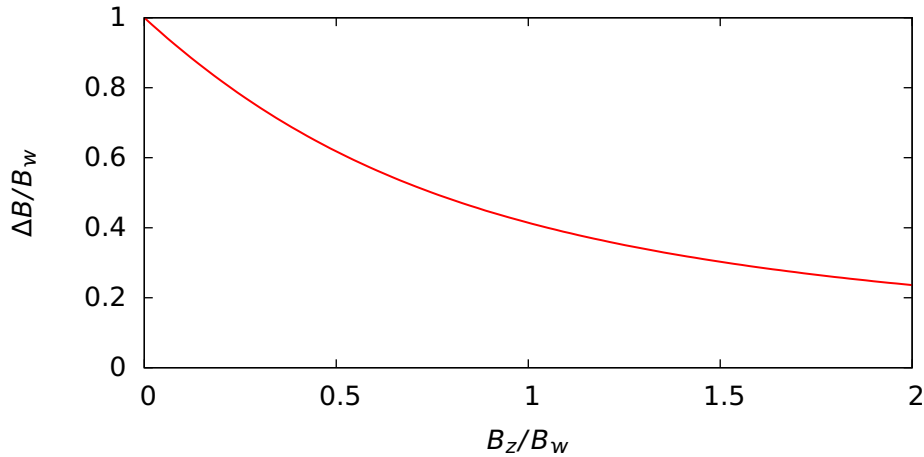}
	\caption{The dependence of the trap depth on the solenoidal field.}
	\label{fig:trapDepth}
	\end{figure}

To create a minimum of magnetic field perpendicular to the axis, a transverse multipole field is used, which can be created through the arrangement of a suitable number of current bars around the axis.
For an ideal multipole magnet of order $l$, the magnet adds a transverse component of magnetic field
	\begin{equation}\label{eq:PerpBField}
		B_\perp \propto r^l.
	\end{equation}
A quadrupole ($l=1$) field is generated using four current bars with the current flowing in alternating directions, a sextupole ($l=2$) with six, an octupole ($l=3$) with eight, and so on.
Assuming that the limiting factor is the multipole strength (not the mirror coils), the depth of the magnetic trap will be the difference in the magnetic field magnitude on the axis (which is just the solenoidal field, $B_z$) and that at the wall radius.
\begin{equation}\label{eq:trapDepth}
	\Delta |\mathbf{B}| = \sqrt{B_z^2 + B_w^2} - B_z.
\end{equation}
Figure \ref{fig:trapDepth} shows $\Delta B$ as a function of $B_z$, where it is clear that for a given $B_w$, the trap depth decreases for larger $B_z$.

Thus, reducing the solenoidal field can improve the probability of trapping neutral atoms, keeping the multipolar field fixed, but adversely affects the cyclotron cooling rate (equation \ref{eq:cyclotronCoolBField}) and confinement of charged particles.
While the first antihydrogen-producing experiments used high magnetic fields (3-5~T), a field of 1~T was chosen in ALPHA as a suitable compromise.
For comparison, the strength of the trap magnets in ALPHA is approximately 1-2~T.
	
When a transverse multipole-based design was considered as a possible trap for antihydrogen atoms, significant experimental and theoretical studies were carried out to identify if the magnetic field would be compatible with the requirement to store the antiprotons and positrons as non-neutral plasmas.
The essential concern was that the transverse magnetic field, which is not present in the simple Penning-Malmberg trap, would reduce the lifetime of the stored plasmas by an unacceptable amount due to an enhanced radial transport mechanism \cite{radialTransportMultipole}.
In addition, the `ballistic loss' mechanism discussed in reference \cite{Joel_Multipole} and treated further in section \ref{sec:ballisticLoss}, restricts the maximum radius of particles stored in such a trap.
Several initial experiments using an electron plasma stored in a quadrupole showed sharp reductions in the stored lifetime
 \cite{GilsonFajans_Quadrupole}, \cite {RouteUltraLowEnergyHBar}.
An obvious solution is to construct the trap while minimising the transverse magnetic field the plasma is exposed to.

\begin{figure}[hbt]
\centering
\input{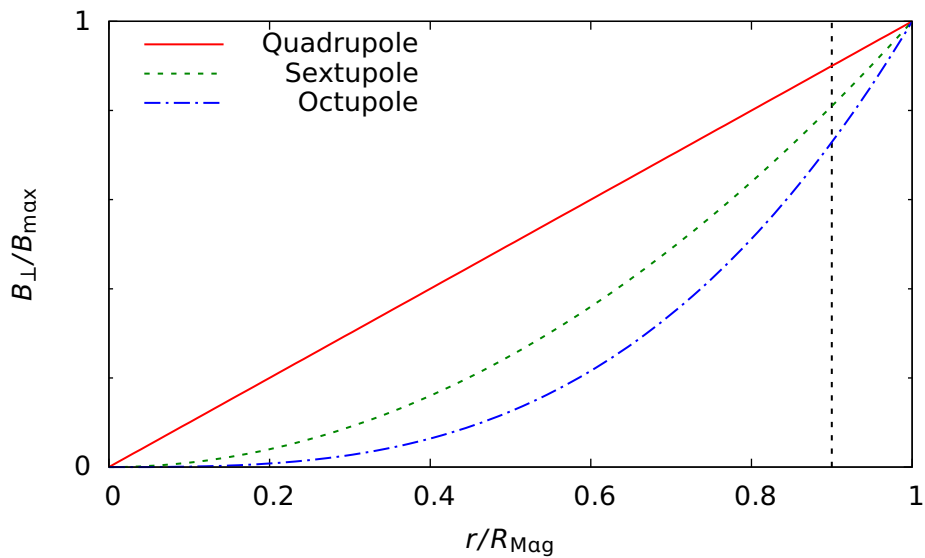}
\caption[The transverse magnetic field as a function of radius for the three lowest-order multipoles]{The transverse magnetic field as a function of radius for the three lowest-order multipoles as a function of radius. The point where the vertical line intersects the field dependence shows that higher-order multipoles produce shallower trap depths.}
\label{fig:compareMultipoleFields}
\end{figure}

The strength of a superconducting magnet is usually limited by the magnetic field at the superconductor surface.
So, for equal strength magnetic fields at the windings (which we can assume are the same radius, $R_\mathrm{Mag}$ for different multipoles), equation \ref{eq:PerpBField} implies that the transverse magnetic field produced by a higher order multipole is much lower close to the axis. 
If the particles are stored near the axis, the effect of the magnetic field is thus minimised.
A comparison of the magnetic field for several orders of multipole magnets is shown in figure \ref{fig:compareMultipoleFields}.
It is clear that near the axis at $r=0$, the higher-order multipoles have a smaller transverse magnetic field.
Near the axis, there is only a  small modification from the normal Penning-Malmberg geometry and plasmas should remain stably trapped.

The vertical line in figure \ref{fig:compareMultipoleFields} is placed at $0.9 R_\mathrm{Mag}$ and represents the edge of the atom trap, usually the surface of the Penning trap electrodes. 
From the points where this line intersects the plots of the magnetic field, it is clear that the higher-order multipoles have a shallower trap depth than lower-order types.
This difference becomes smaller as the distance between the edge of the atom trap and the magnet windings is made smaller.
For an octupole, the trap depth goes as approximately $r^{3}$, so this effect is extremely important, and it is necessary to place the electrodes as close to the magnet windings as possible to make the best use of the magnets.
As a consequence, the electrodes making up the mixing trap have a large inner radius, 22.2~mm and are very thin, approximately 0.5~mm.
Making them twice the thickness (reducing the inner radius by less than 3\%) would have resulted in a trap depth $\sim 8\%$ shallower.

Taking the available information into account, ALPHA designed an octupole magnet for its trap, while the ATRAP experiment opted for a quadrupole \cite{ATRAP_quadrupole}.
Measurements of the particle lifetime in ALPHA \cite{ALPHA_PlasmasMultipole} showed that the lifetime was reduced somewhat (approximately 10\% more antiprotons were lost over 500~s) when the multipole magnet was energised.
However, this is a much longer time than needed to form antihydrogen, and does not impede the experiments.

\subsection{The ALPHA magnets}
ALPHA's  magnetic minimum trap magnets are constructed of wires formed of filaments of Niobium-Titanium (NbTi) alloy embedded in a copper matrix.
NbTi is a type-II superconductor with a critical temperature of 10~K.
To achieve such low temperatures, the magnet is immersed in a liquid helium bath at 4.2~K.
Through a number of considerations and trade-offs, the design of the ALPHA octupole set a maximum current of 1100~A, which results in a surface field of approximately 4~T \cite{ALPHA_Magnet}. 
To maximise the trap depth, as discussed in the preceding section, the magnet is placed as close as possible to the electrodes by directly winding the superconductor onto the vacuum chamber wall. This design produces a transverse magnetic field of 1.8~T at the inner surface of the electrodes, at $r=22.2\;\mathrm{mm}$.
The mirror coils were designed to carry a maximum current of 750~A, and they produce a maximum longitudinal field on the axis of 1.2~T.
Using these values in equation \ref{eq:trapDepth}, we find that the maximum depth of the well is approximately 1~T.
The magnets are not capable of reaching their design current reliably, and are typically used at currents of 900~A for the octupole, and 650~A for the mirror coils.
This gives a trap depth of approximately 0.8~T, or 0.5~K for ground state antihydrogen atoms.

A segment of superconductor that ceases to become superconducting presents a resistance to the current flow, leading to Joule heating. 
The heat produced can spread to the neighbouring superconductor, causing its temperature to rise above the critical temperature, causing a runaway effect known as a `quench'. 
The large currents and energies involved mean that quenches can be quite violent and pose a danger to equipment, possibly destroying the magnet itself.
A particularly spectacular example occurred at the Large Hadron Collider at CERN on September $19^\mathrm{th}$ 2008, causing damage costing in excess of 25 million Swiss Francs and taking almost a year to repair.

\begin{figure}[hbt]
	\centering 

\psset{unit=1in,cornersize=absolute,dimen=middle}%
\begin{pspicture}(-0.028,-0.929167)(3.562722,0.791667)%
\psset{linewidth=0.8pt}%
\psset{linewidth=0.8pt}%
\ifx\MPSTPatchA\undefined{\makeatletter\def\psbezier@ii{\addto@pscode{%
\ifshowpoints true \else false \fi\tx@OpenBezier%
\ifshowpoints\tx@BezierShowPoints\fi}\end@OpenObj}\makeatother%
\global\def\MPSTPatchA{}}\fi%
\psset{arrowsize=1.1pt 4,arrowlength=1.64,arrowinset=0}%
\pscircle[fillstyle=solid,fillcolor=white](0,0){0.028}
\uput{0.501875ex}[r](0.028,0){\rlap{\hbox{$\:$}$ +$\hbox{$\:$}}}
\psline(0,0)(0,0.333333)
\psline(0,0.333333)(1.25,0.333333)
\psline(1.25,0.333333)(1.25,0.733333)
(1.25,0.733333)(1.5,0.733333)
\psline(1.5,0.733333)(1.575,0.733333)
(1.575,0.733333)(1.604167,0.791667)
(1.604167,0.791667)(1.6625,0.675)
(1.6625,0.675)(1.720833,0.791667)
(1.720833,0.791667)(1.779167,0.675)
(1.779167,0.675)(1.8375,0.791667)
(1.8375,0.791667)(1.895833,0.675)
(1.895833,0.675)(1.925,0.733333)
(1.925,0.733333)(2,0.733333)
\psline(2,0.733333)(2.25,0.733333)
\psline(2.25,0.733333)(2.325,0.733333)
(2.325,0.733333)(2.354167,0.791667)
(2.354167,0.791667)(2.4125,0.675)
(2.4125,0.675)(2.470833,0.791667)
(2.470833,0.791667)(2.529167,0.675)
(2.529167,0.675)(2.5875,0.791667)
(2.5875,0.791667)(2.645833,0.675)
(2.645833,0.675)(2.675,0.733333)
(2.675,0.733333)(2.75,0.733333)
\psline(2.75,0.733333)(3,0.733333)
\psline(3,0.733333)(3,0.333333)
\psline(3,0.333333)(3,0.083333)
\psline(3,0.083333)(3,0.008333)
\psbezier(3,0.008333)(3.014583,0.008333)(3.029167,0.008333)(3.04375,0.008333)
(3.072917,0.008333)(3.0875,-0.006833)(3.0875,-0.037167)
(3.0875,-0.0675)(3.067083,-0.082667)(3.02625,-0.082667)
(2.985417,-0.082667)(2.965,-0.075667)(2.965,-0.061667)
(2.965,-0.047667)(2.985417,-0.040667)(3.02625,-0.040667)
(3.067083,-0.040667)(3.0875,-0.059333)(3.0875,-0.096667)
(3.0875,-0.134)(3.067083,-0.152667)(3.02625,-0.152667)
(2.985417,-0.152667)(2.965,-0.145667)(2.965,-0.131667)
(2.965,-0.117667)(2.985417,-0.110667)(3.02625,-0.110667)
(3.067083,-0.110667)(3.0875,-0.129333)(3.0875,-0.166667)
(3.0875,-0.204)(3.067083,-0.222667)(3.02625,-0.222667)
(2.985417,-0.222667)(2.965,-0.215667)(2.965,-0.201667)
(2.965,-0.187667)(2.985417,-0.180667)(3.02625,-0.180667)
(3.067083,-0.180667)(3.0875,-0.199333)(3.0875,-0.236667)
(3.0875,-0.274)(3.067083,-0.292667)(3.02625,-0.292667)
(2.985417,-0.292667)(2.965,-0.285667)(2.965,-0.271667)
(2.965,-0.257667)(2.985417,-0.250667)(3.02625,-0.250667)
(3.067083,-0.250667)(3.0875,-0.265833)(3.0875,-0.296167)
(3.0875,-0.3265)(3.072917,-0.341667)(3.04375,-0.341667)
(3.029167,-0.341667)(3.014583,-0.341667)(3,-0.341667)
\psline(3,-0.341667)(3,-0.416667)
\uput{0.501875ex}[l](2.965,-0.166667){\llap{\hbox{$\:$}$ \mathrm{Magnet}$\hbox{$\:$}}}
\psline(3,-0.416667)(3,-0.666667)
\psline(3,-0.666667)(2.875,-0.666667)
\psline(2.875,-0.666667)(2.8,-0.666667)
\psline(2.45,-0.666667)(2.45,-0.736667)
(2.45,-0.736667)(2.8,-0.736667)
(2.8,-0.736667)(2.8,-0.596667)
(2.8,-0.596667)(2.45,-0.596667)
(2.45,-0.596667)(2.45,-0.666667)
\psline(2.45,-0.666667)(2.375,-0.666667)
\uput{0.501875ex}[u](2.625,-0.596667){\hbox{$\:$}$ \mathrm{Shunt}$\hbox{$\:$}}
\psline(2.375,-0.666667)(0,-0.666667)
\psline(0,-0.666667)(0,-0.333333)
\pscircle[fillstyle=solid,fillcolor=white](0,-0.333333){0.028}
\uput{0.501875ex}[ur](0,-0.305333){\rlap{\hbox{$\:$}$ -$\hbox{$\:$}}}
\psline(0.5,-0.666667)(0.5,-0.841667)
\psline(0.616667,-0.841667)(0.383333,-0.841667)
\psline(0.577778,-0.885417)(0.422222,-0.885417)
\psline(0.55,-0.929167)(0.45,-0.929167)
\psline(0.5,0.333333)(0.5,0)
\psline(0.5,0)(0.5,-0.199482)
\psline(0.5,-0.300518)(0.558333,-0.199482)
(0.558333,-0.199482)(0.441667,-0.199482)
(0.441667,-0.199482)(0.5,-0.300518)
\psline(0.437615,-0.300518)(0.562385,-0.300518)
\psline(0.529167,-0.300518)(0.5875,-0.401554)
\psline(0.5,-0.300518)(0.5,-0.5)
\uput{0.501875ex}[r](0.5875,-0.25){\rlap{\hbox{$\:$}$ \mathrm{SCR}$\hbox{$\:$}}}
\psline(0.5,-0.5)(0.5,-0.666667)
\psline(1.25,0.333333)(1.5,0.333333)
\psline(1.5,0.333333)(1.633333,0.333333)
\psline(1.633333,0.333333)(1.808333,0.508333)
\psline(1.866667,0.333333)(2,0.333333)
\uput{0.501875ex}[d](1.720833,0.302945){\hbox{$\:$}$ \mathrm{IGBT}$\hbox{$\:$}}
\psline(2,0.333333)(2.25,0.333333)
\psline(2.25,0.333333)(2.449482,0.333333)
\psline(2.550518,0.333333)(2.449482,0.391667)
(2.449482,0.391667)(2.449482,0.275)
(2.449482,0.275)(2.550518,0.333333)
\psline(2.550518,0.270948)(2.550518,0.395718)
\psline(2.550518,0.333333)(2.75,0.333333)
\psline(2.75,0.333333)(3,0.333333)
\psline(2.125,0.733333)(2.125,0.433333)
\psline(2.125,0.433333)(2.125,0.391667)
\psarc(2.125,0.333333){0.058333}{90}{270}
\psline(2.125,0.275)(2.125,0.233333)
\psline(2.125,0.233333)(2.125,-0.666667)
\pscircle[fillstyle=solid,fillcolor=white](3.5,-0.166667){0.028}
\uput{0.501875ex}[r](3.528,-0.166667){\rlap{\hbox{$\:$}$ \mathrm{QPS}$\hbox{$\:$}}}
\psline[linestyle=dashed](3.5,-0.166667)(3,0.083333)
\psline[linestyle=dashed](3.5,-0.166667)(3,-0.416667)
\psline[linestyle=dashed](3.1,-0.083333)(3.5,-0.166667)
\psline[linestyle=dashed](3.1,-0.25)(3.5,-0.166667)
\end{pspicture}%
	\caption{A circuit diagram showing the powering system for one of the trap magnets.}
	\label{fig:Magnetcircuit}
	\end{figure}
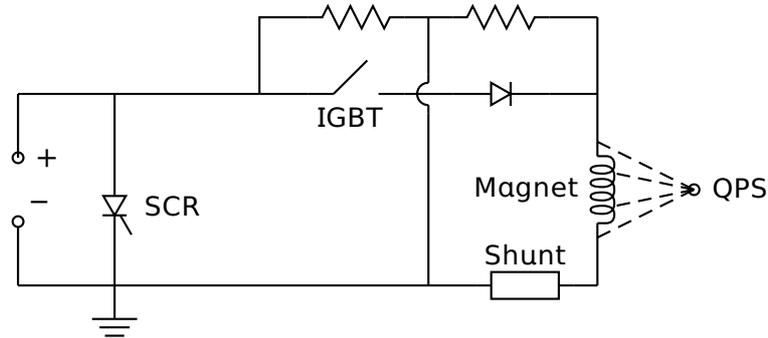

A quench protection system (QPS), which actively detects and responds to quenches by safely extracting the current, is a requisite of any superconducting magnet system of this kind.
Once a quench is identified, it is important to extract the current from the magnet before damage occurs. 
A quench can be detected by the increase in voltage across the magnet generated by the current flow through the resistive section. 
A typical QPS measures the voltages across segments of the magnet, and responds to a voltage increase beyond a defined level by rapidly switching off the magnet and extracting the energy in a safe way.
To achieve this in ALPHA, a Field Programmable Gate Array (FPGA) controller monitors the voltages and determines whether or not a quench has occured with a frequency of order $\sim$ kHz and a high current insulated-gate bipolar transistor (IGBT) is used to quickly force the current through a resistor network, which harmlessly dissipates the energy as heat. 
A diagram of the system used in ALPHA is shown in figure \ref{fig:Magnetcircuit}.

\begin{figure}[hbt]
	\centering
	\input{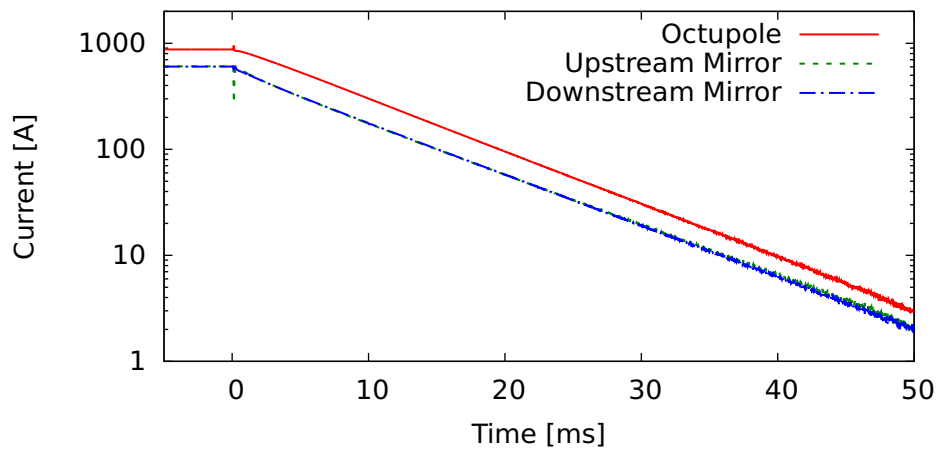}
	\caption[The current decay in the trap magnets during a triggered shutdown]{The current measured in the shunt resistor immediately following a triggered shutdown at $t=0$. The dependence is exponential, with a time constant of 9~ms for the octupole and 8~ms for the mirror coils.}
	\label{fig:quenchTraces}
	\end{figure}

The same system is used to quickly remove the trapping fields when attempting to detect trapped antihydrogen atoms.
To reduce the background due to noise counts and cosmic rays, the magnets have been designed with extremely low inductances so that the current is removed in as short a time as possible. 
The current can be monitored using a shunt resistor connected in series with the magnet, and measurements of the decay (figure \ref{fig:quenchTraces}) show that the time constant is approximately 9~ms.

ALPHA also contains a fourth internal superconducting magnet, not part of the atom trap.
This is a solenoid placed over the catching trap region, which is capable of locally increasing the magnetic field by 2~T, which leads to enhanced antiproton catching and cooling efficiency. 
The magnet is otherwise very similar to the trap magnets.

\section{Cryostat/Vacuum} \label{sec:vacuum}

Plasma physics experiments must be carried out under High Vacuum or Ultra-High Vacuum (UHV) conditions (typically less than $10^{-9}$ mbar) to achieve usable plasma lifetimes.
In the presence of background gas, the particles can recombine to form atoms or ions, or undergo collisions that transfer angular momentum, leading to diffusion, and eventually, loss.
When working with antimatter, it is doubly important to have low pressures, as antimatter particles annihilate on their matter counterparts.

ALPHA has several distinct vacuum chambers. 
The `trap vacuum' is the only truly UHV chamber, and is where the particles are held during experiments. 
This is partially surrounded by a liquid helium bath which contains the atom-trap magnets and the antiproton capture solenoid. 
Outside this is the `Outer Vacuum Chamber' (OVC), which acts as an insulator and eliminates conductive and convective heat flow from the outside world.
There is also another vacuum chamber which acts as an interface between the ALPHA apparatus and the vacuum system of the Antiproton Decelerator.

In addition to providing cooling for the superconducting magnets, the liquid helium reservoir cools the wall of the trap vacuum chamber and acts as a heatsink for the cables electrically connecting the electrodes to the amplifier chains. 
The equilibrium temperature of the electrodes is determined by the balance between the power radiated and conducted onto them and the speed at which the heat sink can remove it. 
The design of the heat sinks have undergone several iterations, and the most recent implementation has resulted in a measured electrode temperature lower than 8~K. 
Cooling the vacuum chamber and the electrodes should reduce the temperature of the stored particles, since in the absence of other heating sources, the particles will come into thermal equilibrium with the surrounding materials.

The cold surfaces in the trap vacuum also help to reduce the gas pressure. 
Gas molecules that strike the walls lose energy can condense or freeze to the surface (`cryopump'), and become effectively removed from the volume. 

To allow access for instruments into the trap vacuum, the trap vacuum chamber extends outside the liquid helium bath.
Since this portion of the chamber is not effectively cryopumped, the gas pressure is higher than in the cold region.
To prevent gas travelling from this region into the trap, an aperture that can be remotely opened and closed is located at the end of the electrode stack (seen in figure \ref{fig:allofit}). 
This aperture is also cooled to a low temperature so that it prevents the transmission of thermal radiation from the room-temperature materials onto the plasmas.

A wide variety of pressure gauges are commercially available, and are used on all of the ALPHA vacuum systems.
However, in the electrode structure space is too limited to admit a pressure gauge capable of measuring in the UHV range.
Nonetheless, it is possible to use the rate of antiproton annihilation as an indicator of the residual pressure in the particle storage region.

From \cite{FeiThesis}, the cross section for annihilation of an antiproton on molecular hydrogen, which is likely to be the dominant process in a cryogenic vacuum, is 
	\begin{equation}\label{eq:pbarAnnihlCrossSection}
		\sigma = 3 \pi a_0^2 \sqrt{\frac{27.2\;\mathrm{eV}}{E}},
	\end{equation}
where $E$ is the centre-of-mass collision energy and $a_0$ is the Bohr radius. When the antiproton thermal velocity is much greater than the velocity of gas atoms, we can assume that $E$ is simply the antiproton kinetic energy. 
This is valid since gas molecules have higher masses than an antiproton, and gas atoms will cool to approximately the temperature of the cryostat walls through collisions, while, as we will see, the antiprotons are significantly warmer.

The density of gas particles, $n_\mathrm{gas}$, the annihilation rate $\Gamma$ and the cross section are related through
	\begin{equation}
		\Gamma = \,^1/_\tau =  n_{gas} v \sigma,
	\end{equation}
where $n_{gas} v$ is the flux of gas particles. \newline
Through a simple change of reference frame, we can take $v$ to be the antiproton velocity, and using $E = \,^1/_2 m_\pbar v^2$
	\begin{equation}
		\tau = \frac{1}{n_{gas} \sigma \sqrt{\frac{2 E}{m_\pbar}}}
	\end{equation}
This leads to 
	\begin{equation}\label{eq:gasDensity}
	\begin{tabular}{l c l}
		$n_{gas}$ & = & $ \left(3 \sqrt{2} \pi a_0^2 \sqrt{\frac{27.2 \mathrm{eV}}{m_\pbar}}\right)^{-1} \frac{1}{\tau}$ \\[8pt]
		&  = & $5.2 \times 10^{8} \; \mathrm{s.cm^{-3}} \;\frac{1}{\tau}$.
	\end{tabular}
	\end{equation}

The value of $\tau$ can be measured by counting the rate of annihilations, $-{\mathrm{d}N}/{\mathrm{d}t}$, while a cloud of antiprotons is held in the apparatus.
The lifetime is then calculated from $-{\mathrm{d}N}/{\mathrm{d}t} = \frac{N}{\tau}$.
This method gives a value of ($10 \pm 3$) hours, where the uncertainty is dominated by the measurements of the relative efficiencies of the detectors used to count the total number of particles and the number that annihilated while being held.
Using this value in equation \ref{eq:gasDensity} gives $n_{gas} = \left( 1.4 \pm 0.4 \right) \times 10^4 \;\mathrm{cm^{-3}}$.

For an ideal gas, the pressure is related to the gas density through
	\begin{equation}\label{eq:idealGas}
		P = n_{gas} k_B T.
	\end{equation}
If we estimate a gas temperature of 10~K, $P = \left(1.9 \pm 0.6 \right) \times 10^{-14} \;\mathrm{mbar}$.

When the aperture used to block the flow of gas from the room-temperature part of the apparatus is opened, the lifetime of the antiprotons falls to ($1.0 \pm 0.3$) hours, implying a gas density of $\left( 1.4 \pm 0.4 \right) \times 10^5 \;\mathrm{cm^{-3}}$, or a pressure $\sim (10^{-13} - 10^{-12})~\mathrm{mbar}$ (depending on the temperature of the gas).

\section{Particle sources}
\subsection{Electrons}

Electrons, while not a constituent of antihydrogen, are used in ALPHA because their short cooling time in a strong magnetic field (section \ref{sec:energyLoss}) allows them to cool antiprotons (see chapter \ref{chap:Cooling}).
Electrons are produced by thermionic emission from a barium-oxide filament integrated into an electron gun, which is mounted outside the main magnet on a movable structure that also holds the MCP/phosphor assembly (section \ref{sec:MCP}).
An electrode placed in front of the filament is biased to $\sim$ -15~V to produce a collimated electron beam, which is then guided by the magnetic field into the catching trap region.
A well is formed to trap some of the particles, which then cool through the emission of cyclotron radiation and form a plasma.

\subsection{Positrons}

Positrons are emitted in many beta radioactive decays.
ALPHA uses the positrons produced in the beta-plus decay of sodium-22 atoms.
\begin{equation}
	^{22}_{11}\mathrm{Na} \: \rightarrow \: ^{22}_{10}\mathrm{Ne} \:+\: ^{0}_{1}e^+ \: + \: \nu_e.
\end{equation}
Sodium-22 has a reasonably short half-life of 2.6 years \cite{CRC_Handbook}, so high activity sources can be produced without a need for a large amount of material.
The half-life is also sufficiently long that the source does not need to be replaced with a high frequency.
The source used in ALPHA had an activity of 2.5 GBq at the time of installation (2007).

\begin{figure}[hbt]
	\centering
	\includegraphics[width=0.9\textwidth]{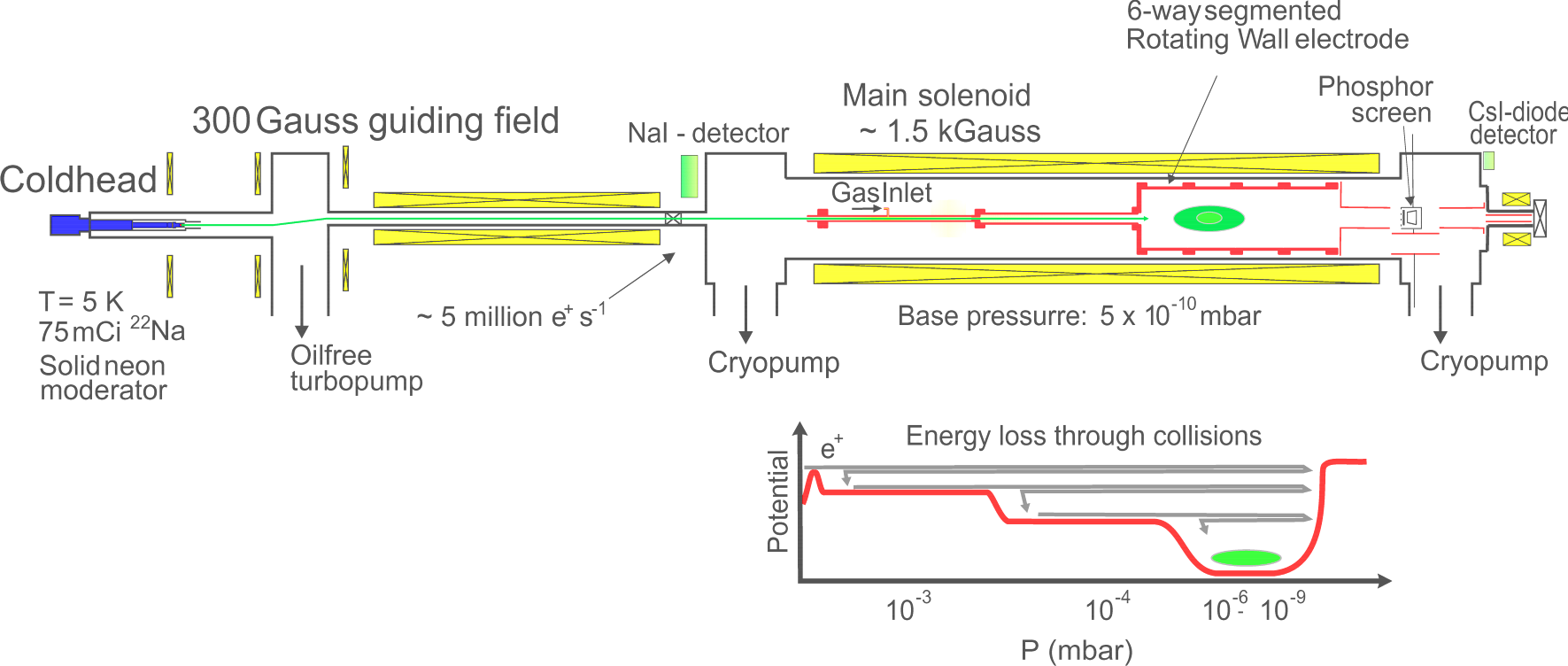}
	\caption[A schematic of the positron accumulator]{A schematic of the positron accumulator, showing the $^{22}\mathrm{Na}$ source at the left. The positrons form a beam (green) into the accumulator, and are collected as a plasma in the final stage.}
	\label{fig:accumulator}
\end{figure}

Since the instantaneous flux from the radioactive source is still quite low, it is necessary to accumulate the emitted positrons as a plasma before use.
To achieve this, a device known as a Surko-type positron accumulator is used \cite{Surko_positronAccumulator}.
A schematic of the accumulator originally built for ATHENA, and also used in ALPHA, is shown in figure \ref{fig:accumulator}.

Positrons from beta decay can have several hundred keV of energy, and are too energetic to be trapped.
In a Surko-type accumulator, a fraction of the positrons emitted from the radioisotope become implanted in a thin layer of material, or `moderator', usually a layer of frozen neon or argon.
Inside the moderator, a small fraction of the positrons lose energy through interactions with the material and are emitted from the surface of the moderator with a few electronvolts of energy \cite{RouteUltraLowEnergyHBar}.

The positrons are guided by a magnetic field into a Penning-Malmberg trap at a magnetic field strength of 0.15 T, where they undergo inelastic collisions with nitrogen gas molecules.
Positrons colliding with the nitrogen molecules are more likely to lose energy by producing an excitation of the nitrogen molecule than annihilate or produce positronium (a bound state of an electron and positron, which is neutral and escapes the trap) \cite{RouteUltraLowEnergyHBar}, so a fraction of the positrons cool into an electric potential well.
To reduce the gas pressure around the well holding the cooled positrons, a pressure gradient is created by introducing the nitrogen gas in a narrow section of the apparatus and differentially pumping through a series of electrodes with larger radii.
The radius of the positron cloud is controlled with the rotating wall technique described further in section \ref{sec:rotatingWall}.

While the positrons are accumulated, the positron accumulator is isolated from the trap vacuum with a mechanical valve to prevent the nitrogen gas contaminating the UHV system in the main trap.
Before transfer, the nitrogen gas is removed using two high-speed pumps, and a pressure sufficiently low for transfer is achieved in about twenty seconds. 
The positrons are transferred as a pulse through a narrow section to further reduce the gas conductance and are trapped with an electric potential in the positron trap, where they cool through radiative emission.

\begin{figure}[hbt]
\centering
%
%
\begin{psfrags}%
\psfragscanon%
%
\psfrag{s13}[b][b]{\color[rgb]{0,0,0}\setlength{\tabcolsep}{0pt}\begin{tabular}{c}Potential\end{tabular}}%
\psfrag{s14}[l][l]{\color[rgb]{0,0,0}\setlength{\tabcolsep}{0pt}\begin{tabular}{l}(a)\end{tabular}}%
\psfrag{s15}[l][l]{\color[rgb]{0,0,0}\setlength{\tabcolsep}{0pt}\begin{tabular}{l}(b)\end{tabular}}%
\psfrag{s16}[l][l]{\color[rgb]{0,0,0}\setlength{\tabcolsep}{0pt}\begin{tabular}{l}(c)\end{tabular}}%
\psfrag{s17}[b][b]{\color[rgb]{0,0,0}\setlength{\tabcolsep}{0pt}\begin{tabular}{c}Potential\end{tabular}}%
\psfrag{s18}[t][t]{\color[rgb]{0,0,0}\setlength{\tabcolsep}{0pt}\begin{tabular}{c}z\end{tabular}}%
\psfrag{s19}[l][l]{\color[rgb]{0,0,0}\setlength{\tabcolsep}{0pt}\begin{tabular}{l}(d)\end{tabular}}%
\psfrag{s20}[t][t]{\color[rgb]{0,0,0}\setlength{\tabcolsep}{0pt}\begin{tabular}{c}z\end{tabular}}%
%
\psfrag{x01}[t][t]{0.3}%
\psfrag{x02}[t][t]{0.35}%
\psfrag{x03}[t][t]{0.4}%
\psfrag{x04}[t][t]{0.45}%
\psfrag{x05}[t][t]{0.5}%
\psfrag{x06}[t][t]{0.55}%
\psfrag{x07}[t][t]{0.3}%
\psfrag{x08}[t][t]{0.35}%
\psfrag{x09}[t][t]{0.4}%
\psfrag{x10}[t][t]{0.45}%
\psfrag{x11}[t][t]{0.5}%
\psfrag{x12}[t][t]{0.55}%
%
\psfrag{v01}[r][r]{-80}%
\psfrag{v02}[r][r]{-60}%
\psfrag{v03}[r][r]{-40}%
\psfrag{v04}[r][r]{-20}%
\psfrag{v05}[r][r]{0}%
\psfrag{v06}[r][r]{20}%
\psfrag{v07}[r][r]{-80}%
\psfrag{v08}[r][r]{-60}%
\psfrag{v09}[r][r]{-40}%
\psfrag{v10}[r][r]{-20}%
\psfrag{v11}[r][r]{0}%
\psfrag{v12}[r][r]{20}%
%
\resizebox{12cm}{!}{\includegraphics[width=0.8\textwidth]{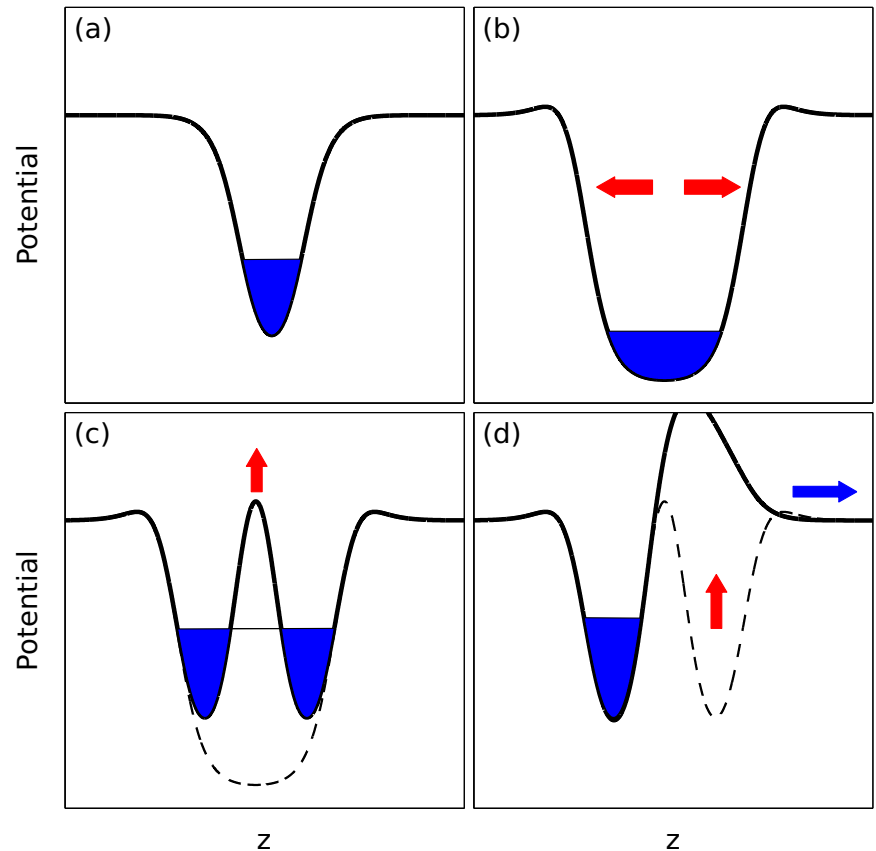}}%
\end{psfrags}%
%

\caption[An illustration of the procedure to reduce the number of particles by half]{An illustration of the procedure to reduce the number of particles by half. The particles are initially in a well like that shown in (a). The well is lengthened (b) and a potential barrier raised to cut the plasma in two pieces (c). One of these halves is discarded (d) and the other is returned to the original well. }
\label{fig:cutting}
\end{figure}

Prior to 2009, this technique produced plasmas containing up to $8\times10^7$ positrons, but due to a hardware failure in this year, only approximately $\mathrm{2.5\times10^7}$ positrons could be recaptured, though this was sufficient for our needs.
The positron number can be tuned using a method of splitting the plasma into two similar plasmas, each with half the total charge, discarding one and keeping the other (see figure \ref{fig:cutting}).
Thus, any number of particles which is a fraction $\left(^1/_2\right)^N$ of the original number can be prepared.
This method has been seen to remain stable at fractions as low as $\mathrm{^1/_{4096}}$.

After transfer, a small population of contaminant ions were identified mixed with the positron plasma.  
The presence of the ions was seen to result in rapid expansion of the positron plasma, so they were separated from the positrons with the application of a fast voltage pulse.
The timing and height of the pulse is arranged so that the positrons can escape the well and be recaptured in a neighbouring one, but the slower ions do not have sufficient time to do the same and remain behind.

\subsection{Antiprotons and the Antiproton Decelerator} \label{sec:antiprotons}
Antiprotons are produced at the Antiproton Decelerator (AD) \cite{AD2004}, \cite{AD2006} through the collision of protons at 26 GeV with an iridium target.
Many different species of particles are produced, and antiprotons are selected based on their mass and charge and captured into a storage ring with an momentum of 3.5 GeV/c.

\begin{figure}[hbt]
	\centering
	\input{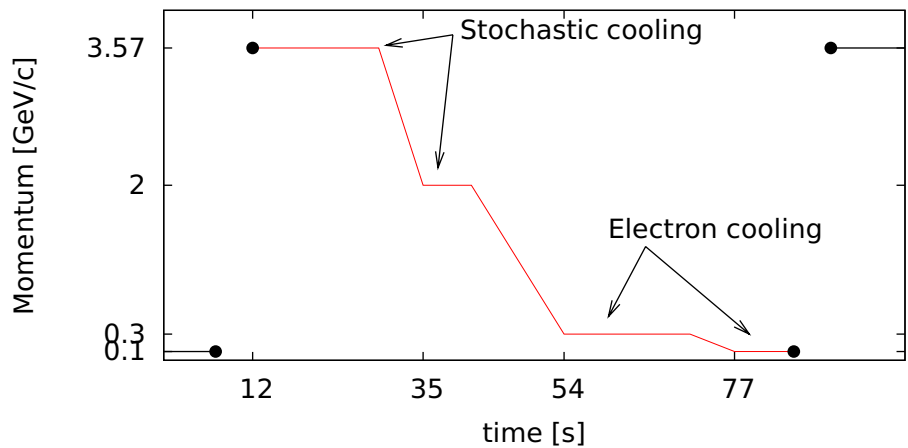}
	\caption[The antiproton momentum as a function of time during an AD cycle]{A visualisation of an AD cycle (adapted from \cite{AD2006}), showing the antiproton momentum as a function of time. The energy of the antiprotons is changed in four steps using radio-frequency power. Stochastic cooling or electron cooling is carried out at the flat portions.}
	\label{fig:ADcycle}
	\end{figure}

The AD cycle, shown in figure \ref{fig:ADcycle}, consists of alternating deceleration and cooling stages. 
Deceleration is achieved by passing the particles thorough a series of resonant radio-frequency (RF) cavities.
With precise timing control, the electric field in the cavity is arranged to decelerate the particles as they pass through.
Deceleration causes the emittance of the beam (a function of the spread of beam momentum and angular divergence) to increase.
To counteract this effect, the deceleration is broken into four stages, and the emittance is reduced after each step. 
Reduction of the emittance is termed `cooling'.

Two techniques are used to perform cooling of the beam: stochastic cooling, which is more effective at higher energies, and electron cooling, which is more effective at lower energies.

Stochastic cooling \cite{stochasticCooling} acts by detecting deviations of particle momenta from the nominal value as they pass a sensor.
These signals can be re-applied as a `kick' to the particles to correct their orbits.
However, the beam is made up of an ensemble of particles, so a pulse that acts to cool one particle can heat another.
It can be shown, fortunately, that for a careful choice of gain, that it is possible to construct a system where the average effect is a net cooling.

Electron cooling \cite{electronCooling} combines the antiproton beam with a cold electron beam over a short length.
The antiprotons transfer energy to the electrons through Coulomb interactions and tend to an equilibrium with the co-streaming electron beam.

After the final cooling stage, the antiprotons have a momentum of 100~MeV/c (equivalent to a kinetic energy of 5.3~MeV). 
They are extracted in one 200~ns long bunch toward an experiment.
The AD usually produces one spill of 3-4 $\times 10^7$ antiprotons every 100~s. 
The deceleration efficiency of the antiprotons captured at the target is in excess of 90\%.

The techniques to capture the antiprotons in a Penning trap and to cool them to cryogenic temperatures are discussed in chapter \ref{chap:Cooling}.
\section{Particle detectors} \label{sec:ParticleDetectors}
	
\subsection{Faraday cup} \label{sec:FaradayCup}

The simplest particle detector in ALPHA is the Faraday cup.
A Faraday cup consists of a piece of conductor -- in ALPHA, the aluminium foil that acts as the final degrader for the antiproton beam (chapter \ref{chap:Cooling}) is also the Faraday cup.
Ideally, a Faraday cup collects the charge of all particles impacting it.
The device has an intrinsic capacitance, so the charge induces a voltage that can be measured using a suitable amplifier.
The capacitance can be determined to high precision, so the amount of collected charge can be measured, which indicates the number of particles.

Some deviations from this simple behaviour can occur when charge is lost from the Faraday cup, either through particles missing the foil, or through the ejection of secondary electrons or other charged particles.
Antiproton annihilations result in several charged daughter particles (section \ref{sec:cosmicAnalysis}), which escape and carry some charge with them, so the antiproton number is not counted using a Faraday cup.
The Faraday cup is used to measure the number of electrons or positrons in a plasma.
ALPHA's Faraday cup is only sensitive to collections of more than $\sim 10^6$ particles, and is used to calibrate other, more sensitive detectors.

\subsection{MCP/phosphor/CCD} \label{sec:MCP}

A microchannel plate (MCP) \cite{MCP} is an array of miniature electron multipliers in the form of microscopic holes machined through a plate of semiconducting material, with a metal deposit on the front and back faces.
In operation, a large potential difference (typically hundreds of volts) is applied across the MCP. 
The impact of a charged particle or high-energy photon on the front face can release secondary electrons from the material. 
The material of the front face is chosen to increase the probability of electron production and the incident particles are typically accelerated to high energies before impact for the same reason.
These electrons are accelerated down the channels by the potential difference, striking the channel walls as they go.
Each further impact releases more secondary electrons, leading to a cascade amplification of the initial charge.
The amount of charge emitted at the back face is an exponential function of the accelerating voltage, and is proportional (up to a saturation point) to the amount of charge incident on the front face.

The MCP used in ALPHA is a type E050JP47 device manufactured by El-Mul Technologies \cite{ElMulCatalog}.
The active face of the MCP is a circle with 41.5~mm active diameter and is covered with an array of holes 12~$\mu$m in diameter spaced 15~$\mu$m in a hexagonal array. 
The device has a gain of ~$8 \times 10^5$ at the maximum rated applied voltage of 1~kV.
The gain behaviour of the MCP was investigated for each of the particle species used in ALPHA over a range of operating parameters \cite{ALPHA_MCP}.

\begin{figure}[hbt]
	\centering
	\includegraphics[width = 0.9 \textwidth, clip = true, trim = 1cm 15.5cm 1cm 1cm]{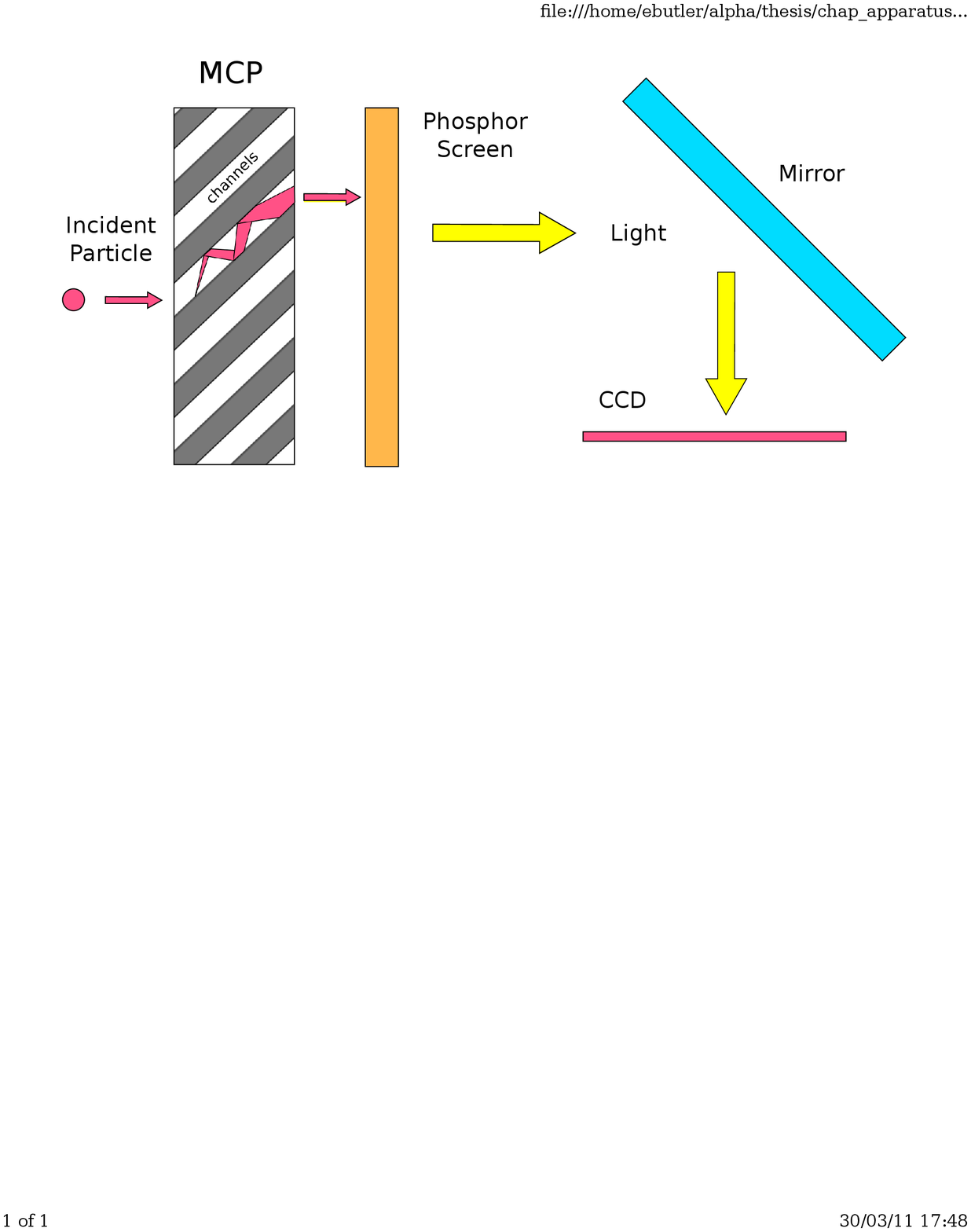}
	\caption[A schematic of the operation of the MCP/phosphor imaging device]{A schematic of the operation of the MCP/phosphor imaging device. An particle incident on the front face of the MCP initiates a cascade of electrons. These strike a phosphor screen, producing light, which is collected by a CCD.}
	\label{fig:MCPdrawing}.
\end{figure}

The electrons from the back of the MCP are further accelerated onto a phosphor screen, where the impacts excite the phosphor atoms, which then decay producing visible light. 
This light can be collected by an imaging device, such as a CCD, to produce a two-dimensional (z-integrated) projection of the particles incident on the MCP.

\begin{figure}[hbt]
\centering
\subfloat[]{ \begin{overpic}[width = 0.3\textwidth]{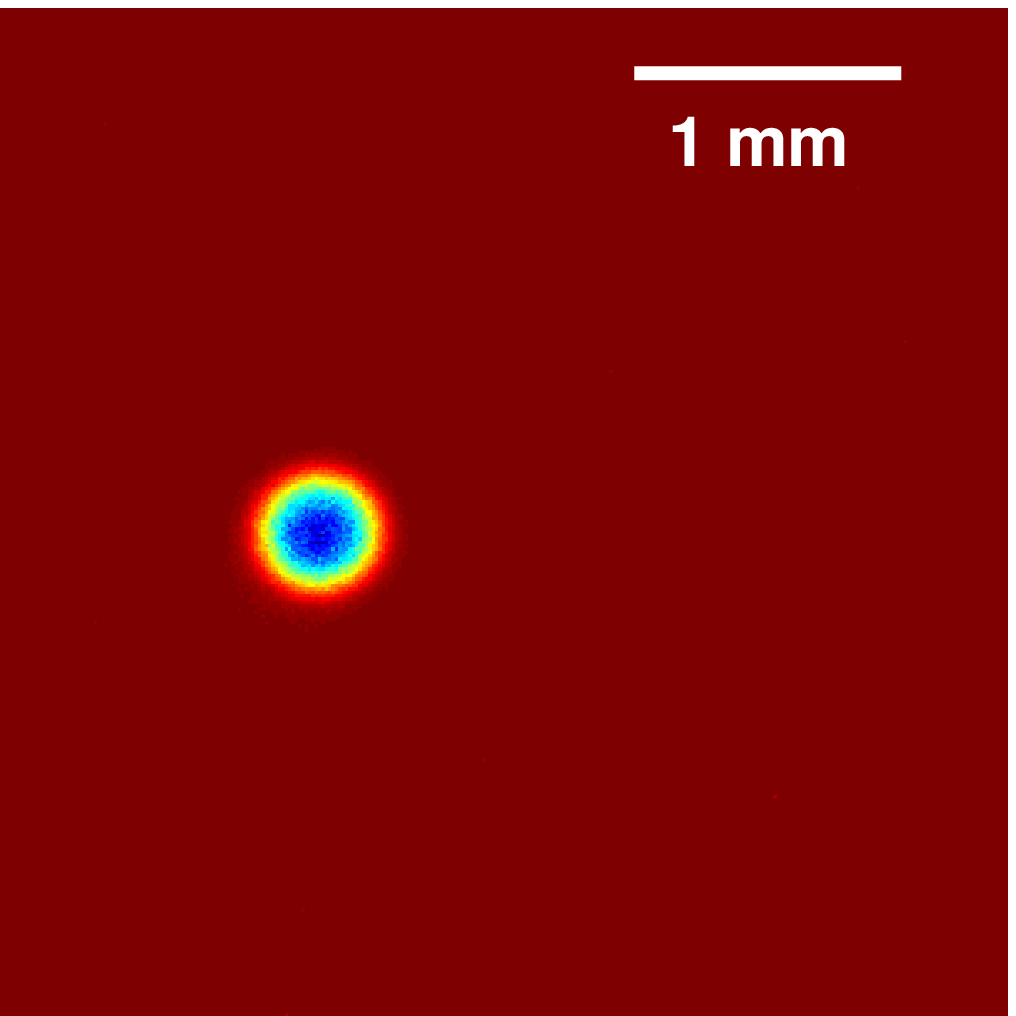} \put(5, 90){\color{white}(a)}\end{overpic}}
\hspace{0.3cm}
\subfloat[]{ \begin{overpic}[width = 0.3\textwidth]{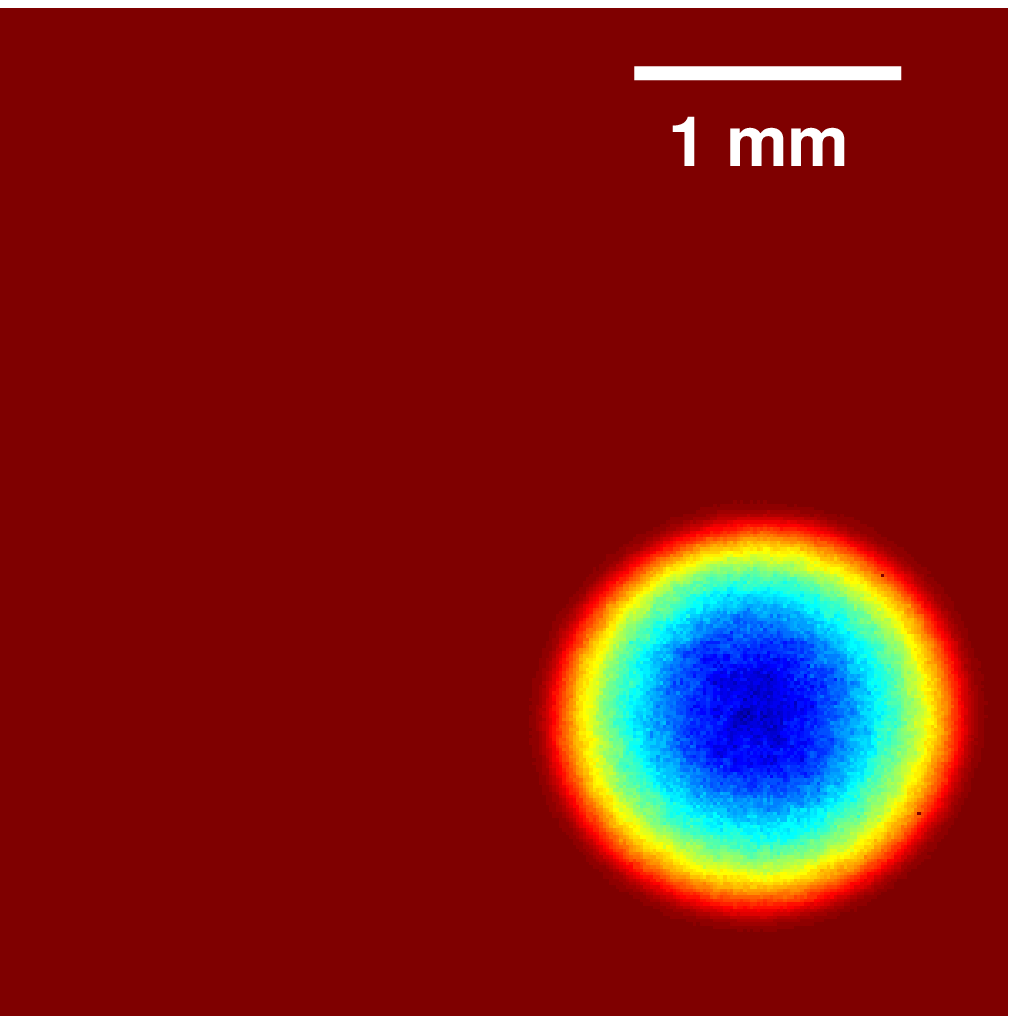} \put(5, 90){\color{white}(b)}\end{overpic}}
\\
\subfloat[]{ \begin{overpic}[width = 0.3\textwidth]{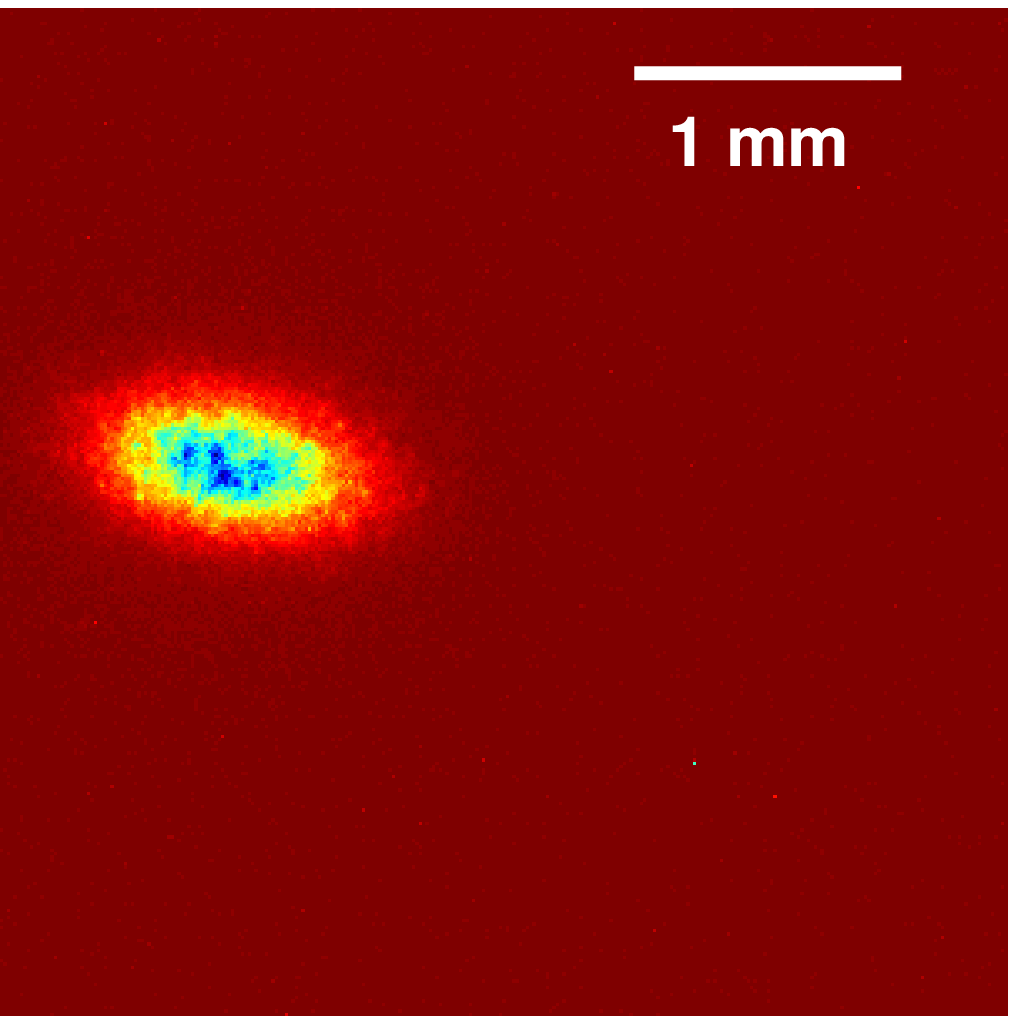} \put(5, 90){\color{white}(c)}\end{overpic}}
\hspace{0.3cm}
\subfloat[]{ \begin{overpic}[width = 0.3\textwidth]{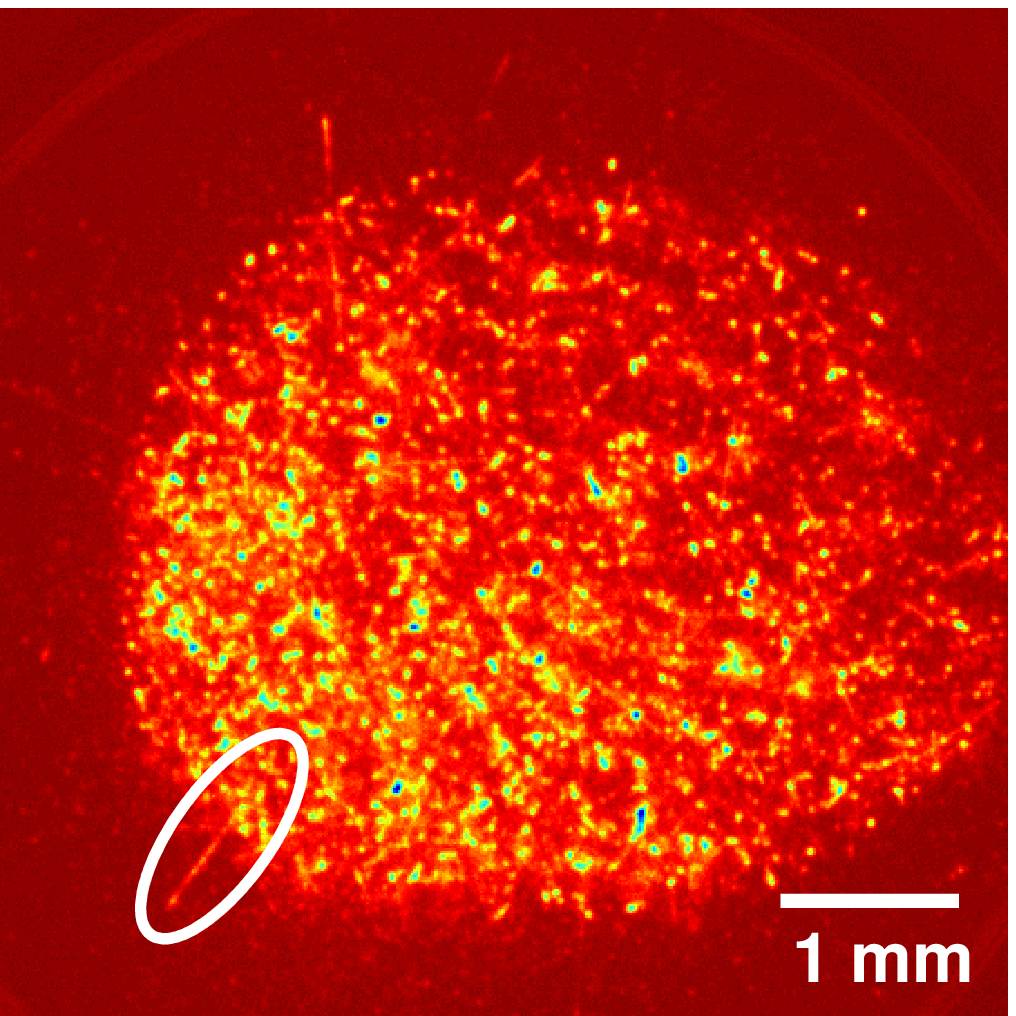} \put(5, 90){\color{white}(d)}\end{overpic}}
\caption[MCP images of electrons, positrons and antiprotons]{MCP images of (a) electrons (b) positrons and (c) antiprotons. At high gain, tracks from antiproton annihilation products can become visible, as in (d), where a `track' left by an annihilation product is circled. The scale bars are the size of the plasmas in a 1~T magnetic field, before extraction.}
\label{fig:exampleMCP}
\end{figure}

A schematic of the operation of the MCP/phosphor device is shown in figure \ref{fig:MCPdrawing}, and some examples of recorded images are shown in figure \ref{fig:exampleMCP}.
The projections of the antiproton clouds are distinctly elliptical in shape, unlike the electron and positron plasmas, which are the expected circular shape.
This is thought to be due to the shape of the magnetic field outside the central homogeneous region (in the `fringing' region) of the external solenoid.
In the fringing field, the strength of the magnetic field falls, and the particles cease to follow magnetic field lines and move in a straight line. 
A slight rotational asymmetry in the magnetic field in this region, due to construction imperfections, can cause the observed deformation of the distribution.
Electron and positrons are not as susceptible to this effect as their lower mass means that they detach from the magnetic field lines at a different point.
The same effect also causes electrons and antiprotons to appear in different positions on the MCP \cite{Joel_Hollow}.

\captionsetup[subfloat]{labelfont={color=white}}

\begin{figure}[b!]
\centering
\input{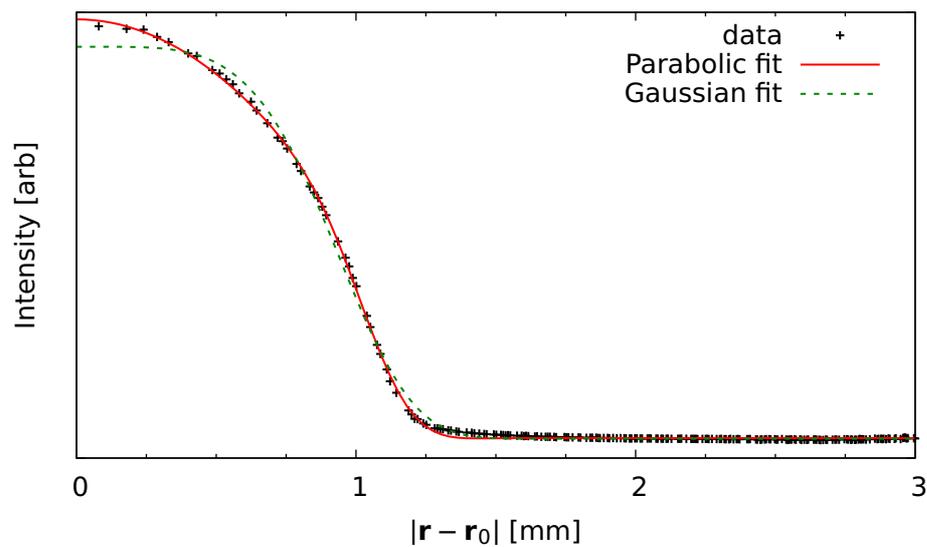}
\caption[A measured intensity profile from the MCP/phosphor/CCD with fits]{A example of a measured intensity profile from the MCP/phosphor/CCD. Intensity is proportional to particle number. The data is fit using a generalised Gaussian (equation \ref{eq:MCPFit1}) or a smoothed projection (equation \ref{eq:MCPFit2}). }
\label{fig:exampleMCPProfile}
\end{figure}

These profiles are fit using a two-dimensional `generalised-gaussian' function, of the form
\begin{equation}
	I(\mathbf{r}) = A\; \mathrm{exp}\left( \left( \frac{\left| \mathbf{r} - \mathbf{r_0}\right|}{B}\right)^n\right),
	\label{eq:MCPFit1}
\end{equation}
where $A$, $B$, $n$ and $\mathbf{r_0}$  are fit parameters.
An example profile, with the fit function, is shown in figure \ref{fig:exampleMCPProfile}.
It is clear that the function from equation \ref{eq:MCPFit1} fits the profile reasonably well, but underestimates the intensity near the centre of the plasma.
This can produce problems when using these functions to calculate the potential and density of the plasma close to the axis (see section \ref{sec:equilSol}).
Guided by knowledge of the physics, an alternative fitting function was developed.

A cold (0~K) plasma in thermal equilibrium will be shaped as a constant density spheroid (see section \ref{sec:plasmaEquilibrium}).
The projection of this function along the z-axis is a function of the form $(C - A r^2)$ with $C, A$ constants.
However, in reality, the plasma is not at 0~K, so the density falls off at the edges of the plasma over the Debye length (section \ref{sec:plasmaEquilibrium}). 
A smoothed step function is applied at the plasma radius to account for this effect.
The complete fit function is of the form
\begin{equation}
I(\mathbf{r}) = (C -A r^2) \;\mathrm{Erfc}(D( |\mathbf{r}-\mathbf{r}_0 | - R)),
\label{eq:MCPFit2}
\end{equation}
where $\mathrm{Erfc}$ is the complementary error function and $A$, $C$, $D$, $\mathbf{r_0}$ and $R$ are fit parameters.
The function is shown in figure \ref{fig:exampleMCPProfile}, where it is obvious that the fit of equation \ref{eq:MCPFit2} describes the plasma more closely than the generalised Gaussian fit.
This function will be used to describe the plasma shape in all of the later analyses in this thesis.

This MCP/phosphor/CCD arrangement has proven to be an exceptionally powerful diagnostic tool for non-neutral plasma physics experiments.
The ease with which it allows the number and spatial distribution of particles to be precisely measured has made many of the advances achieved in ALPHA possible.

It is also possible to measure the charge striking the phosphor screen as a function of time using a capacitively coupled pickup.
By releasing a plasma in a controlled fashion, it is possible to reconstruct the energy spectrum of the plasma particles using the MCP as a charge amplifier (see section \ref{sec:EnergyDumps}).
Because of the very high gains achievable, as few as one or two incident electrons can be detected.

\subsection{Scintillators}\label{sec:Scintillators}

Detection of ionizing radiation using scintillation detectors is a long-established technique, particularly in particle physics.
The active, scintillating material of the detector can be one of a wide range of substances, including a crystal, organic liquid, gas or plastic, but all scintillating materials produce light in response to the passage of ionising radiation. 
The light is collected and converted to an electrical signal by a Photo-Multiplier Tube (PMT).
When the voltage level passes a set threshold, there is considered to be a `count' from the scintillator.
On average, a proton-antiproton annihilation produces three charged pions, which are readily detected by the scintillator/PMT combination.

It is standard practice to place two independent scintillators close together to reduce the sensitivity to electronic noise. 
In this case, a signal from each scintillator within a certain time window is required to register a count.
The scintillators are typically arranged so that the expected trajectory of a particle that strikes one scintillator will also carry it through the other, producing a count in each, while noise, which is uncorrelated, will have a smaller probability of producing counts in coincidence.

Rectangular paddles of scintillator material are placed at three longitudinal positions along the ALPHA trap. 
At each position two pairs of scintillators are placed, one to either side of the apparatus.
Various combinations of counts in coincidence can be used to optimise the signal-to-noise ratio of the measurement.
The most common combinations used are to require a count in at least one of the two scintillator pairs at the same longitudinal position (usually called the `Or' of the pair), and requiring a count in both of the pairs (referred to as the `And').
Because of varying solid angle acceptances, each set of scintillators has a different sensitivity for detecting annihilations in different parts of the trap.
By comparing the relative number of counts, it is possible to make an estimate of the position of the annihilations.

Scintillators are also sensitive to the passage of cosmic rays, which produce a significant background on measurements.
Near the Earth's surface, most cosmic rays come from directly overhead, and for this reason, the scintillators are placed vertically to expose the smallest possible cross section to the downward propagating particles.
The cosmic background rate can also be reduced by requiring counts in coincidence from scintillators that are laterally separated.
The typical background rate for an `Or' of a pair of scintillators is 40 Hz. 
When using an `And', the rate drops to approximately 0.1 Hz. 
However, the sensitivity to annihilations drops by a factor of 15, so in some situations it may be more desirable to trade off the background rate against the detection efficiency.

\subsection{Annihilation vertex detector} \label{sec:siliconDetector}

ALPHA uses a three-layer tracking detector arranged in a cylindrical fashion coaxial with the electrodes (visible in figures \ref{fig:Search_app} and \ref{fig:allofit}). 
Each layer is made up of a number of silicon wafers (`modules'), divided on each side into strips. 
A charged particle passing through the detector loses energy, leaving deposits of charge marking the position at which it passed through each layer.
The strips are oriented in perpendicular directions on either side of the silicon, and can be individually addressed to measure the charge deposited in the material.

The charge is measured using amplifiers and a type-VF48 analogue-to-digital convertor. 
The charge profiles are stored in the MIDAS DAQ system (section \ref{sec:DAQ}), and analysed using  custom-written software.
A strip is identified as `hit' (i.e. a particle passed through it) if the charge deposited exceeds the noise level by a defined amount.
The intersection of two perpendicular hit strips localises the point of passage of a particle in three dimensions.

\begin{figure}[hbt]
	\centering
	\subfloat[]{\includegraphics[width = 0.4\textwidth, clip, trim = 5cm 5cm 5cm 5cm, angle=90]{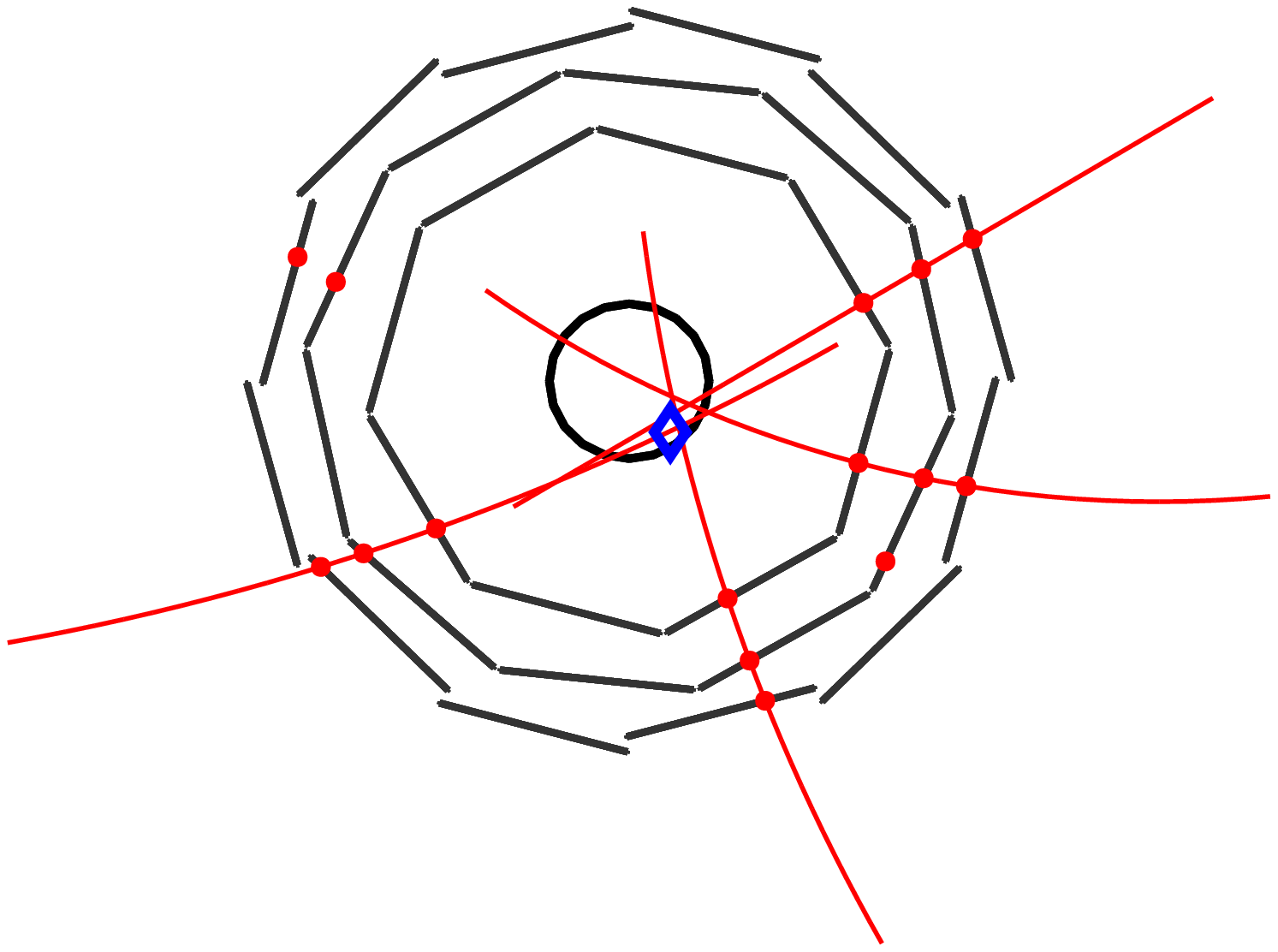}}
	\hspace{0.5cm}
	\subfloat[]{\includegraphics[width = 0.4\textwidth, clip, trim = 5cm 5cm 5cm 5cm, angle=90]{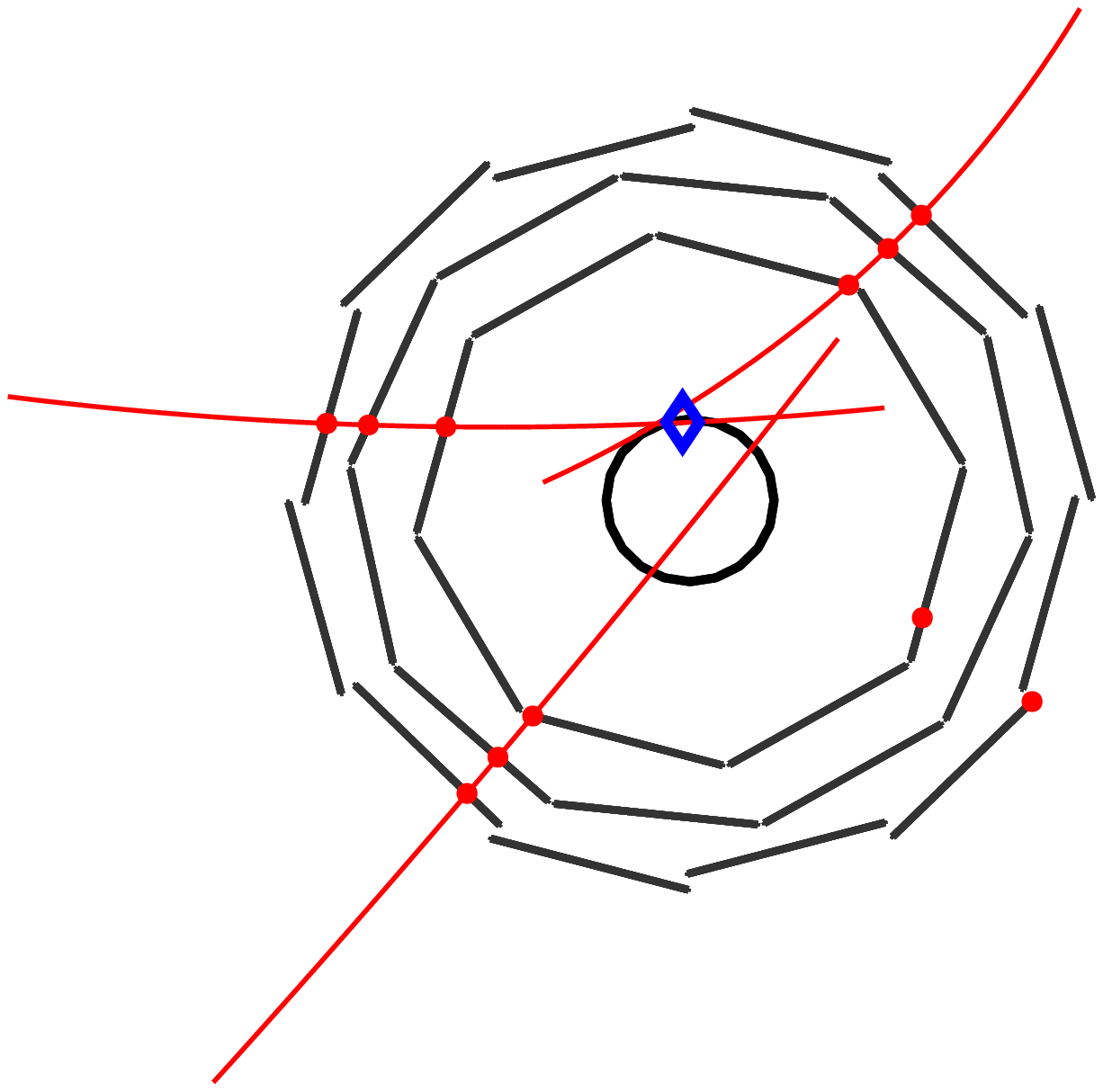}}
	\caption[Examples of reconstructed antiproton annihilations]{Two examples of reconstructed antiproton annihilations. The straight sections are the silicon modules, and the central circle marks the position of the Penning trap electrodes. The red dots mark the positions of hits, the curved lines are the particle tracks and the position of the vertex is shown by the blue diamond. }
	\label{fig:exampleReconstruction}
\end{figure}

If a particle produces hits in all three layers of the detector, a helix is fit to the trajectory, giving a `track'.
The intersection of two or more tracks defines the `annihilation vertex'.
The vertex is an estimate of the position of the antiproton annihilation.
Examples of reconstructed antiproton annihilations can be found in \ref{fig:exampleReconstruction}.
See \cite{RichardFutureThesis} for more details on the reconstruction of antiproton annihilations.

The rate at which the charge profiles can be read out is limited by hardware to around 100~Hz (improved to around 500~Hz since the work reported in this thesis).
The vertex detector can be simultaneously run in a counting mode similar to the scintillators.
In this mode, each module generates a digital trigger when a charge deposit is detected, and a coincidence of triggers is recorded as a count.
Coincidences of different numbers of modules in different positions have varying efficiencies for annihilations and sensitivity to noise.
An often-used channel is called the `Si$>$1' channel, which requires at least two triggers from the innermost layer for a count.
This channel has $(85 \pm 15)\%$ efficiency for annihilations and a background rate (including cosmic rays) of $\sim 4.8~\mathrm{Hz}$.

\begin{figure}[hbt]
\centering
\subfloat[]{\input{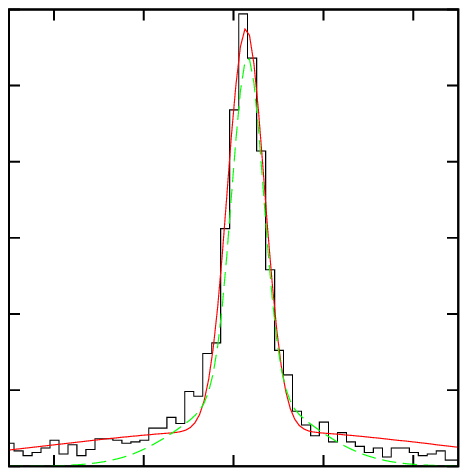}}
\subfloat[]{\input{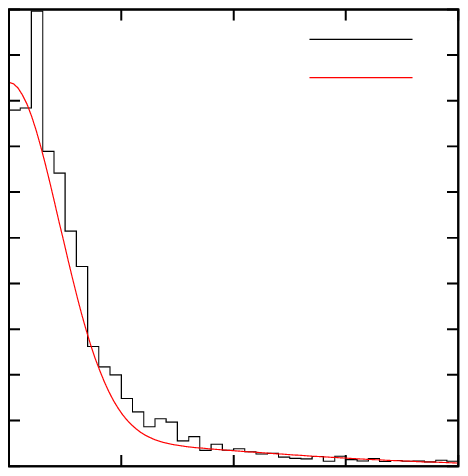}}
\caption[Projections of annihilation vertices from a small antiproton cloud.]{The distribution of annihilation vertices projected onto (a) the $z$ axis and (b) the $r$ axis recorded from a small, trapped antiproton cloud. The solid lines are fits to the data used to parametrise the resolution. }
\label{fig:detectorResolution}
\end{figure}

The resolution of the detector is affected by the amount of material present between the annihilation point and the vertex detector, as the paths of charged particles that scatter can not be accurately reconstructed. 
For this reason, the vacuum and magnet support structures were made from low-density materials to minimise the scattering probability.
The resolution can be measured by holding a small antiproton cloud in the detector and recording the annihilations.
In the limit that the cloud size is much smaller than the resolution, this gives the resolution of the detector.
Figure \ref{fig:detectorResolution} shows the $z$ and $r$ projections of the vertex distributions obtained from such a measurement -- the size of the antiproton cloud was approximately 1~mm in both $z$ and $r$.
The distribution is fit using a function of the form \cite{RichardFitting}
\begin{equation}
N = a\,\mathrm{exp}\left(-\left(\frac{x-b}{c}\right)^2\right) + d\,\mathrm{exp}\left(-\left(\frac{x-e}{f}\right)^2\right),
\end{equation}
which reproduces the shape of the distribution reasonably well.
The width of each of the Gaussian components of this formula for the $z$ distribution in figure \ref{fig:detectorResolution}(a) are 4~mm and 40~mm (each making up 50\% of the total), while the $r$ distribution (figure \ref{fig:detectorResolution}(b)) has 75\% of the data in a Gaussian 4~mm width, and 25\% of the data in a Gaussian 30~mm wide.

The resolution can be improved by placing `cuts' on the data, at some cost to the number of vertices collected.
The cuts used are typically the same as those used for cosmic rejection (section \ref{sec:cosmicAnalysis}).
We make use of the knowledge that annihilations must occur within the Penning trap, so reconstructed vertices far outside the trap radius of 22.2~mm are obviously poorly reconstructed in this direction, and are likely to also be poorly reconstructed in the $z$ direction. 
Restricting the vertices to only those with 40~mm of the trap axis reduces the width of the broader gaussian in the $z$ distribution to $\sim$10~mm.
This function is shown as the dashed line in figure \ref{fig:detectorResolution}(a).
It is clear that it is mostly only vertices with large deviations from the bulk of the distribution that are rejected using this technique, and that the resolution of the detector is improved.
The fits to the resolution of the detector are used in simulations where this is required (chapters \ref{chp:vertex} and \ref{chp:trapping}).

\section{Sequencer} \label{sec:Sequencer}

The wide variety of instruments used in ALPHA are controlled by a large number of independent control systems.
A typical system is composed of custom-written National Instruments (NI) LabVIEW \cite{labview} applications running on a desktop PC with a suitable number of analogue and digital inputs and outputs.
To successfully perform an experiment, it is necessary to link these systems together in a way that is both versatile and robust.

The ALPHA Sequencer consists of an array of digital inputs and outputs controlled by a Field Programmable Gate Array (FPGA) controller (type NI PXI-7811R). 
For versatility, the sequencer is divided into two independent halves, one controlling the catching trap, while the other controls the mixing and positron traps.
At the beginning of every experiment, the FPGA is loaded with a list of digital states that has been prepared by the physicist using a custom graphical user interface (GUI). 
At each time step, digital triggers are issued to the equipment, with each trigger having a meaning specific to that application.
The sequencer is also capable of `waiting' an indefinite amount of time to receive a input trigger before continuing, allowing it to synchronise with external devices.

\begin{figure}[hbt]
\centering

\psset{unit=1in,cornersize=absolute,dimen=middle}%
\begin{pspicture}(0,-4.5)(5,0.25)%
\psset{linewidth=0.8pt}%
\psset{linewidth=0.8pt}%
\ifx\MPSTPatchA\undefined{\makeatletter\def\psbezier@ii{\addto@pscode{%
\ifshowpoints true \else false \fi\tx@OpenBezier%
\ifshowpoints\tx@BezierShowPoints\fi}\end@OpenObj}\makeatother%
\global\def\MPSTPatchA{}}\fi%
\psset{arrowsize=1.1pt 4,arrowlength=1.64,arrowinset=0}%
\psframe(0,-0.25)(1,0.25)
\rput(0.5,0){Editor}
\psline[linestyle=dashed,arrowsize=0.05in 0,arrowlength=2,arrowinset=0]{->}(1,0)(2,0)
\uput{0.501875ex}[u](1.5,0){Sequence}
\psframe(2,-0.25)(3,0.25)
\rput(2.5,0){Main Control}
\psline[linestyle=dashed,arrowsize=0.05in 0,arrowlength=2,arrowinset=0]{->}(3,0)(4,0)
\rput(3.5,0){\shortstack{Sequence\\
XML data}}
\psframe(4,-0.25)(5,0.25)
\rput(4.5,0){MIDAS}
\psline[linestyle=dashed,arrowsize=0.05in 0,arrowlength=2,arrowinset=0]{->}(2.5,-0.25)(2.5,-0.75)
\psline[linestyle=dashed](2.5,-0.75)(4,-0.75)
\uput{0.501875ex}[u](3.25,-0.75){Analogue states}
\psline[linestyle=dashed,arrowsize=0.05in 0,arrowlength=2,arrowinset=0]{->}(4,-0.75)(4,-1.25)
\psframe(3.5,-1.75)(4.5,-1.25)
\rput(4,-1.5){\shortstack{Analogue\\
Controller}}
\psline[linestyle=dashed,arrowsize=0.05in 0,arrowlength=2,arrowinset=0]{->}(2.5,-0.25)(2.5,-0.75)
\psline[linestyle=dashed](2.5,-0.75)(1,-0.75)
\uput{0.501875ex}[u](1.75,-0.75){Digital states}
\psline[linestyle=dashed,arrowsize=0.05in 0,arrowlength=2,arrowinset=0]{->}(1,-0.75)(1,-1.25)
\psframe(0.5,-1.75)(1.5,-1.25)
\rput(1,-1.5){\shortstack{Digital\\
Controller}}
\psline[arrowsize=0.05in 0,arrowlength=2,arrowinset=0]{->}(1.5,-1.5)(3.5,-1.5)
\uput{0.501875ex}[u](2.5,-1.5){Timing triggers}
\psline[arrowsize=0.05in 0,arrowlength=2,arrowinset=0]{->}(1.4,-2.75)(1.4,-1.75)
\uput{0.501875ex}[ur](1.4,-2.25){\rlap{Digital Input}}
\psline[arrowsize=0.05in 0,arrowlength=2,arrowinset=0]{->}(1.35,-2.75)(1.35,-1.75)
\psline[arrowsize=0.05in 0,arrowlength=2,arrowinset=0]{->}(1.3,-2.75)(1.3,-1.75)
\psline[arrowsize=0.05in 0,arrowlength=2,arrowinset=0]{->}(0.6,-1.75)(0.6,-2.75)
\uput{0.501875ex}[ul](0.6,-2.25){\llap{Digital Output}}
\psline[arrowsize=0.05in 0,arrowlength=2,arrowinset=0]{->}(0.65,-1.75)(0.65,-2.75)
\psline[arrowsize=0.05in 0,arrowlength=2,arrowinset=0]{->}(0.7,-1.75)(0.7,-2.75)
\psline[arrowsize=0.05in 0,arrowlength=2,arrowinset=0]{->}(0.75,-1.75)(0.75,-2.75)
\psline[arrowsize=0.05in 0,arrowlength=2,arrowinset=0]{->}(0.8,-1.75)(0.8,-2.75)
\psline[arrowsize=0.05in 0,arrowlength=2,arrowinset=0]{->}(0.85,-1.75)(0.85,-2.75)
\psline[arrowsize=0.05in 0,arrowlength=2,arrowinset=0]{->}(0.9,-1.75)(0.9,-2.75)
\psline[arrowsize=0.05in 0,arrowlength=2,arrowinset=0]{->}(0.95,-1.75)(0.95,-2.75)
\psframe(0.45,-3.25)(1.45,-2.75)
\rput(0.95,-3){\shortstack{Magnet\\
Control}}
\psframe(0.45,-3.875)(1.45,-3.375)
\rput(0.95,-3.625){SIS}
\psframe(0.45,-4.5)(1.45,-4)
\rput(0.95,-4.25){Instruments}
\psline[arrowsize=0.05in 0,arrowlength=2,arrowinset=0]{->}(3.6,-1.75)(3.6,-2.75)
\psline[arrowsize=0.05in 0,arrowlength=2,arrowinset=0]{->}(3.7,-1.75)(3.7,-2.75)
\psline[arrowsize=0.05in 0,arrowlength=2,arrowinset=0]{->}(3.75,-1.75)(3.75,-2.75)
\psline[arrowsize=0.05in 0,arrowlength=2,arrowinset=0]{->}(3.85,-1.75)(3.85,-2.75)
\psline[arrowsize=0.05in 0,arrowlength=2,arrowinset=0]{->}(3.9,-1.75)(3.9,-2.75)
\psline[arrowsize=0.05in 0,arrowlength=2,arrowinset=0]{->}(3.95,-1.75)(3.95,-2.75)
\psline[arrowsize=0.05in 0,arrowlength=2,arrowinset=0]{->}(3.8,-1.75)(3.8,-2.75)
\psline[arrowsize=0.05in 0,arrowlength=2,arrowinset=0]{->}(3.65,-1.75)(3.65,-2.75)
\psframe(3.275,-3.25)(4.025,-2.75)
\rput(3.65,-3){Electrodes}
\psline[arrowsize=0.05in 0,arrowlength=2,arrowinset=0]{->}(4.35,-1.75)(4.35,-2.75)
\psline[arrowsize=0.05in 0,arrowlength=2,arrowinset=0]{->}(4.3,-1.75)(4.3,-2.75)
\psline[arrowsize=0.05in 0,arrowlength=2,arrowinset=0]{->}(4.4,-1.75)(4.4,-2.75)
\psframe(4.175,-3.25)(4.625,-2.75)
\rput(4.4,-3){\shortstack{Instru-\\
ments}}
\end{pspicture}%
\caption[The principal elements of the ALPHA control system]{A flow diagram showing the principal elements of the ALPHA control system.}
\label{sequencerflow}
\end{figure}
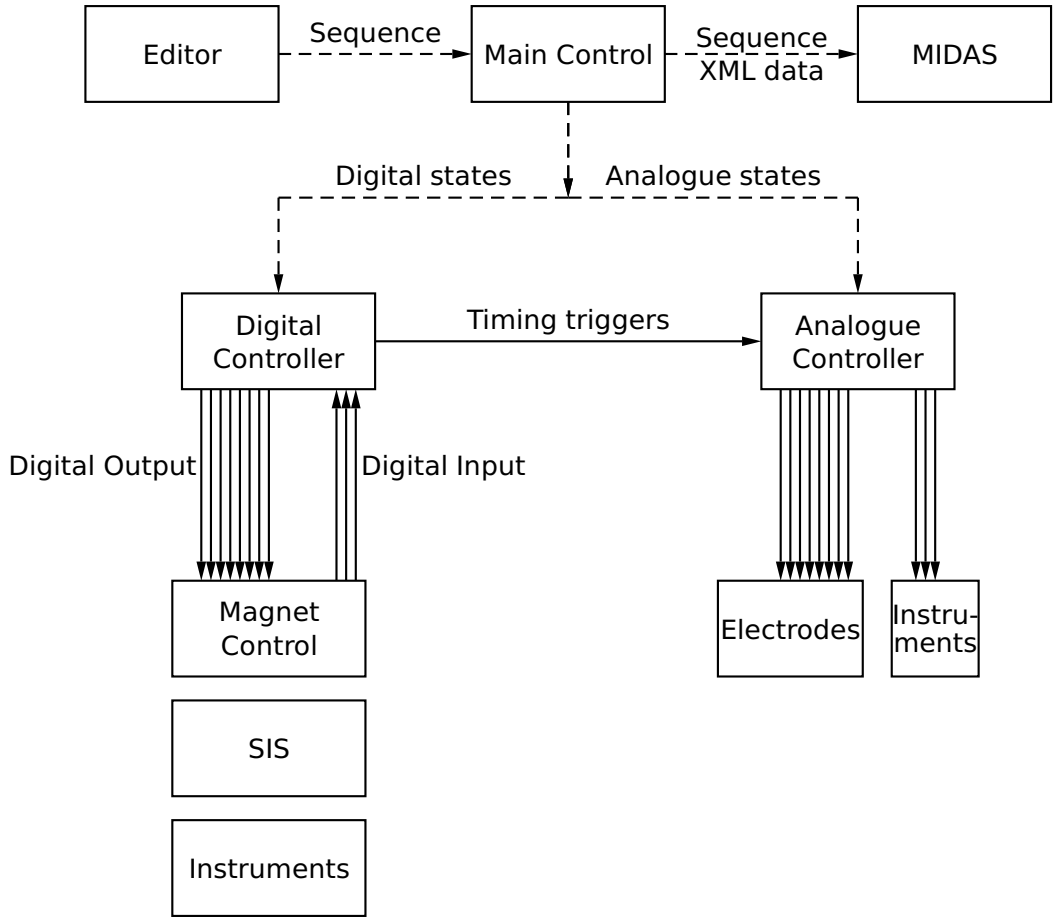

To control the voltages applied to the electrodes, as well as any miscellaneous analogue signals, analogue states are defined using the same GUI. 
The list of analogue states is then loaded into the onboard memory of a type NI PXI-6733 digital to analogue converter.
The transition between states is triggered by a digital signal from the FPGA.

\section{Data acquisition and logging}\label{sec:DAQ}

A large part of the data produced by an experiment takes the form of antiproton annihilation counts produced by scintillation detectors or the vertex detector.
The counts are recorded by a SIS3840 multiscaler module, which produces an array of the counts as a function of time with a specified bin width.
Two such modules are used by ALPHA, capable of recording a total of 64 channels at a continuous rate of 40~kHz.
In addition to the counts from the detectors, signals from the sequencer are also recorded to act as time references for later analysis.

The number of counts in each time bin is stored using the Maximum Integration Data Acquisition System (MIDAS) software \cite{midas}.
MIDAS is a framework jointly developed by the Paul Scherrer Institute (PSI) in Switzerland and TRIUMF in Canada which can accommodate a wide variety of hardware and software data acquisition components and is a powerful recording and data-logging tool.
MIDAS also logs data recorded by the asynchronous external equipment. 
Generically referred to as `environment' data, it includes measurements of temperatures, vacuum pressures, magnetic fields, as well as monitoring of the output currents and voltages of power supplies and amplifiers.
This data can be presented as a time history of each recorded variable over a webserver.

The MIDAS data is post-processed with a custom written analysis suite (called `alphaAnalysis') written in the ROOT \cite{root} extension to C++, which produces data structures called `trees'.
Trees are then accessed in individual, C++ codes written for individual analyses.
This versatile arrangement allows almost any conceivable analysis to be performed.

    \chapter{Physics of Particles and Plasmas in a Penning-Malmberg/Octupole Trap}
\label{chap:trapPhysics}

\epigraph{Science. It works, bitches.}{Randall Munroe, \href{http://xkcd.com/54/}{xkcd.com}}

\section{Non-neutral plasma physics}
\label{sec:plasmaEquilibrium}

In section \ref{sec:penningTrap}, the motion of a single charged particle in a Penning-Malmberg trap was described in detail.
A single particle moves only under the influence of the electric and magnetic fields applied from the trap.
In experiments at ALPHA, between $10^4$ and $10^8$ particles are used, and the particle-particle interactions must be considered, as well as the effects from the trap alone.

Collections of charged particles interact predominantly through the Coulomb force.
At low densities, the effect of the Coulomb interaction is small, and only plays a role when particles make a close approach.
In this scenario, the particles behave much like a gas.
At higher densities, the influence of the Coulomb interaction grows; when the Coulomb force plays the dominant role in determining the particles' dynamics, the particles form a plasma.

In ALPHA, plasmas are almost always made up of particles of the same sign of charge, for which the term `non-neutral' will be exclusively reserved.
In contrast, plasmas containing both positively and negatively charged particles, but carrying a net charge are referred to as `partially-neutralised'.
Non-neutral plasmas, in contrast to neutral or partially neutralised plasmas, can be stably confined whilst in a state of global thermal equilibrium \cite{trappedPlasmasReview}.

The plasma frequency, given by 
\begin{equation}
	\omega_p = \sqrt{\frac{ne^2}{\epsilon_0 m}},
\end{equation}
is a fundamental parameter associated with each plasma.
$1/\omega_p$ is the timescale over which a plasma can react to a small disturbance in the charge distribution, and thus sets the smallest time over which it is possible to observe plasma behaviour.

The Debye length, given by
\begin{equation}\label{eq:debyeLength}
		\lambda_D = \sqrt{ \frac{\epsilon_0 k_B T}{n e^2} }
	\end{equation}
is the distance travelled by a typical particle in a time $1/\omega_p$, and sets the minimum length over which plasma behaviour is observed.
Thus, to be a considered a plasma, the size of a charge distribution must be at least one Debye length.
For this reason, a comparison of the Debye length and the size of the plasma is often used as a rule of thumb when deciding if a collection of charges should be treated as a plasma or not.
	
The Debye length also describes the range over which a single charge can influence the potential - the change in potential induced by a disturbance in the charge distribution falls off as $\mathrm{e}^{-r/\lambda_D}$.
For example, if a test charge is introduced into a plasma, the plasma particles will rearrange themselves so that they counteract the change in potential.
The re-arrangement `screens' the rest of the plasma from the test charge, and this screening effect is a characteristic of plasmas.

The fact that any particle in a plasma interacts with several other particles gives rise to collective effects.
This characteristic can manifest in one of a large number of wave and oscillatory phenomena, e.g. those discussed in \cite{PlasmaModes} and \cite{ElectronAcousticWaves}.
One of the most fundamental collective motions is the centre-of-mass normal plasma mode, in which the plasma as a whole moves back and forth along the magnetic field in its confining potential, in many ways similar to the axial motion of a discrete particle.
An application of plasma modes will be discussed in section \ref{sec:Modes}.

\subsection{Thermal equilibrium}
\label{sec:PlasmaEquilibrium}
The charged particles making up the plasma move according to the electric and magnetic fields in the trap, while the presence of these charges in turn modify the electric fields in and around the plasma.
In thermal equilibrium, the electric potential and the charge distribution must simultaneously satisfy Poisson's equation
\begin{equation} \label{eq:poisson}
	\grad^2 \Phi(r,z) =  - \frac{e n(r,z)}{\epsilon_0},
\end{equation}
and the Maxwell-Boltzmann distribution,
\begin{equation} \label{eq:boltzmann}
	n(r,z) = n_0 \;\mathrm{exp} \left[ \frac{-e \Phi(r,z)}{k_B T} \right].
\end{equation}

\begin{figure}[hbt]
	\centering
	\input{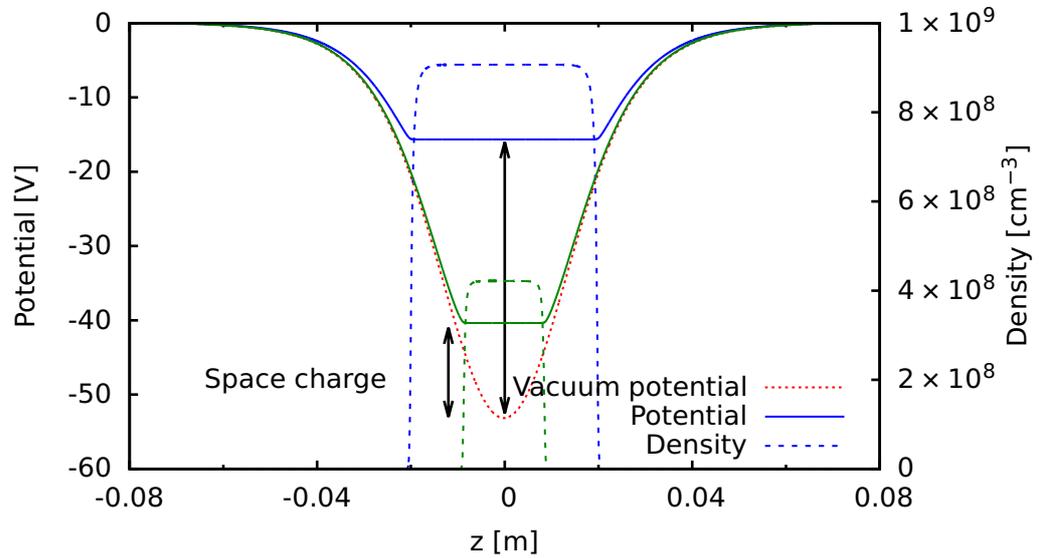}
	\caption[The on-axis potential and density for two example plasmas]{The on-axis potential and density for two example plasmas. The `vacuum potential', i.e. the potential in the absence of charge has been modified so that the potential inside the plasma is flat.}
	\label{fig:sampleEquilSol}
\end{figure}

If at first we only consider the direction parallel to the magnetic field, it is easy to see that an electric field along the plasma will cause a flow of charge.
Charge flows until the potential along a magnetic field line becomes flat.
The level at which the well is flattened is determined by the charge density, and is usually called the `space charge' or `self charge'.
Examples of how the presence of the plasma modifies the trap potential can be seen in figure \ref{fig:sampleEquilSol}.
Equations \ref{eq:poisson} and \ref{eq:boltzmann} imply that the density along a field line is constant along most of the length of the plasma and drops to zero at the ends over a length scale comparable to the Debye length.

For the three dimensional case, full solutions of equations \ref{eq:poisson} and \ref{eq:boltzmann} all have the same general qualities \cite{trappedPlasmasReview}.
The charge density is constant inside a volume defined by a surface of rotation, and then falls exponentially to zero on a length scale of the Debye length (equation \ref{eq:debyeLength}).
The plasma rotates about the cylindrical axis with a frequency determined by a so-called $\mathbf{E} \times \mathbf{B}$ \footnote{$\mathbf{E}$ the traditional symbol for the electric field: $\mathbf{F}$ is used throughout this thesis to avoid confusion with the energy, also denoted $E$.} motion with velocity
\begin{equation}\label{eq:plasmaExB}
	v_r = \omega_r r = \frac{ \mathbf{F} \times \mathbf{B} }{ B^2} = \frac{e n}{2 \epsilon_0} \frac{r}{B_z}.
\end{equation}
It is interesting to note that $\omega_r$ is independent of $r$, indicating that the plasma rotates as a rigid body.

In a quadratic potential (equation \ref{eq:penningFieldsPhi}), the plasma will assume a spheroidal shape - the volume created by rotating an ellipse around one of its axes \cite{trappedPlasmasReview}.
A plasma can either be oblate (pancake shaped) or prolate (cigar shaped), depending on the geometry of the trapping potential and the transverse charge distribution.
In ALPHA, the plasmas are almost exclusively prolate, with length-to-diameter ratios (aspect ratios) exceeding 20 not uncommon.

\subsection{Collisions} \label{sec:collisions}

Plasmas come into thermal equilibrium by redistributing energy between the particles through Coulomb interactions.
`Close' interactions, in which particles' velocities significantly change over the course of the interaction, are considered `collisions'.
For non-neutral plasmas relevant to ALPHA, the approach to thermal equilibrium can be divided into two stages.
First, the velocity distribution of particles bound to the same field line relaxes to the Boltzmann distribution, referred to as a local thermal equilibrium.
The strong magnetic field suppresses transport of particles and energy across the magnetic field, so thermal equilibrium is established across the magnetic field on a longer time scale.
The time taken to come into thermal equilibrium is determined in part by the collision rate \cite{fitzpatrickNotes}
\begin{equation}\label{eq:collisionRate}
	\Gamma_\mathrm{col} \simeq \frac{n\; e^4\; \mathrm{ln}\, \Lambda}{6 \sqrt{2}\;\pi^{3/2}\;\epsilon_0^2\; m^{1/2} \;(k_\mathrm{B}T)^{3/2}},
\end{equation}
where $\mathrm{ln}\, \Lambda$ is a quantity known as the Coulomb logarithm.

Equation \ref{eq:collisionRate} is only strictly valid in the limiting case of no magnetic field.
The collision rate between particles in a strong magnetic field is given by \cite{MagCollisions}
\begin{equation}
	\Gamma_\mathrm{col} \simeq n \bar{v} \bar{b}^2 I\left( \bar{\kappa} \right),
\end{equation}
where $\bar{v}$ is the thermal velocity, $\bar{v} = \sqrt{2 k_\mathrm{B}T/m}$ and $\bar{b}$ is the classical distance of closest approach, $2 e^2/\left( 4 \pi \epsilon_0 k_\mathrm{B} T \right)$.
The 'magnetisation parameter', $\bar{\kappa} = {\bar{b}}/{r_c}$, where $r_c = \bar{v}/\omega_c$ is the cyclotron radius, determines if the effect of the magnetic field is important -- $\bar{\kappa} \ll 1$ implies that the collision rate is close to the no-field case, while $\bar{\kappa} \gg 1$ (a `strongly magnetised' plasma) strongly suppresses the collision rate.
$I (\bar{\kappa})$ is a function that falls rapidly for $\bar{\kappa} > 1$; it is evaluated in \cite{MagCollisions}.

\begin{figure}[bht]
	\centering
	\input{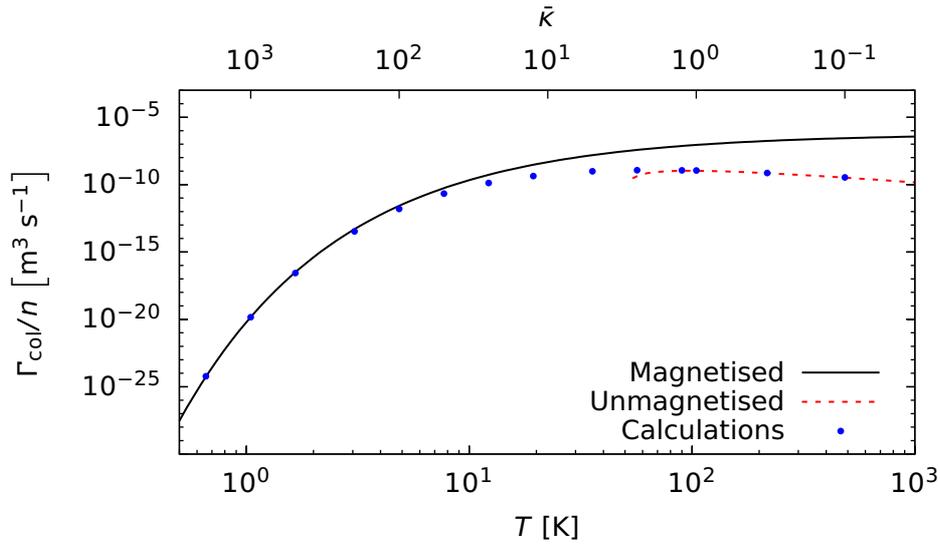}
	\caption[Collision rates as a function of temperature for magnetised and unmagnetised plasmas]{The collision rates as a function of temperature for a plasma using expressions valid in the strongly magnetised ($\bar{\kappa} \gg 1$, low $T$) and unmagnetised ($\bar{\kappa} \ll 1$, high $T$) regimes for an electron plasma in a 1~T magnetic field. The points are the results of calculations in the intermediate regime \cite{MagCollisions}.}
	\label{fig:magCollisions}
\end{figure}

The two asymptotic expressions of $\Gamma_\mathrm{col}/n$ for strongly magnetised and unmagnetised plasmas are plotted in figure \ref{fig:magCollisions}.
The curves deviate from each other by several orders of magnitude in the central, $\bar{\kappa} \sim O(1)$ region, and neither expression gives a good estimate of the collision rate.
The correct values of $I(\bar{\kappa})$ have been calculated numerically \cite{MagCollisions}, and are shown as the discrete points. 
These values agree with each of the asymptotic expressions in their regions of validity, and produce a smooth function over the entire range of $\bar{\kappa}$.

The equilibration along a field line is typically quite rapid, on the order of seconds to fractions of a second, while the radial equilibration time can be quite long, perhaps measured in hours.
There is some evidence that many of the plasmas prepared in ALPHA do not fully relax to global thermal equilibrium on the timescale of our experiments.
Whether or not the plasmas (in particular, their density distributions) are in global thermal equilibrium is not particularly relevant to the formation of antihydrogen, and will not be discussed further.

The global thermal equilibrium configuration for a non-neutral plasma made up of two species with the same sign of charge is different to the case of a one-component plasma.
This is relevant to ALPHA when antiprotons are mixed with an electron plasma for cooling.
In global thermal equilibrium, the heavier species (antiprotons) radially separate from the lower-mass species (electrons), residing in a thin ring just outside the electron plasma at low temperatures \cite{centrifugalsep}.
This effect has recently been observed in ALPHA \cite{Joel_Hollow} and can be of importance when electrons are used to cool antiprotons to very low temperatures (chapter \ref{chap:Cooling}).

\subsection{Calculating the thermal equilibrium distribution}
\label{sec:equilSol}

When conducting a plasma physics experiment, it is desirable, and often necessary, to have precise knowledge of the particle distribution and the electrostatic field in the Penning trap, including the contribution from the plasma itself.
With arbitrary voltages applied to the trap electrodes, and with arbitrary charge distributions, it is possible to obtain this information by solving equations \ref{eq:poisson} and \ref{eq:boltzmann} numerically.

This can be approached using an iterative algorithm with two steps.
Firstly, the electric potential is calculated using equation \ref{eq:poisson} for a guessed distribution of charge.
Using this solution, a new charge distribution is then calculated from equation \ref{eq:boltzmann} and used to find a more accurate description of the electric potential.
The algorithm repeats until the calculated parameters `converge', or remain stable from one iteration to the next.
These potential and density distributions, which simultaneously satisfy the Poisson and Maxwell-Boltzmann equations, are called the self-consistent solution.

A practical implementation of the algorithm is more involved.
For the interested reader, a technical discussion can be found in \cite{fitzpatrickNotes}, though some aspects will be discussed here.

In a Penning trap, the rotational symmetry allows the problem to be reduced to two dimensions, $z$ and $r$.
To calculate the electric potential from a given charge distribution, the problem is separated into sets of longitudinal and transverse equations along rows and columns of a two-dimensional grid.
Boundary conditions are defined along the edges of the grid -- at the upper limit in the transverse direction by the potentials applied to the electrodes, constant along the length of each electrode, and at the lower limit (along the cylindrical axis) by the fact that the potential is symmetric upon reflection at this axis, and therefore the first partial derivative vanishes.
The potential is assumed to tend to a constant value at the edges of the grid in the longitudinal direction, setting the boundary conditions in this direction.
This is equivalent to placing a conducting wall at the end of the electrode structure, which is not present in reality.
To minimise the perturbation to the electric fields near the region of interest, this boundary must be placed as far away as possible, but without causing the calculation volume to grow to an unmanageable size.

The potential at a fixed radius can then be expressed as a truncated Fourier series, and is solved using a numerically efficient matrix equation \cite{fitzpatrickNotes}.
The calculation benefits from the use of the Fourier series, since implementations of the Fast Fourier Transform (FFT) can yield significant reductions in the time needed to complete the operation.
The radial dependence is solved using a simple ordinary differential equation.

An efficient calculation also requires that the solution converges quickly with the number of iterations.
To achieve this, the calculation is first carried out at low temperatures, where the plasma density is more uniform and converges easily.
This solution is then used to seed calculations at successively higher temperatures, where the Debye tails of the plasma contain more of the charge and so are more important.
To ensure stability from one iteration to the next, the calculated density at each iteration is combined with the density from the previous iteration in a weighted average that heavily favours the old density.
Thus, the density is only allowed to evolve slowly, but the elimination of the wild oscillations that otherwise can occur results in an overall more efficient algorithm.

An implementation of this algorithm, based on a modification of a program by F. Robicheaux was developed as a library with a LabVIEW interface application for ALPHA.
By taking experimental inputs of the electrode excitations and measurements of the particle number, temperature and radial profile, the program produces calculations of the two-dimensional density, electric potentials and the electric fields.
These outputs can be further used in calculations of the stored electrostatic energy, the rotational energy, and many others.
This has proven to be an invaluable tool when designing or analysing experiments, and will be used extensively in the analyses later in this thesis.

\section{Simple plasma manipulations}

In practically every plasma Penning trap experiment, including ALPHA, it is essential to be able to manipulate the plasmas to change their position, geometry or energy distribution.

Almost all manipulations take place adiabatically - slowly compared with the time it takes the plasma to react to the changes.
If the plasma is initially in equilibrium, then at all points during an adiabatic process, the plasma remains close to equilibrium.
There are a number of physical quantities that are conserved during an adiabatic process, termed adiabatic invariants, which allow the final state of the plasma after the manipulation to be predicted.

The first invariant states that the magnetic moment of a gyrating particle,
\begin{equation}\label{eq:adiabatic1}
	\mu = \frac{\frac{1}{2} m v_\perp^2}{B} = \frac{E_\perp}{B},
\end{equation}
 is a constant of the motion \cite{fitzpatrickNotes}.

The second invariant, referred to as the longitudinal invariant, is given by
\begin{equation}\label{eq:adiabatic2}
	J = \int_a^b v_\parallel \;\mathrm{d}\mathrm l 
\end{equation}
where the integral is taken between the turning points of the axial motion \cite{fitzpatrickNotes}.

The third invariant states that the magnetic flux enclosed by the magnetron drift of a particle,
\begin{equation}\label{eq:adiabatic3}
	\Phi = \int_S \mathbf{B}\,.\,\mathrm{d}\mathbf{s} = \oint_C \grad \times \mathbf{B}\,.\,\mathrm{d}\mathbf{l},
\end{equation}
where $S$ is the surface enclosed by the closed orbit $C$, is conserved \cite{fitzpatrickNotes}.

\begin{figure}[hbt]
\centering
%
%
\begin{psfrags}%
\psfragscanon%
%
\psfrag{s12}[b][b]{\color[rgb]{0,0,0}\setlength{\tabcolsep}{0pt}\begin{tabular}{c}Potential\end{tabular}}%
\psfrag{s13}[l][l]{\color[rgb]{0,0,0}\setlength{\tabcolsep}{0pt}\begin{tabular}{l}(a)\end{tabular}}%
\psfrag{s14}[l][l]{\color[rgb]{0,0,0}\setlength{\tabcolsep}{0pt}\begin{tabular}{l}(b)\end{tabular}}%
\psfrag{s15}[l][l]{\color[rgb]{0,0,0}\setlength{\tabcolsep}{0pt}\begin{tabular}{l}(c)\end{tabular}}%
%
\psfrag{x01}[t][t]{0.05}%
\psfrag{x02}[t][t]{0.1}%
\psfrag{x03}[t][t]{0.15}%
\psfrag{x04}[t][t]{0.2}%
\psfrag{x05}[t][t]{0.05}%
\psfrag{x06}[t][t]{0.1}%
\psfrag{x07}[t][t]{0.15}%
\psfrag{x08}[t][t]{0.2}%
\psfrag{x09}[t][t]{0.05}%
\psfrag{x10}[t][t]{0.1}%
\psfrag{x11}[t][t]{0.15}%
\psfrag{x12}[t][t]{0.2}%
%
\psfrag{v01}[r][r]{-15}%
\psfrag{v02}[r][r]{-10}%
\psfrag{v03}[r][r]{-5}%
\psfrag{v04}[r][r]{0}%
%
\resizebox{12cm}{!}{\includegraphics{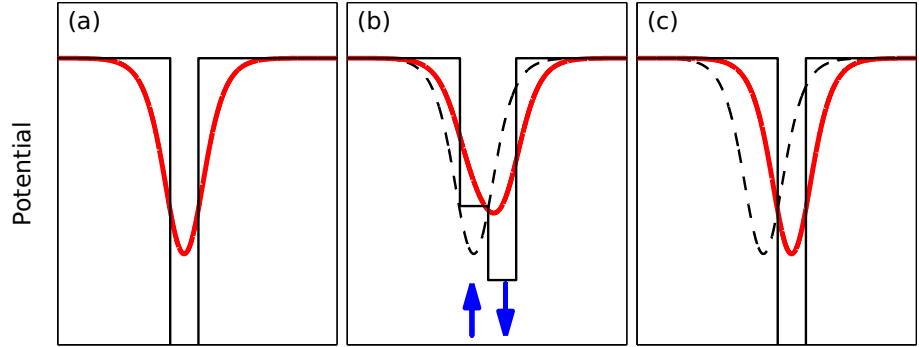}}%
\end{psfrags}%
%

\caption[A simple transfer of particles between electrodes]{A schematic of a simple transfer of particles between neighbouring electrodes.}
\label{fig:transfer}
\end{figure}

Movement of a plasma along the axis of the trap is achieved by altering the potentials of the electrodes.
A simplified scheme to transfer the plasma from one electrode to another is shown in figure \ref{fig:transfer}.
The speed of the transfer is chosen to ensure that the manipulation is firmly in the adiabatic regime.
This ensures that the adiabatic invariant $J$ (equation \ref{eq:adiabatic2}) is conserved, so that if the total length of the plasma remains constant, the average parallel velocity (and thus, the average energy) of the particles does not change.

In most manipulations, the radius of the plasma remains unchanged, since the particles are strongly bound to the magnetic field lines.
When the particles are moved from a region of high solenoidal magnetic field to lower magnetic field, the field lines diverge, and sizes of the particles' orbits increase, in accordance with the third invariant (equation \ref{eq:adiabatic3}).
The plasma radius follows the relationship
\begin{equation}
	\frac{r_p}{r_{p,0}} = \sqrt{\frac{B_{z,0}}{B_z}},
\end{equation}
where the zero subscripts indicate the initial conditions.

There is a also a theorem \cite{trappedPlasmasReview} based on the conservation of the total canonical angular momentum, which states that
\begin{equation}\label{eq:r2Conserve}
	\sum_i r_i^2 = \mathrm{constant},
\end{equation}
where the sum is carried out over the collection of particles.
This is relevant to situations when a plasma is prepared with an unstable transverse density distribution.
The rearrangement of charge must follow this relationship as the plasma evolves to a more stable configuration.
The total canonical angular momentum of the plasma can change through collisions with residual gas particles in the apparatus.
This leads to a slow expansion of the plasma, and in many devices this limits the lifetime of the stored plasmas.

\subsection{Characterising plasmas}
\label{sec:diagnostics}

\begin{figure}[hbt]
\centering
%
%
\begin{psfrags}%
\psfragscanon%
%
\psfrag{s08}[b][b]{\color[rgb]{0,0,0}\setlength{\tabcolsep}{0pt}\begin{tabular}{c}Potential\end{tabular}}%
\psfrag{s09}[l][l]{\color[rgb]{0,0,0}\setlength{\tabcolsep}{0pt}\begin{tabular}{l}(a)\end{tabular}}%
\psfrag{s10}[l][l]{\color[rgb]{0,0,0}\setlength{\tabcolsep}{0pt}\begin{tabular}{l}(b)\end{tabular}}%
%
\psfrag{x01}[t][t]{0.05}%
\psfrag{x02}[t][t]{0.1}%
\psfrag{x03}[t][t]{0.15}%
\psfrag{x04}[t][t]{0.2}%
\psfrag{x05}[t][t]{0.05}%
\psfrag{x06}[t][t]{0.1}%
\psfrag{x07}[t][t]{0.15}%
\psfrag{x08}[t][t]{0.2}%
%
\psfrag{v01}[r][r]{0}%
\psfrag{v02}[r][r]{10}%
\psfrag{v03}[r][r]{20}%
\psfrag{v04}[r][r]{30}%
\psfrag{v05}[r][r]{40}%
\psfrag{v06}[r][r]{50}%
\psfrag{v07}[r][r]{60}%
%
\resizebox{12cm}{!}{\includegraphics{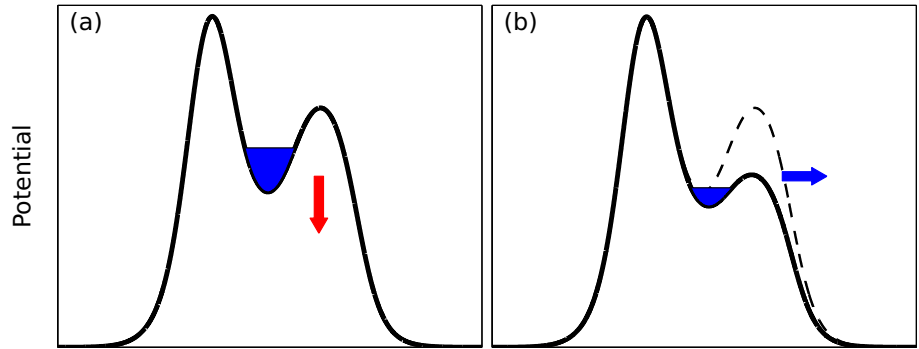}}%
\end{psfrags}%
%

\caption[Extraction of particles from the trap]{Particles are extracted from the trap by lowering the depth of the confining well (a). After they are released (b), the particles are guided by the magnetic field as they escape and strike a detector placed at a potential lower than the initial potential.}
\label{fig:dumpFigure}
\end{figure}

To measure the characteristics of a plasma, it is often necessary to extract, or `dump', the plasma from the trap onto some measurement device.
When a dump is performed, the voltages applied to the electrodes are manipulated to reduce the depth of the well.
Particles escape once the depth of the well falls below their longitudinal energy -- this is shown graphically in figure \ref{fig:dumpFigure}.
Escaping particles strike a detector, and the number of particles escaping at each moment in time is recorded.
For antiprotons, the `detector' is often just a matter object on which they annihilate, and the pions produced are detected using the scintillator/PMTs (section \ref{sec:Scintillators}) or the silicon detector (section \ref{sec:siliconDetector}).
For positrons and electrons, there is often sufficient charge that these particles can be measured using a Faraday cup (section \ref{sec:FaradayCup}).
All of the particles can be detected using the MCP (section \ref{sec:MCP}) as a charge amplifier.

The speed at which a plasma is dumped depends on the aim of the measurement and the capabilities of the instrument used.
For example, when measuring the radial density profile of a plasma using the MCP/Phosphor (section \ref{sec:MCP}), it is necessary to extract the plasma swiftly to prevent redistribution of the particles over the course of the dump.
In this case, the potential is changed as quickly as the hardware (primarily the amplifiers and filters) allows - typically over microseconds.
When determining the number of particles present by counting the number of annihilations, the maximum counting rate of the hardware must not be exceeded, so the dumps are performed more slowly.
For this case, the potentials are changed over tens to hundreds of milliseconds.

\subsection{Particle energy} \label{sec:EnergyDumps}

The energy of the particles is an important parameter in almost every experiment performed.
In the worst case, high-energy particles escape from the trap and are lost, while in other cases, design and interpretation of the experiment hinges on having well-known particle energy distributions.
Production of low-energy antihydrogen necessarily requires use of low-energy particles (particularly antiprotons), so measurements of the particles' energies are vital to the success of ALPHA experiments.

The voltages applied to the electrodes over the course of the dump are changed deterministically by the sequencer (section \ref{sec:Sequencer}), so the well depth is known at all times during the dump.
This allows the number of particles detected over time to be converted to energy distributions.

There are, however, several complicating factors that must be taken into account when determining the well depth for the escaping particles.
First of all is the fact that the well depth is not constant at all radii, but is a minimum on the trap axis.
The typical plasma radii seen in ALPHA are quite small compared to the radius of the trap electrodes (less than 2~mm compared to electrode radii of 16~mm or 22~mm), so this is usually a minor correction, and is ignored in most cases.

\begin{figure}[hbt]
	\subfloat[]{\input{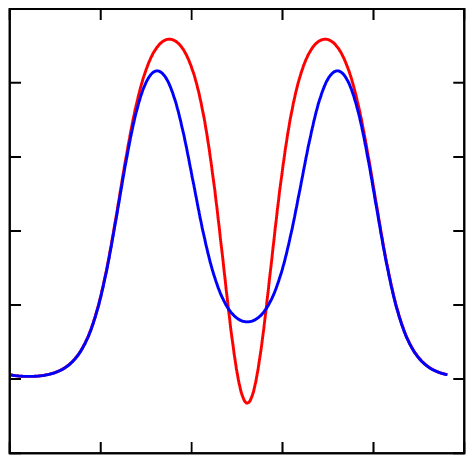}}
	\subfloat[]{\input{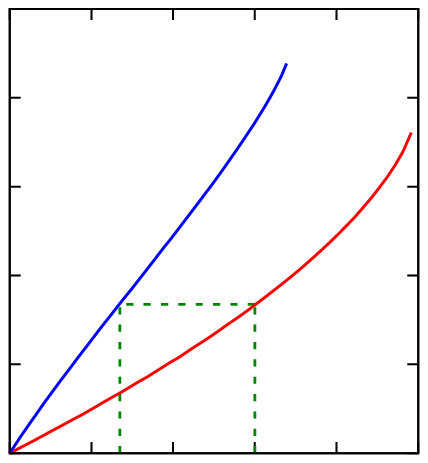}}
	\\
	\subfloat[]{\input{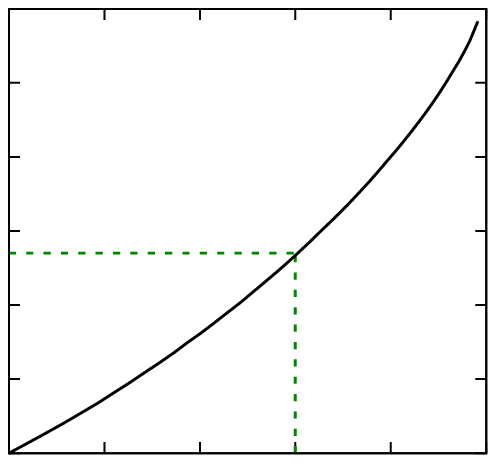}}
\caption[Conservation of the second adiabatic invariant during a change in the shape of the potential.]{An adiabatic change in potential, such as from the red potential in (a) to the blue potential changes the energy of a trapped particle. The second invariant, graphed in (b), is conserved, allowing the relationship between the initial and final energies to be calculated (c).}
\label{fig:adiabatic}
\end{figure}

A second effect comes about due to the conservation of second adiabatic invariant, $J$ (equation \ref{eq:adiabatic2}).
The energy of a particle confined in the initial (red) potential in figure \ref{fig:adiabatic}(a) will change as the potential is changed to the blue potential as a result of work done on it by the electric field.
If the change in the well shape is performed adiabatically, $J$ is conserved.
Calculating $J$ as a function of particle energy $E$ (figure \ref{fig:adiabatic}(b)) allows one to link the energies in the initial and final wells.
An example of this procedure can be seen by following the green lines in figure \ref{fig:adiabatic}(b).
A particle with 6~eV of energy in the initial well has the same value of $J$ as a 2.7~eV particle in the final potential.
Thus, the energy of a 6~eV particle will be reduced to 2.7~eV in this adiabatic manipulation (figure \ref{fig:adiabatic}(c)).

When calculating the correction for a dump, the value of $J$ for particles released from the well (i.e. particles with total energy equal to the instantaneous well depth) as a function of time is determined.
This is compared to $J(E)$ in the initial well to obtain $E(t)$, the initial energy of a particle released at time $t$.
Since the number of particles released at each time is measured, this allows the energy distribution, $N(E)$, to be calculated.

\subsection{Plasma temperature} \label{sec:plasmaTemperature}

A plasma stored in a Penning trap is almost surrounded by the trap electrodes and is exposed to the blackbody thermal radiation emitted from the electrode surfaces.
In the absence of other sources of heat, a plasma will come into thermal equilibrium at the temperature of the electrodes.
In ALPHA, we measure that the temperature of the electrodes reaches a minimum of approximately 8 K, so it might be expected that this is the minimum achievable temperature of the stored particles.
However, there are a number of complicating factors, discussed later.

\begin{figure}[hbt]
\centering
\input{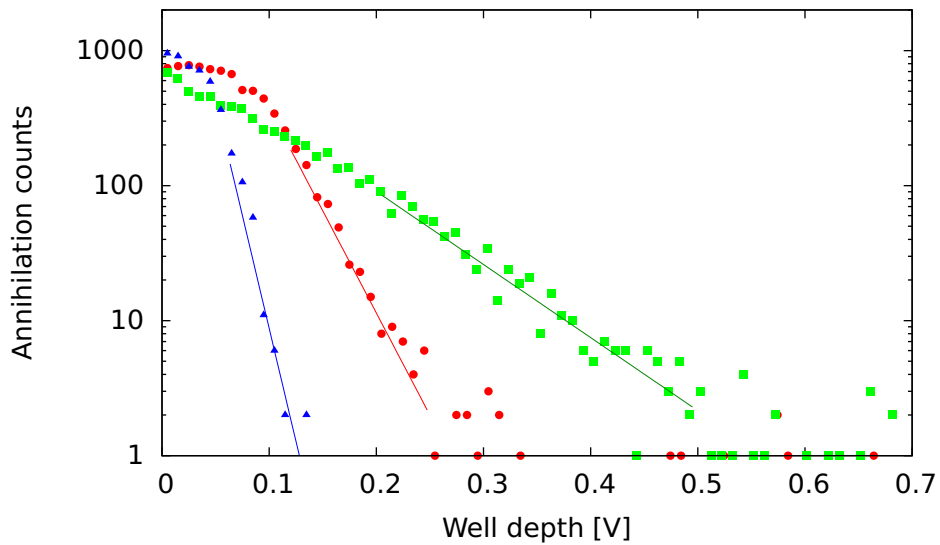}
\caption[Examples of temperature measurements of antiproton distributions]{Examples of temperature measurements of antiproton distributions. The straight lines are fits of the form $A\,\mathrm{exp}\left(-E/k_BT\right)$. The temperatures measured are $(931\pm42)$~K ($\blacksquare$), $(333\pm15)$~K ($\bullet$) and $(151\pm9)$~K ($\blacktriangle$), where the uncertainties are the statistical uncertainties of the exponential fits.}
\label{fig:exampleTemperature}
\end{figure}

The temperature of a plasma can be determined by sampling the Boltzmann distribution of particle energies \cite{TemperatureDiagnostic}.
In a one-dimensional well, the energy distribution will be
\begin{equation}\label{eq:1DBoltzmann}
	N(E) \propto e^{\left(-\frac{E}{k_B T}\right)}.
\end{equation}
To measure the temperature, the particles are slowly  (compared to the axial oscillation frequency) dumped from their confining well; the first (highest-energy) particles to be released will be drawn from the tail of the Boltzmann distribution.
Removing a portion of the plasma disturbs the equilibrium, and as the dump progresses, the loss of particles causes redistribution of energy.
At later times, the measured distribution deviates from the expected Boltzmann form.
We see that the model breaks down after removal of as little as a few thousand particles.
It is important, therefore, that the instrument used to detect the escape of the particles is sensitive to a small number of particles.
Examples of temperature measurements are shown in figure \ref{fig:exampleTemperature}, showing fits to the exponential regions.

For electrons and positrons, the MCP is used to amplify the small amount of charge in the exponential region so that it can be detected electronically, while the sensitivity for antiproton annihilations is high enough that it is easier to count even single particles.

The corrections to the energy diagnostic dumps discussed in section \ref{sec:EnergyDumps} apply equally well to dumps to measure the temperature.
The plasma lengthens as the potentials are manipulated, and conservation of the adiabatic invariant $J$ results in a reduction in the longitudinal energy, effectively cooling the plasma.
The exact consequences of this depend on the speed of the manipulation relative to the collision and thermalisation rates.
In the limit that the dump happens so quickly that no thermalisation occurs, only the temperature of the plasma parallel to the magnetic field will change, and the measured temperature will fall as $\left(L/L_0\right)^2$, where $L$ and $L_0$ are the final and initial plasma lengths.
On the other hand, if complete thermalisation between the parallel and perpendicular degrees of freedom occurs, the temperature will fall as $\left(L/L_0\right)^{2/3}$.
(These scaling laws can be derived starting from the rate equations found in reference \cite{NoiseMeasurements}.)
The real situation lies somewhere between these extremes.

In addition to this correction based on the adiabatic invariant, the changing length of the plasma over the course of the dump also causes the space charge of the plasma to change, which in turn changes the height of the confining potential.
To a rough approximation, the space change $V_s$ is inversely proportional to the plasma length $L_p$ \cite{StevePresentation}, and the measured temperature will scale as approximately 
\begin{equation}
1+\frac{V_s}{L_p}\frac{\mathrm{d}L_p}{\mathrm{d}V},
\end{equation}
where $V$ is the height of the confining potential.

Note that the correction factor due to this effect is greater than one, while the adiabatic correction is less than one, so these effects cancel each other out to some degree.
The value of the correction factor to apply can be calculated for a set of dump and plasma parameters each using a particle-in-cell (PIC) simulation.
The procedure typically finds that the measurement process will yield a temperature higher than the plasma temperature before the dump, by a factor in the range 1.1 - 2, depending on the plasma parameters and dump manipulations.
Denser plasmas typically require larger correction factors.

The temperatures measured in ALPHA using this method are significantly higher than would be expected from assuming that particles come into thermal equilibrium at the temperature of the cryogenic surroundings.
The source of this is not yet identified, but there are some interesting observations which might point to the sources.

\begin{figure}[hbt]
	\centering
	\input{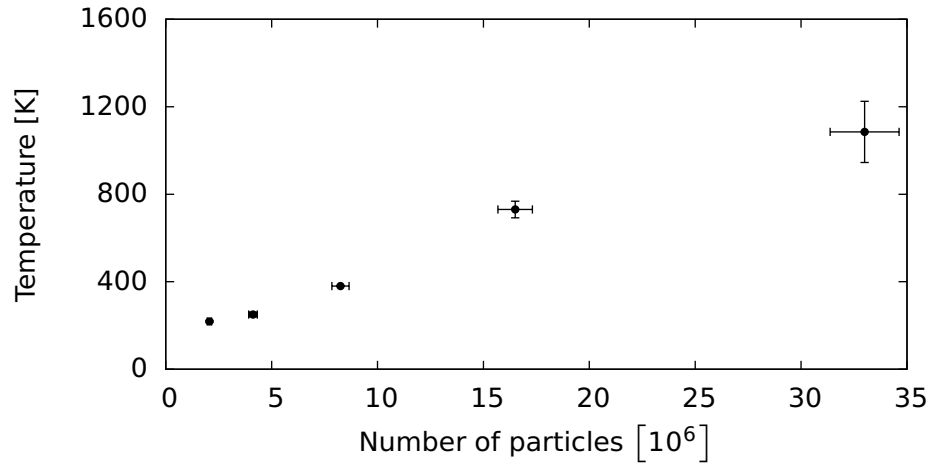}
	\caption[The dependence of the equilibrium temperature on the number of particles]{The dependence of the equilibrium temperature on the number of particles. This example uses a 2.5~mm radius positron plasma.}
	\label{fig:tempNumber}
\end{figure}

The temperature of a plasma is seen to scale with the number of particles it contains, as shown in figure \ref{fig:tempNumber}.
Because of the particle-particle repulsive interactions, the electrostatic potential energy of a particle in a plasma is typically far larger than its (thermal) kinetic energy.
Release of even a small fraction of the potential energy can therefore have a large impact on the plasma's temperature.
Starting from the equation for the energy stored in an electrostatic field, $U = \int \frac{\epsilon_0}{2}\left|\mathbf{F}\right|^2\;\mathrm{d}V$, the average potential energy per particle in an infinite, constant density plasma column with radius $r$ and $N$ particles per unit length, surrounded by a grounded wall at radius $r_w$, can be derived as
\begin{equation}\label{eq:potentialEnergy}
	U = \frac{N e^2}{4 \pi \epsilon_0} \left[  \mathrm{ln}\left( \frac{r_w}{r} \right) + \frac{1}{4}\right].
\end{equation}
Expansion of the plasma due to factors that cannot be completely prevented, such as collisions on residual gas atoms in the trap, will reduce the potential energy, converting it to kinetic energy, a process sometimes referred to as Joule heating.
This effect scales with the number of particles, present, so the observations may imply that this, or a related effect is the source.
However, complementary measurements of the expansion and temperature do not agree well with the predictions of this model \cite{DanielExpansion}.

Another possible reason is the fact that the electrode system cannot be completely enclosed, but has openings to allow particles to be loaded or extracted for diagnostics.
At one end of the ALPHA system, the electrode stack is closed by the antiproton degrader, and at the other by a closable aperture.
These block thermal radiation from the warm parts of the apparatus and are cooled to reduce the amount of radiation they emit.
However, because of the thermal power they are exposed to, they are cooled to only approximately 100~K, which is significantly above the temperature of the electrodes.
The equilibrium temperature will be set by a power-balance relationship, where the rate of radiative energy loss is equal to the power coupled from the environment.
Further cooling may be introduced in a future iteration of the ALPHA device.

Electronic noise, as mentioned in section \ref{sec:ALPHAPenningTrap}, can increase the temperature of a stored plasma \cite{NoiseMeasurements}.
The effect is similar to the oscillating drives that are used to deliberately increase the temperature of a plasma for an experiment(see section \ref{sec:HbarTemperature}).
The amount by which a plasma is heated by background noise will be determined by the amplitude and frequency components of the noise present.
Characterisation and reduction of the effects of electronic noise is an on-going study at ALPHA.

Understanding of the equilibrium temperature of plasmas in a Penning trap like that in ALPHA, and its dependence on the plasma parameters, is not yet complete and is still an active area of research.

\section{Rotating wall}	\label{sec:rotatingWall}

\begin{figure}[hbt]
	\centering

\psset{unit=1in,cornersize=absolute,dimen=middle}%
\begin{pspicture}(-0.796956,-0.8)(1.69282,0.8)%
\psset{linewidth=0.8pt}%
\psset{linewidth=0.8pt}%
\ifx\MPSTPatchA\undefined{\makeatletter\def\psbezier@ii{\addto@pscode{%
\ifshowpoints true \else false \fi\tx@OpenBezier%
\ifshowpoints\tx@BezierShowPoints\fi}\end@OpenObj}\makeatother%
\global\def\MPSTPatchA{}}\fi%
\psset{arrowsize=1.1pt 4,arrowlength=1.64,arrowinset=0}%
\psset{linewidth=2pt}%
\psarc(0,0){0.8}{5}{55}
\psarc(0,0){0.8}{65}{115}
\psarc(0,0){0.8}{125}{175}
\psarc(0,0){0.8}{-175}{-125}
\psarc(0,0){0.8}{-115}{-65}
\psarc(0,0){0.8}{-55}{-5}
\psset{linewidth=1pt}%
\psline(0.69282,-0.4)(1.19282,-0.4)
\psline(1.19282,-0.4)(1.31782,-0.4)
\pscircle(1.44282,-0.4){0.125}
\psarcn(1.401154,-0.4){0.041667}{180}{0}
\psarc(1.484487,-0.4){0.041667}{180}{360}
\psline(1.56782,-0.4)(1.69282,-0.4)
\uput{0.501875ex}[d](1.44282,-0.525){\hbox{$\:$}$ \theta = 0 $\hbox{$\:$}}
\psline(0.69282,0.4)(1.19282,0.4)
\psline(1.19282,0.4)(1.31782,0.4)
\pscircle(1.44282,0.4){0.125}
\psarcn(1.401154,0.4){0.041667}{180}{0}
\psarc(1.484487,0.4){0.041667}{180}{360}
\psline(1.56782,0.4)(1.69282,0.4)
\uput{0.501875ex}[d](1.44282,0.275){\hbox{$\:$}$ \theta = 60^{\circ} $\hbox{$\:$}}
\psline[linestyle=dashed](0,0.8)(0.5,0.8)
\psline[linestyle=dashed](-0.69282,0.4)(-0.39282,0.4)
\psline[linestyle=dashed](-0.69282,-0.4)(-0.39282,-0.4)
\psline[linestyle=dashed](0,-0.8)(0.5,-0.8)
\newgray{fillval}{0.3}
\pscircle[fillstyle=solid,fillcolor=fillval](0,0){0.25}
\psarc[arrowsize=0.0625in 0,arrowlength=2,arrowinset=0]{->}(0,0){0.4}{-25}{25}
\end{pspicture}%
	\caption[A schematic diagram of the apparatus used to apply a rotating-wall field to a plasma]{A schematic diagram of the apparatus used to apply a rotating-wall field to a plasma. The electrode is azimuthally segmented into (in this case, six) sections, and phase-shifted signals are applied to each sector.}
	\label{fig:RWfig}
\end{figure}
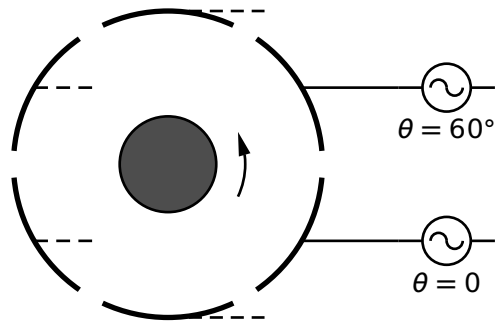

An azimuthal `rotating-wall' electric field is used to manipulate the density of non-neutral plasmas and to increase their confinement time \cite{RotWall}.
The rotating wall field is generated by azimuthally segmenting an electrode in a number of sectors (usually at least four), and applying a sinusoidally varying voltage to each, as shown in figure \ref{fig:RWfig}.
The signal to each sector is phase shifted from its neighbour by an amount $\theta = \,^{2 n \pi} / _N$, where $N$ is the number of sectors, and $n$ is a non-zero integer.
The name of the technique derives from the similarity of this field to that generated by a single, asymmetric electrode rotating about the trap axis.

The rotating wall acts by transferring torque to the plasma, and changing its rotation rate.
The density and rotation rates are linked through equation \ref{eq:plasmaExB}, so increasing the rotation rate also increases the plasma density and shrinks the plasma radius.

According to measurements and analysis of the system \cite{RotWall}, the applied torque is proportional to the difference between the drive frequency and the rotation frequency, $\omega_r$ (equation \ref{eq:plasmaExB}).
The plasma rotation rate increases until the applied torque balances the external torque and the density settles to a constant value.
The first studies of the rotating wall technique only achieved significant compression if the rotating wall frequency stayed tuned to the frequency of a plasma mode - most often the rotation rate itself.
However, later studies identified another regime, the so-called `strong-drive' regime \cite{StrongDriveRW}, where the plasma tended to a state with a rotation rate close to the drive frequency, without the need for tuning to plasma modes.
The strong drive regime manifests itself at large drive amplitudes and has been seen to produce much higher densities than the weak-drive scheme.

Denser, smaller plasmas are much more desirable for antihydrogen experiments.
The antihydrogen recombination rates (section \ref{sec:HbarFormationTheory}) increase with the density of particles, while the perturbing influence of the octupole falls quickly near the axis (section \ref{sec:magneticminimum}).
Antihydrogen atoms that are formed near the axis also experience the deepest magnetic minimum, increasing their chance of being confined.
The rotating wall technique presents a simple, efficient scheme to compress the plasmas, and has been implemented in ALPHA.

Compression of all three particle species has been achieved in ALPHA \cite{ALPHA_RW} making use of the imaging capabilities of the MCP/Phosphor equipment to measure the effect of the rotating wall drive on the radial distributions (section \ref{sec:MCP}).
Compression of electron and positron plasmas is achieved using a constant frequency drive in the MHz range, indicating that we operate in the strong-drive regime.
Densities as high as $\sim~4~\times~10^9~\mathrm{cm^{-3}}$ in a 3~T magnetic field have been achieved.

\begin{figure}[hbt]
	\centering
	\input{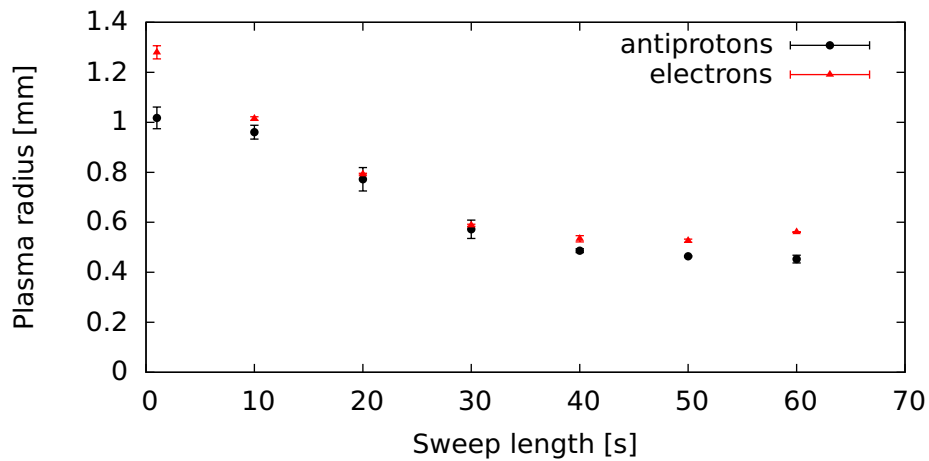}
	\caption{The radius of the electron and antiproton plasmas as they are simultaneously compressed. No compression is seen if antiprotons are compressed alone, implying that the particles are sympathetically compressed.}
	\label{fig:RWpbars}
\end{figure}

Antiprotons by themselves are not seen to respond to a rotating wall field \cite{ALPHA_RW}.
Instead, they are compressed simultaneously with an electron plasma.
Best compression of the antiprotons is achieved if the electron plasma is compressed slowly enough that the sizes of the antiproton and electron clouds remain roughly matched.
Data showing simultaneous compression of a superimposed antiproton and electron plasma is shown in figure \ref{fig:RWpbars}.
Why antiproton compression requires an electron cloud to be present is not completely understood, but one possible explanation is that the rotating wall drive cannot address the antiprotons directly, but does act on the electrons, which collisionally transfer torque to the antiprotons.

Application of a rotating-wall field to a plasma produces a dramatic rise in temperature \cite{StrongDriveRW}, so some form of cooling is required to maintain a steady state, and to recool the particles after the rotating wall procedure.
As has already been mentioned, antiprotons do not efficiently self-cool, but instead are cooled through collisions with cold electrons, so it is advantageous that both species are simultaneously compressed.

\section{Octupole} \label{sec:OctupolePhysics}
In a typical Penning trap, where the magnetic field is solenoidal, charged particles move along straight magnetic field lines, parallel to the cylindrical axis.
With the addition of a transverse multipole field, the field lines become distorted.
The combined magnetic field of a coaxial solenoid and octupole can be written as
	\begin{equation}\label{eq:octupoleField}
		\mathbf{B} = B_z \hat{z} + B_w \left( \frac{r}{r_w} \right)^3 \left[ \hat{r} \mathrm{cos}(4\theta) - \hat{\theta} \mathrm{sin}(4\theta) \right],
	\end{equation}
where $B_w$ is the magnitude of the transverse magnetic field at the radius of the trap wall, $r_w$ \cite{BallisticLossTheory}.

\begin{figure}[hbt]
	\centering
	\includegraphics[width=0.8\textwidth]{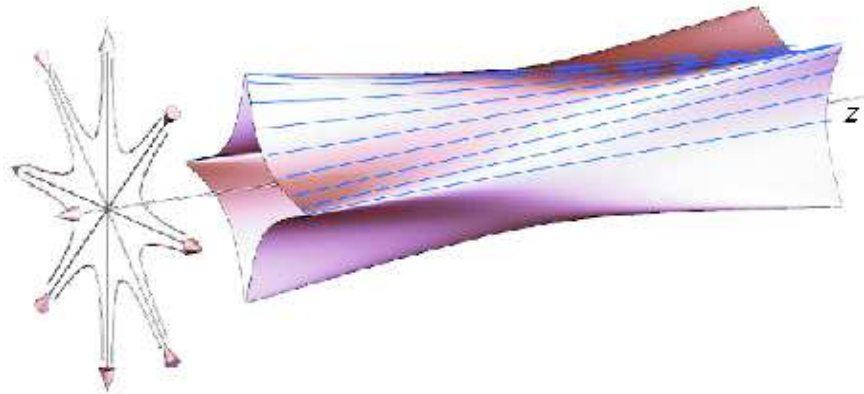}
	\caption[The surface traced out by particles moving in a combined octupole and solenoid magnetic field]{The surface traced out by particles moving in a combined octupole and solenoid magnetic field (from \cite{ALPHA_OctupoleDiagnostic}).}
	\label{fig:octupoleFlutes}
	\end{figure}

As the particles rotate about the trap centre at the magnetron frequency, they sample all of the possible field lines, so it is reasonable to consider the case that produces the maximum effect.
The field lines that propagate outward the fastest follow the equation \cite{BallisticLossTheory}
	\begin{equation}\label{eq:octupoleFieldLines}
	r(z) = \frac{r(0)}{\sqrt{1 - 2 \frac{B_w}{B_z} \frac{r(0)^2}{r_w^2} \frac{z}{r_w} }},
	\end{equation}
where $r(0)$ is the radial position of the particle at $z=0$.
These lines can be visualised by plotting the surface traced out by particles starting from a circle at $z$=0, giving a `fluted' shape with four-fold symmetry, as in figure \ref{fig:octupoleFlutes}.
The structures at either end are rotated by half a flute period (in this case 45$^\circ$) relative to each other.

\subsection{Ballistic loss and the critical radius} \label{sec:ballisticLoss}

Because the multipolar field lines bend outwards in radius, particles following the field lines can be guided towards the surface of the electrodes.
If a particle reaches the electrode wall at $r = r_w$, it will be lost from the trap.
We term this kind of loss `ballistic loss' \cite{BallisticLossTheory}.

Particles confined in a Penning trap have a given axial oscillation length, and for a given length and multipole strength, we can find the minimum initial radius the particle must have at $z=0$ to reach $r=r_w$.
We call this quantity the `critical radius', $r_c$.
Rewriting equation \ref{eq:octupoleFieldLines},
	\begin{equation}
		r_w = \frac{r_c} {\sqrt{1 - 2 \frac{B_w}{B_z} \frac{r_c^2}{r_w^2} \frac{L}{r_w} }},
	\end{equation}
where $L$ is the full length of the particle's axial trajectory.
This can be rearranged to give
	\begin{equation}
		r_c = \frac{r_w}{ \sqrt{1+ \frac{B_w}{B_z}  \frac{L}{r_w} } }.
	\end{equation}
This places an effective maximum radius on the particles that can be confined without loss.

The ballistic loss mechanism can be exploited to produce a destructive measurement of the radial density profile of a stored plasma \cite{ALPHA_OctupoleDiagnostic}.
By holding a plasma in the trap, and slowly increasing the multipole field while monitoring the loss of particles as a function of time,  one obtains the number of particles which exceed the critical radius as a function of time, $N(t)$.
Since $B_w(t)$ is known, the number of particles at a given radius $N(r)$ can be derived if $L$ is known.

\subsection{Diffusion and heating} \label{sec:OctupoleHeating}
Ballistic loss, described in the preceding section, is an example of one of the simplest, and perhaps the most directly destructive, consequences of adding a transverse multipole to a Penning trap.
As well as this phenomenon, a host of more subtle effects exist, notably resonant transport effects.
Studies of these effects using a quadrupole \cite{GilsonFajans_Quadrupole} have shown enhanced radial diffusion of the plasma, and it was predicted that antihydrogen experiments using multipoles, especially those of lower-order, would be hindered by diffusion and the associated plasma heating \cite{Joel_Multipole}.
The fact that the harmful effects are expected to be more pronounced for lower-order multipoles prompted ALPHA to opt for an octupole-based atomic trap when designing the experiment.

\begin{figure}[hbt]
	\centering
	\subfloat[]{\input{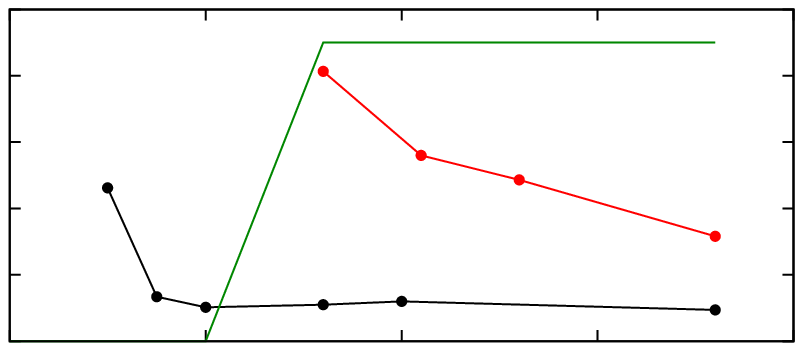}}
	\\
	\subfloat[]{\input{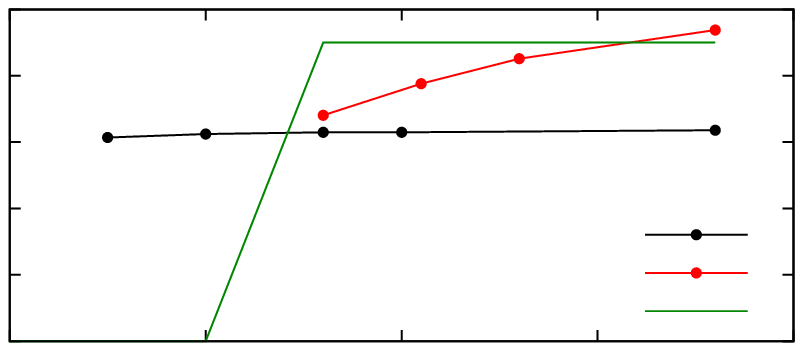}}
\caption[The temperature and radius of an electron plasma, as affected by switching on an octupole field.]{The temperature (a) and radius (b) of an electron plasma, as affected by switching on an octupole field. The octupole is energised over 25s starting at t=60~s. The plasma was made up of approximately $1.6\times 10^7$ electrons with a radius of 1.2~mm and density $\sim 4 \times 10^8 ~\mathrm{cm^{-3}}$.}
\label{fig:octupoleEffect}
\end{figure}

The methods developed to measure the temperature and radial distribution of the particles in the experiment allow for characterisation of the effect of the multipole on the stored plasmas.
The effect of the octupole is measured on the temperature and radius of a stored electron plasma is shown in figure \ref{fig:octupoleEffect}, 
It can be seen that energising the octupole coincides with a large temperature increase, and the plasma begins rapidly expanding.

The degree to which the multipole affects a stored plasma depends on the plasma parameters.
Many of the identified loss and radial transport phenomena are resonant effects, and depend on close matching between frequencies of two of the particles' motions.
For example, if the angle through which a particle rotates in one axial oscillation is close to one period of the multipole field, the particle can experience rapid radial transport.
In this case, resonance occurs between harmonics of the magnetron frequency and the axial oscillation frequency.
One example of such an effect was identified in ALPHA's nested wells (see section \ref{sec:incmix} and \cite{ZeroFreqLoss}).
Empirically, the size of the effect depends on the geometry and density of the plasma, and using the rotating wall technique to prepare plasmas with different parameters, the dependence can be explored.

\begin{figure}[hbt]
	\centering
	\input{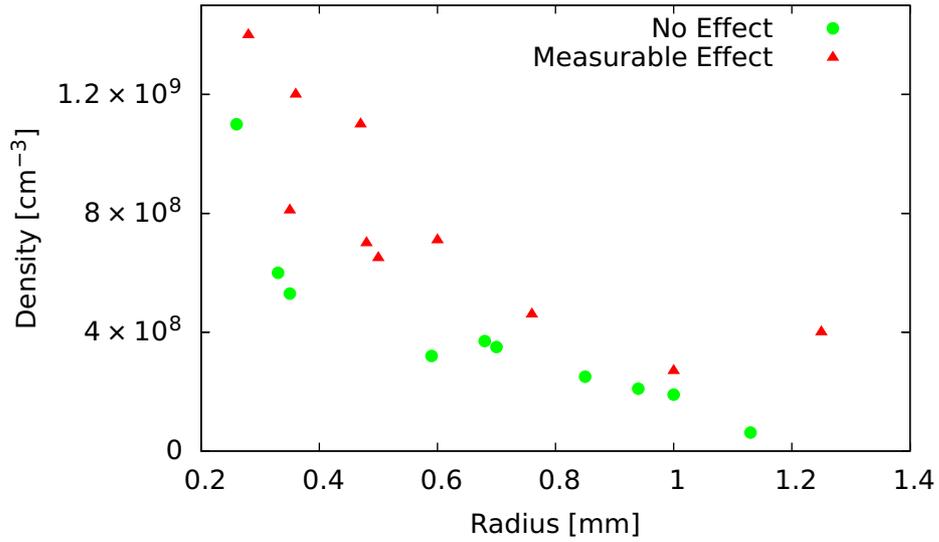}
	\caption[A scatter plot of the effect of the octupole field on a trapped plasma as a function of the density and radius.]{A scatter plot of the effect of the octupole field on a trapped plasma as a function of the density and radius. A rough boundary between regions of `effect' and `no effect' can be identified.}
	\label{fig:octupoleThreshold}
\end{figure}

This can be demonstrated by measuring the effect of the octupole (the most sensitive measurement is on the temperature) on stored plasmas with differing parameters.
By categorising the data points and plotting them on a scatter plot, such as in figure \ref{fig:octupoleThreshold}, the ranges of parameters that remain unperturbed by the octupole can be identified.
However, this plot is likely to be different for different confining potentials.
The threshold effect and scaling is in qualitative agreement with a model of multipole-induced heating and diffusion \cite{Steve_FutureThesis}, but further work is needed to fully understand the behaviour.

\section{Plasma modes} \label{sec:Modes}

\subsection{Theory}

Collective motion, in which the particles move together under the influence of their mutual interactions, is characteristic of plasmas.
In confined non-neutral plasmas, many different classes of collective motion can occur.
In this section, we are interested in the normal, axisymmetric modes, which have been used as diagnostics of the properties of non-neutral plasmas in Penning traps \cite{TinkleGreaves}.

\begin{figure}[hbt]
\centering
\subfloat[]{\includegraphics[width=0.4\textwidth ]{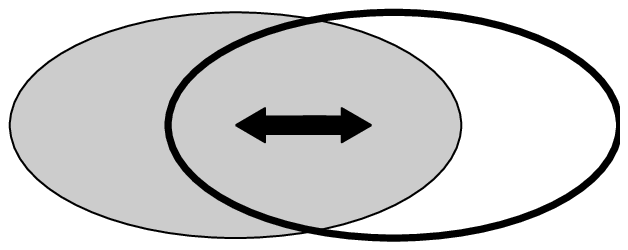}}
\hspace{0.5cm}
\subfloat[]{\includegraphics[width=0.4\textwidth ]{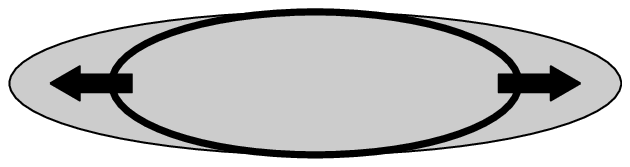}}
\vspace{-0.3cm}
\caption[Illustrations of the oscillatory motions associated with the lowest-order plasma modes.]{Illustrations of the oscillatory motions associated with (a) the first, `sloshing' mode, and (b) the second, `breathing' mode.}
\label{fig:modemodel}
\end{figure}

The lowest-order mode is similar to the axial oscillation of a single particle in the electric potential, described in section \ref{sec:penningTrap}.
The plasma as a whole oscillates in the external potential (figure \ref{fig:modemodel}(a)), giving rise to a centre-of-mass motion known as the `sloshing mode'.
The second-lowest-order mode is characterised by a elongation and compression of the plasma shape parallel to the magnetic field, known as the `breathing' mode (figure \ref{fig:modemodel}(b)).
There are other, higher-order and non-axisymmetric modes that we do not discuss here \cite{TinkleGreaves}.

A model of the behaviour of these modes show a dependence on the plasma parameters, notably the temperature and density \cite{ModesTheory}.
The frequency of first mode is the same as the single-particle oscillation frequency:
\begin{equation}
	\omega_1 = \omega_z,
\end{equation}
and the frequency of the second mode, assuming that the plasma is cold, is given by \cite{TinkleGreaves}
\begin{equation}
	\omega_z = 1 - \frac{\omega_p^2}{\omega_2^2} = \frac{k_2}{k_1} \frac{P_2(k_1) {Q_2^0}'(k_2)}{P_2'(k_1) Q_2^0(k_2)},
	\label{eq:mode2cold}
\end{equation}
where $\omega_p$ is the plasma frequency, $P_2$ and $Q_2$ are the second-order Legendre functions of the first and second kind respectively, $k_1 = \alpha / \sqrt{\alpha^2 -1 +\omega_p^2 / \omega_2^2}$, $k_2 = \alpha/\sqrt{\alpha^2 -1}$, and $\alpha$ is the `aspect ratio', the ratio of the plasma's half-length, $L$ to its radius.

For a finite temperature, the frequency of the second mode is shifted upwards from the cold value \cite{TinkleGreaves}, according to
\begin{equation}
	\omega_{2,h}^2 = \omega_2^2 + 5 \left( 3- g(\alpha) \right) \frac{k_\mathrm{B}T}{m L^2}, 
	\label{eq:modeTemperature}
\end{equation}
where $g(\alpha)$ is a function that describes how the shape of the plasma changes as a function of temperature.

\subsection{Measurements}

This theory led to the development of a technique where frequencies of the plasma modes are measured and used to infer the plasma parameters.
This technique is attractive because it is \textit{non-destructive}, unlike the dumps described in section \ref{sec:diagnostics}, and its application has met with some success in previous experiments: e.g. \cite{TinkleGreaves} and \cite{ATHENA_modes}.

The presence of the plasma  alters the electric field in the trap and induces image charges on the electrodes.
As the plasma oscillates, the electric fields change, causing charge to flow onto and off the electrodes.
This current can be detected as a voltage drop across a resistor and, with suitable amplification, can be measured.
The frequency of the motion can be extracted from the voltage by recording the voltage waveform and applying a Fourier transformation.

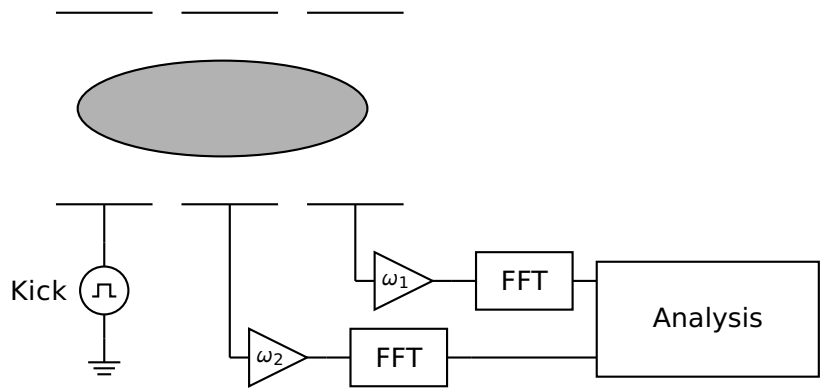
\begin{figure}[hbt!]
\centering

\psset{unit=1in,cornersize=absolute,dimen=middle}%
\begin{pspicture}(-0.1125,-1.45)(3.8375,0.5)%
\psset{linewidth=0.8pt}%
\psset{linewidth=0.8pt}%
\ifx\MPSTPatchA\undefined{\makeatletter\def\psbezier@ii{\addto@pscode{%
\ifshowpoints true \else false \fi\tx@OpenBezier%
\ifshowpoints\tx@BezierShowPoints\fi}\end@OpenObj}\makeatother%
\global\def\MPSTPatchA{}}\fi%
\psset{arrowsize=1.1pt 4,arrowlength=1.64,arrowinset=0}%
\newgray{fillval}{0.7}
\psellipse[fillstyle=solid,fillcolor=fillval](0.75,0)(0.75,0.25)
\psline(1.6875,0.5)(1.1875,0.5)
\psline(1.0375,0.5)(0.5375,0.5)
\psline(0.3875,0.5)(-0.1125,0.5)
\psline(-0.1125,-0.5)(0.3875,-0.5)
\psline(0.5375,-0.5)(1.0375,-0.5)
\psline(1.1875,-0.5)(1.6875,-0.5)
\psline(0.1375,-0.5)(0.1375,-0.7)
\psline(0.1375,-0.7)(0.1375,-0.825)
\pscircle(0.1375,-0.95){0.125}
\psline(0.075,-0.98125)(0.10625,-0.98125)
(0.10625,-0.98125)(0.10625,-0.91875)
(0.10625,-0.91875)(0.16875,-0.91875)
(0.16875,-0.91875)(0.16875,-0.98125)
(0.16875,-0.98125)(0.2,-0.98125)
\psline(0.1375,-1.075)(0.1375,-1.2)
\uput{0.501875ex}[l](0.0125,-0.95){\llap{\hbox{$\:$}$ \mathrm{Kick}$\hbox{$\:$}}}
\psline(0.1375,-1.2)(0.1375,-1.325)
\psline(0.220833,-1.325)(0.054167,-1.325)
\psline(0.193056,-1.35625)(0.081944,-1.35625)
\psline(0.173214,-1.3875)(0.101786,-1.3875)
\psline(1.4375,-0.5)(1.4375,-0.9)
\psline(1.4375,-0.9)(1.5375,-0.9)
\psline(1.8375,-0.9)(1.5375,-0.75)
(1.5375,-0.75)(1.5375,-1.05)
(1.5375,-1.05)(1.8375,-0.9)
(1.8375,-0.9)(1.9375,-0.9)
\rput(1.6875,-0.9){\hspace{-0.2cm}$\scriptstyle{\omega_1}$}
\psline(1.9375,-0.9)(2.0625,-0.9)
\psline(2.5625,-0.9)(2.5625,-0.75)
(2.5625,-0.75)(2.0625,-0.75)
(2.0625,-0.75)(2.0625,-1.05)
(2.0625,-1.05)(2.5625,-1.05)
(2.5625,-1.05)(2.5625,-0.9)
\psline(2.5625,-0.9)(2.6875,-0.9)
\rput(2.3125,-0.9){FFT}
\psline(0.7875,-0.5)(0.7875,-1.3)
\psline(0.7875,-1.3)(0.8875,-1.3)
\psline(1.1875,-1.3)(0.8875,-1.15)
(0.8875,-1.15)(0.8875,-1.45)
(0.8875,-1.45)(1.1875,-1.3)
(1.1875,-1.3)(1.2875,-1.3)
\rput(1.0375,-1.3){\hspace{-0.2cm}$\scriptstyle{\omega_2}$}
\psline(1.2875,-1.3)(1.4125,-1.3)
\psline(1.9125,-1.3)(1.9125,-1.15)
(1.9125,-1.15)(1.4125,-1.15)
(1.4125,-1.15)(1.4125,-1.45)
(1.4125,-1.45)(1.9125,-1.45)
(1.9125,-1.45)(1.9125,-1.3)
\psline(1.9125,-1.3)(2.0375,-1.3)
\rput(1.6625,-1.3){FFT}
\psline(2.0375,-1.3)(2.6875,-1.3)
\psframe(2.6875,-1.4)(3.8375,-0.8)
\rput(3.2625,-1.1){Analysis}
\end{pspicture}%
\caption[A schematic diagram of the modes measurement and detection system.]{A schematic diagram of the modes measurement and detection system. A pulse (`kick') is applied at one end of the plasma to excite the oscillations. The $\omega_1$ signal is acquired on the other end of the plasma, while $\omega_2$ is acquired near the centre. The signals are passed through a Fast Fourier transform and an analysis suite to extract the mode frequencies.}
\label{fig:modesSetup}
\end{figure}

The amplitude of the mode oscillations are quite small in a plasma at rest, so to measure the frequencies it is necessary to excite the motions.
This is done by applying a voltage pulse -- the best results are obtained using a pulse that is composed of frequency components close to the expected mode frequency.
Both the first and second modes can be excited simultaneously by generating a pulse containing both frequency components.
The highest-amplitude signals are induced on electrodes for which the nearby charge density oscillates by the largest amount.
For the first mode, this occurs on the electrodes at the end of the plasma, and on the electrode centred on the plasma for the second mode.
A schematic diagram of the measurement system is shown in figure \ref{fig:modesSetup}.

\begin{figure}[hbt!]
	\centering
	\input{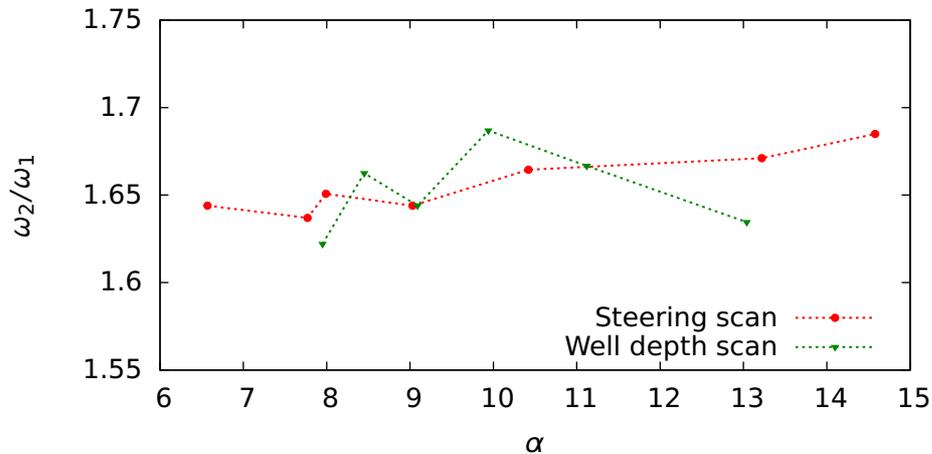}
	\caption{The ratio of first to second mode frequencies as a function of aspect ratio, $\alpha$, for two methods of varying $\alpha$.}
	\label{fig:modesalpha}
\end{figure}

A robust method of producing plasmas with different radii (and thus different values of $\alpha$) is to slightly alter the alignment of the magnetic field of the external solenoid relative to the Penning trap \cite{Modesreport}.
This is thought to alter how electrons emitted from the electron gun enter the Penning trap -- electrons off the axis of the trap will tend to form a radially larger plasma.
This is verified using MCP/phosphor images of the plasmas.
The accessible values of $\alpha$, and the ratio $\omega_2/\omega_1$, are shown in figure \ref{fig:modesalpha}.
It is obvious that the mode frequency does not change significantly over the measurement range, which is problematic when developing a diagnostic technique.

\begin{figure}[hbt!]
\centering
\input{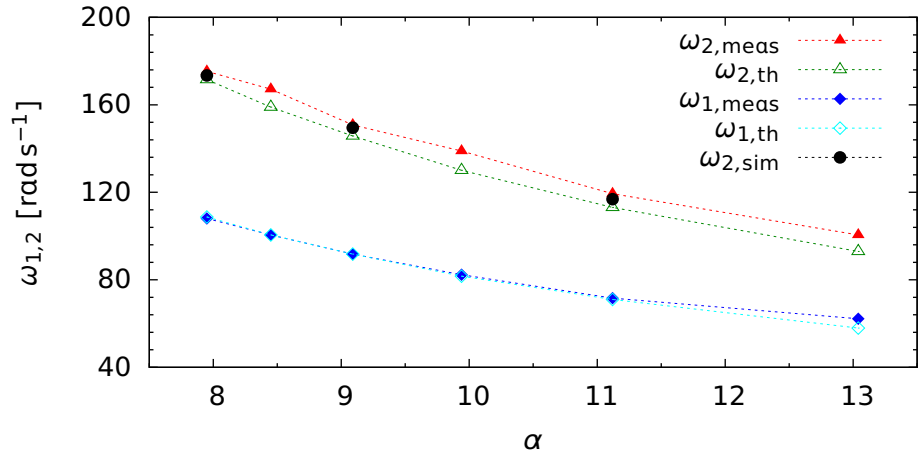}
\caption[The frequency of the first ($\omega_1$) and second ($\omega_2$) mode as a function of the aspect ratio $\alpha$,  obtained by varying the confining potential.]{The frequency of the first ($\omega_1$) and second ($\omega_2$) mode as a function of aspect ratio obtained by varying the confining potential. Shown are the measurements (meas), the values obtained from equation \ref{eq:mode2cold} (th) and the predications from a PIC simulation (sim).}
\label{fig:modesalpha2}
\end{figure}

It is also possible to vary $\alpha$ by altering the shape of the potential well to change the length of the plasma, keeping the radius constant.
This is a less-ideal measurement, as the first mode also depends on the shape of the potential well, but is validated by the data in figure \ref{fig:modesalpha}, which shows that the ratio $\omega_2/\omega_1$ is constant for differently-prepared plasmas with the same value of $\alpha$.
The results of this measurement, compared to the predictions for the measured plasma geometry from \cite{ModesTheory} are shown in figure \ref{fig:modesalpha2}.

It is clear that the measured mode frequency is offset by $\sim 1~\mathrm{MHz}$ above the prediction from theory.
This change could be due to a temperature difference, but equation \ref{eq:modeTemperature} requires a plasma temperature in excess of 10,000~K, which is implausibly high.
Note that when these measurements were being performed, a direct temperature diagnostic (section \ref{sec:plasmaTemperature}) had not been developed to a degree to which it could make useful comparisons.

The theory described above assumes that the plasma is spheroidal in shape -- i.e. in  global thermal equilibrium.
The approach to global thermal equilibrium is slow in strong magnetic fields, and in ALPHA equilibration is thought to occur on a timescale of hours, so in these measurements, it is possible, even likely, that the plasmas were not completely spheroidal.
It is possible to predict the mode frequencies of non-equilibrium plasmas by simulating motion of the plasma particles in a particle-in-cell (PIC) computer simulation.
This was performed using code made available by F. Robicheaux.
The PIC code allows any distribution of particles to be used, so non-equilibrium plasmas can be studied.
Only a limited number of simulations could be carried out because of the computational requirements - several points are shown in figure \ref{fig:modesalpha2}.
The simulated points lie closer to the measured points than the theoretical predictions, thus indicating that taking account of the effect of the plasma density distribution can produce a better agreement between our understanding of the plasma modes and reality.

\subsection{Conclusions}

The aim of this study was to produce a diagnostic system that could non-dest\-ruct\-ively measure plasma parameters such as the density and temperature using the plasma modes.
It was found that the influence of the shape of the plasma could mimic a large temperature change.
This undermines the utility of the diagnostic, as a (destructive) measurement of the plasma density profile is needed to make a good estimate of the temperature from the mode frequencies.

A second consideration is the fact that measuring the mode frequencies requires adding energy to the plasma to excite the oscillations.
This can, in turn, increase the plasma's temperature.
Plasma temperature is one of the factors that influence the probability of producing antihydrogen atoms with trappable kinetic energies, so it is not desirable to introduce possible sources of heating.
This precludes using the modes diagnostic to make measurements on the plasmas used for antihydrogen production, and further development was not included in the physics programme.
Later work \cite{Tim_modes}, not included here, has begun to use the modes diagnostic in conjunction with the direct temperature diagnostic (section \ref{sec:plasmaTemperature}).

\section{Autoresonant manipulations}
\label{sec:autoresonance}

As has already been mentioned, trapped non-neutral plasmas undergo several oscillatory motions.
As with any oscillator, energy can be added or extracted from the system by applying a periodic force at the frequency of the oscillation.
In the example of the simple pendulum, or that of a child's playground swing, pushing on the oscillator at the oscillation frequency increases the energy and amplitude of the motion.
The force imparted by each individual push need not be very strong, but the cumulative effect over many oscillations can drive the oscillator to high energy.

\begin{figure}[hbt]
	\centering
	\subfloat[]{\input{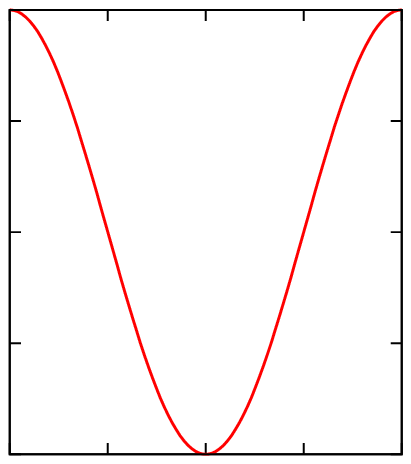}}
	\subfloat[]{\input{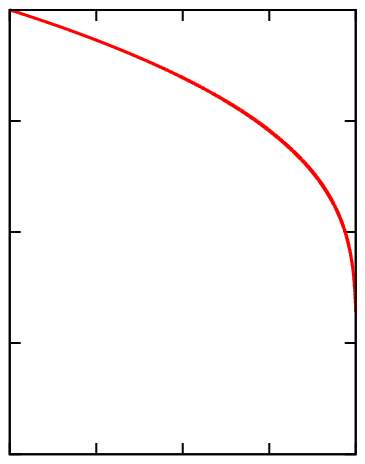} \label{fig:pendulumFreq}}
\caption[A plot of the potential for a simple pendulum and the oscillation frequency as a function of energy.]{A plot of (a) the potential for a simple pendulum and (b) the oscillation frequency $f$ as a function of energy. The parameters have been made dimensionless by scaling by the pendulum mass $m$, length $L$, and the gravitational acceleration $g$. The frequency depends on the oscillator energy, decreasing with increasing energy.}
\label{fig:pendulum}
\end{figure}

The case in which the strength of the drive is small compared to the oscillator's energy is called the weakly-driven oscillator.
To illustrate this, we will use the example of the simple pendulum, which moves in the potential shown in figure \ref{fig:pendulum}(a).
In the limit of small oscillations, the simple pendulum is a linear oscillator, with frequency approximately independent of the amplitude.
At higher amplitudes, this is not true, and the pendulum must be treated as a non-linear oscillator; as shown in figure \ref{fig:pendulum}(b), the frequency is a decreasing function of the oscillator energy.
Thus, continuing to apply the drive at the same fixed frequency does not result in efficient energy transfer.
Though each impulse applied by the drive can change the oscillator energy, this occurs in a stochastic, or random way.
Instead, if the frequency of the drive is changed to match the oscillator frequency, and stays resonant with the oscillator, energy is efficiently transferred.

However, suppose the drive is so small (and there is some damping present) that the oscillator does not gain energy from the drive.
If the frequency of the \textit{drive} is changed by a small amount, it is observed that the oscillator frequency will tend to change so that the drive is still resonant.
The amplitude and frequency are inextricably linked, however, so the change in frequency is accompanied by a corresponding change in the energy.
This tendency of non-linear oscillators to remain in resonance with their drives is known as autoresonance (AR) \cite{FirstAR}, \cite{Joel_AR}.

Autoresonance is a general principle and has been described in many systems, ranging from planetary systems to atoms and Bose-Einstein condensates.
In cyclotron accelerators, particles are accelerated by the application of swept frequency drives taking advantage of autoresonance.
Autoresonance occurs for any change in the oscillator parameters, not just when a swept frequency drive is applied.
For instance, if the length of our simple pendulum slowly changes, the amplitude of the oscillation will also change so that the frequency (locked to the drive) remains constant.

In ALPHA, we are principally interested in applying autoresonance to non-neutral plasmas, using drives that couple to collective oscillations of the plasmas.
These include the axisymmetric modes already discussed, as well as others, such as the diocotron mode, where the plasma is displaced from the cylindrical axis of the trap and orbits the axis with a frequency given by the size of the displacement.
The autoresonant excitation of this mode in an electron plasma has previously been achieved \cite{Joel_AR}.

The most important application in ALPHA is the control of the longitudinal energy by exciting the axial centre-of-mass motion (the `sloshing mode') \cite{ALPHA_AR}.
Before establishing autoresonance, the plasma is initially essentially at rest, and the oscillation frequency is close to the linear frequency defined by the confining potential.
Autoresonance is established by applying an electric field at a frequency higher than this frequency and sweeping it to lower frequency.
As the drive passes through the linear frequency, the oscillation can become locked to the drive, and remain so as the frequency is swept further.

\begin{figure}[hbt]
	\centering
	\subfloat[]{\input{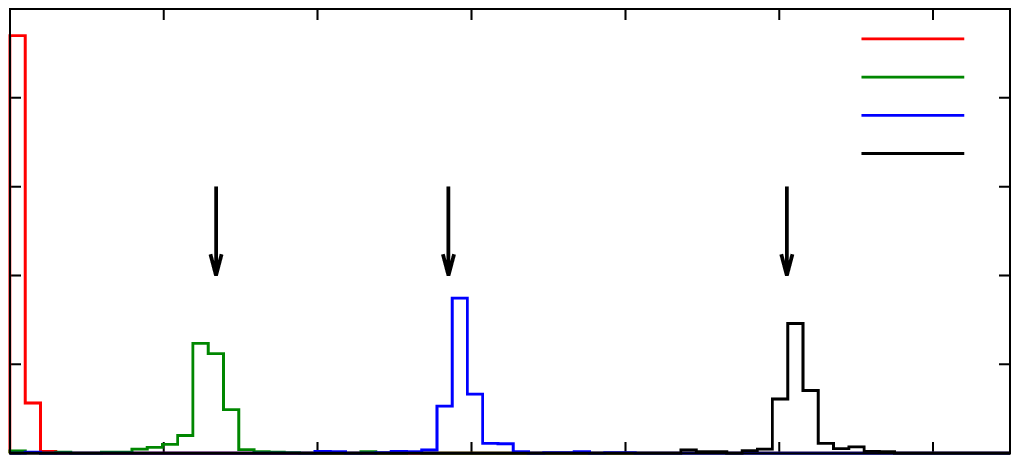}\label{fig:AR_EnergyFreqDistributions}}
	\\
	\subfloat[]{\input{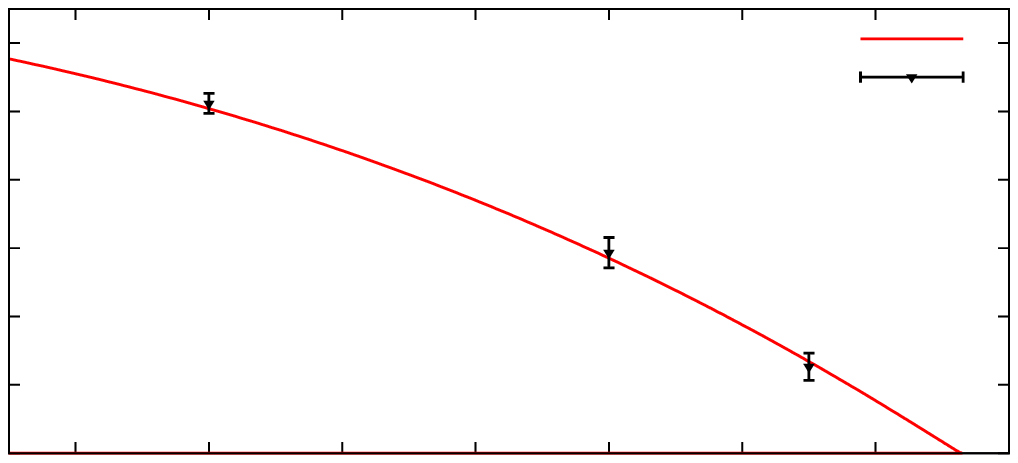}\label{fig:AR_EnergyVSFreq}}
\caption[The energy distributions of antiproton clouds after application of autoresonance sweeps with varying final frequencies.]{The energy distributions of antiproton clouds after application of autoresonance sweeps with varying final frequencies $f_2$. The vertical arrows indicate the energy predicted by calculating the energy-dependent bounce frequency, shown in (b).}
\label{fig:ARresults}
\end{figure}

At the end of the sweep, the drive is switched off, and the energy of the particles is determined by the final frequency of the drive.
This model was tested for a cloud of antiprotons, by exciting their axial oscillation with a 500~mV sinusoidal drive swept at $\alpha = \frac{\mathrm{d}f}{\mathrm{d}t} = $20~MHz/s.
The results are shown in figure \ref{fig:ARresults}, and it is clear that the antiprotons have moved to an energy where the oscillation frequency matches the final frequency of the drive.

\begin{figure}[bht]
	\centering
	\subfloat[]{\input{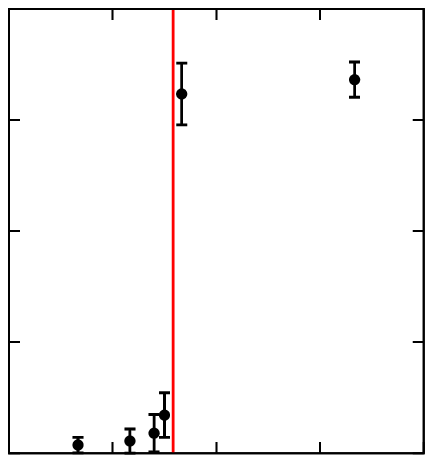}}
	\subfloat[]{\input{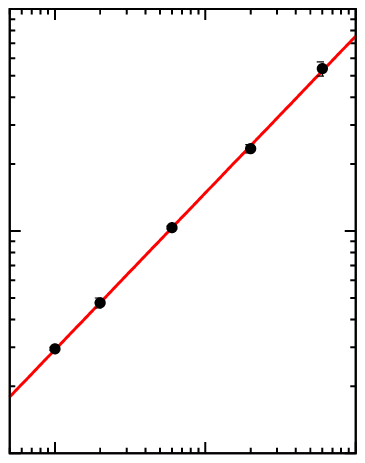}}
	\caption[The threshold behaviour of an autoresonant drive.]{(a) The final energy of an antiproton distribution as a function of the sweep amplitude (420~kHz -- 300~kHz at a rate of 20 $\mathrm{MHz\:s^{-1}}$). A clear threshold is visible at V = 48~mV. (b) The critical amplitude $V_c$ as a function of the	sweep rate $\alpha$. The dependence is well described by a power law - $V_c \propto \alpha^{0.70 \pm 0.01}$.}
	\label{fig:AR_Vc}
\end{figure}

Autoresonance does not occur for all drives.
Instead, there is an observed threshold behaviour on the strength of the drive.
Below a critical amplitude, the oscillation does not lock to the drive and the drive excites the distribution in a non-resonant way.
A mathematical treatment of the problem \cite{Joel_AR} yields the expression for the critical amplitude
\begin{equation}
V_c \propto \alpha^{3/4}.
\label{eqn:alpha3/4}
\end{equation}
This is illustrated in figure \ref{fig:AR_Vc}, which shows the mean energy of the final, excited distribution, as a function of the sweep amplitude.
For sweeps with amplitudes below $V_c$, the distribution will tend to not be excited, but above $V_c$, the mean energy moves to the value determined by the end-point of the frequency sweep.
The critical amplitude $V_c$ is taken to be mid-way between the highest amplitude for which the distribution energy did not follow the sweep and the lowest amplitude for which it did.
By measuring $V_c$ for a number of values of $\alpha$, the power-law scaling (equation \ref{eqn:alpha3/4}) could be confirmed (figure \ref{fig:AR_Vc}(a)).
In figure \ref{fig:AR_Vc}(b), the data points are fit with a power law, finding an exponent of $0.70 \pm 0.01$.
This is in reasonable accord with the expected value of 0.75 from the theory.

A perfect oscillator will have a sharp transition from no effect of the autoresonant sweep to full effect.
However, a real non-neutral plasma does not behave as a perfect oscillator, and a transition region is observed where the energy distribution is modified, but not excited in the way that a fully autoresonant sweep does.
Some of this can be explained by including the random thermal motion of the plasma particles, which adds `noise' to the oscillator, and the effect of the repulsive space charge fields.
A theoretical treatment of these considerations \cite{Barth_AR} shows how they affect the behaviour of the plasma when a drive is applied.
The thermal motion essentially inserts a degree of randomness in the initial phase difference between the plasma motion and the drive.
A large phase difference means that it is less likely for the plasma to become resonantly locked to the drive and vice-versa.
The width of the transition thus depends on the temperature of the plasma and rate of the frequency sweep $\alpha$, scaling as $\left(\alpha T \right)^{1/2}$.

Somewhat counter-intuitively, the analysis also found that the width scaled \textit{inversely} with the strength of the self-fields, or in other words, the charge density.
This is due to the possibility of the formation of a `macro-particle' -- where the particles are highly localised in phase space.
In this state, the distribution will tend to either be locked to the autoresonant drive or not, without a transition region.
Stronger self-fields help the formation of a macro-particle, explaining the scaling with density.

A narrower transition width is desirable when using autoresonance as a particle manipulation tool, because the final distribution produced will not be sensitive to small changes in the initial conditions if the sweep is far enough from the transition.
This means that particles are prepared with as low a temperature and as high a density as possible.

The physics of autoresonance is far richer than has been described here and is still under active study.
At ALPHA, autoresonance is used as a tool to change particle energies in a well-defined and reproducible way.
A particularly important application of autoresonance has been to remove particles from their confining well entirely.
This has been used as an injection scheme to produce spatial overlap of antiprotons and positrons in a nested well, as described in section \ref{sec:AR_mixing}.

    \chapter {Antiproton Cooling} \label{chap:Cooling}

\epigraph{In science one tries to tell people, in such a way as to be\\understood by everyone, something that no one ever knew before.\\But in the case of poetry, it's the exact opposite!}{Paul Dirac}

At the Antiproton Decelerator (AD), antiprotons are produced through the collisions of protons and nuclei at an energy of $\sim$ 26~GeV. 
The antiprotons produced are captured into the AD at an energy of $\sim$ 3.6 GeV -- fourteen orders of magnitude higher than the depth of the magnetic traps to be used to confine antihydrogen atoms.
A wide variety of techniques are needed to reduce the energy by this large factor so that antihydrogen atoms with trappable energies can be produced.

The first stage of cooling occurs in the AD, already described in section \ref{sec:antiprotons}.
At the extraction stage, the AD produces a bunch of approximately $3-4 \times 10^7$ antiprotons at 5.3~MeV, a reduction in energy of a factor of 1000.
Antiprotons are then captured in a Penning trap at $\sim$ 4~keV, with an efficiency of $\sim 10^{-3}$.

Of more relevance to this work are the techniques used to cool the antiprotons when stored in the Penning trap.
The first of these is collisional cooling via electrons, which has produced antiproton temperatures as low as a few hundred Kelvin, (a reduction in energy of around $10^5$) with almost unit efficiency. 
This technique will be described in detail in section \ref{sec:ElectronCooling}.
Lastly, evaporative cooling (described in section \ref{sec:EVC}) can reduce the temperature by at least another order of magnitude, with an efficiency of a few percent.

An antiproton temperature of a few Kelvin is low enough that a significant fraction of the antihydrogen atoms formed from this distribution will have a trappable kinetic energy.

\section{Antiproton catching} \label{sec:PbarCatching}

After extraction from the AD, the antiprotons must be confined in a Penning trap. 
It is exceedingly difficult to produce wells that can trap particles of such high energies, so the first energy reduction is performed by passing the antiprotons through a 218~$\mu$m thick aluminium foil (`degrader').
As the antiprotons pass through the degrader, they lose energy through interactions with the foil material.
As a result, the energy of the initially mono-energetic beam from the AD is shifted downwards and the width of the distribution increases dramatically.
The thickness of the degrading foil has been chosen so that the average energy lost by an antiproton is equal to the initial energy of the beam.
In this model, around half of the antiprotons stop in the degrader and annihilate.
The antiprotons that escape the degrader have a broad energy distribution ranging from zero to $\sim$ 5~MeV.

\begin{figure}
	\centering
%
%
\begin{psfrags}%
\psfragscanon%
%
\psfrag{s12}[b][b]{\color[rgb]{0,0,0}\setlength{\tabcolsep}{0pt}\begin{tabular}{c}Potential\end{tabular}}%
\psfrag{s13}[l][l]{\color[rgb]{0,0,0}\setlength{\tabcolsep}{0pt}\begin{tabular}{l}(a)\end{tabular}}%
\psfrag{s14}[l][l]{\color[rgb]{0,0,0}\setlength{\tabcolsep}{0pt}\begin{tabular}{l}(b)\end{tabular}}%
\psfrag{s15}[l][l]{\color[rgb]{0,0,0}\setlength{\tabcolsep}{0pt}\begin{tabular}{l}(c)\end{tabular}}%
%
\psfrag{x01}[t][t]{0}%
\psfrag{x02}[t][t]{0.05}%
\psfrag{x03}[t][t]{0.1}%
\psfrag{x04}[t][t]{0.15}%
\psfrag{x05}[t][t]{0.2}%
\psfrag{x06}[t][t]{0.25}%
\psfrag{x07}[t][t]{0}%
\psfrag{x08}[t][t]{0.05}%
\psfrag{x09}[t][t]{0.1}%
\psfrag{x10}[t][t]{0.15}%
\psfrag{x11}[t][t]{0.2}%
\psfrag{x12}[t][t]{0.25}%
\psfrag{x13}[t][t]{0}%
\psfrag{x14}[t][t]{0.05}%
\psfrag{x15}[t][t]{0.1}%
\psfrag{x16}[t][t]{0.15}%
\psfrag{x17}[t][t]{0.2}%
\psfrag{x18}[t][t]{0.25}%
%
\psfrag{v01}[r][r]{0}%
\psfrag{v02}[r][r]{1000}%
\psfrag{v03}[r][r]{2000}%
\psfrag{v04}[r][r]{3000}%
\psfrag{v05}[r][r]{4000}%
%
\resizebox{12cm}{!}{\includegraphics{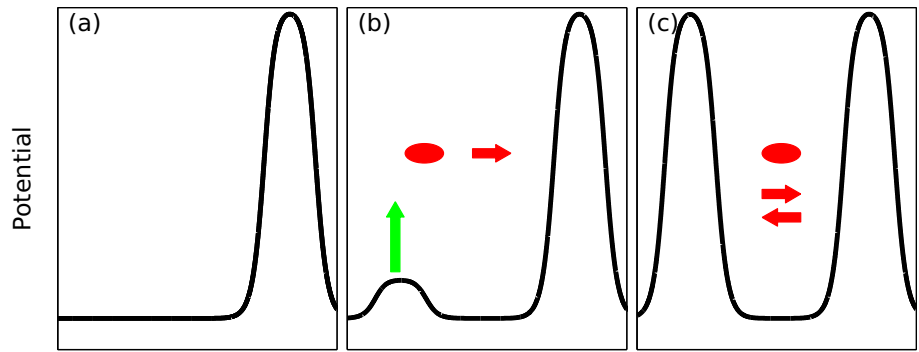}}%
\end{psfrags}%
%

	\caption[A schematic of the procedure to capture antiprotons from the AD.]{(a) Before the antiproton beam arrives, a reflecting potential is erected. (b) As the antiprotons enter the apparatus, the gate electrode is triggered and begins to rise. (c) The antiprotons are then trapped between the two electrodes.}
	\label{fig:pbarCatching}
\end{figure}

Before the arrival of the antiproton beam from the AD, a $\sim$ 4~keV blocking potential is erected using a high voltage electrode 225~mm after the degrading foil (`HVB' in figure \ref{fig:ALPHAelectrodes}).
Most of the 5~MeV antiprotons pass over this barrier, but the low-energy tail of the antiproton distribution is reflected back towards the degrader.
However, in the time the antiprotons take to reach the high voltage electrode and return, a second high voltage electrode, placed immediately after the degrader, is energised (`HVA' in figure \ref{fig:ALPHAelectrodes}).
The antiprotons are trapped in the well formed by these electrodes.
This procedure is shown graphically in figure \ref{fig:pbarCatching}, and was first performed in 1986 at LEAR at CERN \cite{pbarCapture}.

The number of antiprotons caught is observed to be much higher if a stronger solenoidal magnetic field is used \cite{MattThesis}.
However, the neutral trap depth (equation \ref{eq:trapDepth}) is reduced for higher solenoidal fields.
To balance these two effects, ALPHA opted to place a smaller solenoid inside the external magnet, only over the catching trap section.
This solenoid is capable of adding 2 T to the axial magnetic field, and can be switched on and off as necessary in approximately 5~s.
It is thought that the higher magnetic field compensates for the divergence of the beam caused by scattering through the degrading material, though the magnitude of the effect does not agree well with simulations, and the process is still not completely understood.
Under optimum conditions, approximately $8 \times 10^4$ antiprotons can be captured per AD cycle.

\begin{figure}[H]
	\centering
	\input{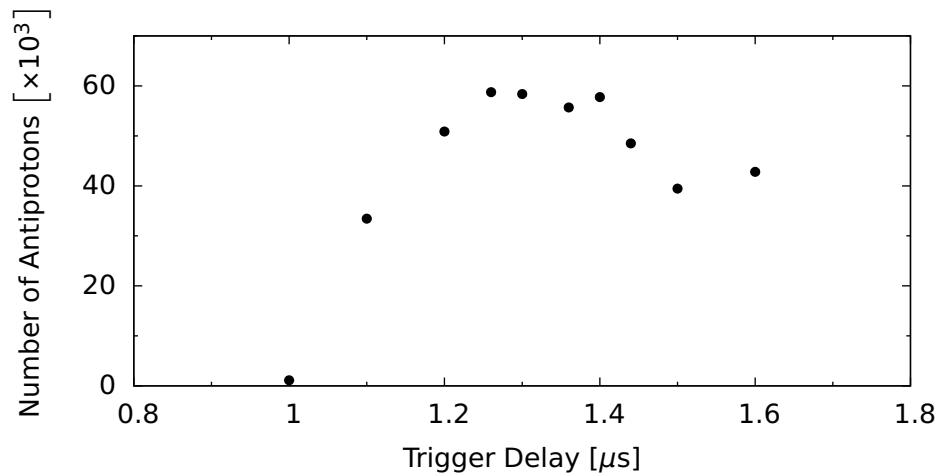}
	\vspace{-10pt}
	\caption{The number of antiprotons caught as a function the delay of the catching trigger.}
	\label{fig:closingCurve}
	\end{figure}

The antiprotons are moving at approximately $4 \times 10^5 \;\mathrm{m.s^{-1}}$ at 4~keV, and transit the distance between the high-voltage electrodes (18.6~cm) in less than 1~$\mathrm{\mu}$s.
The short time scale requires fine control of the timing for the switch to apply the `gate' high voltage.
The time at which the switch is triggered is defined as a delay  from a warning signal from the AD control system that antiproton extraction is imminent.
The optimum value for this delay is determined experimentally by measuring the number of antiprotons captured over a range of settings. 
An example of a measured distribution is shown in figure \ref{fig:closingCurve}, where the optimum value is approximately 1.3~$\mu$s.

\begin{figure}[H]
	\centering
	\mtotex{chap_apparatus/HVCircuit} 

\psset{unit=1in,cornersize=absolute,dimen=middle}%
\begin{pspicture}(-0.159722,-0.8875)(3.191261,0.35)%
\psset{linewidth=0.8pt}%
\psset{linewidth=0.8pt}%
\ifx\MPSTPatchA\undefined{\makeatletter\def\psbezier@ii{\addto@pscode{%
\ifshowpoints true \else false \fi\tx@OpenBezier%
\ifshowpoints\tx@BezierShowPoints\fi}\end@OpenObj}\makeatother%
\global\def\MPSTPatchA{}}\fi%
\psset{arrowsize=1.1pt 4,arrowlength=1.64,arrowinset=0}%
\psline(0,0)(0,-0.125)
\pscircle(0,-0.25){0.125}
\rput(0,-0.1875){$-$}
\rput(0,-0.3125){$+$}
\psline(0,-0.375)(0,-0.5)
\uput{0.501875ex}[l](-0.125,-0.25){\llap{\hbox{$\:$}$ \mathrm{HV}$\hbox{$\:$}}}
\psline(0,-0.5)(0,-0.7)
\psline(0,-0.7)(0,-0.825)
\psline(0.083333,-0.825)(-0.083333,-0.825)
\psline(0.055556,-0.85625)(-0.055556,-0.85625)
\psline(0.035714,-0.8875)(-0.035714,-0.8875)
\psline(0,0)(0,0.2)
\psline(0,0.2)(0.6,0.2)
\pscircle[fillstyle=solid,fillcolor=white](0.6,0.2){0.02}
\psline(0.6,0.2)(0.6,0)
\pscircle[fillstyle=solid,fillcolor=white](0.6,0){0.02}
\psline(0.6,0)(0.3,0)
\psline(0.3,0)(0.3,-0.125)
(0.3,-0.125)(0.341667,-0.145833)
(0.341667,-0.145833)(0.258333,-0.1875)
(0.258333,-0.1875)(0.341667,-0.229167)
(0.341667,-0.229167)(0.258333,-0.270833)
(0.258333,-0.270833)(0.341667,-0.3125)
(0.341667,-0.3125)(0.258333,-0.354167)
(0.258333,-0.354167)(0.3,-0.375)
(0.3,-0.375)(0.3,-0.5)
\psline(0.3,-0.5)(0.6,-0.5)
\pscircle[fillstyle=solid,fillcolor=white](0.6,-0.5){0.02}
\psline(0.6,0)(0.6,-0.225)
\psline(0.516667,-0.225)(0.683333,-0.225)
\psline(0.516667,-0.275)(0.683333,-0.275)
\psline(0.6,-0.275)(0.6,-0.5)
\uput{0.501875ex}[r](0.683333,-0.25){\rlap{\hbox{$\:$}$ \scriptstyle{\mathrm{C_{store}}}$\hbox{$\:$}}}
\psline(0.6,-0.5)(0.6,-0.7)
\pscircle[fillstyle=solid,fillcolor=white](0.6,-0.7){0.02}
\psline(0.6,-0.7)(0,-0.7)
\psline(0.6,0.2)(0.9,0.2)
\psline(0.9,0.2)(0.9,0.2)
\psline(1.4,0.2)(1.4,0.35)
(1.4,0.35)(0.9,0.35)
(0.9,0.35)(0.9,0.05)
(0.9,0.05)(1.4,0.05)
(1.4,0.05)(1.4,0.2)
\psline(1.4,0.2)(1.4,0.2)
\uput{0.501875ex}[d](1.15,0.05){\hbox{$\:$}$ \scriptstyle{\mathrm{Behlke}}$\hbox{$\:$}}
\psline(0.9,0.2)(1.066667,0.2)
\psline(1.066667,0.2)(1.191667,0.325)
\psline(1.233333,0.2)(1.4,0.2)
\psline(1.4,0.2)(1.55,0.2)
\psline(1.55,0.2)(1.675,0.2)
(1.675,0.2)(1.695833,0.241667)
(1.695833,0.241667)(1.7375,0.158333)
(1.7375,0.158333)(1.779167,0.241667)
(1.779167,0.241667)(1.820833,0.158333)
(1.820833,0.158333)(1.8625,0.241667)
(1.8625,0.241667)(1.904167,0.158333)
(1.904167,0.158333)(1.925,0.2)
(1.925,0.2)(2.05,0.2)
\psline[arrowsize=0.05in 0,arrowlength=2,arrowinset=0]{->}(1.658579,0.058579)(1.941421,0.341421)
\uput{0.501875ex}[d](1.8,0.058579){\hbox{$\:$}$ \mathrm{R_{block}}$\hbox{$\:$}}
\psline(2.05,0.2)(2.2,0.2)
\psline(2.2,0.2)(2.3,0.2)
\pscircle[fillstyle=solid,fillcolor=white](2.3,0.2){0.02}
\psline(2.3,0.2)(2.473205,0.1)
\pscircle[fillstyle=solid,fillcolor=white](2.473205,0.1){0.02}
\pscircle[fillstyle=solid,fillcolor=white](2.3,0){0.02}
\psline(2.473205,0.1)(2.573205,0.1)
\psline(2.3,0)(2.3,-0.15)
\psline(2.3,-0.15)(2.3,-0.275)
\pscircle(2.3,-0.4){0.125}
\rput(2.3,-0.4){LV}
\psline(2.3,-0.525)(2.3,-0.65)
\psline(2.3,-0.65)(2.3,-0.7)
(2.3,-0.7)(0,-0.7)
\psline(2.473205,0.1)(3.073205,0.1)
(3.073205,0.1)(3.073205,-0.05)
\pscircle[fillstyle=solid,fillcolor=white](3.073205,-0.05){0.02}
\psline(3.073205,-0.05)(2.773205,-0.05)
\psline(2.773205,-0.05)(2.773205,-0.175)
(2.773205,-0.175)(2.814872,-0.195833)
(2.814872,-0.195833)(2.731538,-0.2375)
(2.731538,-0.2375)(2.814872,-0.279167)
(2.814872,-0.279167)(2.731538,-0.320833)
(2.731538,-0.320833)(2.814872,-0.3625)
(2.814872,-0.3625)(2.731538,-0.404167)
(2.731538,-0.404167)(2.773205,-0.425)
(2.773205,-0.425)(2.773205,-0.55)
\psline(2.773205,-0.55)(3.073205,-0.55)
\pscircle[fillstyle=solid,fillcolor=white](3.073205,-0.55){0.02}
\psline(3.073205,-0.05)(3.073205,-0.275)
\psline(2.989872,-0.275)(3.156538,-0.275)
\psline(2.989872,-0.325)(3.156538,-0.325)
\psline(3.073205,-0.325)(3.073205,-0.55)
\uput{0.501875ex}[r](3.156538,-0.3){\rlap{\hbox{$\:$}$ \scriptstyle{\mathrm{Electrode}}$\hbox{$\:$}}}
\psline(3.073205,-0.55)(3.073205,-0.7)
(3.073205,-0.7)(0,-0.7)
\end{pspicture}%
	\caption{A simplified schematic of the catching voltage switching circuit.}
	\label{fig:HVcircuit}
	\end{figure}
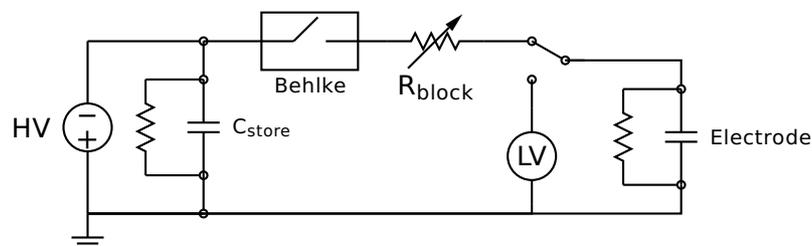

Specialised circuitry, shown in figure \ref{fig:HVcircuit}, must be employed to switch the high voltages to the electrodes in such a short time interval.
It is also required that the electrodes can be controlled with a low voltage (LV) amplifier when not in use for capturing antiprotons.
The design is similar to that discussed in \cite{MattThesis}, but has been re-engineered to alleviate problems with electronic noise.

The main component is a fast MOSFET-based switch (Behlke type HTS-161-03-LC) \cite{Behlke}.
Before the switch is triggered, a 4 nF bank of storage capacitors ($\mathrm{C_{store}}$) in parallel with the power supply is charged to provide enough current to quickly charge the electrode.
A blocking resistor is placed on the electrode side of the switch to minimise reflections from impedance transitions in the path.
Together with the resistance ($R$) of the cable and the capacitance ($C$) of the electrode, this defines an ideal characteristic rise time given by
	\begin{equation}
		\tau = \left(R_\mathrm{block}+R_\mathrm{cable} \right) \times \left(C_\mathrm{electrode} + C_\mathrm{cable} \right).
	\end{equation}
As it is preferable to have as sharp a voltage change as possible, it seems desirable to make $R_\mathrm{block}$ as small as possible.
However, this competes with the damping of reflections and oscillations.
As usual, it is easiest to measure any effect experimentally.
Several values of $R_\mathrm{block}$ were tried, and the optimum value was found to be around 50-60 $\Omega$.
	
\begin{figure}[hbt]
	\centering
	\input{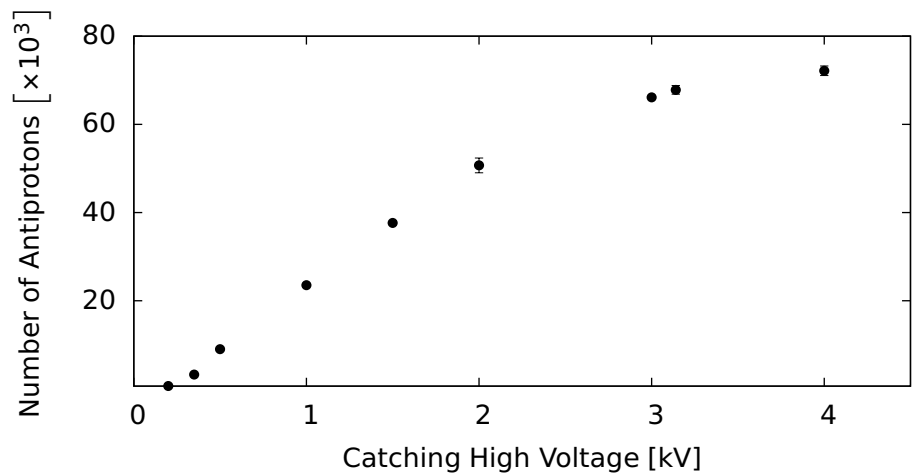}
	\vspace{-10pt}
	\caption{The number of antiprotons caught as a function of the catching voltage, at a constant delay of 1.2~$\mathrm{\mu s}$.}
	\label{fig:catchingCurve}
	\end{figure}

The number of antiprotons caught is a function of the level of the voltages applied to the catching electrodes.
The energy distribution of the antiprotons after passing through the degrader can be measured by varying the voltage levels on the electrodes and measuring the number of antiprotons captured at each setting.
An example measurement is shown in figure \ref{fig:catchingCurve}. 
It is notable that the slope of the distribution becomes more shallow at higher energies, indicating a diminishing return on applying higher catching voltages.

The voltages applied are limited by the size of the electric fields generated in the apparatus.
A large electric field can cause ionisation of residual gas atoms, forming a path of current carriers that rapidly discharges the electrode, potentially causing damage to other equipment, as well as destroying the confining well and allowing the antiprotons to escape.
To minimise the chance of such an occurrence, the catching electrodes are separated from their neighbours on ceramic spacers, and the electrodes are designed to avoid sharp edges, which produce the highest electric fields.

At the end of this procedure, we are left with a cloud of antiprotons  with energies around 4~keV.
Antiprotons at 4~keV are first of all, difficult to manipulate, since using voltages in the keV range requires special design considerations, and secondly, they are at far too high an energy to produce cryogenic antihydrogen atoms.

\section{Electron cooling} \label{sec:ElectronCooling}

Charged particles in a magnetic field lose energy radiatively, but since the cyclotron cooling time for antiprotons is unacceptably large, approximately 84 years in a 3~T field (section \ref{sec:energyLoss}), it is not feasible to cool antiprotons in this way.
Instead, the antiprotons are combined with an electron plasma to form a two-component non-neutral plasma.
The electrons act as a sink for the antiprotons' energy, which is transferred through elastic collisions \cite{pbarCooling}.
The electrons can radiate the energy away far faster than can the antiprotons -- recall that their cooling time in 3~T is approximately 0.43~s.
The timescale for cooling is thus set by the rate of energy transfer between the two species.

\begin{figure}[hbt]
	\centering
%
%
\begin{psfrags}%
\psfragscanon%
%
\psfrag{s12}[b][b]{\color[rgb]{0,0,0}\setlength{\tabcolsep}{0pt}\begin{tabular}{c}Potential\end{tabular}}%
\psfrag{s13}[l][l]{\color[rgb]{0,0,0}\setlength{\tabcolsep}{0pt}\begin{tabular}{l}(a)\end{tabular}}%
\psfrag{s14}[l][l]{\color[rgb]{0,0,0}\setlength{\tabcolsep}{0pt}\begin{tabular}{l}(b)\end{tabular}}%
\psfrag{s15}[l][l]{\color[rgb]{0,0,0}\setlength{\tabcolsep}{0pt}\begin{tabular}{l}(c)\end{tabular}}%
%
\psfrag{x01}[t][t]{0.05}%
\psfrag{x02}[t][t]{0.1}%
\psfrag{x03}[t][t]{0.15}%
\psfrag{x04}[t][t]{0.2}%
\psfrag{x05}[t][t]{0.05}%
\psfrag{x06}[t][t]{0.1}%
\psfrag{x07}[t][t]{0.15}%
\psfrag{x08}[t][t]{0.2}%
\psfrag{x09}[t][t]{0.05}%
\psfrag{x10}[t][t]{0.1}%
\psfrag{x11}[t][t]{0.15}%
\psfrag{x12}[t][t]{0.2}%
%
\psfrag{v01}[r][r]{-100}%
\psfrag{v02}[r][r]{0}%
\psfrag{v03}[r][r]{100}%
\psfrag{v04}[r][r]{200}%
\psfrag{v05}[r][r]{300}%
\psfrag{v06}[r][r]{400}%
\psfrag{v07}[r][r]{500}%
%
\resizebox{12cm}{!}{\includegraphics{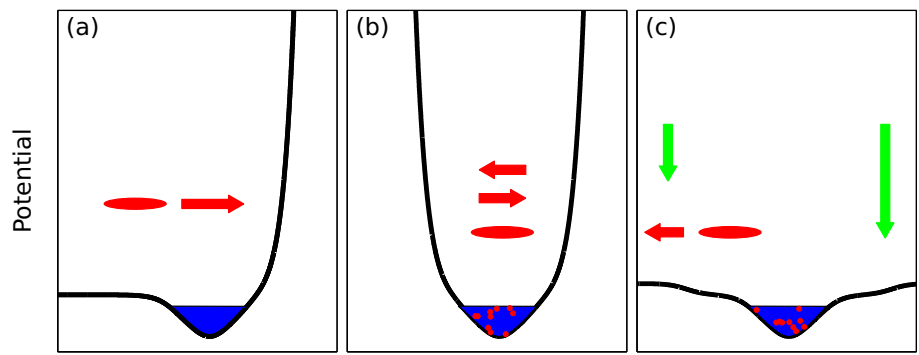}}%
\end{psfrags}%
%

	\caption[A schematic showing the use of an electron plasma to cool captured antiprotons.]{(a) Before the antiprotons arrive, an electron plasma is placed between the high voltage electrodes. (b) While the antiprotons are trapped in the high-voltage well, they collide and transfer energy to the colder electrons. As they lose energy, the antiprotons accumulate in the electron plasma. (c) After a predetermined amount of time, the high voltage well is removed. Uncooled antiprotons escape axially, while the cold antiprotons remain confined with the electron plasma.}
	\label{fig:coolingFig}
\end{figure}

To cool the antiprotons captured from the AD, approximately $10^7$ electrons are placed into a `dimple' well between the high voltage electrodes prior to the arrival of the antiproton beam (see figure \ref{fig:coolingFig}).
Once confined between the high voltage electrodes, the antiprotons pass back and forth through the electron plasma, colliding with the electrons and losing energy.

\begin{figure}[hbt]
	\centering
	\input{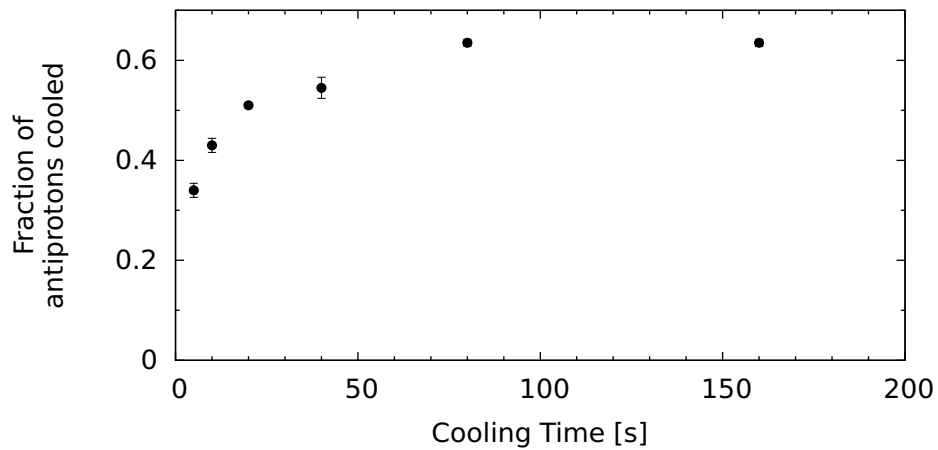}
	\caption{The fraction of antiprotons cooled as a function of the interaction time with the electron plasma.}
	\label{fig:coolingCurve}
\end{figure}

After a certain time, the high-voltage electrodes are discharged -- any antiprotons that have cooled into the dimple well remain confined, but high-energy particles escape.
The fraction of cooled antiprotons can be determined by comparing the number of annihilations detected when the high voltage is removed to the number detected when the dimple well is dumped.
The optimum cooling time is determined experimentally by varying the length time for which the antiprotons and electrons are allowed to interact.
Example data is shown in figure \ref{fig:coolingCurve}, where it can be seen that it requires several tens of seconds to cool a spill of antiprotons to the point where the high voltage well can be switched off.
The catching and cooling procedure can be repeated a number of times to accumulate a large number of antiprotons (a process known as `stacking').

Along with the cooling time, the degree of overlap between the electron plasma and the antiproton distribution will determine the fraction of antiprotons cooled.
The antiprotons, once captured, are bound to the solenoidal magnetic field lines and move at a fixed radius in the Penning trap.
The radius of the distribution is defined by the radial profile of the AD beam and the amount of divergence introduced by passage through the degrader.
An electron plasma with a large radius can interact with, and thus cool, more of the antiprotons compared to a plasma with a radius smaller than the antiproton distribution.

Though one might think that cooling as large a fraction as possible is the most desirable situation, this is not always the case.
Recall that in equilibrium at low temperatures, the antiprotons form a hollow ring outside the electron plasma (section \ref{sec:plasmaEquilibrium}), so the size of the antiproton cloud will be at least the size of the electron plasma.
The density of antiprotons close to the axis of the trap is just as, if not more, important than the total number of particles when producing antihydrogen (see section \ref{sec:TrappingConsiderations}), so a trade-off is usually made between the radial size of the antiprotons (and electrons) and the number of antiprotons cooled.
Although the rotating-wall technique (section \ref{sec:rotatingWall}) can compress the electron-antiproton cloud to smaller radius, in practice, it is seen to perform poorly on diffuse plasmas.
A denser final antiproton distribution can be produced by sacrificing some cooling efficiency for a smaller cloud.
Typically a good tradeoff is made by using a 0.6~mm electron plasma, which cools approximately 50\% of the antiprotons caught.

The stored electron/antiproton plasma will continue to radiatively cool, eventually reaching an equilibrium defined by the rate of cyclotron cooling and the power of heating sources.

\subsection{Electron removal}

Before mixing the antiprotons with positrons to produce antihydrogen, the electrons and antiprotons must be separated.
If the electrons were allowed to remain, they could potentially deplete the positron plasma by forming an electron-positron atom, positronium (Ps), or destroy antihydrogen atoms through the charge exchange process
\begin{equation}
	\mathrm{\Hbar + e^- \; \rightarrow \; \pbar + \mathrm{Ps}}.
\end{equation}
Injecting the electrons along with the antiprotons into the positron plasma also disturbs the equilibrium of the positron plasma, causing an increase in temperature.
This effect will scale with the amount of charge introduced, so it can be minimised by removing the electrons and only injecting the antiprotons.

\begin{figure}[hbt]
	\centering
%
%
\begin{psfrags}%
\psfragscanon%
%
\psfrag{s12}[b][b]{\color[rgb]{0,0,0}\setlength{\tabcolsep}{0pt}\begin{tabular}{c}Potential\end{tabular}}%
\psfrag{s13}[l][l]{\color[rgb]{0,0,0}\setlength{\tabcolsep}{0pt}\begin{tabular}{l}(a)\end{tabular}}%
\psfrag{s14}[l][l]{\color[rgb]{0,0,0}\setlength{\tabcolsep}{0pt}\begin{tabular}{l}(b)\end{tabular}}%
\psfrag{s15}[l][l]{\color[rgb]{0,0,0}\setlength{\tabcolsep}{0pt}\begin{tabular}{l}(c)\end{tabular}}%
%
\psfrag{x01}[t][t]{0.05}%
\psfrag{x02}[t][t]{0.1}%
\psfrag{x03}[t][t]{0.15}%
\psfrag{x04}[t][t]{0.2}%
\psfrag{x05}[t][t]{0.05}%
\psfrag{x06}[t][t]{0.1}%
\psfrag{x07}[t][t]{0.15}%
\psfrag{x08}[t][t]{0.2}%
\psfrag{x09}[t][t]{0.05}%
\psfrag{x10}[t][t]{0.1}%
\psfrag{x11}[t][t]{0.15}%
\psfrag{x12}[t][t]{0.2}%
%
\psfrag{v01}[r][r]{-60}%
\psfrag{v02}[r][r]{-50}%
\psfrag{v03}[r][r]{-40}%
\psfrag{v04}[r][r]{-30}%
\psfrag{v05}[r][r]{-20}%
\psfrag{v06}[r][r]{-10}%
\psfrag{v07}[r][r]{0}%
%
\resizebox{12cm}{!}{\includegraphics{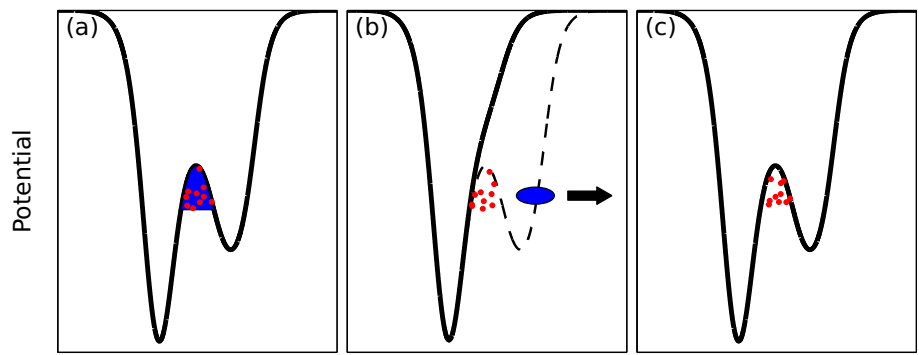}}%
\end{psfrags}%
%

	\caption[A schematic of the `ekicking' procedure.]{A schematic of the `ekicking' procedure. In (a), electrons (blue) and antiprotons (red) are mixed. When one side of the confining well is removed (b), the electrons escape. The well is re-established (c) before the antiprotons can leave.}
	\label{fig:ekickSimple}
	\end{figure}

The electrons are removed by taking advantage of the large difference in masses of the electron and antiproton.
This `ekicking' technique is illustrated in figure \ref{fig:ekickSimple}.
A fast voltage pulse, typically around 100~ns, is applied to one side of the well, briefly removing the confining barrier and allowing the electrons to escape.
The antiprotons, which move much more slowly, cannot escape before the barrier is re-established.
The height and timing of the voltage pulse are empirically determined, with feedback from diagnostic dumps, to achieve the best results.
Each pulse only removes a fraction of the electrons, and so the procedure must be repeated a number of times.

This process is by design non-adiabatic for the antiprotons, and even though they do not escape, the antiprotons can be heated.
In particular, the removal of the large space charge associated with the electron plasma results in a significant change to the potential shape. 
This leaves some antiprotons with a large amount of potential energy, which is quickly converted to a high temperature through collisions.
The result is that the antiprotons are left very warm, with a temperature approximately 1~eV (11,000~K).

\begin{figure}[tb]
	\centering
%
%
\begin{psfrags}%
\psfragscanon%
%
\psfrag{s23}[b][b]{\color[rgb]{0,0,0}\setlength{\tabcolsep}{0pt}\begin{tabular}{c}Potential\end{tabular}}%
\psfrag{s24}[l][l]{\color[rgb]{0,0,0}\setlength{\tabcolsep}{0pt}\begin{tabular}{l}(a)\end{tabular}}%
\psfrag{s25}[l][l]{\color[rgb]{0,0,0}\setlength{\tabcolsep}{0pt}\begin{tabular}{l}(b)\end{tabular}}%
\psfrag{s26}[l][l]{\color[rgb]{0,0,0}\setlength{\tabcolsep}{0pt}\begin{tabular}{l}(c)\end{tabular}}%
\psfrag{s27}[l][l]{\color[rgb]{0,0,0}\setlength{\tabcolsep}{0pt}\begin{tabular}{l}(d)\end{tabular}}%
\psfrag{s28}[b][b]{\color[rgb]{0,0,0}\setlength{\tabcolsep}{0pt}\begin{tabular}{c}Potential\end{tabular}}%
\psfrag{s29}[l][l]{\color[rgb]{0,0,0}\setlength{\tabcolsep}{0pt}\begin{tabular}{l}(e)\end{tabular}}%
\psfrag{s30}[l][l]{\color[rgb]{0,0,0}\setlength{\tabcolsep}{0pt}\begin{tabular}{l}(f)\end{tabular}}%
%
\psfrag{x01}[t][t]{0.05}%
\psfrag{x02}[t][t]{0.1}%
\psfrag{x03}[t][t]{0.15}%
\psfrag{x04}[t][t]{0.2}%
\psfrag{x05}[t][t]{0.05}%
\psfrag{x06}[t][t]{0.1}%
\psfrag{x07}[t][t]{0.15}%
\psfrag{x08}[t][t]{0.2}%
\psfrag{x09}[t][t]{0.05}%
\psfrag{x10}[t][t]{0.1}%
\psfrag{x11}[t][t]{0.15}%
\psfrag{x12}[t][t]{0.2}%
%
\psfrag{v01}[r][r]{-60}%
\psfrag{v02}[r][r]{-50}%
\psfrag{v03}[r][r]{-40}%
\psfrag{v04}[r][r]{-30}%
\psfrag{v05}[r][r]{-20}%
\psfrag{v06}[r][r]{-10}%
\psfrag{v07}[r][r]{0}%
\psfrag{v08}[r][r]{-60}%
\psfrag{v09}[r][r]{-50}%
\psfrag{v10}[r][r]{-40}%
\psfrag{v11}[r][r]{-30}%
\psfrag{v12}[r][r]{-20}%
\psfrag{v13}[r][r]{-10}%
\psfrag{v14}[r][r]{0}%
%
\resizebox{12cm}{!}{\includegraphics{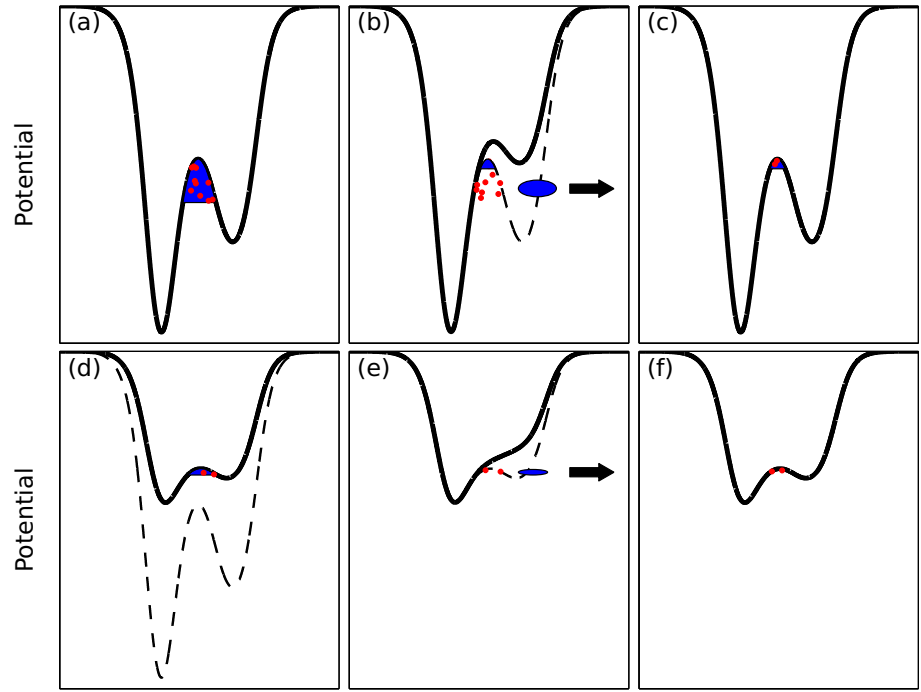}}%
\end{psfrags}%
%

	\caption[A schematic of the two-stage ekicking procedure.]{A schematic of the two-stage ekicking procedure. This process begins (a) in the same way as the one-step process shown in figure \ref{fig:ekickSimple}, but a small number of electrons remains after the first 'ekick' (b), and re-cool the antiprotons (c). The depth of the well is reduced (d), and the remainder of the electrons are removed (e). This leaves antiprotons with significantly lower energy than the one step process.}
	\label{fig:softEkicks}
	\end{figure}

In order to combat these effects, a multi-stage electron removal technique was developed, illustrated in figure \ref{fig:softEkicks}.
In the first stage, more than 95\% of the electrons are removed by the application of between one and three pulses in quick succession.
This violent process leaves the plasma in an extremely hot state ($> 10^3~\mathrm{K}$).
The antiprotons re-cool on the remaining electrons, and a final pulse train is applied to remove the last of the electrons.
Because a smaller number of electrons is removed on the final step, the temperature of the resultant antiproton cloud is lower, but still around 200~K.

\begin{figure}[hbt]
	\centering
	\input{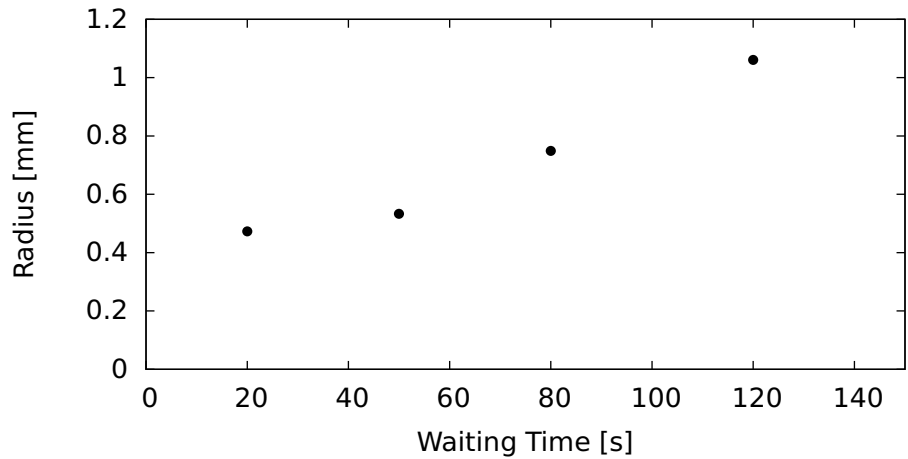}
	\caption{The expansion of the antiproton/electron cloud following extraction of part of the electron plasma}
	\label{fig:ekickWaitExpand}
	\end{figure}

The voltage pulse used to remove the particles is most effective close to the axis, and so this is where the electrons are preferentially removed from, resulting in a hollow charge profile after the first ekick pulses.
Such a charge profile is unstable and tends to redistribute itself to fill in the centre while the particles are re-cooling.
Since charge moves inward, conservation of total canonical angular momentum (equation \ref{eq:r2Conserve}) requires that some particles also move to higher radii, with the result that the antiproton density can fall quite dramatically (see figure \ref{fig:ekickWaitExpand}).
Once all the electrons have been removed, the radial redistribution essentially stops since the antiproton-antiproton collision rate is more than an order of magnitude lower than the antiproton-electron rate.

The time between the first and second stages is then a trade-off between the time given to allow the antiprotons to cool to a lower temperature and the time for the expansion to have a significant effect.
Leaving more electrons in the second stage can cool the antiprotons at a faster rate, but the redistribution also speeds up (since it is related to the collision rate). 

\subsection{Conclusions}

The few-hundred Kelvin temperatures achieved represent an improvement of around two orders of magnitude in temperature from the simple ekicking procedure.
Though the temperature is still far higher than the depth of the magnetic minimum, there is a non-negligible fraction of antiprotons in the low energy tail below 0.5~K (at 200~K, $10^{-4}$ of the antiprotons are below 0.5~K), so it is not unreasonable to expect that it is possible to use these antiprotons to produce a small number of antihydrogen atoms with Kelvin-scale energies.

Having colder antiproton is almost always better, however.
Electron cooling, though extremely successful in cooling the antiprotons the five orders of magnitude from $10^3~\mathrm{eV}$ to $10^{-2}~\mathrm{eV}$, does not seem capable of achieving the cryogenic temperatures necessary to produce large numbers of antihydrogen atoms at energies of order 1~K.
In the next section, we discuss a further cooling technique that can achieve much colder antiproton temperatures.

\section{Evaporative cooling of antiprotons} 
\label{sec:EVC}

\subsection{Introduction} \label{sec:EVCintro}

Evaporative cooling of particles is a technique that was first applied to neutral atoms by physicists working towards Bose-Einstein condensates in the 1990s \cite{EVC_Sodium}, \cite{EVC_Hydrogen}.
The technique operates on the principle that it is possible to selectively remove particles with large kinetic energies from an ensemble of trapped particles.
This can be achieved by slightly reducing the strength of the confining fields, so that the highest-energy particles simply escape the trap.
The average kinetic energy of the trapped portion is reduced, and the particles re-equilibrate to a distribution with a lower temperature.
Evaporative cooling is now a fundamental technique in ultra-cold atomic and condensed matter physics, and has been used to achieve temperatures as low as 450~pK \cite{recordTemperature}.

In contrast to neutral atoms, before the work described in this section, evaporative cooling had not been demonstrated for charged species.
A related technique, in which charged ions stored in an EBIT device were cooled using a sacrificial species with smaller charge to carry away the energy \cite{EVC_in_EBIT1}, \cite{EVC_in_EBIT2}, has been used for some time.

\subsection{Theoretical description} \label{sec:EVCTheory}

In a Penning-Malmberg trap, charged particles are strongly bound to the magnetic field lines, and do not tend to escape across the field lines, in the transverse direction.
The confining well is defined by the electric potential along the $z$ axis (parallel to the magnetic field) produced by the trap electrodes (and any nearby charge distributions) and evaporating particles in this situation escape in the $z$ direction.
Therefore, to first order, the problem is one-dimensional, and only the energies of the particles in the $z$ direction are important.

The energy of a collection of particles in global thermal equilibrium will follow a Maxwell-Boltzmann distribution.
If the particles are confined in a well with finite depth, the Maxwell-Boltzmann distribution will be truncated above the depth of the well, and particles that are scattered into the truncated portion are removed from the distribution.
The particles are said to have `evaporated' and it is from this that the process takes its name.
The average energy of the population still trapped is lower than the starting distribution, and the particles will re-distribute energy amongst themselves to produce a distribution with lower temperature.

The rate at which evaporative cooling proceeds is determined by the rate at which particles are lost and the amount of energy carried away by an evaporating particle.

We define the evaporation timescale, $\tau_{ev}$, using the expression
\begin{equation}
\frac{\mathrm{d}N}{\mathrm{d}t} = -\frac{N}{\tau_{ev}}.
\label{eq:EVC_DNDT}
\end{equation}

The `excess' energy of an evaporating particle, the amount of energy above the average energy, can be characterised as $\alpha k_\mathrm{B} T$.
This leads to an expression for the rate of change of temperature of
\begin{equation}
\frac{\mathrm{d}T}{\mathrm{d}t} = - \frac{\alpha}{\tau_{ev}} T.
\label{eq:EVC_DTDT}
\end{equation}

$\tau_{ev}$ is the rate at which particles are scattered into the loss region.
This is a function of the rate of collisions $\tau_{col}$, and the fraction of the Maxwell-Boltzmann distribution extending above the top of the well. 
This latter quantity can be parametrised using $\eta = \frac{U}{k_B T}$, the ratio of the well depth to the temperature.
$\tau_{ev}$ and $\tau_{col}$ are related through the expression \cite{ALPHA_EVC}
\begin{equation} \label{eq:EVC_tau_ev_col}
	\frac{\tau_{ev}}{\tau_{col}} = \frac{\sqrt{2}}{3} \eta e^\eta,
\end{equation}
and $\alpha$ can be written as \cite{ketterleReview}
\begin{equation}
	\alpha = \frac{\eta + \kappa}{\delta + 3/2} - 1,
\end{equation}
where $\kappa$ is the energy (divided by $k_\mathrm{B}T$) by which an evaporating particle `overshoots' the top of the well, and $\delta$ is a constant that describes the well shape, ($\simeq 1/2$ for experiments in ALPHA) \cite{ALPHA_EVC}.

We can relate the change in temperature and the number of particles lost by combining equations \ref{eq:EVC_DNDT} and \ref{eq:EVC_DTDT} to give
\begin{equation}
\frac{\mathrm{d}T}{\mathrm{d}N} = \alpha \frac{T}{N}.
\label{eq:EVC_DTDN}
\end{equation}

\subsection{Measurements} \label{sec:EVCMeasurements}

Evaporative cooling was first applied in ALPHA to antiprotons.
The antiproton dominates the  kinetic energy of an antihydrogen atom, so an obvious improvement in the number of trappable antihydrogen ought to be achieved by reducing the antiproton temperature.

Before carrying out evaporative cooling, a cloud of antiprotons was placed into a 1.5~V deep potential well.
The antiproton density was approximately $8 \times 10^6~ \mathrm{cm^{-3}}$, so the collision time, $\tau_{col}$ is of order $10~\mathrm{ms}$.
The temperature of the antiprotons was approximately 1,000~K, so the value of $\eta \simeq 17$ implies an evaporation timescale, $\tau_{ev} > 10^6~\mathrm{s}$, large enough that any evaporation can be neglected.

The depth of the well was then slowly reduced, ramping the voltage applied to one of the electrodes linearly in time, allowing the highest-energy antiprotons to escape -- this technique is sometimes referred to as `forced evaporative cooling'.
As the well depth is reduced, the number of particles and their temperature falls.
We studied the dependence of the temperature on the speed of the cooling ramp and the depth of the final cooling well.
The well depth is not a linear function of time, so a single value of `ramp rate' does not exist.
We instead parametrise the experiments in terms of the time taken to reach the final well depth from the initial 1.5V deep well.
This introduces a small systematic error as the initial portions of the cooling ramps are not identical for each of the measurements, but the effect is small enough to not be important for this discussion.

\begin{figure}
	\centering
	\input{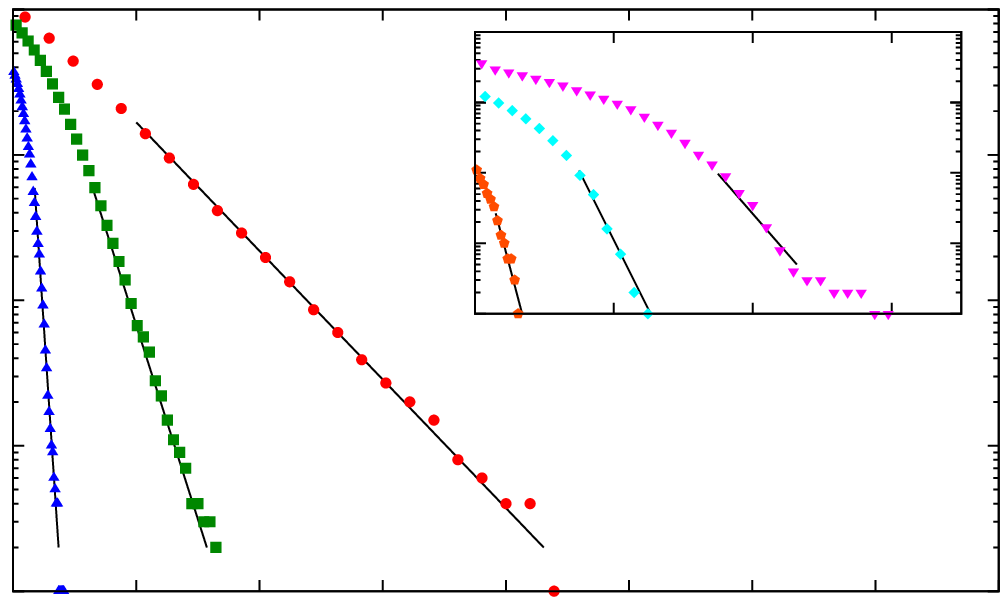}
	\caption[Temperature measurements of antiprotons after evaporative cooling.]{A series of temperature measurements of antiprotons after different amounts of evaporative cooling. The indicated temperatures are from the exponential fits to the data (shown as the straight lines), corrected by the factor derived from the PIC code. In descending order of temperature, the well depths for these measurements were 1480~mV, 433~mV, 108~mV, 69~mV, 36~mV and 10~mV. Uncertainties are of the order of 10\%; the coldest temperature is $(9 \pm 4)$~K.}
	\label{fig:EVCTraces}
\end{figure}

\begin{figure}
	\centering
	\input{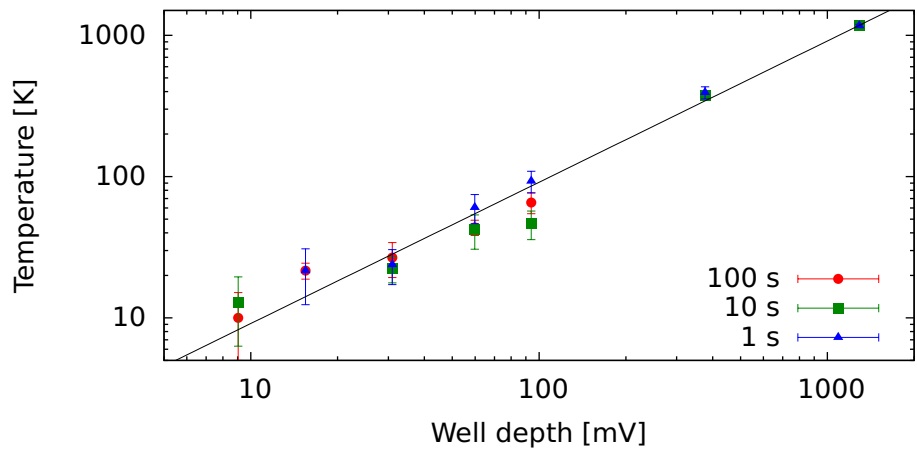}
	\caption{The temperature of the antiproton distribution as a function of well depth for three different ramp times. The straight line is a linear fit through the origin, with slope ($\eta$) $\simeq$ 12.}
	\label{fig:EVCTimeCompare}
\end{figure}

During each cooling ramp, the number and temperature of the antiprotons remaining in the well was measured at a number of well depths.
Example temperatures are shown in figure \ref{fig:EVCTraces} and are plotted against the confining well depth in figure \ref{fig:EVCTimeCompare}.
The temperatures have been corrected  by the factor discussed in section \ref{sec:plasmaTemperature}.
It is clear that the temperature is a decreasing function of the well depth, as expected.
The form of the dependence is quite close to linear, with a slope ($\eta$) $\sim 12$.
To first order, the temperature achieved is independent of the speed of the ramp, except for ramps that are faster than a certain threshold -- the measured temperatures change by only around $20\%$, even with a two-order-of-magnitude change in the cooling time.

We can explain this behaviour from equation \ref{eq:EVC_tau_ev_col}, where the key feature is that $\tau_{ev}$ is an exponential function of $\eta$.
If we consider a distribution that is initially so hot compared to the (fixed) well depth (i.e. low $\eta$) that evaporation occurs copiously, it is clear that the temperature will rapidly fall.
As the temperature falls, $\eta$ rises and evaporation slows.
At some point, $\tau_{ev}$ becomes large compared to the relevant experiment time -- at this point, evaporation has effectively stopped.
The exponential dependence on $\eta$ dominates the expression, so this occurs at approximately the same value of $\eta$, even for different values of $\tau_{col}$ (if for example, there are different densities of antiprotons).
This means that the particles' temperature is determined almost exclusively by the confining well depth.

The critical value of $\eta$ can be estimated from the above equations and the expression for the collision rate (equation \ref{eq:collisionRate}).
When the antiprotons are `cold', $\tau_{col}$ is on the order of $100~\mu \mathrm{s}$, and if we consider $\tau_{ev} = 10~\mathrm{s}$ `stopped', then this procedure yields a value of $\eta \sim 9$.
Given the simplicity of the model, this is in reasonable agreement with the measured slope (12) in figure \ref{fig:EVCTimeCompare}.

At the shallowest well depths, this relationship begins to break down, as other factors come into play -- the particles' space charge, or noise-induced fluctuations in the confining potential, for instance.

\subsection{Cooling efficiency}

\begin{figure}[hbt!]
	\centering
	\input{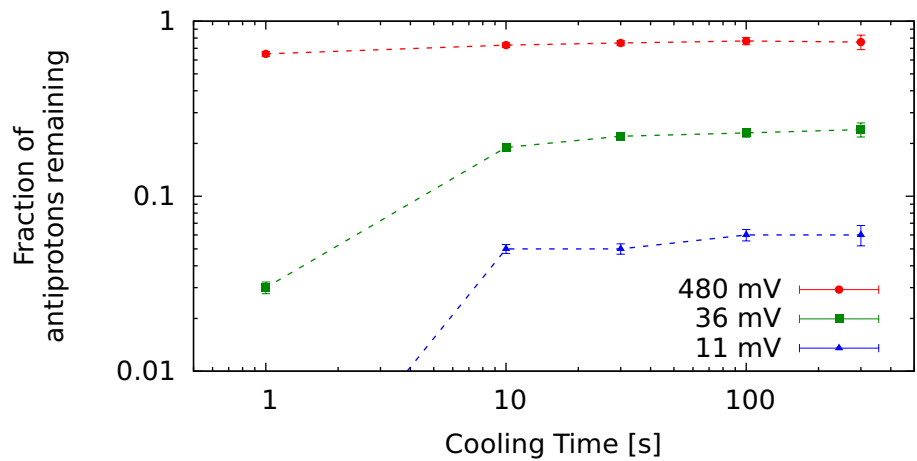}
	\caption{The number of antiprotons remaining at three different well depths, as a function of the cooling time.}
	\label{fig:EVCFracRemaining}
\end{figure}

One of the quantities that describes the performance of evaporative cooling is the cooling efficiency, or the fraction of particles that must be removed to achieve a given temperature.
It can be seen from figure \ref{fig:EVCFracRemaining}, that, as long as the cooling ramp takes place over a long enough time (more than 10~s, in this case), that the efficiency is approximately independent of the cooling time.
However, for shorter cooling times, the number of antiprotons that survive to the end of the ramp is significantly smaller.

From equation \ref{eq:EVC_DTDN}, it is clear that, in a similar fashion to the temperature, the efficiency is determined primarily by $\eta$, and as we have discussed, $\eta$ is an essentially constant parameter for a wide range of plasma parameters.

\begin{figure}[hbt!]
	\centering
	\input{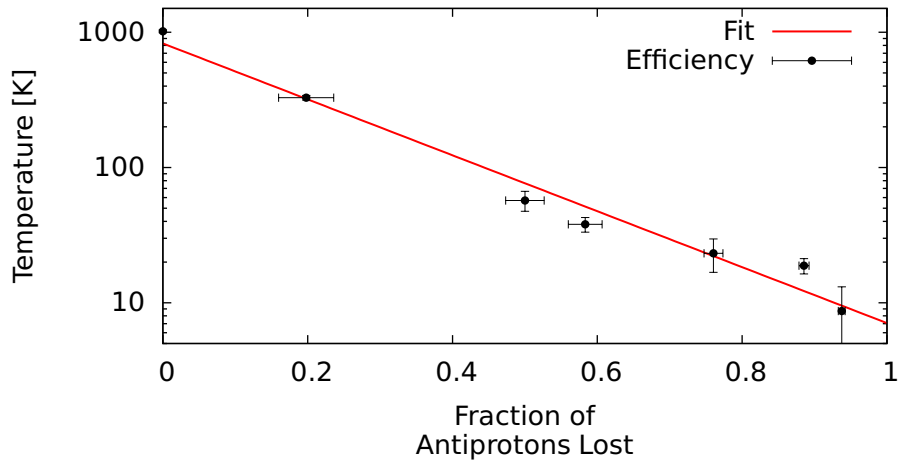}
	\caption[The temperature of the antiproton distribution as a function of the number of particles lost from the well.]{The temperature of the antiproton distribution as a function of the number of particles lost. The line is an empirical fit to the data. }
	\label{fig:EVCCoolingEff}
\end{figure}

Data showing the efficiency can be seen in figure \ref{fig:EVCCoolingEff}.
The data is approximately described by a function $T/T_0 = \mathrm{exp}\left( -\frac{1}{0.20}\frac{N_0-N}{N_0}\right)$.
Over the range of applicability, this implies that each factor of ten drop in temperature requires loss of $\sim 46\%$ of the particles.
The minimum achievable temperature in this model (when $N=0$) is $\sim$ 7~K, in good agreement with the minimum observed temperature of $(9 \pm 4)~\mathrm{K}$.

\subsection{Plasma size} \label{sec:EVCSize}

\begin{figure}[t]
	\centering
	\input{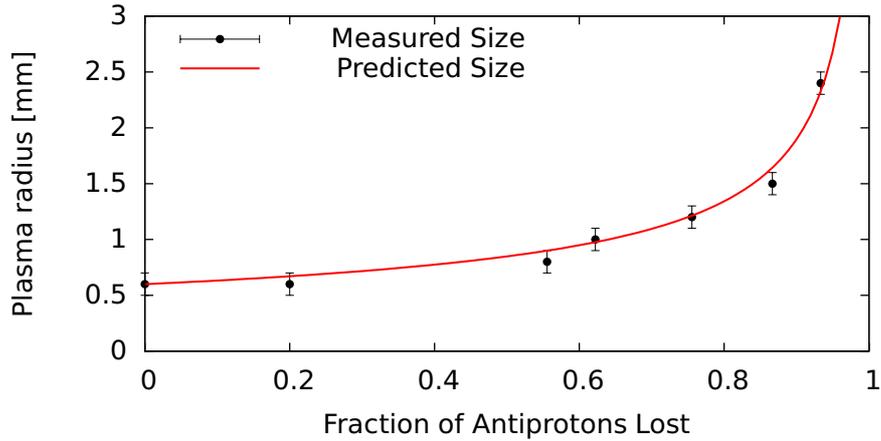}
	\caption[The transverse size of the antiproton cloud as a function of the number of antiprotons lost from the well.]{The transverse size of the antiproton cloud as a function of the number of antiprotons lost from the well. The data is compared to the predictions of equation \ref{eqn:r2conserve}. }
	\label{fig:EVCSize}
\end{figure}

The radial size of the plasma was also measured using the MCP/phosphor device. 
The measurements, shown in figure \ref{fig:EVCSize}, show a dramatic increase in radial size as the plasma evaporatively cools.

This effect stems from the fact that most of the charge  lost from the plasma  escapes from close to the axis.
This occurs because the potential well is modified by the electric charge of the plasma, and is a minimum on the axis.
Since the probability of evaporation scales exponentially with the well depth, most of the charge will be lost from the centre of the plasma.

A plasma that has been `hollowed' by the removal of charge near the centre is not stable and the particles will redistribute to return to an equilibrium configuration.
In doing so, the total angular momentum is conserved, which implies that the mean squared radius, $\langle r^2 \rangle$, is constant (equation \ref{eq:r2Conserve}).
Thus, since some charge must move inwards to replace the charge that has escaped, other particles must move outwards to conserve $\langle r^2 \rangle$.

Assuming that the evaporating particles all leave on the axis, at $r=0$, we can predict the increase in radial size, knowing of the number of particles lost.
This leads to the relationship between the radius and the fraction of particles lost:
\begin{equation} \label{eqn:r2conserve}
	\frac{r}{r_0} = \sqrt{\frac{N_0}{N}},
\end{equation}
where $r_0$ is the radius of the plasma when there are $N_0$ particles.
This model is shown in figure \ref{fig:EVCSize}, and is seen to follow the data very well.

In reality, the particles do not escape exactly from the centre of the plasma, but from an area approximately the size of the Debye length \cite{GormThesis}.
Thus, this effect is most important for collections of particles in the plasma regime, where the Debye length is smaller than the size of the plasma.
All but the warmest distributions discussed here have $\lambda_D$ less than the plasma radius.
 
\subsection{Limits on evaporative cooling}

The factors that limit evaporative cooling can be separated into the technical and the more fundamental physics issues.

As we have seen, the temperature of the particles follows a simple relationship with the confining well depth.
It might therefore be expected that the lowest temperature achievable is set by the minimum well depth that can be constructed.

Constructing very shallow potential wells is  quite technically challenging.
Firstly, electronic noise, which is present to some extent on all electrical systems, causes fluctuations in the voltages applied to the Penning trap electrodes.
This can cause the confining barrier to the side of the potential well to be lowered, allowing some of the particles to escape.
These particles will carry away a smaller amount of energy than an evaporating particle, thus reducing the efficiency of the evaporative cooling.
The stability of the amplifier chain over the course of the experiment can be important.
Small drifts in the amplifier gain or offset due to, for example, temperature changes, can influence the depth of the well confining the particles.
To reduce this effect, an engineering study must be made to improve the noise and stability of the amplifier chain.
Ultimately, however, it is not possible to remove all noise from a real system, or to make a perfectly stable system, so the amplifier performance will set one of the limits on the lowest temperatures achievable.

As the wells become very shallow, the accuracy of our knowledge of the electric potential close to the axis becomes important.
Without considering space charge effects, the leading sources of error are likely to be related to how well the geometry of the trap electrodes is known, and to what precision the electrodes have been machined.
The accuracy of the calculation is estimated to be on the order of one part in $10^5$ \cite{Chukman_elog9961}.
A cooling well must be offset from ground to ensure that an evaporating particle will completely leave the trap, and in the measurements described here, are offset by $\sim$ 10~V.
This yields an uncertainty on the well depth of $\sim 1~\mathrm{mV}$.
Setting well depths more accurately than this cannot be done just from calculations of the potential, but may be possible if the well depths are calibrated experimentally.

When particles are present, the degree to which their electric fields deform the well is important.
If the space charge of the particles is not negligible compared to the depth of the well, it must be taken into account.
The space charge can be calculated by solving the Poisson-Boltzmann equation, as discussed in section \ref{sec:PlasmaEquilibrium}
This calculation depends on prior measurements of the number of particles and their transverse density distribution to be accurate.
The reproducibility of these numbers is typically on the order of 5-10\%, and the space charge for antiproton plasmas in the typical ALPHA operating range are $\sim$ 10~mV, so this introduces an uncertainty on the well depth of order 1~mV.
Development of improved stability of the number and geometry of the particles is ongoing.

The physics effects stem from changes in $\tau_{col}$ as cooling proceeds.
If $\tau_{col}$ increases by too much, $\eta$ (and thus $\alpha$) must be reduced so that $\tau_{ev}$ remains reasonably small.
A smaller value of $\alpha$ reduces the cooling efficiency (equation \ref{eq:EVC_DTDN}).

$\tau_{col}$ increases during evaporative cooling because it is a function of temperature and density.
The density falls because the antiproton number is decreasing, and secondly because the cloud radius is expanding.
As discussed in \cite{GormThesis}, $\tau_{col}$ falls as a function of temperature, until the plasma undergoes a transition from magnetised to unmagnetised.
As already mentioned in section \ref{sec:collisions}, when the plasma becomes magnetised, the collision rate drops precipitously.
For antiprotons in a 1~T magnetic field, the magnetisation parameter, $\bar\kappa = 1$ corresponds to a temperature of $\sim 8.5~\mathrm{K}$.
It is notable, but probably somewhat coincidental, that this temperature is quite close to the minimum achieved temperature in the evaporative cooling measurements described above -- ($9 \pm 4$)~K.

The last effect is expansion-induced heating of the plasma because of the plasma's radial expansion.
The radial redistribution of charge converts some of the electrostatic potential energy to kinetic energy -- from equation \ref{eq:potentialEnergy}, the potential energy $U$ is a decreasing function of $r$.
Expansion therefore introduces a heating term in competition with evaporative cooling which increases the minimum achievable temperature.
The magnitude of the heating term has been calculated to be equivalent to a $\sim$ 5~mK temperature rise for each antiproton lost \cite{ALPHA_EVC}.
This power is quite low compared to the observed value of $\mathrm{d}T/\mathrm{d}N$, so it is not likely to impose a limit on the temperature above $\sim 0.5~\mathrm{K}$.

\subsection{Evaporative cooling on other species}
\label{sec:LeptonEVC}

The evaporative cooling techniques described in this section are not restricted to antiprotons -- they can also be applied to other species.
Directly relevant to ALPHA is the possibility of reducing the temperature of the positron plasma used to produce antihydrogen -- why this is important will be discussed in chapter \ref{chp:hbar}.

\begin{figure}[h]
	\centering
	\subfloat[]{\input{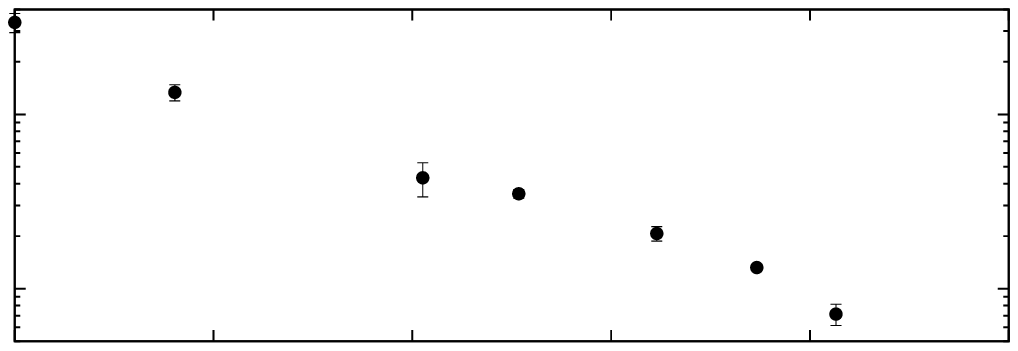}}\\
	\subfloat[]{\input{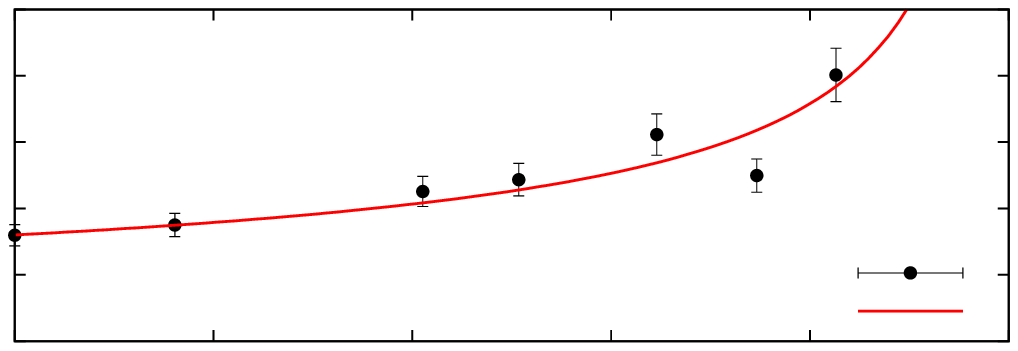}}
	\caption[The temperature and radius of a positron plasma during evaporative cooling.]{The temperature (a) and radius (b) of an evaporatively-cooled positron plasma as a function of the number of positrons lost. The line in (b) is the prediction of the size using equation \ref{eqn:r2conserve}.}
	\label{fig:posEVC}
\end{figure}

The experimental procedure is almost identical to that followed for antiprotons.
Some example measurements of the positron temperature and radius as a function of the number of positrons remaining is shown in figure \ref{fig:posEVC}.

The densities of positron plasmas are typically in the range $10^7 - 10^8~\mathrm{cm^{-3}}$, much higher than the antiproton densities used.
Combined with the lower positron mass, this gives values of $\tau_{col}$ of $\leq 10^{-5}~\mathrm{s}$.
When compared to the $\sim 10^{-4}~\mathrm{s}$ antiproton collision time, we see that evaporative cooling of the positron plasmas can take place over much shorter times.
The shorter times were in fact found to be \textit{necessary}, as the positron plasma was found to spontaneously re-heat over the course of a few seconds after the end of the cooling ramp.
The fact that this is seen could be linked to the much shorter cyclotron cooling time of the positron, which means that it comes into equilibrium with the external radiation field much faster than an antiproton.

Because of the higher positron density, the space charge of the particles fills a large fraction (sometimes more than 90\%) of the vacuum well depth.
It is thus more difficult to determine the height of the confining barrier -- producing a self-consistent solution to Poisson's equation and the Maxwell-Boltzmann distribution can yield some insight, but the precision of the calculation, and how well the inputs have been determined, makes its usefulness marginal.
The final well depth is chosen experimentally by measuring the particle temperature at various depths.

Positrons enter the magnetised regime at a higher temperature than antiprotons ($\bar{\kappa} = 1$ when T $\simeq$ 100~K at B = 1~T), thus it would be expected that it is not possible to achieve as low temperatures with positrons as with antiprotons.
In fact, the lowest temperatures seen with evaporatively cooled positrons in ALPHA are of order 10-20~K.
The higher collision rate in a positron plasma is suspected to compensate for the magnetisation to some degree.
The behaviour of evaporative cooling acting on numerous lepton plasmas is still under investigation.

\subsection{Conclusions}

Evaporative cooling is a promising technique for cooling antiprotons to lower temperatures than those achievable using electron cooling alone.
The physics behind the process has been studied extensively -- the reader hungry for more details is directed to reference \cite{GormThesis}.

Evaporative cooling is also a new tool in ion-trap physics in general and promises to open the door to study of phenomena not previously accessible.

The combined cooling techniques cool keV-range antiprotons to energies of the order meV, a six order-of-magnitude reduction.
We now turn to the reason for producing cold antiprotons -- to use them to produce cold antihydrogen atoms.

    \chapter{Producing Antihydrogen}
\label{chp:hbar}

\epigraph{Any sufficiently advanced technology is indistinguishable from magic.}{Arthur C. Clarke}

\section{Antihydrogen formation processes} \label{sec:HbarFormationTheory}

At the most basic level, the formation of an antihydrogen atom is the conversion of an initially unbound antiproton and positron pair into a single bound state.
Some mechanism must exist to carry away the excess energy and momentum, be it a third particle, a photon, or a changing field.
The different methods of doing this have very different characteristics and behaviours, which are important to understand and will be described in the following sections.

The most efficient method of producing antihydrogen atoms (in terms of number of atoms produced) demonstrated so far is the interaction of antiprotons with a plasma of positrons.
Combination of clouds or plasmas of antiprotons and positrons has become commonly known as `mixing'.

One of the most important results of a mixing experiment is the rate of antihydrogen formation, typically quoted as a rate per antiproton present.
In ATHENA, rates on the order of $4\times 10^{-2}~\mathrm{s^{-1}}$ were observed \cite{ATHENA_highRate}.
This is similar to the usual rates seen in ALPHA experiments, but rates as high as $1~\mathrm{s^{-1}}$ have been observed for brief periods.
These rates will be compared to the rates predicted by theoretical models of some of the possible antihydrogen formation schemes in a mixing experiment, which are outlined below.

\subsection{Radiative recombination}

A positron incident on an antiproton can be captured into a bound state with the emission of a photon, the time-reversed analogue of the photoionisation process.
Both processes have been extensively studied and are well understood.
A full theoretical treatment can be found in \cite{bethesalpeter}, and an overview of the semi-classical theory is outlined below.

The cross section for capture of a positron into a state with principal quantum number $n$ is given by \cite{bethesalpeter}
\begin{equation}
	\sigma_{RR,n} = \frac{0.48 \pi}{\epsilon_0}\, \frac{e^2 \hbar^2}{m_e^2c^3} \frac{\nu_1^2}{\nu(\nu-\nu_n)} \frac{1}{n^3},
	\label{eq:RadiativeCapture}
\end{equation}
where $\nu$ is the frequency of the emitted photon, $\nu_1$, and $\nu_n$ are the frequencies of photons with just enough energy to ionise an atom in the ground state and the state with principal quantum number $n$ respectively.
Alternatively, this can be written in terms of the kinetic energy of the positron in the rest frame of the antiproton, $E_e$,  as \cite{RouteUltraLowEnergyHBar}
\begin{equation}
	\sigma_{RR,n} = 2 \times 10^{-22}\;\mathrm{cm}^2\;\frac{\mathcal{E}_0}{n E_e\left(1+n^2 E_e/\mathcal{E}_0\right)},
\end{equation}
where $\mathcal{E}_0$ is the binding energy of the ground state (13.6~eV).

An important feature to note  is the inverse dependence of the cross-section on $n$, implying that there is a preference for capture into strongly bound states.
This means that the antihydrogen atoms produced through this method tend to be resistant to reionisation by the electric fields in a Penning trap or through collisions with other particles (see section \ref{sec:fieldionisation}).

The total rate of formation through radiative recombination can be found by integrating the expressions for $\sigma_{RR,n}$ over the distribution of positron energies and all possible quantum states.
This treatment produces an expression for the formation rate per antiproton, $\Gamma_{RR}$, in a thermal distribution of positrons with density $n_e$ and temperature $T$, where the antiproton is at rest \cite{Gabrielse_AntihydrogenProductionTrappedPlasmas}:
\begin{equation}
	\Gamma_{RR} = 3 \times 10^{-11}\;\sqrt{\frac{4.2~\mathrm{K}}{T}}\frac{n_e}{\mathrm{cm^{-3}}}~\mathrm{s^{-1}}.
	\label{eq:GammaRR}
\end{equation}

For a positron density of $10^7~\mathrm{cm^{-3}}$ and a speculative, `best-case' positron temperature of 4.2~K (the temperature of the cooling liquid helium bath), $\Gamma_{RR}$ is approximately $3 \times 10^{-4}~\mathrm{s^{-1}}$ per antiproton.
For the higher temperatures seen in ALPHA $(T \simeq 120~\mathrm{K})$, equation \ref{eq:GammaRR} predicts a rate of approximately $6 \times 10^{-5}~\mathrm{s^{-1}}$ per antiproton.
It should be noted that these rates are much lower than the measured antihydrogen formation rates.

In a magnetic field, the dependence of the cross section for radiative recombination on the positron's kinetic energy is modified by approximately $h e B/ 2 \pi m_e$ \cite{Francis_TopicalReview}.
For a 1~T magnetic field, as that used in ALPHA, this term is approximately equal to $1.3~\mathrm{K} \times k_\mathrm{B}$, negligible compared to the thermal spread of the positron energies, so it can safely be ignored.

As well as the spontaneous process described above, radiative recombination can be stimulated by the presence of another photon.
In principle, therefore, it should be possible to increase the rate of radiative recombination by increasing the photon density, allowing for \textit{stimulated} radiative recombination.
The increase in production rate should be approximately \cite{Francis_TopicalReview}
\begin{equation}
\frac{1}{8\pi h c} \frac{I\lambda^3}{\delta \nu},
\end{equation}
where $I$ is the laser intensity, $\lambda$ is the laser wavelength and $\delta \nu$ is the spread of the laser frequency.
An experiment using a carbon dioxide laser ($\lambda=11~\mu\mathrm{m}$, $I = 160~\mathrm{W\,cm^{-2}}$) was performed by ATHENA to search for this effect \cite{ATHENA_laser}, but did not identify any increase in the rate.
This null result, in addition to the fact that the observed antihydrogen formation rates are much higher than the rates calculated from equation \ref{eq:GammaRR} is strong evidence that the dominant formation process is not through radiative recombination.

\subsection{Three-body recombination}

In the context of antihydrogen production, `three-body recombination' refers to the process where two positrons scatter in the field of an antiproton, with one becoming bound, and the other carrying away the excess energy.

In the conceptually simplest analysis, the steady-state, equilibrium rate of formation is given by \cite{Francis_TopicalReview}
\begin{equation}
	\Gamma_{TBR} = C n_e^2 v b^5,
	\label{eq:GammaTBR1}
\end{equation}
where $b$ is the classical distance of closest approach (the distance where the energy associated with the Coulomb interaction equals the positron's thermal energy, $b = e^2/\left( 4 \pi \epsilon_0 k_\mathrm{B} T \right)$, and $C$ is a numerical constant.
The equation can be understood as the flux of positrons into the cross section $b^2$, $\sim n_evb^2$, times the probability of finding another positron in the same volume, $\sim n_eb^3$.

The constant C includes the dependence on the magnetic field, which is much more important for three-body recombination than for radiative recombination.
In the limit $B \rightarrow 0$, $C \simeq 0.76$ \cite{TBR_Beq0}, while in the strong field limit, $B \rightarrow \infty$, $C \simeq 0.07$ \cite{TBR_Binf}, an order of magnitude reduction.
In the range of the original ATHENA and ATRAP experiments, $C \simeq 0.1$ \cite{Francis_TBR}.

We can substitute for some of the variables in equation \ref{eq:GammaTBR1} to obtain the rate as a function of the experimentally measured parameters -- the positron temperature $T$ and density $n_e$ -- and obtain \cite{Gabrielse_AntihydrogenProductionTrappedPlasmas}
\begin{equation}
	\Gamma_{TBR} = 8 \times 10^{-12}\,C\,\left(\frac{4.2~\mathrm{K}}{T}\right)^{9/2} \left(\frac{n_e}{\mathrm{cm^{-3}}}\right)^2\;\mathrm{s}^{-1}.
	\label{eq:GammaTBR}
\end{equation}

This equation predicts a value of $\Gamma_{TBR}$ per antiproton of approximately $80~\mathrm{s^{-1}}$, assuming a positron density of $10^7~\mathrm{cm^{-3}}$, a temperature of 4.2~K and $C=0.1$.
At the higher, 120~K temperatures, $\Gamma_{TBR}$ falls to $\sim 2 C \times 10^{-4}~\mathrm{s^{-1}}$, a huge change in rate driven by the $T^{-9/2}$ dependence, and brings the predicted rate closer to, or even lower than the radiative rate.

It should be mentioned that a third term must also be included when calculating the total rate.
The term stems from the competing collisional excitation and radiative de-excitation of atoms after they are formed \cite{RouteUltraLowEnergyHBar}, and is given by 

\begin{equation}
	\Gamma_{coll} = 3 \times 10^{-10}\,\left(\frac{4.2~\mathrm{K}}{T}\right)^{2.18} \left(\frac{n}{\mathrm{cm^{-3}}}\right)^{1.37}\;\mathrm{s}^{-1}.
\end{equation}
At 4~K and $10^7~\mathrm{cm^{-3}}$, $\Gamma_{coll}$ has a value of approximately $1~\mathrm{s}^{-1}$, while for a temperature of 120~K, it is approximately  $8 \times 10^{-5}~\mathrm{s}^{-1}$.

\subsection{Temperature dependence} \label{sec:HbarTemperature}

\begin{figure}[h]
\centering
\input{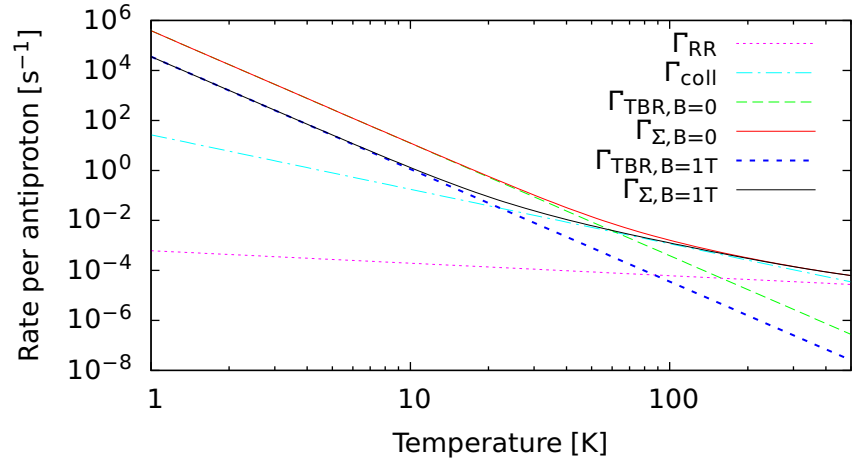}
\caption[Antihydrogen formation rates.]{The antihydrogen formation rates for the radiative recombination (RR), three-body recombination (TBR), and collisional (Coll) processes. The TBR rate is shown for both the magnetic-field free (B=0) and B=1~T cases, as is the total rate ($\Gamma_{\Sigma}$). }
\label{fig:HbarRates}
\end{figure}

The scaling of the antihydrogen production rate with the positron plasma's temperature for the production mechanisms described in the previous sections are plotted in figure \ref{fig:HbarRates}. 

\begin{figure}[hbt!]
\centering
\input{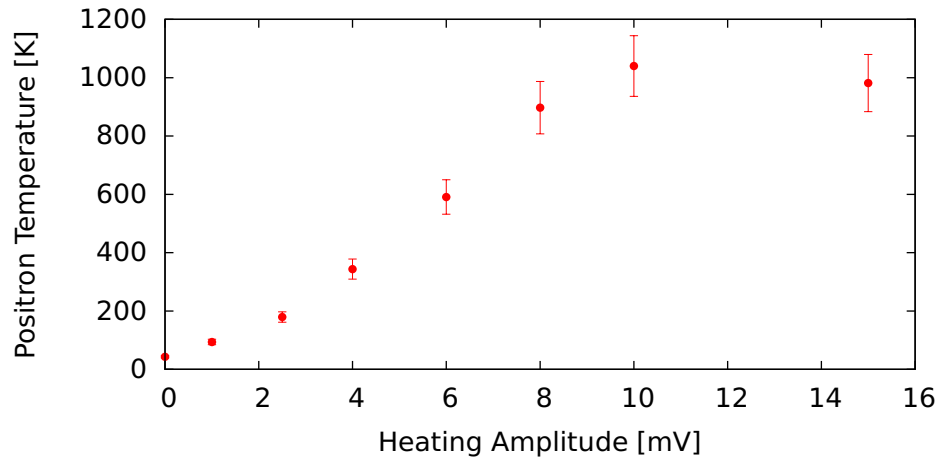}
\caption[The positron plasma temperature as a function of the amplitude of a heating drive.]{The positron plasma temperature as a function of the amplitude of the heating drive. At high temperatures, the plasma begins to evaporatively cool (section \ref{sec:EVC}). The point measured at 15~mV was accompanied by positron loss.}
\label{fig:HeatingAmpl}
\end{figure}

The temperature scaling has been experimentally investigated using a pulsed radio-frequency source to heat the positron plasma .
A trapped plasma can be heated using a radio-frequency electric field to excite the axial centre-of-mass motion of the plasma (the sloshing mode described in section \ref{sec:Modes}).
The kinetic energy transferred from the drive is quickly redistributed to other components of the particles' motion, and the temperature is increased.
To transfer energy efficiently, the drive is repeatedly swept through a frequency band centred on the frequency of the sloshing mode.
The amplitude of the drive can be altered to vary the temperature achieved -- an example measurement of the plasma temperature as a function of drive amplitude is shown in figure \ref{fig:HeatingAmpl}.

\begin{figure}[b!]
\centering
\input{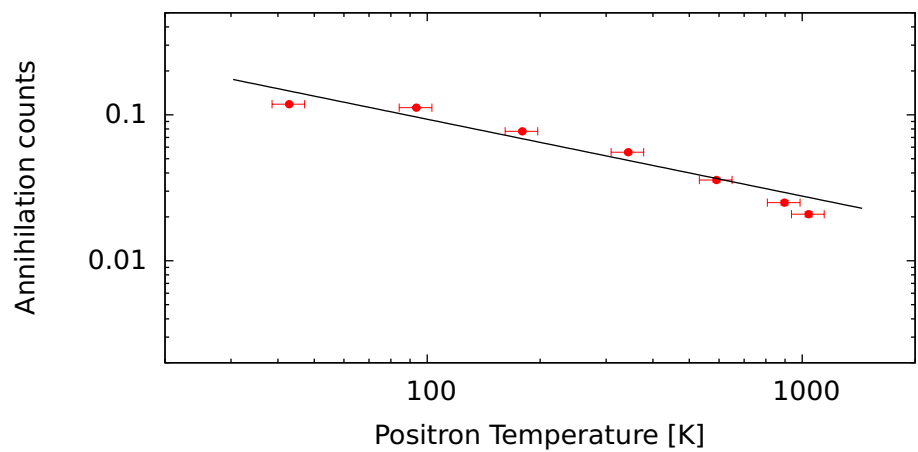}
\caption[The number of annihilation counts as a function of the positron plasma temperature.]{The number of annihilation counts per antiproton during a 1~s mixing period as a function of the positron temperature. The line is a fit of the form $a\,T^{-b}$, with the best-fit value of $b = 0.5 \pm 0.1$. The highest-temperature point from figure \ref{fig:HeatingAmpl} is not shown.}
\label{fig:HeatingRate}
\end{figure}

The number of annihilation counts (indicative of antihydrogen formation) recorded during a 1~s mixing experiment at each temperature is shown in figure \ref{fig:HeatingRate}.
The result of the power-law fit to the data yields an exponent of $-0.5 \pm 0.1$, which is incompatible with the predictions of the simplified theory discussed above.
A previous measurement monitored the antihydrogen signal as the positron plasma radiatively cooled after a burst of radio-frequency heating \cite{ATHENA_OnOff}.
This investigation found  a temperature dependence of $T^{-1.1 \pm 0.5}$ over a temperature change of 1500~K, which is in approximate agreement with the data shown here.
The form of the dependence of the formation process on temperature is not well-understood, and is an active area of study.

Heating the positron plasma to suppress antihydrogen formation can act as a `null' experiment to show that a signal is due to antihydrogen, as none of the other parameters of the plasmas or particle dynamics are affected (Though the positron plasma's Deybe length changes somewhat, altering the density distribution, it remains small compared to the size of the plasma).
This was first established in \cite{ATHENA_Nature} and also used in ALPHA \cite{ALPHA_Nature}.

There is clear disagreement between the size and scaling of the antihydrogen production rates predicted in figure \ref{fig:HbarRates} and those observed experimentally.
In the experiments, high rates are observed at warmer temperatures than the predicted scaling implies is possible.
A possible cause of this is the fact that the three-body recombination process in the experiments does not completely resemble the equilibrium process described above.
Instead of antihydrogen formation occurring in an infinite, uniform density positron plasma, in experiments, the formation process is interrupted or `arrested' when the atom leaves the plasma \cite{FrancisSimulationsHbarFormation}.
This can significantly affect the rate of formation and the binding energies of the atoms.
This will be discussed in more detail in section \ref{sec:NestedWell}.

\subsection{Binding energies}
\label{sec:HbarBE}

The distribution of binding energies of atoms produced through three body recombination is an equilibrium between collisional excitation and collisional and radiative de-excitation.
In global thermal equilibrium, the distribution can be determined from the general form of the Boltzmann distribution 
\begin{equation}
	N(\mathcal{E})  \propto g(\mathcal{E})\, \mathrm{exp}\left( - \frac{\mathcal{E}_0 - \mathcal{E}}{ k_\mathrm{B}T} \right),
	\label{eqn:Boltzmann}
\end{equation}
where $g(\mathcal{E})$ is the density of states with binding energy $\mathcal{E}$, and $\mathcal{E}_0$ is the binding energy of the ground state, 13.6~eV.
$g(\mathcal{E})$ can be determined from the distribution of Bohr energy levels $\mathcal{E} = \mathcal{E}_0 / n^2$, and the degeneracy of states with principal quantum number $n$, $P(n) = 2n^2$.
For these purposes, one can treat $n$ as a continuous variable, and this leads to 
\begin{align}
	g(\mathcal{E}) & =  P(n) \,/\, \frac{\mathrm{d}\mathcal{E}}{\mathrm{d}n} \notag \\
	& \propto  n^2 \,/\, n^{-3} \\
	& \propto  \mathcal{E}^{-5/2}. \notag
\end{align}
Using equation \ref{eqn:Boltzmann}, we obtain that
\begin{equation}
	N(\mathcal{E})  \propto \mathcal{E}^{-5/2} \mathrm{exp}\left( - \frac{\mathcal{E}_0 - \mathcal{E}}{ k_\mathrm{B}T} \right).
\end{equation}

Antihydrogen atoms formed through three-body recombination are initially bound by approximately the kinetic energy of the positrons.
Inelastic collisions between the atom and other positrons changes the binding energy, either up or down.
Monte-Carlo simulations of the process show the presence of a `bottleneck' at a binding energy a few times the positrons' thermal kinetic energy \cite{bottleneck}.
Atoms bound by less than this amount are fragile, and are susceptible to reionisation by further collisions.
If atoms make it to binding energies deeper than the bottleneck, they are resistant to ionisation and will tend to be long-lived.
In arrested mixing, many of the atoms do not have enough time to pass through the bottleneck before leaving the positron plasma, and are bound by only a few times $k_\mathrm{B}T$.

At the moment, no complete model exists that fully describes antihydrogen formation during antiproton-positron mixing.
It is likely that the system is simply too complex for analytic techniques to be effective, though partial or computational approaches (see \cite{Francis_TopicalReview} for an overview) have been successful in describing some observations -- for example, the measured binding energy distribution in \cite{ATRAP_velocityReinterp}.

\subsection {Charge transfer}

A conceptually different antihydrogen production mechanism involves transfer of a positron from an atom of positronium, the bound state of an electron and a positron, to an antiproton \cite{Mike_positronium}.
This method has been demonstrated by the ATRAP experiment, \cite{ATRAP_chargeXchange} and the AEgIS collaboration \cite{AEGIS} also plans to make use of it to produce antihydrogen for gravitational measurements.
Antihydrogen production via positronium charge exchange has the advantage that it is, in principle, possible to produce positronium in defined atomic states, resulting in antihydrogen atoms in selected states.
Positronium can be produced either through charge exchange reactions of positrons with atoms that have been laser-excited into specific internal states \cite{ATRAP_chargeXchange}, or by collisions of positrons with a specially-prepared surface \cite{RouteUltraLowEnergyHBar}.
This method has not been used in ALPHA, so it will not be discussed further.

\subsection{Other processes}

As well as the antiproton-positron and antiproton-positron-positron reactions discussed above, processes involving different combinations of particles are also possible.
For instance, the three-body interaction $\pbar + \pbar + e^+ \rightarrow \Hbar + \pbar$, is possible, but is suppressed by the large ratio of positron density to antiproton density used in the experiments.

Some more exotic schemes for producing antihydrogen have also been discussed in the literature. 
These include using a pulsed electric field to dynamically modify the potential around an antiproton to trap a positron \cite{PulsedFieldRecombination} and interacting antiprotonic helium with positrons \cite{Hbar_pbarHe1} \cite{Hbar_pbarHe2}. Antiprotonic helium is the metastable bound state of a helium nucleus, an antiproton, and an electron.
Such schemes have not been the focus of much experimental effort, and are only mentioned here for completeness.

\section{The nested well}
\label{sec:NestedWell}

\begin{figure}[htb]
\centering
\input{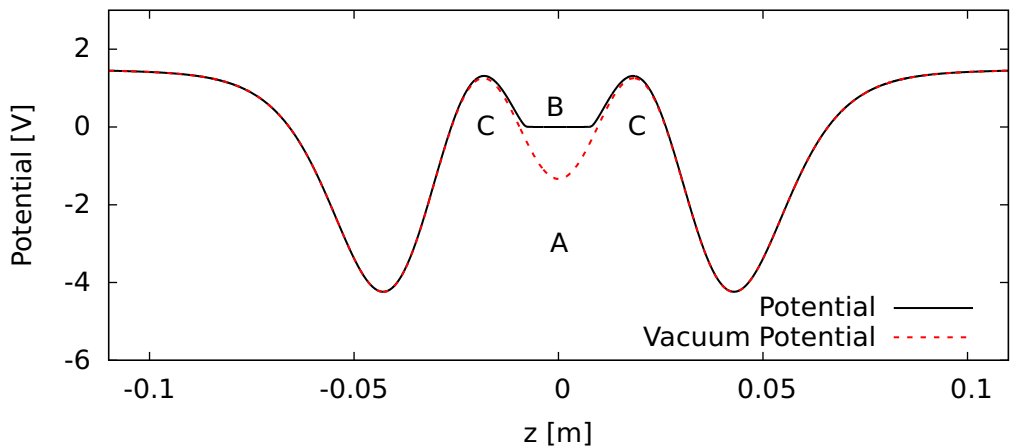}
\caption[An example of a nested well.]{An example of a nested well for antihydrogen production, showing the different regions, marked A, B and C, that are described in the text. The curve marked `potential' shows how the vacuum potential is deformed when a positron plasma is placed in the inner well.}
\label{fig:nested}
\end{figure}

To efficiently produce antihydrogen in a Penning trap, antiprotons and positrons must be confined in the same volume of space.
This is not trivial, since a potential well that confines species of one charge will repel species of the opposite charge.
However, using a `nested well' arrangement \cite{Gabrielse_AntihydrogenProductionTrappedPlasmas}, simultaneous confinement is possible.
The nested well, an example of which is shown in figure \ref{fig:nested}, comprises a long, outer well for species of one sign of charge (marked as A), with a shorter inner well of the opposite sign in the centre (B).
To either side of the inner well, the outer well extends into deeper `side-wells', marked as C.

Both polarities of the well are possible, with either positive (positrons) or negative (antiprotons) particles in the centre.
The kinetic energy of the antihydrogen atom will be dominated by the kinetic energy of the antiproton from which it is formed, so it may be advantageous to place the antiprotons in the centre, where they do not have to be excited in the mixing procedure.
However, the lighter positrons cool quickly in the magnetic field (3.9~s in a 1~T field or 0.43~s in the 3~T fields used in the original ATHENA experiments), so the particles can interact for only a short amount of time before the positrons would cool into the side wells.
In addition, collisions between positrons in the outer well would quickly transfer kinetic energy between the parallel and perpendicular degrees of freedom.
This would cause the average parallel energy to reduce, trapping the positrons in the side wells more quickly, and would increase the temperature of the distribution, reducing the antihydrogen production rate.
For these reasons, most -- if not all -- experiments have used the configuration where the positrons are placed in the centre of the well, and we will concentrate on this case.

To induce the antiprotons and positrons to interact, the antiprotons must have enough kinetic energy relative to the bottom of the side wells to pass through the positron plasma.
Once this is achieved, the antiprotons oscillate in the region marked `A' in figure \ref{fig:nested}.
The process of preparing antiprotons with an appropriate energy distribution is called 'injection' of the antiprotons.

\begin{figure}[bt]
\centering
\includegraphics[width=\textwidth]{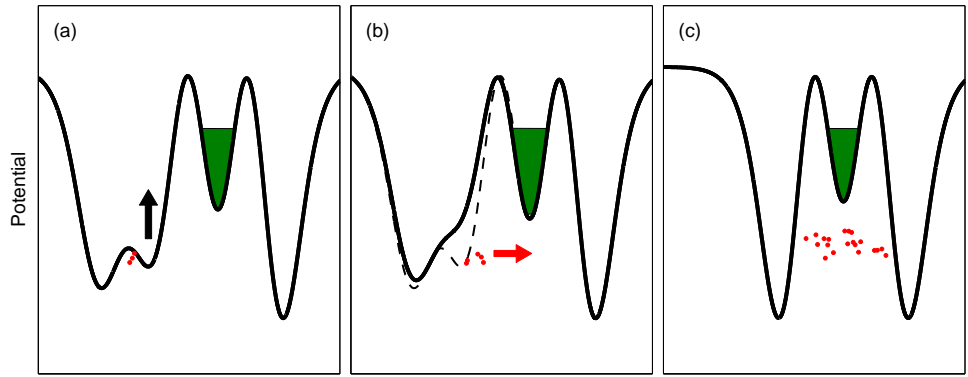}
\caption[An illustration of the procedure to inject antiprotons into the positron plasma with high energy.]{An illustration of the procedure to inject antiprotons into the positron plasma with high energy. The antiprotons (red dots) and positron plasma (green) are prepared in wells similar to those shown in (a). The antiprotons are released into the plasma in (b). In (c) the antiprotons move back and forth in the outer well, passing through the positron plasma.}
\label{fig:ATHENAinject}
\end{figure}

The simplest method to achieve energetic antiprotons is to release them into the nested well from a higher potential, as illustrated in figure \ref{fig:ATHENAinject}.
This is the scheme used in the first antihydrogen production experiments \cite{ATHENA_Nature}.
Typical injection energies ranged from volts to tens of volts.

As the antiprotons pass through the positrons, they exchange energy via the Coulomb interaction, and the high-energy antiprotons are cooled by the positrons.
In ATHENA, it took the antiprotons a few tens of milliseconds to have their energy reduced from approximately 30~eV above the positron space charge level to close to thermal equilibrium with the positrons as they passed through the plasma \cite{ATHENA_Dynamics}.
This timescale agrees with numerical calculations of the average `drag' force on an antiproton passing through a positron column, with a cooling timescale of approximately 4~ms predicted for ATHENA-like parameters \cite{Francis_slowing}.
These results imply that antihydrogen is formed by antiprotons that have reached, or are at least close to, thermal equilibrium with the positron plasma.

However, later analysis of the shape of the spatial annihilation distribution could not explain the measurements without assuming that the antihydrogen velocity distribution was non-isotropic, and characterised by separate temperatures in the degrees of freedom parallel and perpendicular to the magnetic field \cite{ATHENA_SpatialDistribution}.
The motion parallel to the magnetic field was characterised by a higher temperature, associated with a higher energy, and implied that the antiprotons formed antihydrogen before thermalising with the positron plasma.
For this reason, new antihydrogen production techniques had to be identified to produce antihydrogen at the cryogenic temperatures needed to effect trapping.

The ATRAP collaboration also measured the axial velocity of the antihydrogen atoms using an oscillating electric field to selectively ionise slow-moving atoms \cite{ATRAP_velocity}.
Their measurements indicated that the atoms were moving approximately twenty times faster than would be expected if they were in thermal equilibrium with the positron plasma.
However, some later reanalysis attributed this result to `a surprisingly effective charge-exchange mechanism', where antihydrogen atoms moving along the axis (the only atoms that were detected) underwent a charge-exchange interaction with a fast-moving antiproton in the side-wells of the nested trap, resulting in a high-energy atom \cite{ATRAP_velocityReinterp}.
Thus, it is possible, though not demonstrated, that the antiprotons and positrons were in thermal equilibrium during recombination.

\section{Detection of antihydrogen atoms}

\subsection{Annihilation vertex imaging} \label{sec:hbarimaging}
The first cold antihydrogen atoms were identified using a position-sensitive detector to search for annihilation signals from antiprotons and positrons.
This used a detector similar in many respects to the ALPHA annihilation detector described in section \ref{sec:siliconDetector}, but also included a component sensitive to the gamma rays produced in a positron annihilation.
The observation of positron and antiproton annihilations from the same point in space at the same time was indicative of the presence of an antihydrogen atom to a high degree of certainty.

\begin{figure}[b!]
	\hspace{-0.5cm}
	\subfloat[]{\input{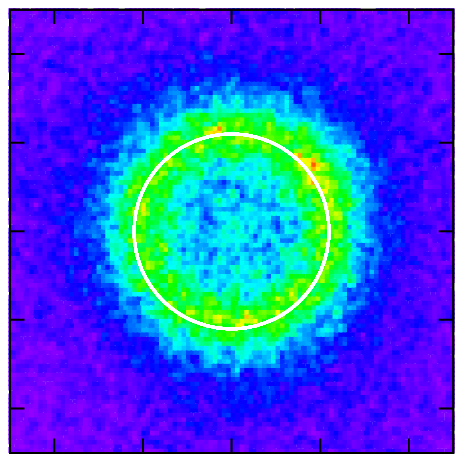}}
	\hspace{-1cm}
	\subfloat[]{\input{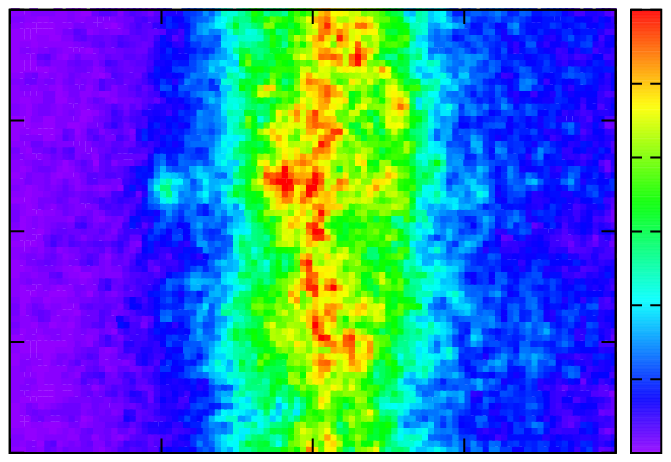}}
	\caption[Antiproton annihilation vertices measured during antihydrogen production.]{Colour-density maps of antiproton annihilation vertices measured during antihydrogen production, shown in (a) $x-y$ and (b) $z-\phi$ projections. The white circle in (a) marks the inner surface of the Penning trap electrodes, at r=22.5~mm.}
	\label{fig:hbarvertexdist}
\end{figure}

This scheme suffered from the low efficiency of gamma photon detectors -- typically only 20\% of positron annihilations were detected \cite{ATHENA_highRate}.
This limited the overall rate at which antihydrogen atoms were identified.
The fraction of antiproton annihilations detected, however, can be much larger -- up to 80\% .
ATHENA established that antiproton annihilations alone were reliable indicators of antihydrogen atoms if they produced an azimuthally smooth `ring' of annihilations at the inner surface of the Penning-Malmberg trap electrodes \cite{ATHENA_Imaging}.
This occurs because a neutral antihydrogen atom is not influenced by the electric or magnetic fields of the Penning-Malmberg trap and travels in a straight line until it encounters matter --  usually the trap electrodes, and annihilates.
The cylindrical symmetry of the experiment ensures that no azimuthal direction is preferred.

An example of this type of distribution, measured in ALPHA, is shown in figure \ref{fig:hbarvertexdist}.
The distribution is not completely smooth, and is contaminated by, for example reconstructed cosmic rays passing through the detector, annihilations in the centre of the ring corresponding to antiproton annihilations on residual gas atoms in the trap, and the escape of bare antiprotons.
However, the dominant component is the uniform ring structure.

\begin{figure}[hbt]
	\hspace{-0.5cm}
	\subfloat[]{\input{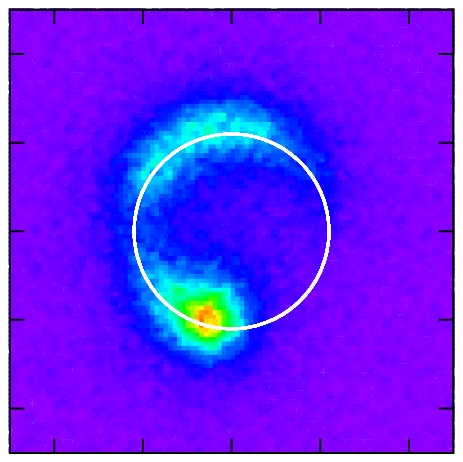}}
	\hspace{-1cm}
	\subfloat[]{\input{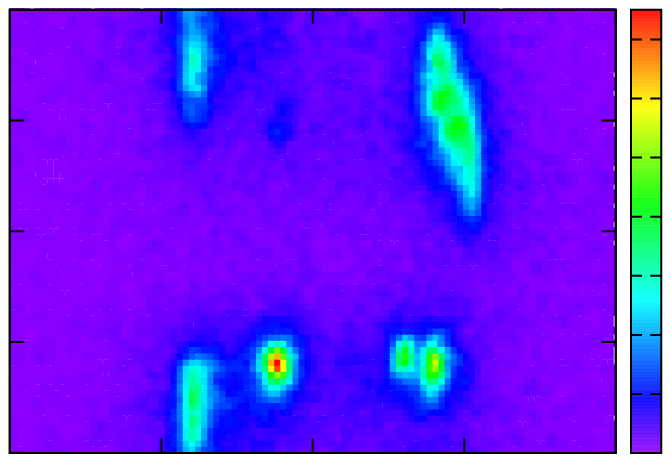}}
\caption[Antiproton annihilation vertices from the escape of bare antiprotons.]{Colour-density maps of antiproton annihilation vertices measured during the escape of an unstable antiproton cloud, shown in (a) $x-y$ and (b) $z-\phi$ projections.}
\label{fig:hotspots}
\end{figure}

In contrast, it was demonstrated that uncombined, or `bare', antiprotons became unconfined from the trap in very localised areas, called `hot-spots'.
This is interpreted to be due to the influence of imperfections in the trap construction. 
Small misalignments or manufacturing errors can result in electric fields that break the cylindrical symmetry of the trap.
These imperfections will, in general, not be azimuthally symmetric and tend to distort the antiproton trajectories so that they preferentially first intersect the electrode wall at a certain point.
An example of the vertex distribution obtained from antiproton escape is seen in figure \ref{fig:hotspots}, where a cloud of trapped antiprotons has been deliberately destabilised to demonstrate the effect.
The symmetry-breaking electric fields do not act on the neutral antihydrogen, and the imperfections do not have a significant effect on the `ring'.

Annihilation vertex imaging is a powerful technique that allows extremely sensitive discrimination between antiproton annihilations and any background contributions -- this is discussed in detail in chapter \ref{chp:trapping}. 
Analysis of the spatial annihilation distributions can also give valuable insight into the physical processes at work. 
An analysis making use of the vertex distributions can be found in chapter~\ref{chp:vertex}.

At high rates of antihydrogen production, where annihilations occur faster than the maximum rate at which the imaging detector can be read-out, the measured rate saturates, and is not useful when determining the absolute or relative numbers of antihydrogen atoms produced in a given experiment.
In such cases, we have found that it is acceptable to use the digital trigger used to signal the detector hardware that data is available as a proxy for vertices.
This signal has a maximum rate in excess of 200~kHz, whereas the detector can only be read out at around 100-500~Hz.
However, this introduces the possibility that some of the counts are due to background processes, such as cosmic rays or uncombined antiprotons leaving the trap.
Still, the fraction of annihilations that are reconstructed can be assumed to be a representative sample of the trigger signals and the spatial distribution can still be used to confirm the presence of antihydrogen, and to measure any other contributions to the signal.

\subsection{Field-ionisation}
\label{sec:fieldionisation}
A second method of identifying antihydrogen takes advantage of the fact that the binding force between the positron and antiproton can be overcome by an external electric field.

\begin{figure}[hbt]
\centering
\input{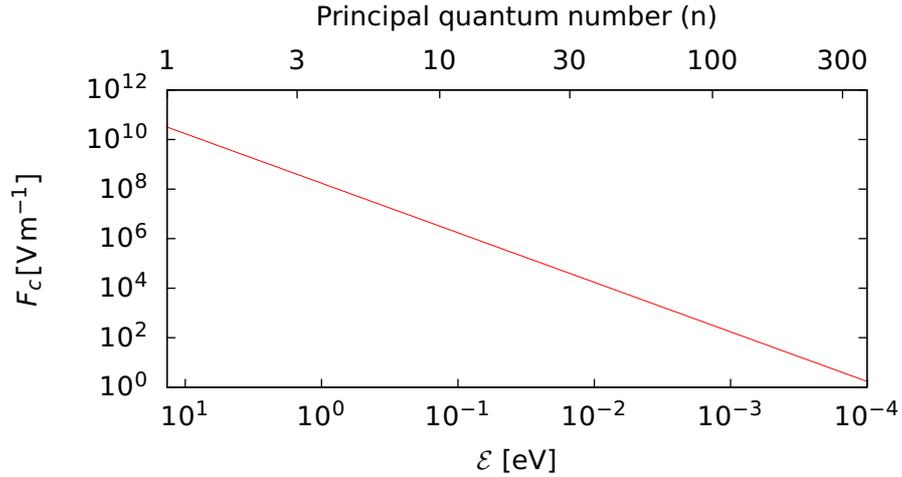}
\caption[The critical electric field required to ionise an atom as  a function of the binding energy.]{The critical electric field required to ionise an atom as  a function of the binding energy. The principal quantum number (for the same binding energy in the absence of electric and magnetic fields) is also shown.}
\label{fig:critfield}
\end{figure}

The classical electric potential around an antiproton in the presence of an electric field $\mathbf{F} = F \hat{z}$, and in a region free of magnetic field, is \cite{bethesalpeter}
\begin{equation}
	\Phi = - \left(\frac{e}{4 \pi \epsilon_0} \right) \frac{1}{r} - F z,
	\label{eqn:classicalPotential}
\end{equation}
where $r$ is the separation between the nucleus (antiproton) and the positron.
The saddle point of the potential has height
\begin{equation}
\Phi_{s} = - 2 \sqrt{\frac{e F}{4 \pi \epsilon_0}}.
\end{equation}
Thus, at binding energies lower than $-e\Phi_s$ , the atom can be ionised.
This defines a minimum, or critical, electric field $F_c$ beyond which the atom may be ionised, given by
\begin{equation}
F_c = \frac{\pi \epsilon_0}{e^3}\mathcal{E}^2,
\label{eq:Fc_epsilon_au}
\end{equation}
where $\mathcal{E}$ is the binding energy.
The critical field as a function of the binding energy is shown in figure \ref{fig:critfield}.
For binding energies ranging from the $k_\mathrm{B}T$ scale expected from three-body recombination to the antihydrogen ground state, the values of critical field range from the extremely strong -- $10^{10}~\mathrm{V\,m^{-1}}$, to relatively weak $\sim 10^2~\mathrm{V\,m^{-1}}$.
The electric field in the ALPHA Penning trap is typically of order $10^3$ -- $10^4~\mathrm{V\,m^{-1}}$ over most of the volume of the trap.

The binding energy for an atomic state with principal quantum number $n$ (in the absence of electric and magnetic fields) is 
\begin{equation}
 \mathcal{E}_n = \left(\frac{m_e e^4}{8 \epsilon_0^2 h^2}\right) \, \frac{1}{n^2},
\end{equation}
allowing equation \ref{eq:Fc_epsilon_au} to be written in terms of $n$ as
\begin{equation}
F_c =  \left( \frac{\pi m_e^2 e^5}{64 \epsilon_0^3 h^4} \right)  \frac{1}{n^4}.
\label{eq:Fc_n_au}
\end{equation}

This simplified treatment overlooks an important point -- in an electric field, the principal quantum number $n$ is not a good quantum number, but is replaced by the parabolic quantum numbers $n_1$ and $n_2$ \cite{Gallagher}.
$n_1$ and $n_2$ are non-negative integers and are related to $n$ through
\begin{equation}
	n = n_1 + n_2 + \left| m \right| + 1, 
\end{equation}
where $m$ is the magnetic quantum number.
The treatment described above corresponds to the special case of $n_1 \simeq n_2 \simeq n/2$ and $m \simeq 0$, which is a so-called `central' state, where the positron wavefunction is centred on the nucleus.
The other cases correspond to states in which the atom is polarised, i.e. the positron is preferentially localised to one side of the nucleus.
This tends to increase the critical field from the value in equation \ref{eq:Fc_n_au}.
For the most extreme 'blue-shifted' atoms, the critical field can be increased by up to around a factor of two \cite{Gallagher}.
The atoms produced in mixing in ALPHA will most likely be a random mixture of states with different values of $n, n_1, n_2$ and $m$, significantly complicating any analysis of the states.

The classical treatment described above also only considers ionisation when it is classically allowed.
Quantum mechanically, the positron can tunnel through a potential barrier of finite height, and the atom can be ionised at fields lower than the classical critical field.
However, when the electric field is close to the classical critical value, the probability of ionisation in one atomic unit of time is close to unity \cite{bethesalpeter}.
At lower electric fields, ionisation is exponentially suppressed, as expected by the exponential dependence of the tunnelling probability on the height and width of the barrier \cite{Griffiths}.
Thus, the behaviour is not much different from the classical treatment.

In a magnetic field, the atom's internal motion becomes more complex.
Some discussions of the effect can be found in \cite{HydrogenMagneticField}, \cite{FieldI_StrongMag} and \cite{Svante_JPhysB}.
Of interest to us is the fact that the potential (equation \ref{eqn:classicalPotential}) is modified to include a dependence on the speed of the atom across the magnetic field.
This can be interpreted as a `motional' Stark shift, or as an electric field in the rest frame of the atom;
\begin{equation}
 \mathbf{F}_{mot} =  \mathbf{v} \times \mathbf{B}.
 \label{eqn:motStarkShift}
\end{equation}
This is simply due to the opposing Lorentz forces on the positron and antiproton, averaged over time.
This term combines with the electrostatic field in the Penning trap and can increase the probability that an atom will be ionised.
In general, the motion of the positron around the antiproton in a magnetic field is not quite as simple as this.
In certain situations, the positron can remain bound to the antiproton, even though it has negative binding energy \cite{Gajewski}.
The analysis of this effect is far more involved, and we will ignore it.

The electric field required to ionise an atom in the ground state is of the order $3 \times 10^{10}~\mathrm{V\,m^{-1}}$, far  higher than can be achieved in Penning trap type experiments.
However, the experiments take advantage of the fact that most of the atoms produced through three-body recombination are in highly-excited, weakly bound, states that can be ionised by the $\sim 10^4~\mathrm{V\,m^{-1}}$ electric fields found in the Penning trap.

\begin{figure}[bt]
\centering
\input{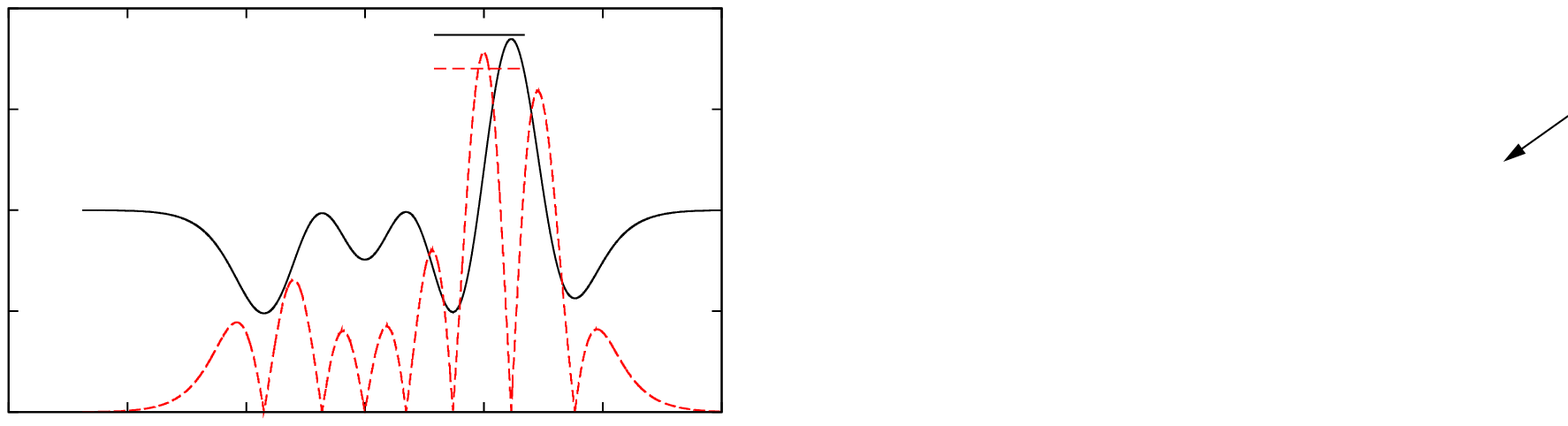}
\caption{An example of a nested trap with an ionisation well.}

\label{fig:ionisationWell}
\end{figure}

When using field ionisation as a detection technique, a potential is constructed with a region of high electric field inside a well that can confine the antiprotons after ionisation.
This kind of configuration is called an `ionisation well' - an example is shown in figure \ref{fig:ionisationWell}.
Antihydrogen is most likely to be ionised at the points where the electric fields are highest.
Notice that the potential has been constructed so that an antiproton placed at one of these points will be confined in an electrostatic well.
In principle, positrons can also be captured by reversing the well polarity, but small numbers of antiprotons are far more readily detectable than positrons.
As antihydrogen atoms pass through the potential and are ionised, antiprotons accumulate in the ionisation well.
At the end of the mixing cycle, the ionisation well is emptied, and the annihilations of the antiprotons released are detected.
The ionisation well is constructed so that it is impossible for bare antiprotons escaping the nested well to become trapped.

\begin{figure}[hbt]
\centering
\input{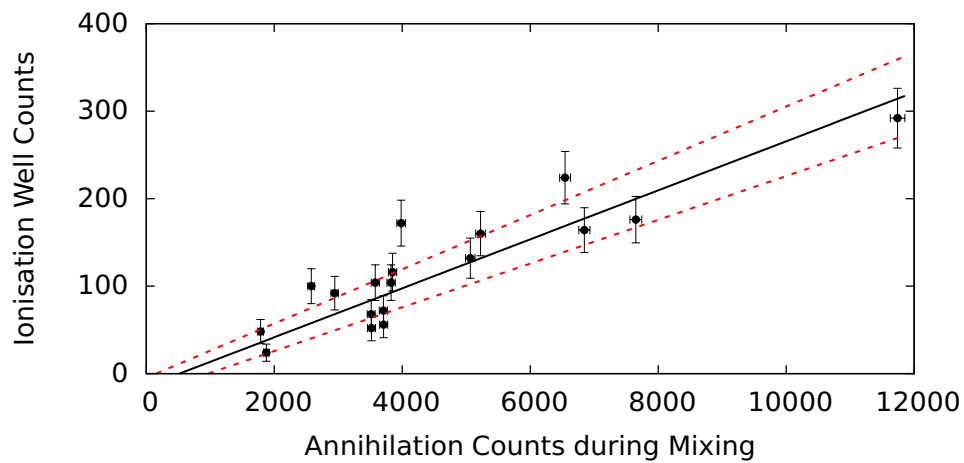}
\caption[The number of annihilation counts during mixing compared to the number of ionisation well counts.]{The number of recorded annihilation counts during mixing compared to the number of ionisation well counts for a series of mixing measurements. The counts have been corrected for the different efficiencies of the detectors. The data is approximately described by a straight line fit - shown as the solid line, with the one-sigma uncertainty region bounded by the dashed lines.}
\label{fig:IW_vs_triggers}
\end{figure}

This method is powerful in that the cosmic background can be all but eliminated by ejecting all of the antiprotons from the ionisation well and counting their annihilations in a short period of time.
However, it also suffers from a much lower sensitivity, since the flux of antihydrogen atoms passing through the ionisation well is only a small fraction of the total.
For example, in experiments with ionisation wells in ALPHA, tens of counts were recorded in the ionisation well while thousands of annihilation vertices can be detected by the imaging detector.
The data shown in figure \ref{fig:IW_vs_triggers} was produced from a series of mixing measurements where the number and density of antiprotons and positrons, and the interaction time, was varied to test the relationship between the number of counts in the ionisation well and the number of annihilation counts recorded by the silicon detector.
The slope of the line is  $\left(2.7~\pm 0.3\right)\%$, or $\left(3.3~\pm 1.1\right)\%$ when scaled by the detector efficiency.
The fraction of the total solid angle covered by the ionisation well was between 1.8\% and 5\% (depending on the positions of formation and ionisation), so these results are in good agreement.
The fact that counting the annihilation triggers during mixing and measuring the number of antiprotons captured in an ionisation well produce consistent results supports the use of both methods as indicators of antihydrogen formation.

In addition to the low sensitivity, use of an ionisation well is not compatible with attempting to trap antihydrogen atoms, when the electric fields must be reduced to prevent ionisation of otherwise trappable atoms.
For these reasons, ALPHA primarily relies on counting and imaging the antiproton annihilations during mixing to confirm antihydrogen production.

\section{Antihydrogen in a nested trap}
\label{sec:NestedTrap}
In contrast to the equilibrium formation processes discussed in section \ref{sec:HbarFormationTheory}, antihydrogen formed in a nested trap is in contact with the positron plasma for only a very short time after formation.
While the atom is inside the positron plasma, atom-positron collisions change the binding energy and velocity, and when it leaves, this influence stops.
This effect was first analysed computationally by F. Robicheaux \cite{FrancisSimulationsHbarFormation}.
His analysis found that the typical binding energy was much smaller than the few $k_\mathrm{B}T$ kinetic bottleneck mentioned in section \ref{sec:HbarBE} as a consequence of the reduced number of collisions.

In a nested trap, an antihydrogen atom can meet one of a number of fates.
The simplest is that the antihydrogen atom, unconfined by the Penning trap, drifts outwards until it encounters a matter nucleus and annihilates - this typically happens on the surface of the Penning trap electrodes, the first solid matter encountered by an atom moving outwards.

The antihydrogen atom can succumb to ionisation - either through field ionisation (section \ref{sec:fieldionisation}), collision with another particle, or by absorbing a photon.
If this occurs at a position where the antiproton's trajectory and energy will take it through the positron plasma again, the antiproton has a second chance to form an antihydrogen atom, and the process can repeat.
The change in the radial position of the antiproton between formation and subsequent ionisation can be thought of as a random-walk-type process.
With each iteration, the antiproton will, on average, move radially outwards, and the radial distribution of antiprotons can become significantly altered \cite{Svante_JPhysB}.
In the most extreme case, this means that antiprotons accumulate near the edge of the positron plasma, and only antihydrogen atoms formed near the edge can escape the positron plasma.

If the atom is ionised outside the nested trap, or at a point of high electric potential, the antiproton will exit the Penning trap and annihilate. 
When the magnetic field is uniform, the antiproton will be guided to the axial ends of the electrode structure.
If the atom is ionised inside the nested trap, but at high radius, so that it will not pass through the positron plasma, the antiproton remains confined and isolated in the nested trap.
This can also happen if the atom is ionised at low energy in the side wells.

\section{Incremental mixing} \label{sec:incmix}

\begin{figure}[h!]
\centering
\includegraphics[width = \textwidth]{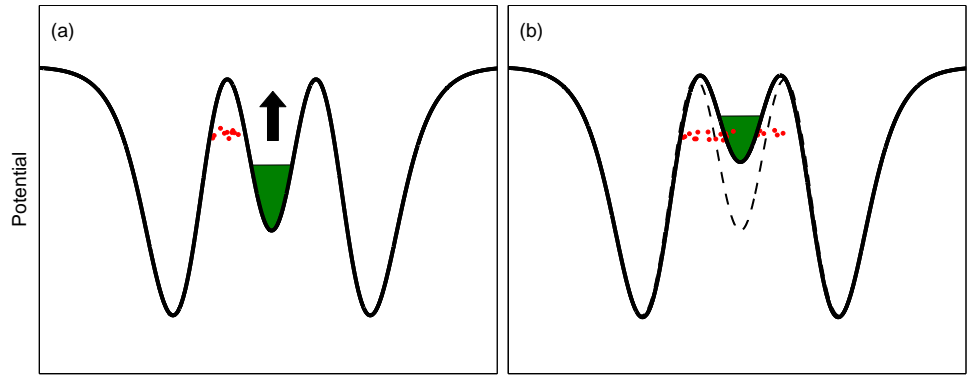}
\caption[A schematic of incremental mixing.]{A schematic of incremental mixing. In (a), the antiprotons (the red dots) are prepared at a well-defined energy in one of the side wells. As the positron well is moved in (b), the antiprotons enter the positron plasma with a minimum of longitudinal energy.}
\label{fig:incrementalMixing}
\end{figure}

`Incremental mixing' is the name given to a modified version of the nested trap mixing scheme.
It is intended to introduce the antiprotons to the positron plasma with lower kinetic energy than the original procedure.
Instead of initially giving the antiprotons enough energy to cross the positron plasma, the antiprotons are placed in one of the side wells, at an energy where they are out of contact with the positron plasma.
Then, the central potential separating the positrons and antiprotons is reduced until the antiprotons can escape the side well and transit the positron plasma.
This procedure is shown in schematic form in figure \ref{fig:incrementalMixing}.

If the change in potential is performed slowly with respect to the antiproton bounce time (of order $1-10 ~\mathrm{\mu s}$, which is a condition easily met experimentally), the antiprotons will be released with small longitudinal kinetic energy inside the positron plasma, and should form antihydrogen with correspondingly low kinetic energy.

\begin{figure}[b!]
\centering
\input{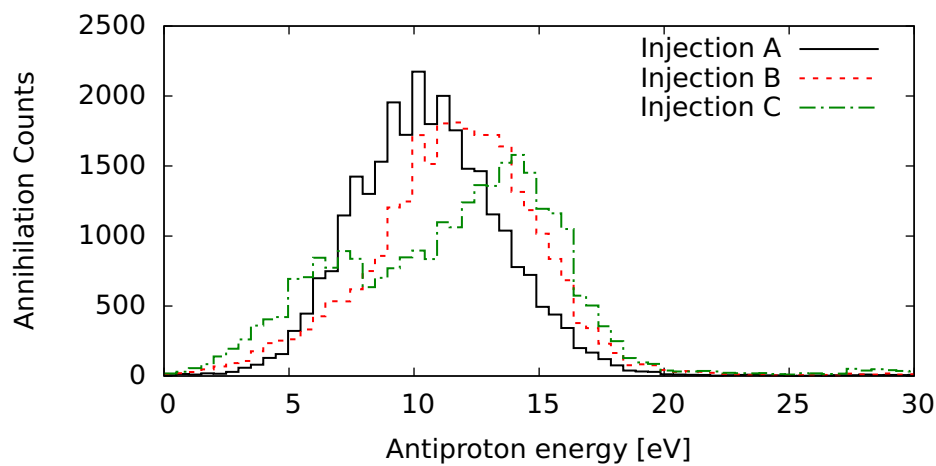}
\caption{The measured energy distributions for three different preparations (injections A, B and C) of antiprotons for incremental mixing.}
\label{fig:IncMixInitial}
\end{figure}

Before the antiprotons are placed in the side well, they are confined in a neighbouring potential, much like the one shown in figure \ref{fig:ATHENAinject}.
The energy with which they are injected is controlled by choosing the difference in potential of this well and the nested potential.
The energy is ultimately determined by the dynamics of the manipulation, so the procedure is usually optimised using the measured energy distribution as feedback.
Three example injected distributions are shown in figure \ref{fig:IncMixInitial}.
The antiproton distributions have a large energy spread, but the peak antiproton energy moves by several electron-volts.

\begin{figure}[t!]
\centering
\input{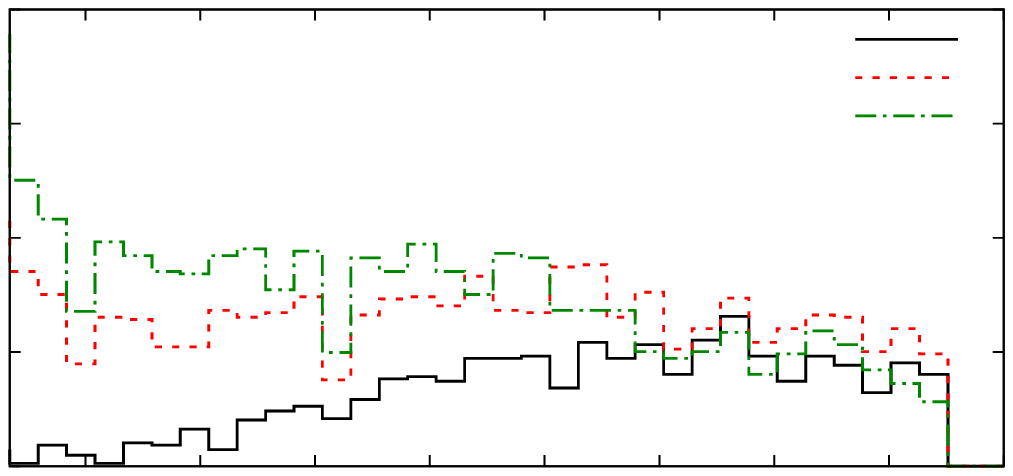}
\caption[The annihilation counts detected during incremental mixing for the .]{The annihilation counts detected as the depth of the antiproton well was reduced in 10~s for the injected distributions shown in figure \ref{fig:IncMixInitial}. The x-axis has been reversed so that time moves from left to right.}
\label{fig:IncMixMixing}
\end{figure}

The different energy distributions are expected to be reflected in the time dependence of antihydrogen formation.
Higher energy antiprotons enter the positron plasma at earlier times, so we expect antihydrogen production (signalled by annihilation counts) to begin earlier.
This is clearly seen in figure \ref{fig:IncMixMixing} -- the lowest-energy antiprotons begin producing antihydrogen at a later time.

\begin{figure}[b!]
\centering
\input{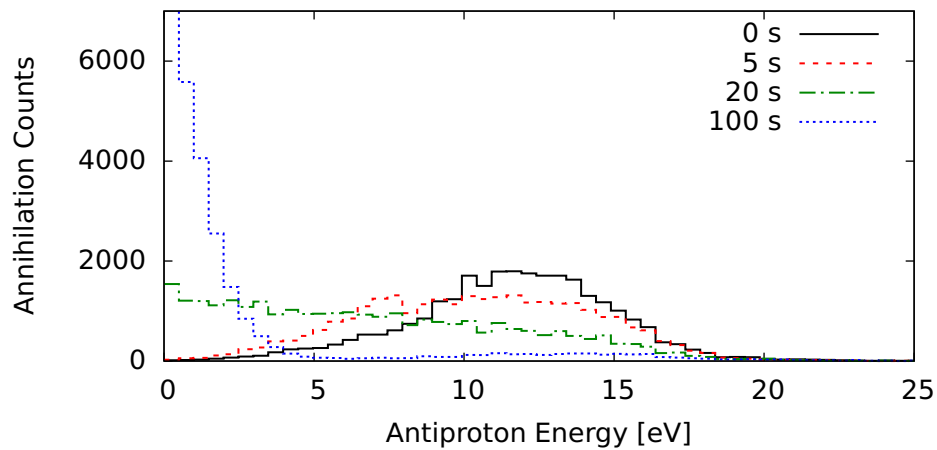}
\caption[The evolution of the antiproton energy distribution over 100~s.]{The evolution of the injected antiproton energy distribution over 100~s, showing the transition from an excited distribution to an equilibrium one. }
\label{fig:IncMixInitialWait}
\end{figure}

Immediately after injection, the antiprotons do not have an equilibrium energy distribution -- they have a lot of the energy in the parallel direction, and little in the perpendicular degrees of freedom.
Energy transfer through collisions will act to equalise the energy in each degree of freedom.
This is demonstrated in figure \ref{fig:IncMixInitialWait} -- after injection, the antiproton energy distribution evolves towards the equilibrium Maxwell-Boltzmann distribution.
The time scale over which this occurs can be measured by dumping the antiprotons from their injected configuration at a defined time after injection.
It can be clearly seen that the energy distribution evolves by spreading to lower energies.
After 20~s, the low-energy tail extends to the bottom of the well, and after 100~s, the distribution is well-described by an exponential, indicating that it has reached thermal equilibrium.

If we assume that the total energy is divided equally between the three degrees of freedom  during equilibration, and if the perpendicular degrees of freedom are initially cold, we expect that the parallel energy after equilibration will be approximately one-third of the initial energy.
For the distributions shown in figure \ref{fig:IncMixInitialWait}, the initial mean energy is approximately 12~eV.
Thus, we expect that the mean parallel energy will be $\sim$ 4~eV after thermal equilibrium is reached.
This is in rough agreement with 3~eV, the mean measured energy of the distribution after 100~s.

This redistribution has two deleterious effects on the antihydrogen production \linebreak[3] scheme.
Firstly, energy is transferred to the perpendicular degree of freedom.
Thus, even though the scheme has been designed so that the antiprotons enter the positron plasma with minimal parallel kinetic energy, there is significant energy in the perpendicular degrees of freedom.
The total kinetic energy of the antiproton when it forms antihydrogen can therefore be significantly higher than the initial antiproton temperature would suggest. 
Secondly, the redistribution causes antiprotons to fall deeper into the side wells of the nested well (figure \ref{fig:nested}), where they do not have sufficient energy to enter the positron plasma.
These antiprotons cannot then form antihydrogen, and are essentially lost from the system.

In principle, it is possible to overcome this effect by moving the positron well quickly compared to the time scale for thermalisation of the antiprotons.
Figure \ref{fig:IncMixInitialWait} shows that the thermalisation timescale is on the order of seconds or longer, while the antiproton bounce time (the well must deform slowly with respect to this in order that the antiprotons enter with little parallel kinetic energy) is $1-10~\mu \mathrm{s}$. 
However, experimentally, it was seen that ramping the well quickly (faster than over 10~s or so) resulted in a dramatic decrease in the number of antihydrogen atoms formed.

There is no definitive explanation for this behaviour, but there are a number of hypotheses.
Moving the positron well could increase the temperature of the positron plasma, suppressing the formation rate (section \ref{sec:HbarFormationTheory}).
If this is the case, making the movement more slowly would allow the plasma time to lose energy through cyclotron radiation, maintaining a lower temperature and a higher antihydrogen formation rate.

It may also be that the antihydrogen formation rate is much slower in this scenario than expected -- the measured positron temperature was of the order of several hundred Kelvin, so this is not unlikely.
If the formation rate is comparable to, or even slower than, the antiproton equilibration rate, most of the antiprotons will not form antihydrogen before being collisionally driven to energies too low to enter the positron plasma.

The amount of kinetic energy transferred to the perpendicular degree of freedom can be minimised by making the side wells as shallow as possible.
The antiprotons then need to have only a small amount of energy above the bottom of the side wells to enter the positron plasma and the rate of energy transfer from the parallel to perpendicular degrees of freedom is lower.
However, this competes with the need to maintain reliable confinement of the positron plasma, and the shallowest wells used were approximately 500~mV deep.

Incremental mixing is also susceptible to a novel antiproton loss mechanism, which was reported in \cite{ZeroFreqLoss}.
The mechanism only occurs in the presence of a transverse multipole magnetic field -- in ALPHA's case, when the octupole magnet is energised.
In the octupole, the particles move along the magnetic field lines, which can be calculated from equation \ref{eq:octupoleField} for the ideal octupole -- importantly, the field lines are not parallel to the axis of the Penning trap.
Thus, a particle moves inwards and outwards in the $r$ direction as it oscillates axially.
The precise amount by which the particle travels in $r$ in a single oscillation depends on the azimuthal angle $\phi$.
The field lines are not constant in $\phi$, so the particle also rotates about the trap axis as it moves.
Ignoring any other effects, the particle would remain bound to the same field line, so the size of the radial excursions would remain constant.

However, recall that a particle in a Penning trap also rotates across the magnetic field lines due to the magnetron motion (section \ref{sec:penningTrap}).
This rotation rate will be determined by the radial electric field.
We can think of this rotation in terms of the particle following different field lines on the two half-oscillations -- the part in which it moves towards positive $z$ and that towards negative $z$.
Thus, over a complete oscillation in $z$, the particle will not necessarily return to the same radial position.

In most cases, the change in radius will be averaged out over many oscillations.
However, there is an exception -- if the change in $\phi$ over an oscillation is a multiple of a period of the magnetic field, i.e,
\begin{equation}
\Delta \phi = n \frac{2 \pi}{m} \;;\; n = 0, \pm 1, \pm 2, ...,
\label{eq:resonantLoss}
\end{equation}
where $m = 4$ for an octupole, the particle can find itself on a trajectory which moves rapidly outwards in radius.
A particle with a value of $\Delta \phi$ satisfying equation \ref{eq:resonantLoss} can be transported from close to the axis to the electrode wall in just a few axial oscillations.
Because the process requires close matching between the average rotation frequency and one of the frequencies in equation \ref{eq:resonantLoss}, this is a resonant loss process.
Not all particles are lost, since the transport depends on the value of $\phi$ when the resonant condition is met.

Since $\Delta \phi$ depends on the electric field in the trap, and the electric field changes during the incremental mixing procedure as the positron well is moved, $\Delta \phi$ for an individual particle slowly changes.
At some point, $\Delta \phi$ reaches one of the resonant values and the particle can be lost.
Simulations carried out (see \cite{ZeroFreqLoss} for details) indicate that, in ALPHA, the loss process corresponds to $n=0$ in equation \ref{eq:resonantLoss}, that is, the angle through which a particle rotates over a single oscillation is zero.
This is made possible by the fact that, in a nested trap,  the radial component of the electric field changes sign, so on one portion of the trajectory the particle rotates in one direction, and in the opposite direction on the rest.

The presence of this process, and the antiproton loss that accompanies it, poses a significant problem for the success of incremental mixing as an efficient antihydrogen production technique.
As well as the simple fact that the lost antiprotons cannot be used to form antihydrogen, their annihilations contaminate the antihydrogen signal.

Despite these drawbacks, incremental mixing is a significant improvement over the original nested-well mixing scheme.
The antiprotons never have kinetic energies of many tens of electron-volts characteristic of the ATHENA mixing schemes, but instead form antihydrogen with energies on the order of 1~eV.

\section{Autoresonant injection scheme}
\label{sec:AR_mixing}

Autoresonance (discussed in detail in section \ref{sec:autoresonance}) was implemented in ALPHA to precisely control the antiproton parallel energy.
It can be used in place of the potential manipulations discussed above (and shown in figure \ref{fig:ATHENAinject}) to prepare the antiprotons in an excited longitudinal energy distribution prior to ramping the central potential.
Using this method, it is possible to produce much narrower antiproton energy distributions than from the standard potential manipulations (figure \ref{fig:IncMixInitial}).
However, the antiprotons still collisionally redistribute energy in the same way, so any benefit is lost in a short amount of time.

\begin{figure}[b!]
\centering
\input{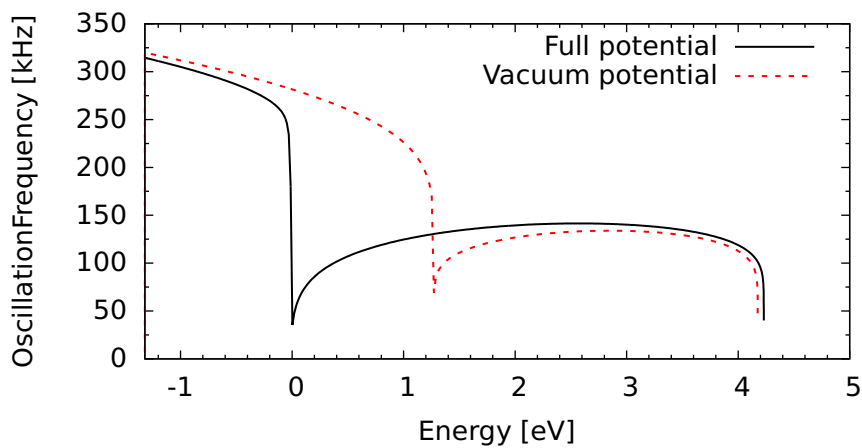}
\caption[The antiproton axial oscillation frequency as a function of energy in a nested well.]{The antiproton axial oscillation frequency as a function of energy, for the potential shown in figure \ref{fig:nested}.}
\label{fig:nestedBounce}
\end{figure}

The two-well structure of the nested trap gives rise to a discontinuity in the oscillation frequency-energy dependence for antiprotons; an example frequency-energy curve showing this behaviour is shown in figure \ref{fig:nestedBounce}.
This occurs at the point where antiprotons have just enough energy to leave the side well and pass through the positron plasma.
Here, the antiprotons' orbit changes length, at least by a factor of two.
Correspondingly, the axial oscillation frequency changes by a large amount.
Because this takes place suddenly -- from one bounce to the next -- the antiprotons do not stay in resonance with the drive and stop gaining energy coherently.
Thus, in the best case, the antiprotons gain just enough energy to enter the positrons and no more.

This injection scheme suffers less from the redistribution of energy between the degrees of freedom since it is not necessary to ramp the central potential, so antihydrogen production and energy redistribution start almost simultaneously.
In contrast, recall that in the incremental mixing scheme, antihydrogen production started some time after the antiprotons had been placed in the excited distribution, while the central potential was moved.
Redistribution still causes the antiprotons to be collisionally driven into the side wells.

\begin{figure}[b!]
\centering
\input{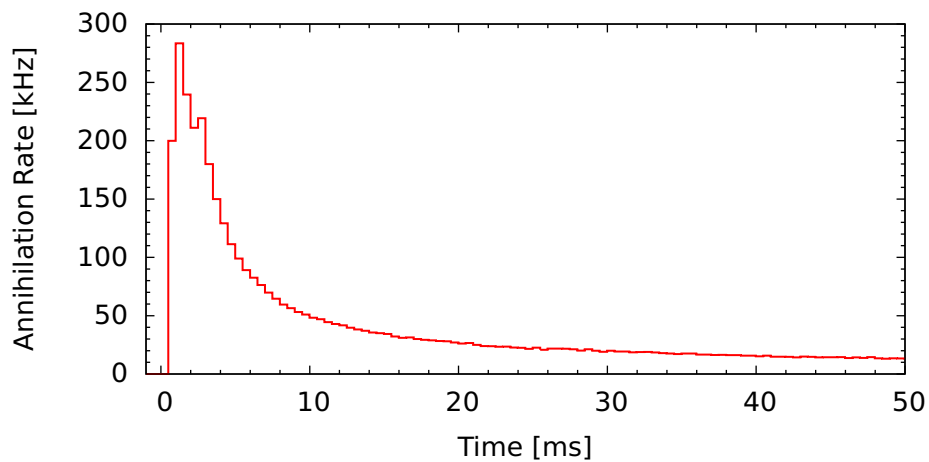}
\caption[The annihilation rate time-structure after autoresonant injection.]{The annihilation rate time-structure after autoresonant injection at t=0. The trace was produced by averaging approximately 200 separate measurements.}
\label{fig:ArMixing}
\end{figure}

The annihilation rate as a function of time, shown in figure \ref{fig:ArMixing}, reflects this behaviour.
It is not necessary to sweep the frequency to zero -- typically entering the vertical part of the curve is sufficient, and sweeping from 350~kHz to 200~kHz over 1~ms was found to perform well, and was used to produce this data.
The annihilation rate abruptly rises shortly after t=0, the time at which the frequency drive begins, and the antiprotons are injected into the positron cloud.
The high rate of annihilations is sustained for only $\sim$10~ms, and falls by a factor of 10 within 50~ms.
This fall may be attributable to the collision-driven redistribution.
Calculating the collision rate between antiprotons in this situation is not an analytically tractable problem.
We can use the approximate expressions in \cite{PlasmaBible} to obtain a crude order--of--magnitude estimate of $10^{1}$ -- $10^{2}$~Hz.
This timescale appears to agree with the observed decline in the annihilation rate.

Autoresonant injection causes the temperature of the positron plasma to increase -- in the example here, from an initial temperature $\sim 30~\mathrm{K}$ to a final temperature of $\sim 70~\mathrm{K}$). 
The exact mechanism of the temperature change is not known -- some hypotheses will be discussed later.
The positron plasma is already hot 10~ms after injection, the minimum length of time needed to perform a temperature analysis, so the timescale of the temperature change is also unknown.
A change in the positron temperature should be associated with a change in the antihydrogen production rate, so this is a further possible explanation for the rapid fall in the production rate.
It is probable that both of the effects just mentioned combine to produce the observed annihilation rate-time dependence.

\begin{figure}
\centering
\input{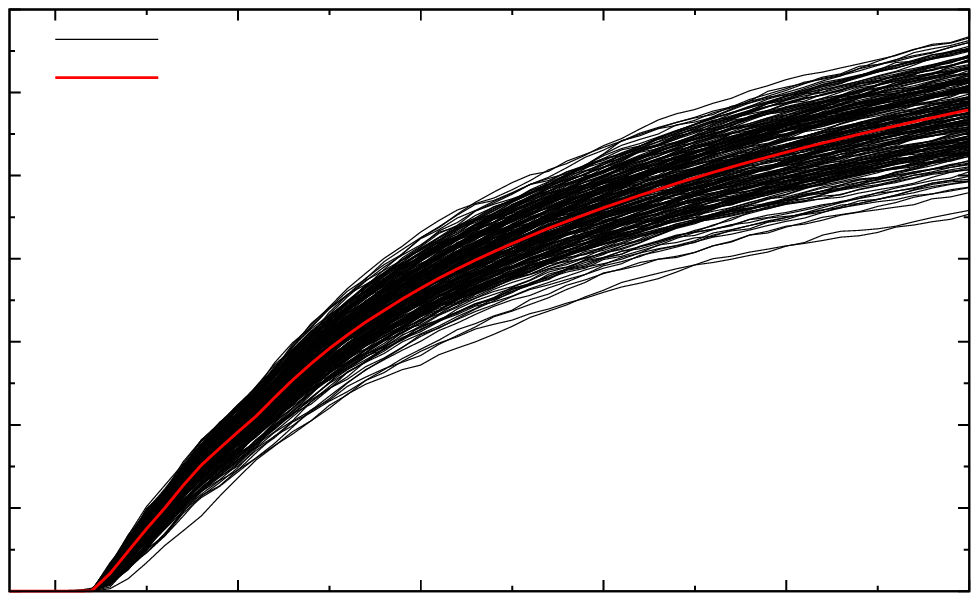}
\caption[The integrated number of annihilation counts after an autoresonant injection sweep.]{The integrated number of annihilation counts after an autoresonant injection sweep commencing at $t$=0. The thin, black lines show the traces from individual experiments, while the heavy, red line shows the average of these traces.}
\label{fig:ARInjection}
\end{figure}

The details of the injection procedure are also informative to study.
The integrated annihilation counts, corresponding to antihydrogen production, over the first 10~ms after the injection sweep are shown in figure \ref{fig:ARInjection}.
The traces rise abruptly above the background at $(0.5 \pm 0.1)$~ms, marking the onset of antihydrogen production.

It is possible to measure the number of antiprotons injected over time independently of the antihydrogen production rate.
To do this, the nested well is modified so that antiprotons that pass through the positron plasma are not confined and escape the well.
In the experiments described here, the rightmost barrier in figure \ref{fig:nested} was removed, allowing the injected antiprotons to strike the MCP (section \ref{sec:MCP}), positioned along the positive $z$ axis.
Special care must be taken to ensure that the shapes of the left-hand side well (where the antiprotons are stored prior to injection) and the positron plasma are not changed, so that the response of the antiprotons to the frequency sweep is not altered.
After the frequency sweep is completed, any antiprotons remaining in the nested trap, including those that have not been injected, are ejected by manipulating the potentials to release them from the well to follow the injected antiprotons.

\begin{figure}[hbt]
\centering
\input{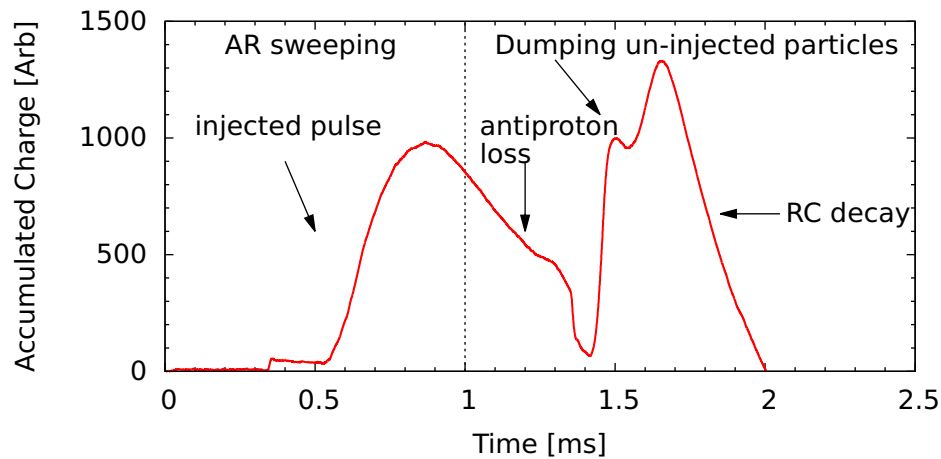}
\caption[The time profile of antiprotons autoresonantly injected into a nested well.]{The time profile of antiprotons autoresonantly injected into a nested well. The detailed features are described in  the text.}
\label{fig:AR2MCP}
\end{figure}

An example of the measured injection profile is shown in figure \ref{fig:AR2MCP}.
This experiment measures the charge of the antiprotons, amplified using the MCP, by measuring the voltage across a capacitor. 
Pulses of antiprotons thus appear as vertical edges, with the height of the edge proportional to the number of antiprotons. 
The first peak is due to antiprotons passing through the positron plasma and escaping the nested well.
The second pulse occurs when the uninjected antiprotons are ejected from the nested well by performing a dump.
The relative heights of the peaks gives the fraction of antiprotons injected by the drive.
In this case, approximately 50\% of the total number of antiprotons were successfully injected.

The charge stored in the capacitor drains through a resistor, with a characteristic RC decay that can be measured from the shape of the curve after the second peak.
Between the peaks, the voltage is given by a combination of the RC decay and a continuation of the antiproton loss, possibly caused by collisional energy transfer between antiprotons that were excited by the autoresonant drive, but not injected in the first pulse.

There is excellent agreement between the time of the rising edge in figure \ref{fig:AR2MCP} and the onset of antihydrogen production (figure \ref{fig:ARInjection}).
This indicates that antihydrogen production begins as soon as antiprotons are introduced to the positron plasma.

It is also interesting that the antiprotons are injected relatively early in the frequency sweep (the entire first rising edge is contained within 0.5~ms - 0.8~ms). 
As mentioned before, antiprotons that have acquired enough energy to pass through the positron plasma are no longer locked to the autoresonant drive and do not gain energy resonantly.
They still experience the changing electric field, but at a random phase with each oscillation. 
They can stochastically, or randomly, gain energy while the drive remains energised.
This additional energy can be transferred to the positron plasma through collisions and is a possible source of the increase in the positron temperature discussed above.
If this is the case, shutting off the drive a portion of the way through the frequency sweep should result in a lower positron temperature, without significantly reducing the number of antiprotons injected.
A lower positron temperature would probably produce a larger number of antihydrogen atoms, as well as at a potentially lower temperature.
This aspect is still under study in ALPHA -- the initial experiments indicate that interrupting the autoresonant drive does in fact reduce the size of the temperature change, though possibly at the cost of injecting a smaller fraction of the antiprotons and reducing the yield of antihydrogen atoms.

\section{Antihydrogen production in a transverse multipole magnetic field}

To trap antihydrogen atoms, the mixing procedure must be carried out inside the magnetic atom trap.
The effects of the inhomogeneous magnetic fields of the trap, principally the transverse components due to the octupole, could potentially disrupt the mixing process to such a degree that it inhibits antihydrogen production.
The effects of the octupole on the stored plasmas have already been described in section \ref{sec:OctupolePhysics} -- typically there is an associated temperature increase, and possibly enhanced radial diffusion.
These reduce the antihydrogen production rate, as expected from the discussion in section \ref{sec:HbarFormationTheory}.

In addition, the inhomogeneous magnetic fields act on the antiprotons liberated from field-ionised antihydrogen atoms.
In a simplified picture, if an antihydrogen atom is ionised outside the critical radius (discussed in section \ref{sec:OctupolePhysics}) for the antiproton, the magnetic field will guide the antiproton into the electrode surfaces, where it will annihilate.
Thus, the annihilation signal during mixing, normally made up solely of antihydrogen atoms, will be contaminated by some annihilations of bare antiprotons.
Recall that individual antihydrogen and antiproton annihilations are indistingushable using ALPHA's silicon detector.
The impact of this feature and its interpretation will be discussed in greater detail in section \ref{sec:annihilationsMultipole}.

Aside from this, antihydrogen production in the magnetic trap is much the same as production in a solenoidal magnetic field.
ALPHA demonstrated this capability in experiments reported in \cite{ALPHA_HbarInOct}.

\section{Conclusions}

Production of antihydrogen via injection of antiprotons into a positron plasma is by now a well-established technique.
The detection of antihydrogen atoms, whether through annihilation vertex imaging or field-ionisation analysis has also been convincingly demonstrated.
All indications are that the antihydrogen atoms produced using the original techniques are at a much higher temperature than is needed to trap some of them in a magnetic-minimum trap.
The techniques described in this chapter have been tailored to introduce the antiprotons to the positron plasma with little kinetic energy, and therefore reduce the antihydrogen temperature.
In the following chapters, we will examine how to estimate the antihydrogen temperature and apply these injection schemes to attempts to produce trapped antihydrogen.

    \chapter {Spatial Distribution of Antihydrogen Annihilations}

\epigraph{It was almost as incredible as if you fired a 15-inch shell\\at a piece of tissue paper and it came back and hit you.}{Ernest Rutherford}

\label{chp:vertex}

The annihilation vertex detector (described in detail in section \ref{sec:siliconDetector}) is an extremely useful tool to identify antiproton annihilations.
Study of the spatial distribution of the annihilations yields insight into the properties of the antihydrogen atoms formed and the important processes affecting antihydrogen atoms and antiprotons in the experiment.
For instance, it has already been discussed in section \ref{sec:hbarimaging} how the transverse distribution provides strong evidence that antiproton annihilations observed during positron-antiproton mixing are due to the escape of antihydrogen atoms.

In this chapter, we will compare the projection of the annihilation distribution onto the $z$-axis to predictions for the distributions produced from computer simulations.
The shape and features of the $z$-projection depend on the properties of the antihydrogen atoms produced from antiproton-positron mixing -- specific examples being their velocity distribution and their binding energy distribution.
The probability of trapping the antihydrogen atoms depends critically on these parameters, and measurements help to inform the development of the mixing techniques.
When the transverse multipole magnetic field is applied, the antihydrogen annihilation signal is contaminated with the loss of antiprotons freed from field-ionised atoms, but examination of the annihilation distribution allows the two components to be separated \cite{ALPHA_HbarInOct}.

This analysis is an empirical and phenomenological study of the annihilation distributions.
We attempt to find a model for the system that produces an annihilation distribution that reproduces the measurements.
This approach can help to give insights about the basic physical mechanisms, even though it does not necessarily take account of all of the underlying physics.

The annihilation distribution is modelled taking a Monte-Carlo-like approach.
This technique is a widely used and powerful computational technique, and is especially useful when a `result' can be deterministically computed for a given set of inputs, but for which the inputs are characterised by a distribution instead of a single value.
In this analysis, the initial conditions (position, velocity, binding energy) of a single atom are drawn from defined distributions and the result -- its contribution to the annihilation distribution -- is computed.
This is repeated for many different atoms to produce a distribution of annihilation positions, to be compared to the experimental measurements.

\section{Annihilation distribution in a uniform magnetic field}
While an unmodified (i.e. without a superimposed multipole) Penning-Malmberg trap performs very well as a trap for charged particles, the magnetic and electric fields (and their gradients) are far too small to exert forces of any significance on an antihydrogen atom.
Thus, when formed, an atom simply moves away from the point of combination in a straight line with the combined momentum of the antiproton and positron.
If it continues along this trajectory, the atom eventually encounters a matter nucleon and annihilates.
The first solid matter object usually encountered is the inner wall of the Penning trap electrodes, so this is where the majority of the annihilations are concentrated.

Since the atom is moving in a straight line, the position of the annihilation is determined by the starting location $\mathbf{r_0} = \left(r_0, \phi_0, z_0\right)$ and the direction of the initial momentum vector $\mathbf{\hat p}$.
The experiment is symmetric in $\phi$, so the problem can be reduced to two dimensions - the longitudinal velocity $v_z$ and the transverse velocity $v_r$.

As the mass of the antiproton is 1836 times the mass of the positron, the momentum of the antihydrogen atom is almost entirely determined by the momentum of the antiproton before combination.
The initial velocity of the antiproton perpendicular to the magnetic field can be considered to be made up of two components - the thermal velocity, taken from a one-dimensional Maxwell-Boltzmann distribution, and the magnetron, or $\mathbf{E}\times \mathbf{B}$, motion given by equation \ref{eq:plasmaExB}.
The antiproton's motion also has a cyclotron component, but the length-scale of this motion is so small, and the time-scale so short compared the time taken for an atom to cross the trap, that the motion averages out and can be ignored.
The velocity parallel to the magnetic field has only a thermal contribution.
Recall from section \ref{sec:PlasmaEquilibrium} that, in equilibrium, the positron plasma rotates about the axis of the Penning trap as a rigid rotor at the magnetron frequency.
Antiprotons inside the positron plasma rotate about the trap axis with the same rotational velocity.
So, the motion of the antiprotons can be also thought of as thermal motion in the rotating frame of reference in which the positron plasma is at rest.

An analysis of the spatial annihilation distribution as a function of the antihydrogen velocity was originally applied to data gathered during antihydrogen production data by ATHENA \cite{ATHENA_SpatialDistribution}.
This analysis concluded that the measured annihilation spatial distribution was not consistent with an isotropic velocity distribution (i.e. the data was not consistent with $T_\parallel = T_\perp$).
Instead, it was found that the longitudinal velocity was a factor of 2-10 times higher than the transverse velocity.
Since these experiments used antiprotons that were initially introduced to the positrons with many electron-volts of energy, and cooled as they formed antihydrogen, this was interpreted to mean that antihydrogen formation occurred before the antiprotons had thermalised with the positron plasma.
In contrast, the antihydrogen production schemes used in ALPHA do not introduce antiprotons with large longitudinal kinetic energies. 
Thus, it is more likely that the annihilation distribution will be closer to the case with $T_\parallel = T_\perp$.
The arguments for this point have already been made in section \ref{sec:NestedWell}.

Another effect on the spatial annihilation distribution, not considered in \cite{ATHENA_SpatialDistribution}, but mentioned in later works, such as \cite{Svante_JPhysB}, is the possibility of ionisation of the antihydrogen atom between the point of formation and impact with the electrode wall. 
This will be explored in section \ref{sec:vertex_fieldionisation}.

\subsection{Data samples}

\begin{figure}
	\centering
	\input{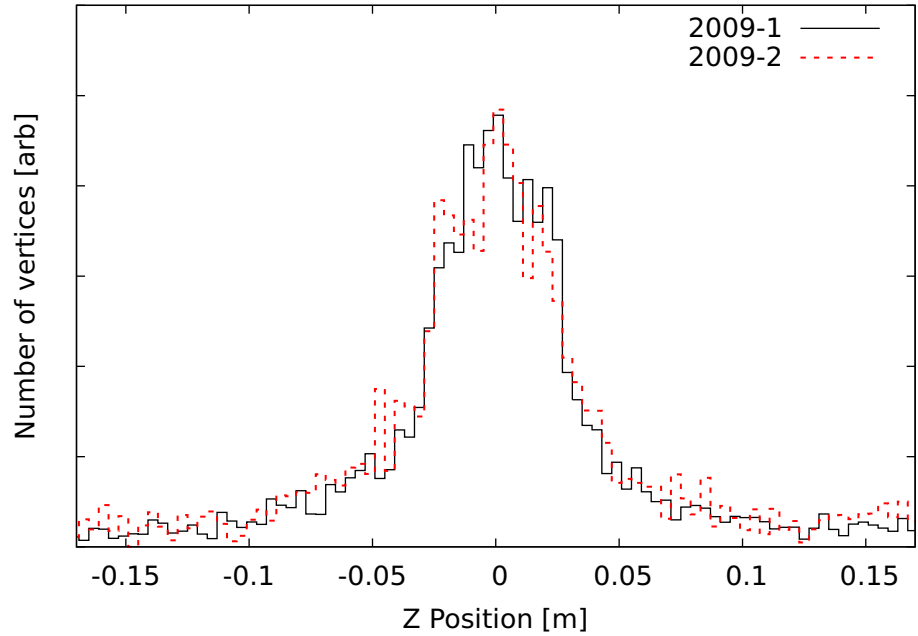}
	\caption[The samples of annihilation vertices used in this study.]{Two data samples used for this study. The distributions have been normalised to have the same area, and have been corrected for the response function of the detector.}
	\label{fig:Isotrop_measured}
\end{figure}

Comparisons with simulations were carried out for two sets of antihydrogen production data; the corresponding $z$-distributions are shown in figure \ref{fig:Isotrop_measured}.
Each data set was collected using nested wells of different construction and with positron plasmas with different characteristics (listed in table \ref{table:datasets}).
The number of vertices collected determines the statistical uncertainties associated with the distribution - the total number of vertices depends on the length of the mixing cycle and the number of runs performed.
The number of vertices available for this analysis is somewhat smaller than is desirable, but is the best data available.
The use of two independent data sets helps to validate any conclusions made.

The positron temperature was measured during a mixing cycle by releasing the positron plasma from a well and measuring the tail of the Boltzmann distribution, as described in section \ref{sec:plasmaTemperature}.
The temperature is corrected based on the results of simulations of the temperature measurement process to produce a good estimate of the positron plasma temperature.

\begin{table}[h]
\centering
\begin{tabular}{| l | l | l | l |}
\hline
\textbf{Data set} & \textbf{Positron density} & \textbf{Temperature} & \textbf{Number of Vertices}\\
\hline
2009-1 &  $1.8 \times 10^8~\mathrm{cm^{-3}}$& 220~K & $1.3 \times 10^4$\\
\hline
2009-2 & $7 \times 10^7~\mathrm{cm^{-3}}$ &  140~K & $5.0 \times 10^3$\\
\hline
\end{tabular}
\caption[The parameters describing the measured annihilation distributions.]{The parameters describing the data sets to be used in this analysis.}
\label{table:datasets}
\end{table}

\subsection{Modelling the distribution}

For this analysis, the conditions that must be known or measured \textit{a priori} are the number of positrons and their radial density profile, the potentials applied to the electrodes, and the magnetic field.
With this data, the Poisson-Boltzmann equations \ref{eq:poisson} and \ref{eq:boltzmann} can be solved numerically to produce the two-dimensional positron density and  electric field.
The equations also require the temperature of the positron plasma, but in reality, a change in temperature modifies only the details of the density and potential in the Debye tails of the  plasma, and is not important here.

The position at which the antihydrogen atom is formed, $\mathbf{r_0}$, is randomly picked from the positron density distribution.
This assumes that the antihydrogen atoms are formed uniformly throughout the plasma; this assumption is not necessarily justified and will be examined later.

The initial thermal velocities are drawn randomly from a defined distribution. 
The velocities of a thermalised distribution will follow a Gaussian distribution in each degree of freedom such that density of atoms with velocities between $v$ and $v+\mathrm{d}v$ is given by
\begin{equation}
	N\left(v\right) = \mathrm{exp}\left(-\frac{m_{\Hbar} v^2}{2k_\mathrm{B} T}\right) \mathrm{d}v.
	\label{eq:N_MaxBolt}
\end{equation}

The $\mathbf{E}\times \mathbf{B}$ velocity, $v_{E\times B}$ is calculated using the size of the transverse component of the electric field, $F_r$, at $\mathbf{r_0}$, from the solution to the Poisson-Boltzmann equations, using equation \ref{eq:plasmaExB}.
The total transverse velocity, $v_r$ is then simply the vector sum, at a random angle, of the transverse component of the thermal velocity and $v_{E\times B}$.

The $z$-coordinate of position where the atom's trajectory intersects the wall is calculated simply from
\begin{equation}
	z_{ann} =\left(r_{ann} - r_0\right)\frac{v_z}{v_r} + z_0,
	\label{eq:z_annihilation}
\end{equation}
where the annihilation position, $\mathbf{r}_{ann}$  is  $\left(z_{ann}, r_{ann}\right)$, the initial position, $\mathbf{r}_0$, is $\left(z_0, r_0\right)$, and the velocity, $\mathbf{v}$, is $\left(v_z, v_r\right)$.
The simulation assumes that all annihilations occur at the electrode surface, which fixes $r_{ann}$.

To analyse the data, the range over which annihilations occur is divided into `bins'.
Each annihilation adds a count to the bin appropriate to its position.
To mimic the experimental observations, the annihilation distribution produced is convolved with the resolution function of the vertex detector (section \ref{sec:siliconDetector}).
A smooth distribution is produced by simulating at least $10^5$ atoms.
The measured and simulated distributions are normalised to have equal areas to allow  direct comparisons to be made.

When comparing the simulated and measured annihilation distributions, a quantitative measure of the level to which they agree is very useful.
To this end, we use the chi-squared test statistic, a commonly-used technique when comparing measured and theoretical frequency distributions. 
It is defined as
\begin{equation}
	\Chi^2 = \sum_i \frac{\left(N_{meas}-N_{sim}\right)^2}{N_{meas}},
\end{equation}
where $N_{meas}$ is the number of counts in bin number $i$ in the measured distribution, and $N_{sim}$ is the corresponding number in the simulated distribution.
Each term in the expression is essentially the (squared) number of standard deviations of uncertainty that the measured and simulated distributions differ by, where the uncertainty is given by $\sqrt{N_{meas}}$.

By varying the input parameters of the distribution, and finding the parameters that minimise the value of $\Chi^2$, the `best-fit' input parameters can be found.
For two distributions that are drawn from the same parent distribution -- i.e that exactly correspond -- the value of $\Chi^2$ will be expected to be close to the number of degrees of freedom in the fit.
The number of degrees of freedom ($ndf$) is simply the number of bins less the number of fit parameters.
The statistical uncertainty on the best-fit parameters can be evaluated by considering the range over which $\Chi^2$ increases (i.e. the fit worsens) by an amount that is a function of the confidence level and the number of degrees of freedom \cite{Chi2Uncertainty}.
Because the models used here will not perfectly represent the measurements, the value of $\Chi^2/{ndf}$ will tend to be somewhat larger than one.
However, we can still compare the relative sizes of $\Chi^2$ to compare the quality of the fit. 

\subsection{The influence of the temperature}\label{sec:VertexTemperature}

If the antiprotons are in thermal equilibrium with the positron plasma, the thermal velocities in the longitudinal and transverse directions will be identical, and if the $\mathbf{E}\times \mathbf{B}$ velocity is ignored, the annihilation distribution will simply be the projection of a circle onto a line, convolved with the shape of the positron plasma and the resolution of the detector.
When the $\mathbf{E}\times \mathbf{B}$ rotation is included, the shape of the distribution will be set by the relative sizes of the thermal velocity and the $\mathbf{E}\times \mathbf{B}$ velocity.
At high temperatures, the thermal velocity dominates, and the distribution approaches the isotropic case.
At lower temperatures, the ratio of transverse to longitudinal velocity increases, and the distribution becomes more sharply peaked about the centre of the positron plasma \cite{ATHENA_SpatialDistribution}.

\begin{figure}[H] 	
	\centering
	\input{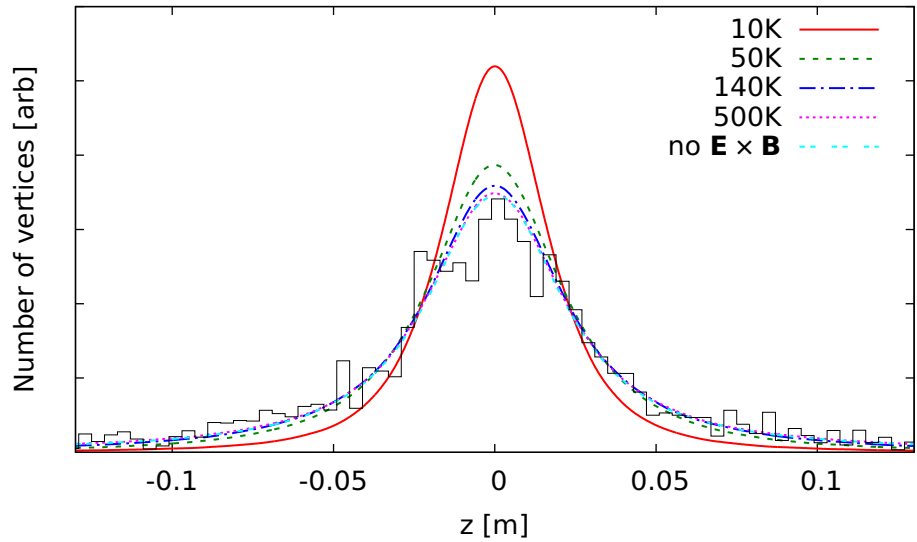}
	\caption[Simulated annihilation distributions, assuming thermal equilibrium.]{Simulated distributions, using the 2009-2 positron plasma assuming equal temperatures in the antiproton parallel and perpendicular motions of 10~K, 50~K, 140~K, and 500~K. These are compared to the distribution neglecting the $\mathbf{E}\times \mathbf{B}$ velocity and the measured distribution.	}
	\label{fig:isotropCompare}
\end{figure}

This is illustrated in figure \ref{fig:isotropCompare}.
It can be seen that the distribution at high temperatures converges towards the distribution ignoring the $\mathbf{E}\times \mathbf{B}$ velocity.
The $\mathbf{E}\times \mathbf{B}$ velocity at the surface of this positron plasma is approximately 600 $\mathrm{m\,s^{-1}}$, equal to the thermal velocity of a 15~K antiproton distribution.
This sets the range over which the $\mathbf{E}\times \mathbf{B}$ velocity is dominant.

\begin{figure}[t!]
	\centering
	\input{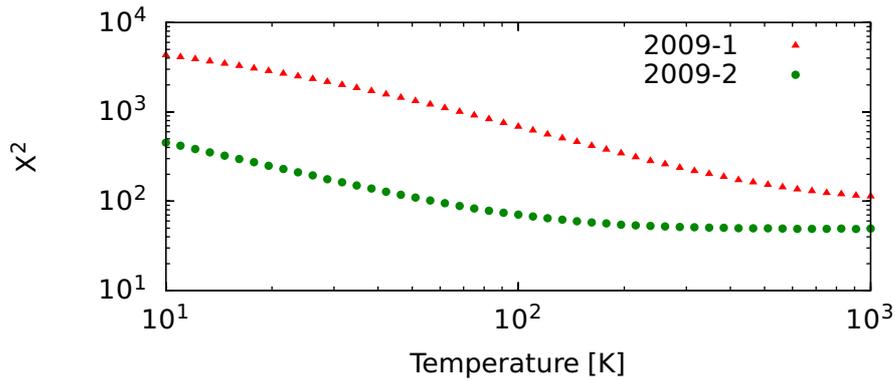}
	\caption{$\Chi^2$ for an isotropic fit to the data, as a function of the temperature.}
	\label{fig:isotropFit}
\end{figure}

The only free parameter in this model is the antiproton temperature.
To determine the temperature that produces the simulated distribution that best matches the data, a series of curves were generated at different temperatures.
The values of $\Chi^2$ as a function of temperature for this data is shown in figure \ref{fig:isotropFit}.
$\Chi^2$ decreases monotonically until reaching a `corner' and then takes on a more-or-less constant value.
This reflects the interplay of the thermal contribution to the velocity of the atom and the portion due to the $\mathbf{E}\times\mathbf{B}$ velocity, discussed above.
The corner in the 2009-2 data is seen at several times the temperature corresponding to the maximum $\mathbf{E}\times\mathbf{B}$ velocity, which is consistent with this model, while the corner in the 2009-1 set occurs later, as expected from the higher density of positrons (and thus, larger $\mathbf{E}\times\mathbf{B}$ velocity) in this experiment.

\subsubsection{Formation Position}
\begin{figure}[b!]
\centering
\input{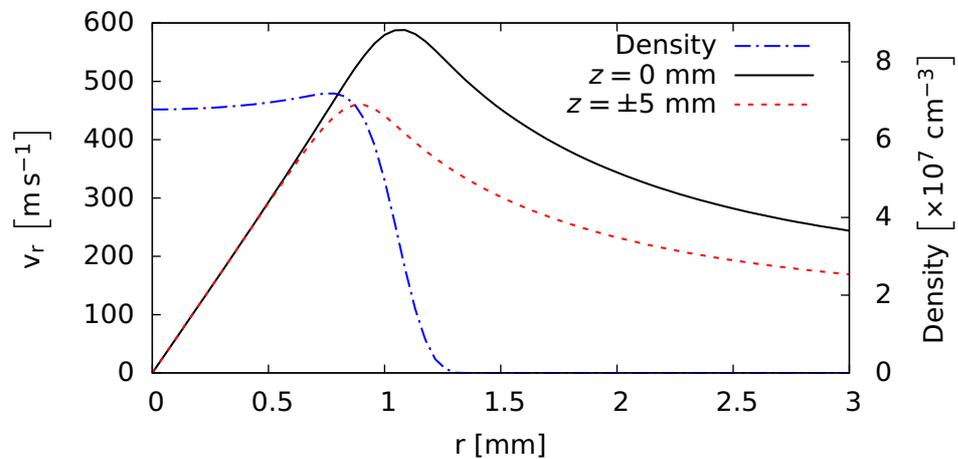}
\caption[The $\mathbf{E}\times\mathbf{B}$ component of the transverse velocity.]{The $\mathbf{E}\times\mathbf{B}$ component of the transverse velocity for the 2009-2 positron plasma, shown at the longitudinal centre of the plasma (z=0) and offset by 5~mm.}
\label{fig:isotrop_vr}
\end{figure}

The $\mathbf{E} \times \mathbf{B}$ contribution to an atom's velocity is determined by the electric field at the position of formation.
The velocity as a function of radius is shown in figure \ref{fig:isotrop_vr} for two longitudinal positions, and clearly increases with radius inside the positron plasma.
The initial trajectory, and thus the annihilation position, is also dependent on the position at which the atom forms.
The spatial distribution of antihydrogen formation should therefore also be considered.

In section \ref{sec:NestedTrap}, the process of radial redistribution of the antiprotons through repeated antihydrogen formation and ionisation was discussed.
Antihydrogen formed near the plasma edge will, on average, spend less time inside the plasma and thus will be more likely to leave the plasma without being collisionally ionised.
If the mean distance travelled before being ionised is small compared to the size of the plasma, most of the observed antihydrogen atoms would have originated close to the edge of the positron plasma.

On the other hand, if the radial density profile of the injected antiprotons has a small width compared to that of the positrons, and the ionisation probability is low, a significant fraction of the detected antihydrogen atoms will have been formed close to the axis.
In the extreme case, all of the antihydrogen will be formed at $r=0$, while a more realistic case would be to consider the radial density profile of the injected antiprotons.
The radial density profile of the antiprotons is measured using the MCP/Phosphor device (section \ref{sec:MCP}), and has a width of $\sim$0.7~mm, which is slightly smaller than the positron plasma, which is approximately 1~mm wide.

\begin{figure}[htb]
\centering
\input{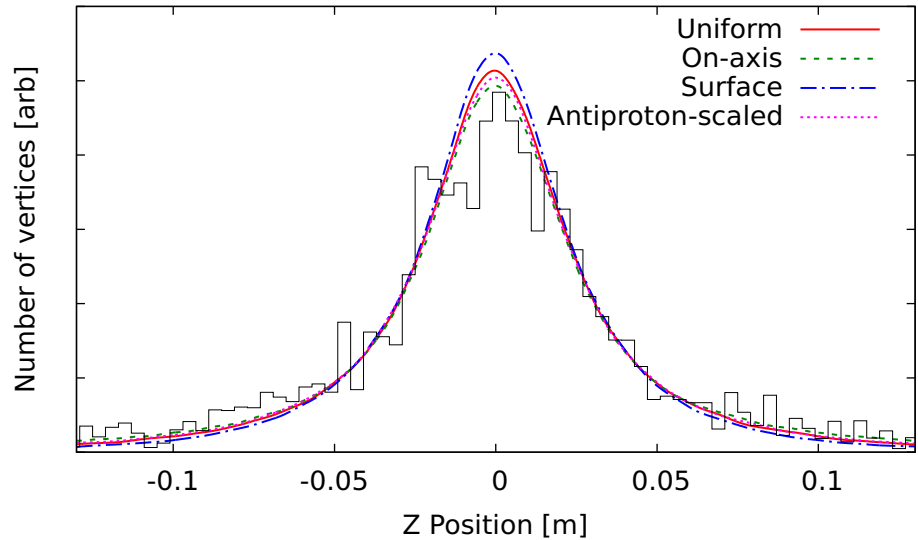}
\caption[A comparison of the annihilation distributions for different formation spatial distributions.]{A comparison of the annihilation distributions obtained where antihydrogen formation occurs uniformly throughout the 2009-2 positron plasma with an isotropic temperature of 140~K, with the cases where antihydrogen formation is concentrated on the axis of the Penning trap, concentrated at the surface of the plasma and with a distribution that follows the measured antiproton density.}
\label{fig:formationSpatDist}
\end{figure}

Figure \ref{fig:formationSpatDist} shows the `normal' uniform spatial distribution with simulated distributions for the two most extreme cases - formation entirely on the axis of the trap, and formation solely at the surface of the positron plasma. 
Also shown is the distribution obtained when the antihydrogen production at a given radius is scaled by the antiproton density at that radius (measured using the MCP/Phosphor device).

To avoid having too many free parameters in the model, it is best to select a single formation distribution for use in the simulations to follow.
In reality, there is little difference between the models, so the choice is not overwhelmingly important.
The rest of the simulations will assume that the probability of formation at a given position is proportional to the positron density (which is almost uniform inside the plasma), and  weighted in the $r$ direction by the measured antiproton density profile.

\subsubsection{Anisotropic temperatures}

The model curve at 140~K shown in figure \ref{fig:isotropCompare} reproduces the approximate shape of the antihydrogen distribution.
This supports the hypothesis that the antiprotons are in thermal equilibrium with the positron plasma when they form an antihydrogen atom, but we can further test this hypothesis by imagining a non-equilibrium distribution of the antiproton velocities and comparing this annihilation distribution to the data.
Following the analysis in \cite{ATHENA_SpatialDistribution}, we use an antiproton distribution that has evolved to independent equilibria in the parallel and perpendicular degrees of freedom.
This is a plausible scenario, since the rate of energy transfer, and thus the speed at which equilibrium is reached, along the magnetic field is faster than energy transfer across the magnetic field.
The velocity distribution will be characterised by the perpendicular temperature $T_\perp$ and the ratio $\kappa = T_\parallel/T_\perp$.

\begin{figure}[htb]
	\centering
	\input{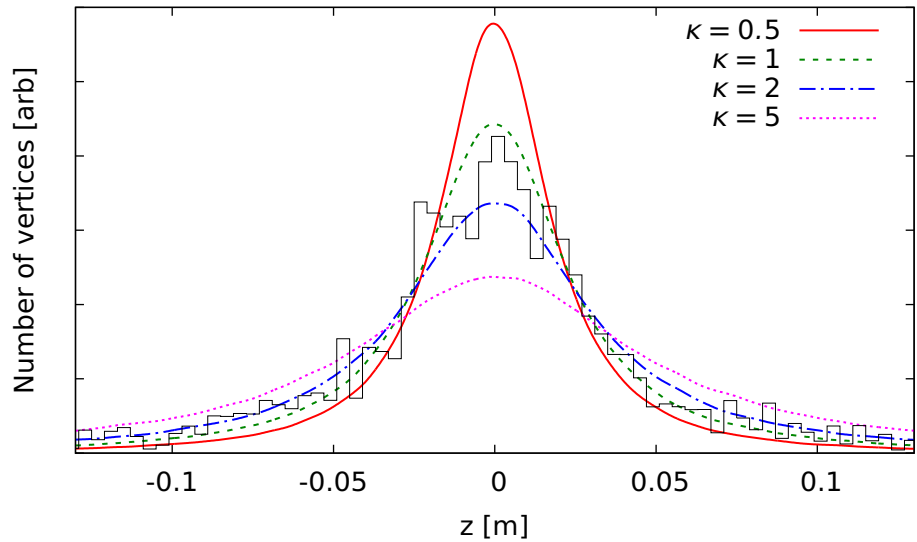}
	\caption[Example annihilation distributions with varying $\kappa$.]{A comparison of modelled annihilation distributions for 2009-2 positron plasma and electric potential, varying $\kappa$ and holding $T_\perp$ at 140~K. }
	\label{fig:compareKappa}
\end{figure}

A series of simulated curves illustrating the effect of different values of $\kappa$ at constant $T_\perp$ are shown in figure \ref{fig:compareKappa}.
It can be seen that values of $\kappa>1$ extend the tails to either side of the positron plasma, while reducing the relative size of the central peak, and achieves a better match between the simulated and measured distributions.
For the data shown in \ref{fig:compareKappa}, it is clear that the value of $\kappa$ that best describes the data for $T_\perp$ = 140~K is roughly between 1 and 2.

\begin{figure}[bt!]
	\centering
	\input{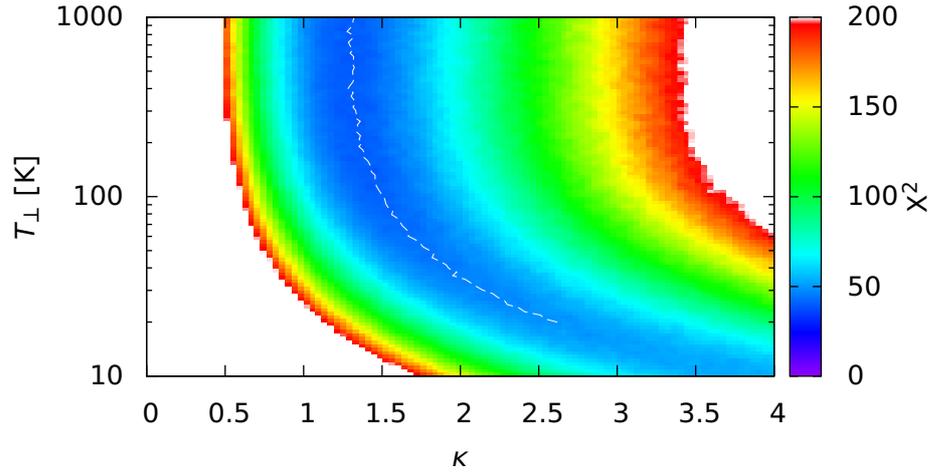}
	\caption[$\Chi^2$ as a function of $\kappa$ and $T_\perp$.]{A two-dimensional colour map showing the variation of $\Chi^2$ with $\kappa$ and $T_\perp$ for the 2009-2 data set. The white regions are beyond the scale of the colour map. The white line indicates the best-fit values of $\kappa$ for each value of $T_\perp$.}
	\label{fig:kappaScan2D}
\end{figure}

\begin{figure}[bt!]
	\centering
	\subfloat[]{\input{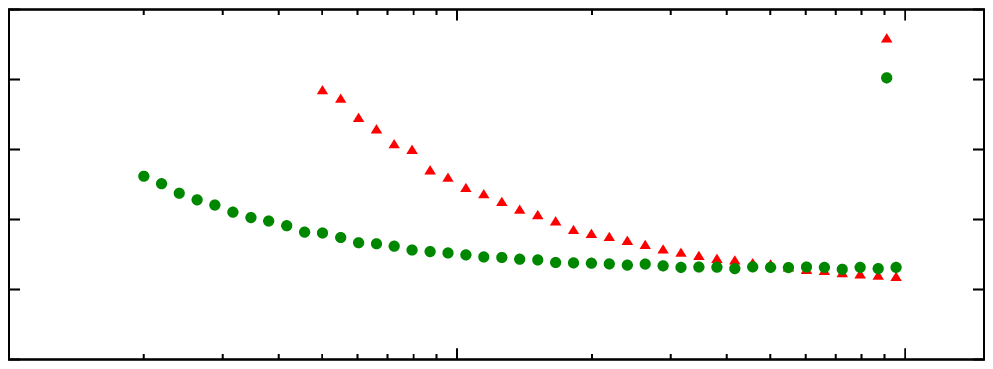}}\\
	\subfloat[]{\input{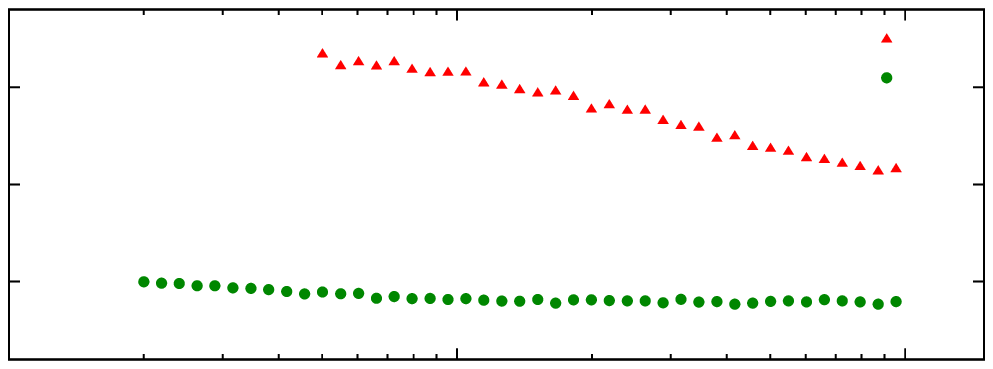}}
	\caption[The best fit values of $\kappa$ and $\Chi^2$ as a function of $T_\perp$.]{(a) the value of $\kappa$ that minimises $\Chi^2$ at a given value of $T_\perp$ and (b) the corresponding values of $\Chi^2$.}
	\label{fig:kappaScan}
\end{figure}

This model now has two free parameters - $\kappa$ and $T_\perp$, so the fitting procedure must be extended to find the `best-fit' parameters.
Figure \ref{fig:kappaScan2D} shows a two-dimensional colour map plot of $\Chi^2$ for different pairs of $\kappa$ and $T_\perp$.
It is clear that for each value of $T_\perp$, there is a value of $\kappa$ for which $\Chi^2$ is minimised.
These values are indicated by the white line in figure \ref{fig:kappaScan2D} and are shown as a function of $T_\perp$ for both data sets in figure \ref{fig:kappaScan}.
The data qualitatively reproduces the features of the previous experiment, reported in \cite{ATHENA_SpatialDistribution}.
At low temperatures, a large value of $\kappa$ is required to fit the data, while at higher temperatures, the best-fit value decreases and tends towards a constant value at very high temperatures.
The values of $\kappa$ at high $T_\perp$ for the two data sets are $1.2 \pm 0.1$ (2009-1) and $1.3 \pm 0.1$ (2009-2), much closer to the equilibrium case of $\kappa = 1$ than the minimum value of $\kappa = 2.3$ reported in \cite{ATHENA_SpatialDistribution}.
This is a strong indication that the antiprotons thermalise with the positron plasma before forming antihydrogen.

From figure \ref{fig:kappaScan}, we see that there is no global minimum value of $\Chi^2$, which prevents this method predicting the temperature.
We therefore make use of the measured positron temperature to allow us to estimate $\kappa$.
From the way the experiment is set up, it is reasonable to assume that the antiprotons can only be \textit{more} energetic in the direction parallel to the magnetic field than in the transverse direction.
The antiproton temperature measured before injection is also higher than the positron temperature.
Thus, it reasonable to take the positron temperature as the minimum value possible value of $T_\perp$.
From figure \ref{fig:kappaScan2D}, we see that if $T_\perp$ is in fact higher than this minimum, the value of $\kappa$ will be closer to 1.

\begin{figure}[bt!]
	\centering
	\subfloat[]{\input{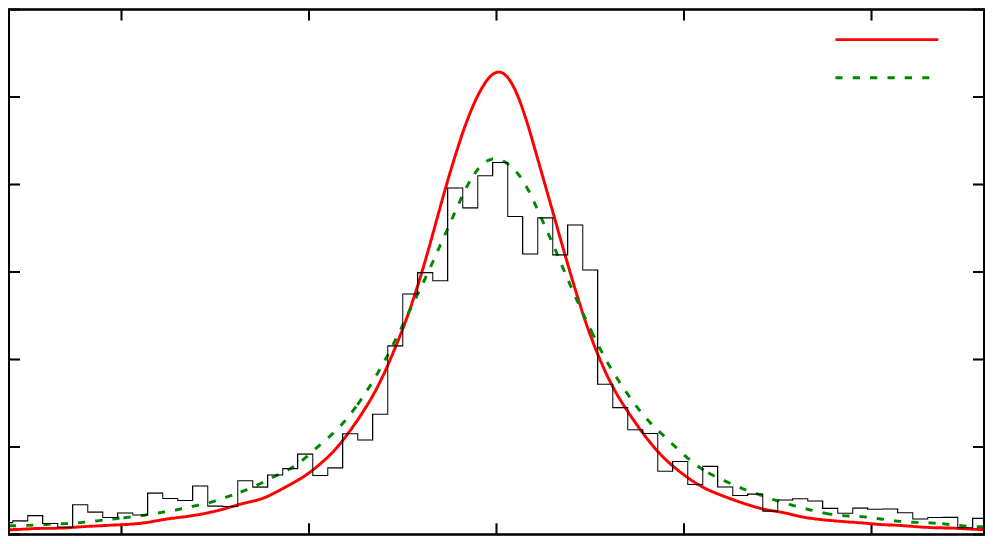}}\\
	\subfloat[]{\input{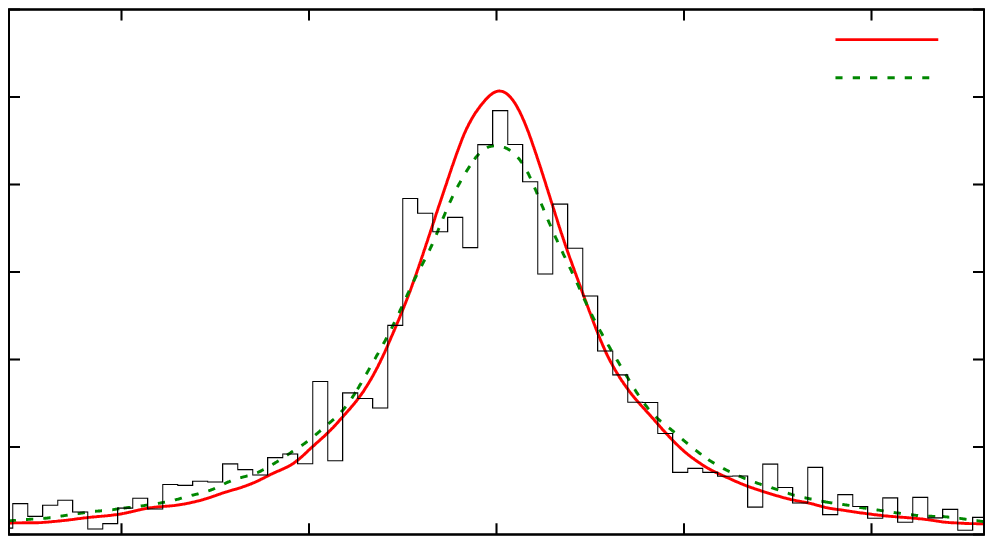}}\\
	\caption[A comparison of the best fit and isotropic distributions.]{A comparison of the best fit and isotropic distributions for (a) the 2009-1 and (b) the 2009-2 data. $T_\perp$ is 220~K in (a) and 140~K in (b).}
	\label{fig:bestKappa}
\end{figure}

In the 2009-1 data set, the temperature (see table \ref{table:datasets}) is approximately 220~K. 
For this temperature, $\Chi^2$ is minimised for $\kappa = 1.7 \pm 0.1$.
Similarly, the temperature for the 2009-2 data set was approximately 140~K, and $\Chi^2$ is minimised for $\kappa = 1.4 \pm 0.1$.
The simulated distributions for these parameters are compared to the measured distributions in figure \ref{fig:bestKappa}.

The uncertainties quoted here are the $1\sigma$ statistical errors calculating using the method outlined in \cite{Chi2Uncertainty}.
There are also systematic errors, which are harder to estimate.
A rough idea of the influence of systematic errors can be obtained by examining the value of $\Chi^2/ndf$.
Recall that for a fit function that describes the data to within the measurement uncertainties, $\Chi^2/ndf \simeq 1$.
For the data presented here, $ndf = 33$, and $\Chi^2/ndf$ for the 2009-1 and 2009-2 points discussed above are $\sim 4.1$ and $\sim 1.2$ respectively.
The corresponding values of $\Chi^2/ndf$ for the isotropic ($\kappa=1$) case are $\sim 9.5$ and $\sim 1.8$, respectively.
At face value, this implies that the model has produced a good fit to the 2009-2 data set, but a poorer one for the 2009-1 data.
A possible explanation for this can stem from the stronger electric fields present in the Penning trap for the 2009-1 experiment.
These electric fields can ionise atoms in flight to the wall and significantly alter the annihilation distribution, as will be seen in the next section.

\section{Annihilation distribution in ionising fields}

\label{sec:vertex_fieldionisation}

The analysis of the antihydrogen annihilation distribution measured in ATHENA \cite{ATHENA_SpatialDistribution} assumed that the antihydrogen atoms travelled from their point of formation to the surface of the Penning-trap electrodes on a straight line without incident.
This neglects the possibility that antihydrogen atoms can be field-ionised (section \ref{sec:fieldionisation}).

The fate of a field-ionised antiproton has already been discussed in section \ref{sec:NestedTrap}.
In a purely solenoidal magnetic field, if the atom is ionised in a region where the antiproton is confined axially by electric fields, the antiproton will remain confined in the trap, otherwise, it will exit the trap through the ends.
In either case, no annihilation is identified to record the formation of the original antihydrogen atom.

The possibility of field-ionisation is irrelevant if the antihydrogen atoms are tightly bound enough that they cannot possibly be ionised by the size of the electric fields found in the Penning trap.
However, the theoretical description of three-body recombination, simulations, and experimental work, including use of field-ionisation as an antihydrogen detection technique, provide overwhelming evidence that the antihydrogen atoms are in states weakly enough bound to be susceptible to ionisation by the fields found in the Penning trap.

\begin{figure}[bt]
	\centering
	\input{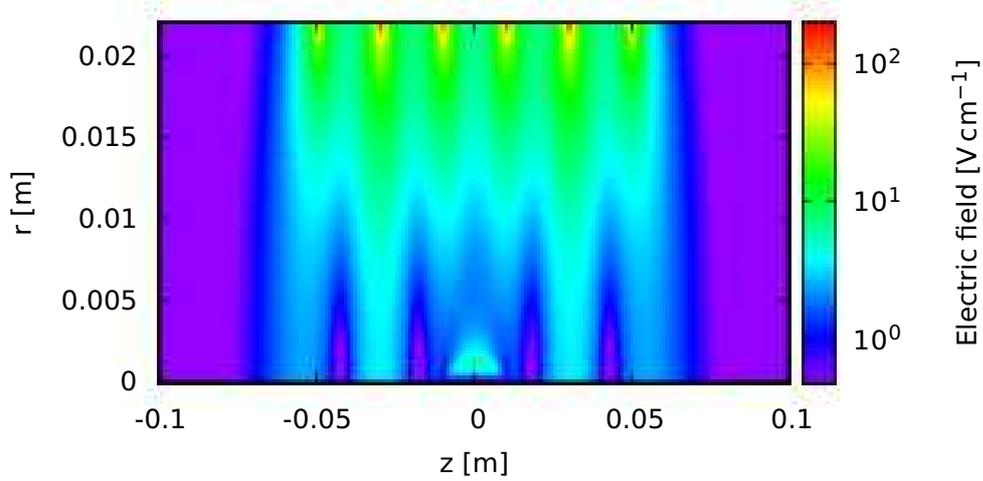}
	\caption[The electric field magnitude in the Penning trap.]{A two-dimensional colour-map showing the electric field magnitude in the Penning trap.}
	\label{fig:example_Efield2D}
\end{figure}

\begin{figure}[bt]
	\input{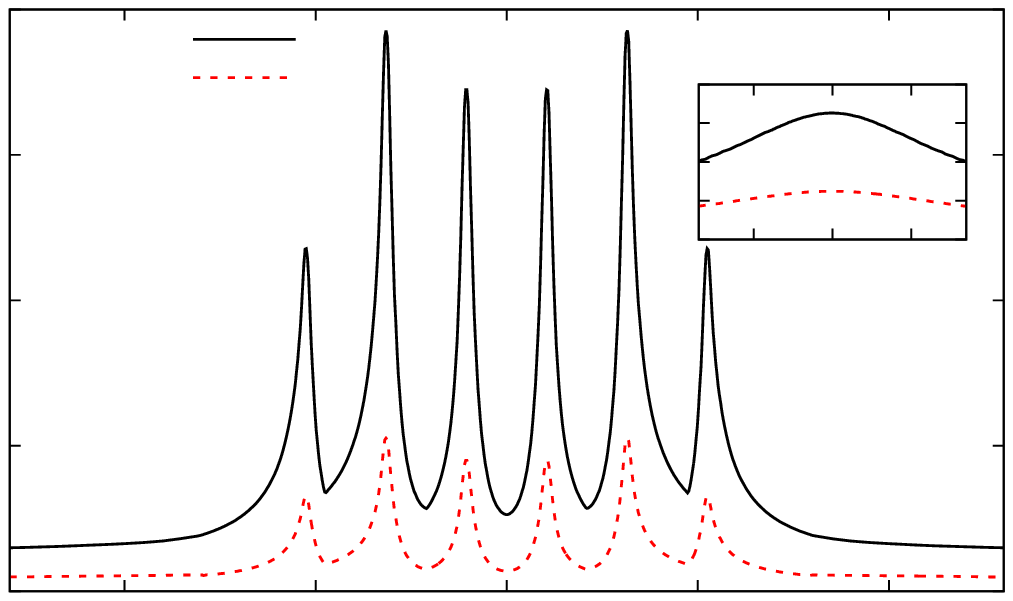}
	\centering
	\caption[The magnitude of the strongest electric field experienced by an atom moving from the centre of the trap to the electrode walls.]{The magnitude of the strongest electric field experienced by an atom moving from the centre of the trap to the electrode walls, as a function of the position where the trajectory intersects the electrode walls. The inset shows the maximum field experienced by the time the atom has just left the positron plasma.}
	\label{fig:Efield_theta}
\end{figure}

It could also be that the atoms have the same probability to be ionised regardless of the path they take between formation and annihilation.
Since the electric fields in the Penning trap are not uniform, this is patently not true.
Figure \ref{fig:example_Efield2D} shows the two-dimensional electric field magnitude in the trap configuration for the 2009-2 data set.
In figure \ref{fig:Efield_theta}, the maximum electric field experienced by an atom formed at the centre of the trap and annihilating on the trap wall as a function of the expected $z$-coordinate of the annihilation position is shown for both of the data sets.
From these graphs, one can clearly see that the electric field strength and thus, the probability of ionisation, is trajectory-dependent.

The shape of the electric field distribution shown in figure \ref{fig:Efield_theta} is made up of two components.
The first is a broad peak centred over the positron plasma that results from the electric fields induced by the plasma's charge. 
This contribution has been separated and is shown in the inset.
The large, spiked features near the centre of the distribution are due to the large electric fields induced between Penning trap electrodes set to different voltages.
It is notable that the size of the highest electric fields changes by almost a factor of four between the two data sets.
This is as a result of shallower nested traps being used for later experiments.

\subsection{Modelling the distribution}
Ionisation is included in the previous simulation by assigning each atom a binding energy randomly picked from a defined distribution.
The distribution of binding energies is not known \textit{a priori}.
However, a power-law-like dependence has been observed in some simulations \cite{Svante_JPhysB} and experiments \cite{GabrielseReview}, so using a similar distribution seems reasonable.
In a power-law distribution, the number of atoms with binding energy between $\mathcal{E}$ and $\mathcal{E}+\mathrm{d}\mathcal{E}$ follows a power-law dependence -- i.e.
\begin{equation}
	\mathrm{d}N(\mathcal{E}) = \mathcal{E}^{-a} \,\mathrm{d}\mathcal{E},
	\label{eq:dN_epsilon}
\end{equation}
where $\mathrm{d}N$ is the number of particles in the interval $\left(\mathcal{E}, \mathcal{E}+\mathrm{d}\mathcal{E}\right)$.
Use of a power law has the advantage of being parameterised by only one variable, thus reducing the dimensionality of the overall parameter space and requiring a smaller amount of computational effort.
The power-law distribution must be bounded to have a finite integral.
The lower limit is chosen to be low enough that no atom with lower binding energy can leave the positron plasma without being ionised by the internal electric fields. 
The value of this limit depends on the positron plasma used -- the typical value for the parameters used here is on the order of 1~meV.
At the upper limit, the binding energy of the ground state of hydrogen, -13.6~eV, was chosen, since no atom can be more strongly bound.
The precise value of the upper limit is not as important as the lower limit, since the number of atoms at deep binding energies is suppressed for positive values of $a$, and thus deeply bound atoms do not have a large impact on the annihilation distribution.

The electric field at each point along the straight-line trajectory from the formation position to the predicted impact position is compared to the electric field required to ionise the atom, calculated from equation \ref{eq:Fc_epsilon_au}.
As previously discussed, this equation only applies to central states of the atom, but the effect of non-central states will be discussed in due course.
We consider ionisation to have occurred when the binding energy falls to zero, thus if the electric field exceeds the ionisation threshold, the atom is discarded (recall that some bound states with \textit{positive} energy are possible, but ignored).
An atom that survives to reach the electrode surface is considered to have annihilated, and its impact position is added to the annihilation distribution.

If the atom is ionised close to the electrode surfaces, the antiproton's cyclotron orbit could still intersect the electrodes, and the particle would still annihilate.
For this reason, any atoms that are ionised closer than 0.2~mm to the electrode surfaces are included in the annihilation distribution.
Altering this threshold slightly in either direction does not have a significant impact on the distributions.

The change in binding energy from the motional Stark shift due to the motion of the atom across the magnetic field is also included.
The degree to which the motional effect plays a role depends on the velocity distributions of the antihydrogen atoms.
For $v_r = 1,000~\mathrm{m\,s^{-1}}$, (typical of a 140~K distribution) the size of the motional Stark shift in 1~T is $10^3~\mathrm{V\,m^{-1}}$.
The experiments have electric fields a few times $10^3~\mathrm{V\,m^{-1}}$, so the effect is non-negligible, and must be taken into account.

\begin{figure}[hbt]
	\centering
	\input{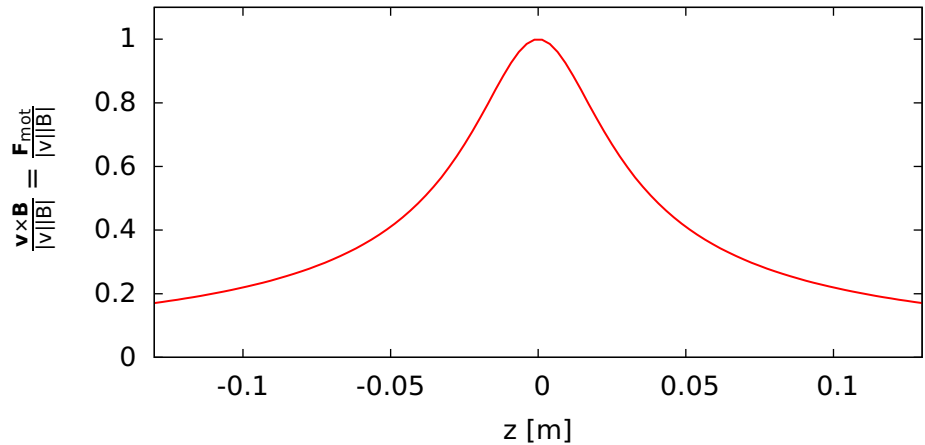}
	\caption[The size of the induced motional Stark shift.]{The size of the induced motional Stark shift, normalised to the atom's transverse speed and the strength of the magnetic field, as a function of the annihilation position, for an atom starting at (0,0).}
	\label{fig:motStarkShift}
\end{figure}

Similar to the static electric field, the motional Stark shift has an angular dependence, and can have an impact on the annihilation distributions.
From equation \ref{eqn:motStarkShift}, atoms moving perpendicular to the magnetic field experience a stronger effective electric field than those moving parallel to the field.
The dependence on the annihilation position is shown in figure \ref{fig:motStarkShift}.

\subsection{Results}

\begin{figure}[b!]
	\centering
	\input{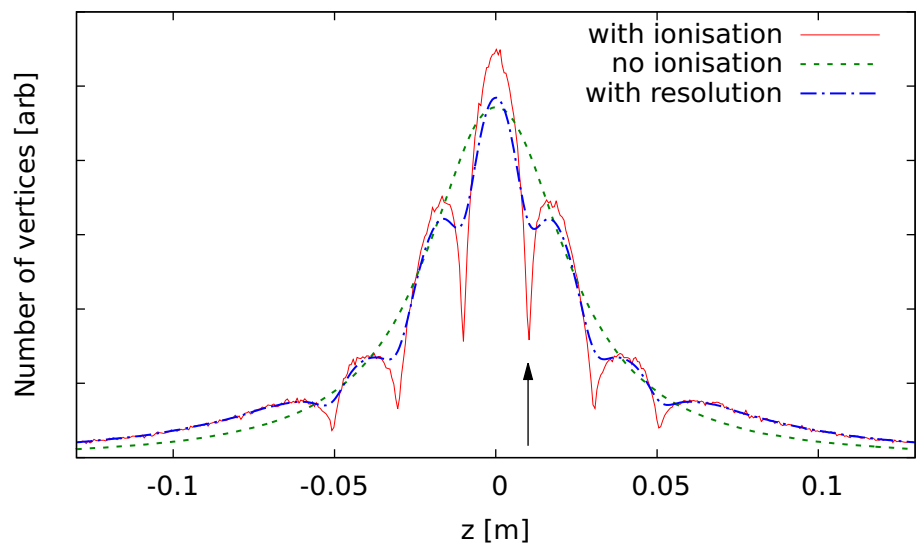}
	\caption[An example of the annihilation distribution obtained when field ionisation is taken into account.]{An example of the annihilation distribution obtained when field ionisation is taken into account. In the red curve, the response function of the detector has been made infinitely sharp to make the details of the distribution visible. This distribution is generated using a 2009-2 positron plasma with an isotropic temperature distribution at 140~K and a binding energy distribution proportional to $\mathcal{E}^{-2}$.}
	\label{fig:exampleEfield}
\end{figure}

An example of the effect of field ionisation on the vertex distribution is shown in figure \ref{fig:exampleEfield}. 
The resolution of the vertex detector obscures much of the detail, and has been suppressed in one of the graphs.
The most striking feature is the presence of regions where the number of annihilations is reduced, such as the region indicated by the arrow.
These correspond to positions where the electric field (figure \ref{fig:Efield_theta}) is high, and a large fraction of the atoms have been ionised.
When convolved with the resolution of the detector, the prominence of the features is significantly reduced.

\begin{figure}[t!]
	\centering
	\input{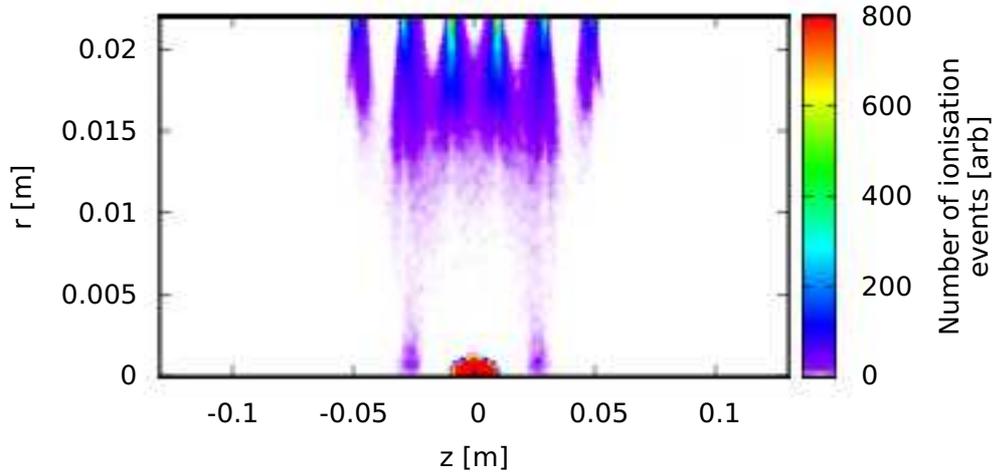}
	\caption[The ionisation positions of atoms.]{A colour-density map of (z,r) positions of atoms when ionised by the electric fields in the 2009-2 configuration. The colour-bar has been scaled so that atoms inside the positron plasma saturate the scale.} 
	\label{fig:ionPos}
\end{figure}

Figure \ref{fig:ionPos} shows the ($z,r$) positions of the atoms at the moment when the total binding energy falls to zero.
In this simulation, a large fraction of the atoms are ionised inside the positron plasma, but the limits of the colour-map have been chosen to suppress this feature.
It is clear that the positions of the ionisation events closely correspond to the regions of high electric field strength in the Penning trap shown in figure \ref{fig:example_Efield2D}.
Apart from the atoms that ionise inside the positron plasma (which anyway have a second chance to form antihydrogen), the majority of the atoms are ionised at high radius.

\begin{figure}[b!]
	\centering
	\input{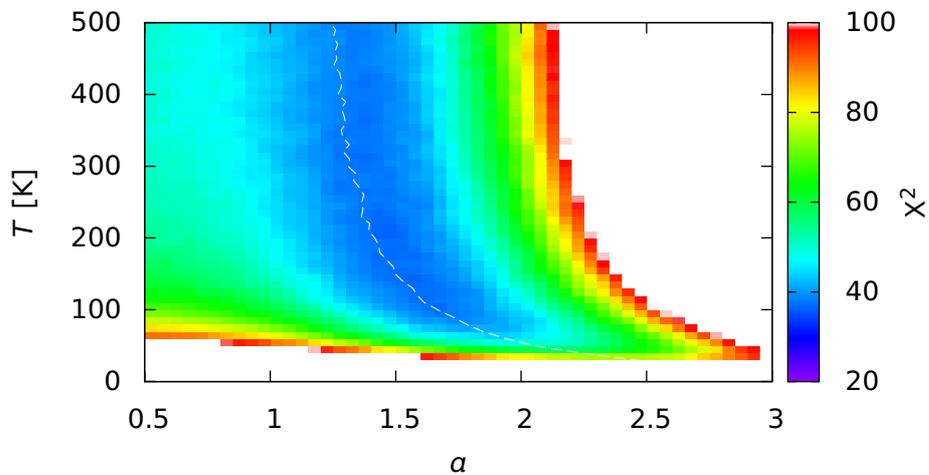}
	\caption[$\Chi^2$ as a function of $a$ and $T$.]{A two-dimensional colour map showing the variation of $\Chi^2$ with $a$ and $T$ for the 2009-2 data set. The white regions are beyond the scale of the colour map. The white line indicates the best-fit values of $a$ for each value of $T$.}
	\label{fig:statesScan2D}
\end{figure}

We will again use $\Chi^2$ to compare the simulated and the measured distributions, but a thermalised velocity distribution ($\kappa = 1$) will be used.
Thus two free parameters remain - the temperature $T$, and the exponent $a$ specifying the form of the power-law distribution of states.
We then proceed as before, varying the values of $T$ and $a$.
The two-dimensional map of $\Chi^2$, similar to figure \ref{fig:kappaScan2D}, for the 2009-2 distribution is shown in figure \ref{fig:statesScan2D}.

\begin{figure}[hbt]
	\centering
	\subfloat[]{\input{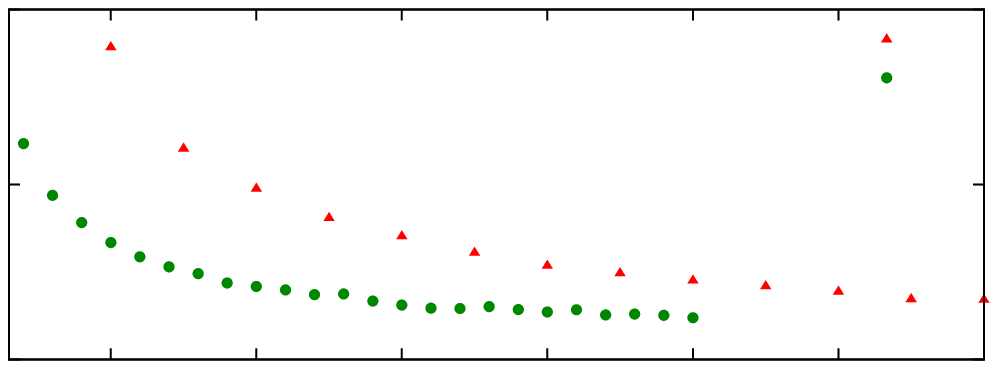}}
	\\
	\subfloat[]{\input{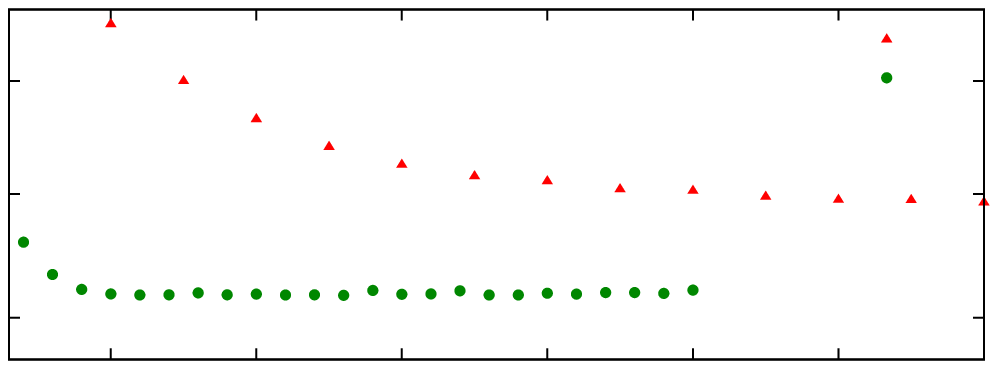}}
	\caption[The best-fit values of $a$ and $\Chi^2$ as a function of $T$.]{(a) the value of $a$ that minimises $\Chi^2$ at a fixed value of temperature, as a function of the choice of temperature. (b) the $\Chi^2$ values corresponding to the points in (a).}
	\label{fig:statesFit}
\end{figure}

\begin{figure}[tb!]
	\centering
	\subfloat[]{\input{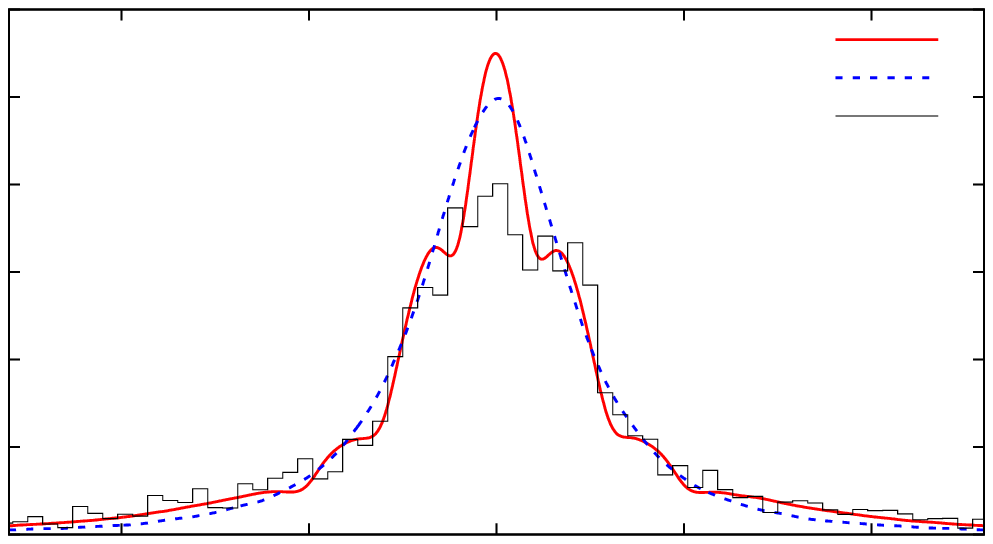}}
	\\
	\subfloat[]{\input{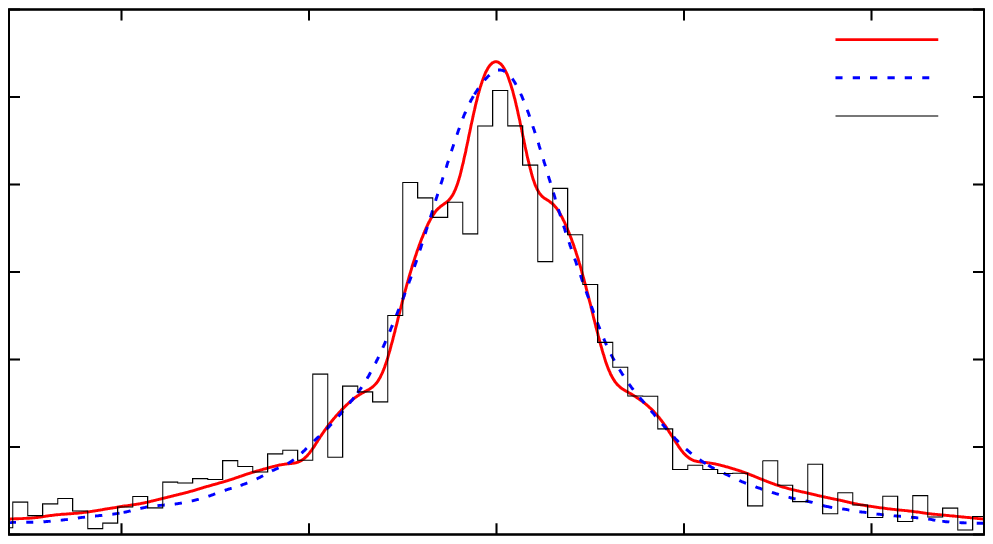}}
	\caption[A comparison of annihilation distributions for the best-fit binding energy distributions and the no-ionisation cases]{A comparison of annihilation distributions from the best-fit binding energy distributions and the no-ionisation cases for (a) the 2009-1 and (b) the 2009-2 data. $T$ is 220~K in (a) and 140~K in (b).}
	\label{fig:bestStatesFit}
\end{figure}

Similar to the case for $\kappa$, it is evident that for a given value of $T$, there is a value of $a$ that minimises $\Chi^2$.
These values, are plotted in figure \ref{fig:statesFit}.
There is a very shallow minimum of $\Chi^2$ at approximately 150~K, which agrees well with the measured positron temperature.
The best-fit value of $a$ at this temperature is $1.5 \pm 0.1$.
For the 2009-1 data, also plotted in figure \ref{fig:statesFit}, there is no global minimum of $\Chi^2$.
The best-fit value of $a$ for $T$ = 220~K is $1.9 \pm 0.1$.
The distributions produced using these parameters are shown in figure \ref{fig:bestStatesFit}.
Combining the results of both experiments and making the reasonable assumption that the antihydrogen formation processes are similar in both, the 95\% confidence interval for $a$ extends between 1.3 and 2.1.

The distributions obtained reproduce the narrower central peak flanked by `shoulders' to either side.
Similar features appear in the same positions in the measured annihilation distributions, suggesting that the collected data shows direct signs of the influence of the field-ionisation of antihydrogen atoms.
The central peak in the 2009-1 simulation, and to a lesser extent, also in the 2009-2 data, is systematically larger than the data.
The minimum $\Chi^2/ndf$ values for the data sets are $\sim 5.6$ (2009-1) and $\sim 1.1$ (2009-2).

\begin{figure}[h]
\centering
\input{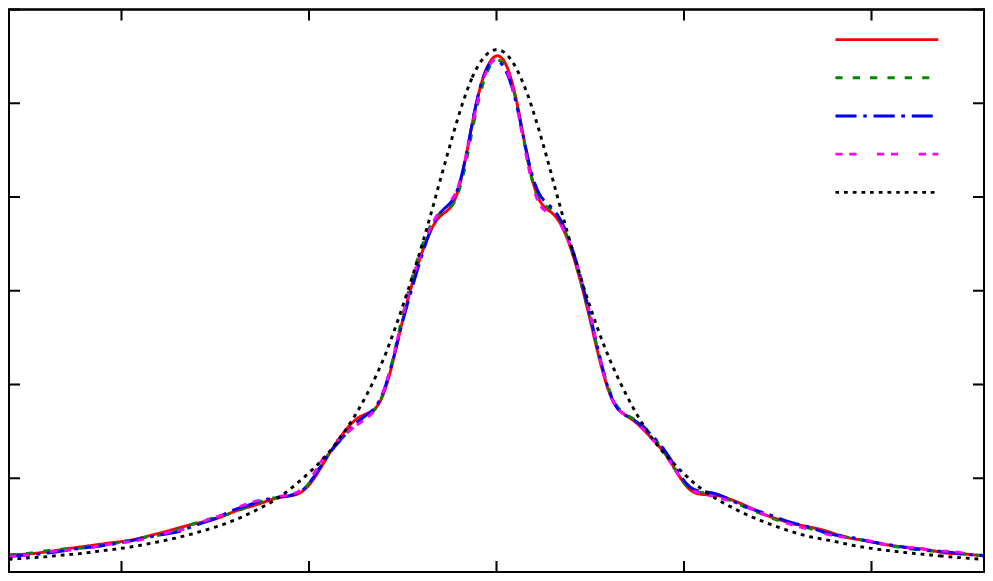}
\caption[The distributions produced by modifying the expression for the critical field.]{The distributions produced by modifying the expression for the critical field $F_c$ (equation \ref{eq:Fc_n_au}) to $s F_c$, for different values of $s$. The distributions are for 2009-2 case, with T=140~K and a = 1.5. Also shown is the distribution ignoring the ionising effect for these parameters.}
\label{fig:starkAComparison}
\end{figure}

As has already been mentioned, the expression for the critical field (equation \ref{eq:Fc_n_au}) is only valid for central atomic states.
We investigate the dependence of the atom polarisation (discussed in section \ref{sec:fieldionisation}) by altering the critical field by a constant value.
Guided by the analysis of \cite{Gallagher}, we change the critical field by a factor of two to four, which should encompass the most extreme changes possible.
The distributions produced for these changes are shown in figure \ref{fig:starkAComparison}.
The distributions for different values of $s$ (the modification factor, $F_c \rightarrow s F_c$) are practically indistinguishable on this graph, but differ significantly from the simulation where field ionisation is suppressed.
The same is true of simulations using the 2009-1 parameters.
This is strong evidence that any polarisation of the atoms does not affect the results of this analysis.
Performing the $\Chi^2$ minimisation fit for $a$ using different values of $s$ produces the same result to within the statistical error.

This lack of dependence can be explained by considering that most of the atoms do not ionise at the maximum values of electric field along their trajectories, but at some lower field that they encounter first.
So an atom that survives ionisation because of a larger value of $s$ is still likely to be ionised at a point further along its trajectory.
Because the position at which an atom ionises is not relevant to this analysis, only whether or not it actually ionises, very similar annihilation distributions are produced.
Any small changes to the annihilation distribution that are made are concealed by the detector's resolution (section \ref{sec:siliconDetector}).

Thus, this effect cannot be the source of the disagreement in the 2009-1 data.
Another possibility is that the $\kappa$ is not exactly 1, but is slightly larger, corresponding to incomplete thermalisation.
Adding a third parameter to the model would make the task of fitting the parameters excessively computationally demanding, so it is not done.
A simple study of changing $\kappa$ while keeping $T$ and $a$ fixed at their best-fit values indicates that a value of $\kappa$ between 1.2 and 1.5 would produce slightly better, but still not complete, agreement.

A potentially important effect that is neglected by these simulations is radiative decay of the atom from the weakly-bound, excited states.
This means that atoms become more strongly bound as they move towards the electrode surfaces from the positron plasma.
One analysis \cite{Francis_cascade} expressed the decay rate as the time required for 50\% of an ensemble of atoms to reach the ground state, as a function of the initial state.
We can use this information to produce a rough estimate of the important time scale -- around 5~ms for atoms bound by $\sim 10~\mathrm{meV}$.
Atoms from the 140~K distribution have a thermal velocity of approximately $1500~\mathrm{m\,s^{-1}}$, and cross the distance from the positron plasma to the electrode surfaces in of order $100~\mu s$, several times smaller than our estimate of the radiative decay time.
So, to first approximation, it seems that the effect is small enough to be neglected.
A full simulation of the effect would be too computationally demanding to implement, so it is not included in this analysis.

Several works, notably \cite{Svante_JPhysB} and \cite{BassAntihydrogen}, have discussed how collisions between antihydrogen atoms and positrons change the atoms' binding energy, possibly even ionising the atom.
Even though collisional effects are likely to be very important, they are not included in the simulation, as the computational complexity and cost is prohibitive.
The binding energy distributions used in this analysis will therefore correspond to the distributions of the atoms that are not collisionally ionised, at the point where they leave the positron plasma.
This process might also have an angular dependence, since the length of time the atom is inside the positron plasma depends on the trajectory it takes.
This would further complicate this analysis.

\begin{figure}[hb!]
	\centering
	\input{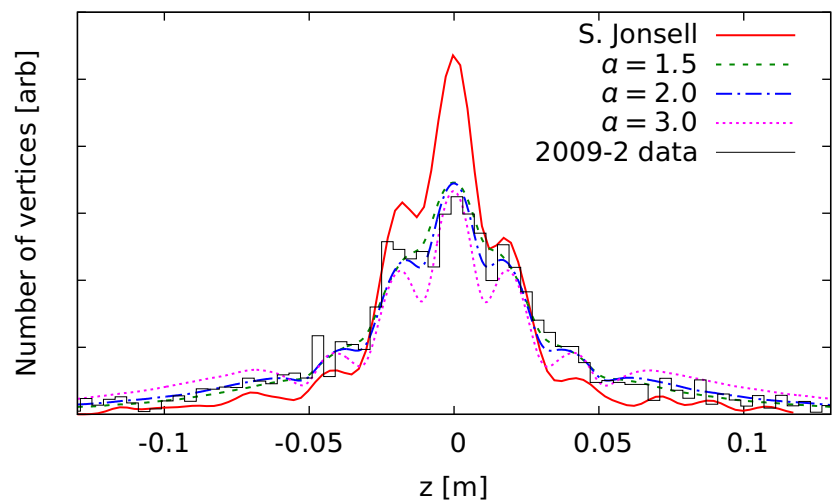}
	\caption[A comparison of the simulated distributions to a more complete simulation.]{A comparison of a distribution produced using a full treatment of the positron and antiproton motion compared to several variations of results from this study and to the measured distribution.}
	\label{fig:svanteCompare}
\end{figure}

A complete simulation, in which the full internal (classical) motion of the positrons and antiprotons is simulated, can be found in reference \cite{Svante_JPhysB}.
An example distribution, produced using the positron plasma used for the 2009-2 data set \cite{SvanteElog}, supplied by S. Jonsell, is shown in figure \ref{fig:svanteCompare}, compared to the distribution produced from this study and the measured distribution.
Also shown are a number of variations of this study's simulation, changing the value of $a$.
It appears that the distribution from the simulation of the full internal motion does not reproduce the measured distribution particularly well.
The number of annihilations in the centre of the distribution is systematically higher than the data and the number of annihilations in the tail is systematically lower.

Jonsell's distribution is not reproduced by this study, even if the binding energy distribution is altered substantially.
It does, however, contain the same essential features of a narrow central peak and regions where the antihydrogen atoms are preferentially ionised.
It would have been expected that Jonsell's simulation would better reproduce the measured annihilation distribution, as it simulates the system in full, without making simplifying assumptions that we have made here.
Why this is not the case is unknown, and may indicate that there is some as-yet unidentified process at work.
It should be remembered that the annihilation distributions produced from our study has an artificial advantage over Jonsell's in that $a$ is an adjustable parameter that can be arbitrarily changed to give good agreement with the measured distributions without necessarily having a good physical basis. 
A similar parameter does not exist in Jonsell's treatment.

\section{Annihilation distribution in a multipole magnetic field}

\label{sec:annihilationsMultipole}

In a uniform magnetic field, the antiprotons produced from field-ionised antihydrogen atoms do not hit the electrode walls, and either remain confined in the Penning trap or are guided by the magnetic field through ends of the electrode stack.
However, when a transverse multipole field such as from the octupole used in ALPHA to produce part of the magnetic minimum trap is applied, the trajectories of the antiprotons are modified and can impact the electrode wall.
These antiprotons are indistinguishable to the vertex detector from the antiprotons bound in antihydrogen atoms, and so will modify the measured annihilation distribution.

\begin{figure}[htb]
\centering
\input{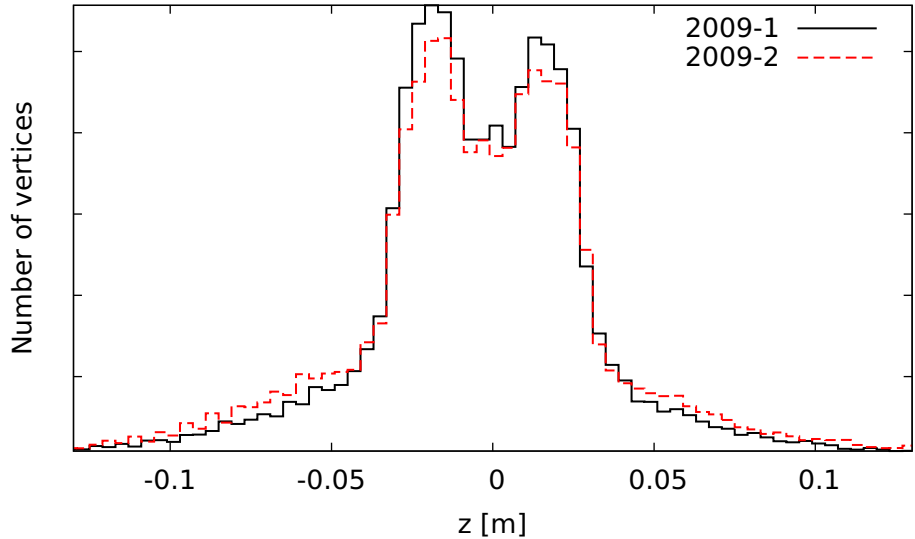}
\caption[Measured antiproton annihilation distributions when the octupole is energised]{The antiproton annihilation distributions when the octupole is energised. There are $4.0 \times 10^4$ vertices in the 2009-1 data set and $2.1 \times 10^4$ in the 2009-2 data set. The distributions have been normalised to have the same area.}
\label{fig:hbarInOct_measured}
\end{figure}

The distributions measured when the octupole is energised to its full strength (a field strength at the surface of the Penning trap electrodes of 1.55~T), are shown in figure \ref{fig:hbarInOct_measured}.
The most striking difference between these distributions and the uniform-field distributions shown in figure \ref{fig:Isotrop_measured} is the appearance of two peaks to either side of the positron plasma.

It is important to separate the effect of the octupole's magnetic field on the antihydrogen formation process and the dynamics outside the positron plasma.
The antihydrogen formation region, defined by the size of the positron plasma, is confined to a region within 1~mm of the trap axis.
In this region, the magnetic field changes by at most 1.4 parts in $10^4$ when the octupole is energised. 
This seems to justify the approximation that there is no effect on the formation process.
We should be wary, though, since it is known that some non-neutral plasmas can be affected by energising the octupole, even though the change in field strength is seemingly minor (section \ref{sec:OctupoleHeating}).
However, measurements of the influence on the positron plasmas used to produce this data showed only a small (10-20\%) increase in temperature, so we will make the assumption that the velocity and binding energy distributions of the antihydrogen produced are identical to those in the uniform field case, and the change in the observed annihilation distribution is solely due to new effects pertaining to field-ionisation and the motion of the antiprotons after field ionisation.

\begin{figure}[hbt]
	\centering
	\input{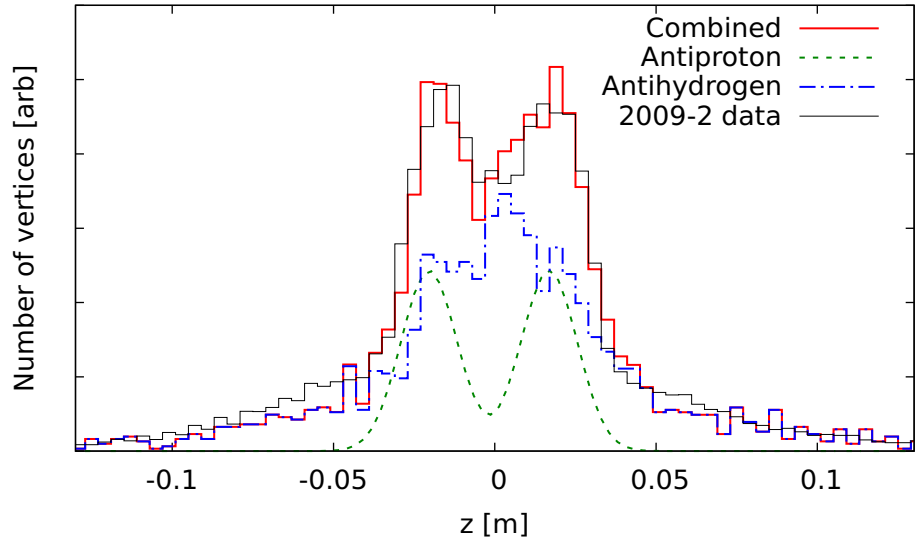}
	\caption[The measured annihilation distributions compared to a fit.]{The annihilation distribution measured with the octupole energised compared to a linear combination of the distribution measured with uniform magnetic field and Gaussian peaks.}
	\label{fig:GaussianPeaks}
\end{figure}

The presence of the peaks has been interpreted to be due to the escape of antiprotons liberated from field-ionised antihydrogen atoms \cite{ALPHA_HbarInOct}.
These antiprotons are subject to the ballistic loss mechanism discussed in section \ref{sec:ballisticLoss}.
The shape of the annihilation distribution was modelled by adding two Gaussian functions to the distribution measured at uniform field.
The distribution is fit using the centres, width and area of the peaks and the area of the uniform-field distribution as free parameters.
As shown in figure \ref{fig:GaussianPeaks}, this treatment produces a reasonable match with the measured distribution.
Even so, the choice of Gaussian functions to model the peaks is arbitrary, as there is no reason to expect that antiproton ballistic loss produces a distribution with this shape.
Using the simulation techniques developed so far, we can study how well these shapes are reproduced, and how the size and shape of the peaks depend on the experimental parameters.

\subsection{Modelling the distribution}

In the simulation discussed in section \ref{sec:vertex_fieldionisation}, an antiproton liberated from the ionisation of an antihydrogen atom was discarded, since, in the solenoidal magnetic field it would not annihilate on the electrode surfaces.
In an extension to the previous simulation, in this section we calculate the antiprotons' trajectories after ionisation and determine their influence on the measured annihilation distribution.

The magnetic field of the octupole is assumed to follow that for the ideal octupole (equation \ref{eq:octupoleField}).
This is a reasonable model for the magnetic field near the centre of the octupole, but breaks down near the ends.
Most of the annihilations occur close to the centre of the octupole, so the approximations needed will not reduce the accuracy of the simulations significantly.
The model does not include the changing $z$-component of magnetic field due to the mirror coils, which is also negligible near the centre of the trap.
The nature of the octupole field breaks the cylindrical symmetry of the apparatus, and requires introduction of an azimuthal angle, $\phi$. 
This is randomly assigned from a uniform distribution in $\left(-\pi,\pi\right)$.

If an atom is ionised, the resulting antiproton is constrained to move along the magnetic field lines - i.e. $\hat{v} = \pm \hat {B}$.
It is allowed to move until it either reaches the electrode surface, where it annihilates, leaves through the ends of the trap, or encounters an electric potential high enough to confine it.
An antiproton confined in the Penning trap will execute a magnetron motion and will rotate about the z-axis and may come onto an unconfined trajectory even though initially trapped. 
This is taken into account by allowing confined antiprotons to sample all angles within a $\pi/2$ sector -- taking advantage of the fourfold symmetry of the octupole.

Recall that we model the motional Stark shift as an electric field given by $\mathbf{F}_\mathrm{mot} = \mathbf{v} \times \mathbf{B}$.
Since $\mathbf{B}$ now has transverse components, the inclusion of this effect becomes slightly more complex.
The magnitude of total effective electric field is given by the sum of $\mathbf{F}_\mathrm{mot}$ and the electric field $\mathbf{F}$ present in the trap, and is
\begin{equation}
	F_\mathrm{eff} = \sqrt{ \left| \mathbf{F}\right|^2 + \left| \mathbf{v} \times \mathbf{B} \right|^2 - 2 \mathbf{F}\cdot(\mathbf{v \times B})}.
\end{equation}
As before, we assume that the internal motion of the positron is fast enough that the maximum value of $F_\mathrm{eff}$ is the appropriate one to take.

\subsection{Results}

\begin{figure}
	\centering
	\input{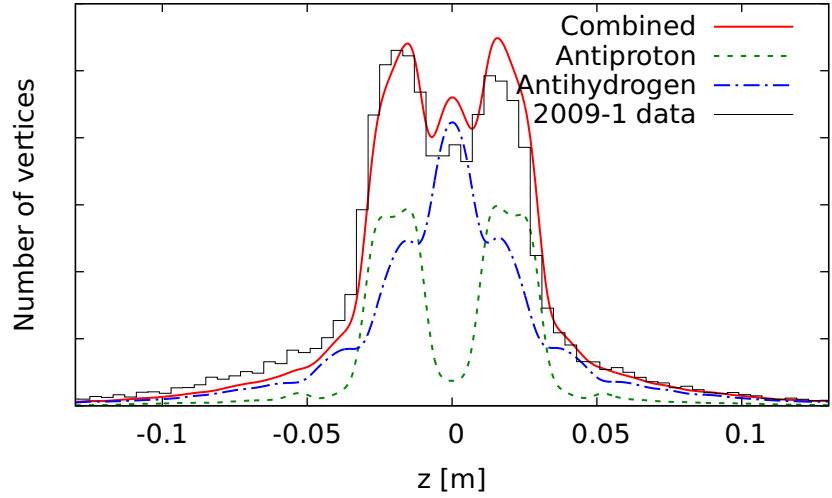}
	\caption[An example distribution produced by the simulation when the octupole magnet is energised]{An example distribution produced by the simulation when the octupole magnet is energised. The distribution was generated using the 2009-1 parameters, with $T=$200~K and $a=$2.0. The dotted and dot-dashed curves show the contributions to the annihilation distribution from antihydrogen annihilations and antiprotons liberated through field-ionisation. }
	\label{fig:exampleHornsSim}
\end{figure}

An illustrative example of the output from this simulation, using the 2009-1 data set with the parameters found above, is shown in figure \ref{fig:exampleHornsSim}.
It is clear that the positions and approximate sizes of the `horn' features are well-reproduced.
By splitting the distribution into the antihydrogen and antiproton components -- the particles that reach the electrode surface with and without ionising, respectively, we can clearly see that the horn structures are formed predominantly from antiprotons liberated from field ionisation.
This validates the hypothesis made in \cite{ALPHA_HbarInOct}.

In a similar fashion to the last analysis, we perform a two-dimensional scan of temperature $T$ and the exponent $a$ of equation \ref{eq:dN_epsilon}.
The value of $\kappa$ is kept fixed at 1.
There are significantly more vertices contained in these data sets as a result of a larger number of mixing cycles being performed.
For this reason, the bin size has been reduced and there are 64 degrees of freedom in the fit, which should be remembered when comparing values of $\Chi^2$.

\begin{figure}
	\centering
	\subfloat[]{\input{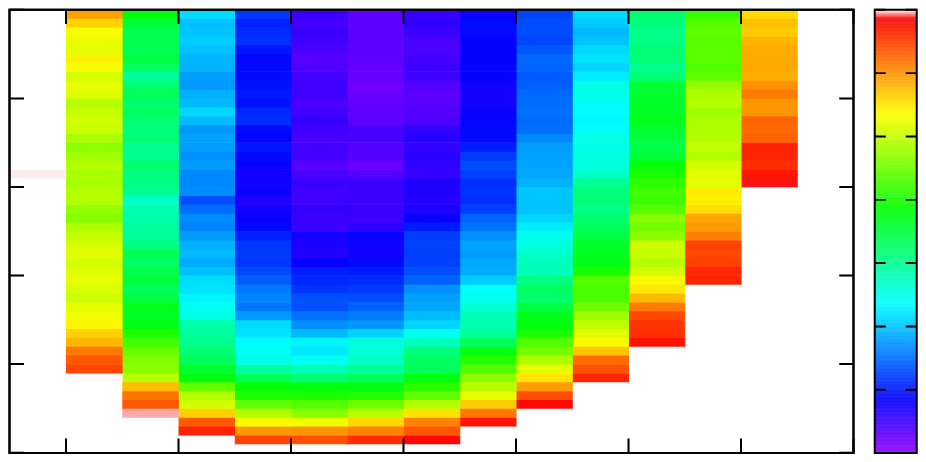}}
	\\
	\subfloat[]{\input{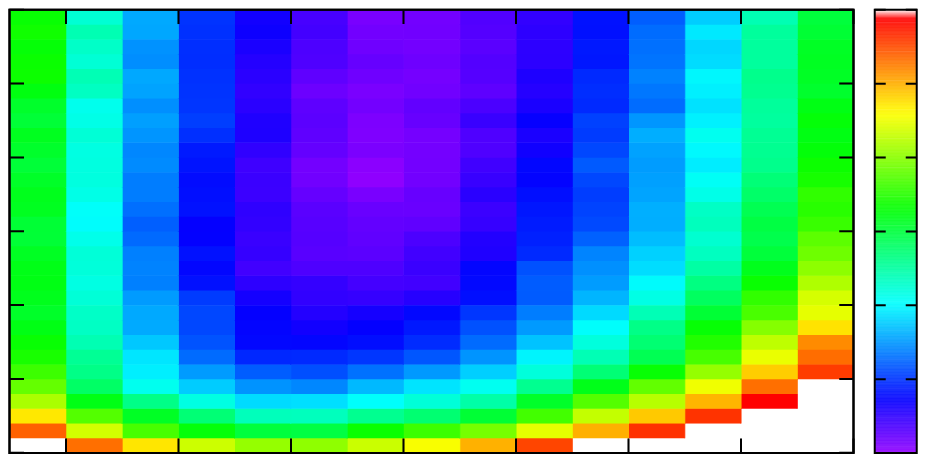}}
		\caption[$\Chi^2$ as a function of $a$ and $T$.]{A two-dimensional colour map showing the variation of $\Chi^2$ with $a$ and $T$ for the (a) 2009-1 and (b) 2009-2 data sets with the octupole magnet energised. The white regions are beyond the scale of the colour map.}
	\label{fig:statesOct2D}
\end{figure}

\begin{figure}
	\centering
	\input{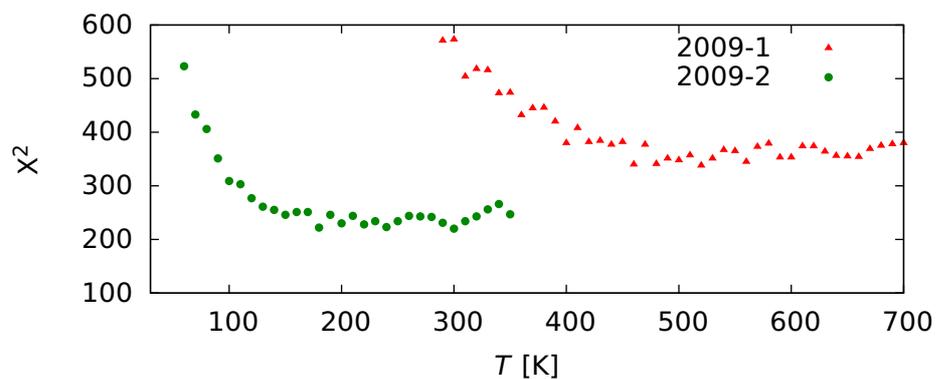}
	\caption{$\Chi^2$ as a function of $T$ for fixed $a=2.0$.}
	\label{fig:temp_constA}
\end{figure}

Figure \ref{fig:statesOct2D}(b) shows the two dimensional colour-map plot of $\Chi^2$ for different values of $T$ and $a$, using the 2009-2 data set. 
For each value of $T$, there is a value of $a$ that minimised $\Chi^2$, but the reverse is not true.
There is no global minimum value of $\Chi^2$.
The best-fit value of $a$, which is roughly constant for $T \gtrsim 100~\mathrm{K}$ (see figure \ref{fig:temp_constA}), is $2.0 \pm 0.2$, and the minimum value of $\Chi^2$ is $\sim 250$.

\begin{figure}
	\centering
	\input{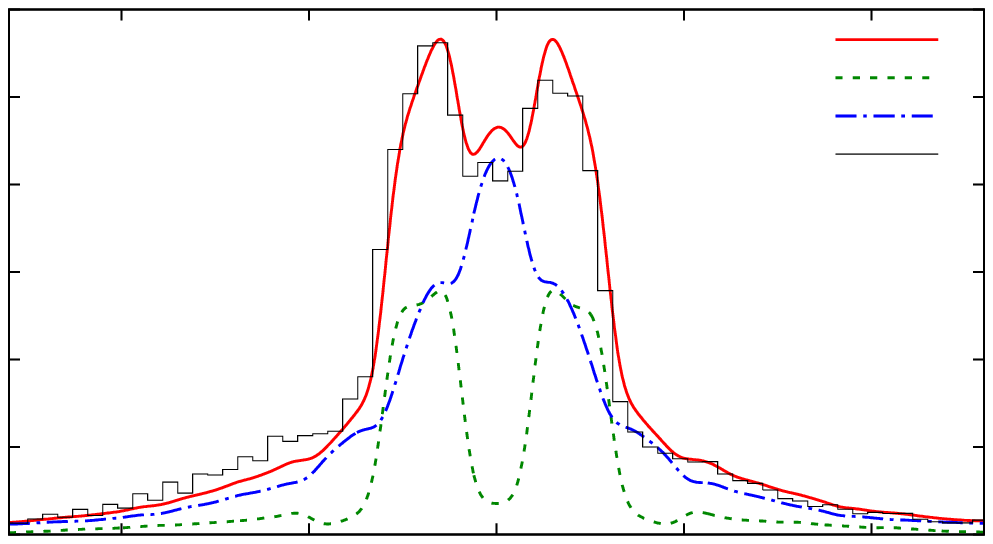}
	\caption[The simulated distribution for the 2009-2 configuration.]{The simulated distribution for the 2009-2 configuration with $T$=140~K, $a$=2.0.}
	\label{fig:octSimCompare}
\end{figure}

Roughly the same conclusions can be made from the 2009-1 data set, plotted in figure \ref{fig:statesOct2D}(a).
However, the downward trend in $\Chi^2$ as a function of $T$ for constant $a$ begins at a higher temperature, as might be expected from the measured positron temperatures.
This is shown in figure \ref{fig:temp_constA}.
This may indicate that although fitting to the vertex distributions in this way does not provide a very good absolute estimate of the antihydrogen temperature, it is sensitive to a temperature change.
The best-fit distributions, showing the contributions from antihydrogen impacts and antiproton impacts are shown in figure \ref{fig:exampleHornsSim} (for the 2009-1 data), and in figure \ref{fig:octSimCompare} for the 2009-2 data.

\begin{figure}[p]
\centering
\subfloat[]{\input{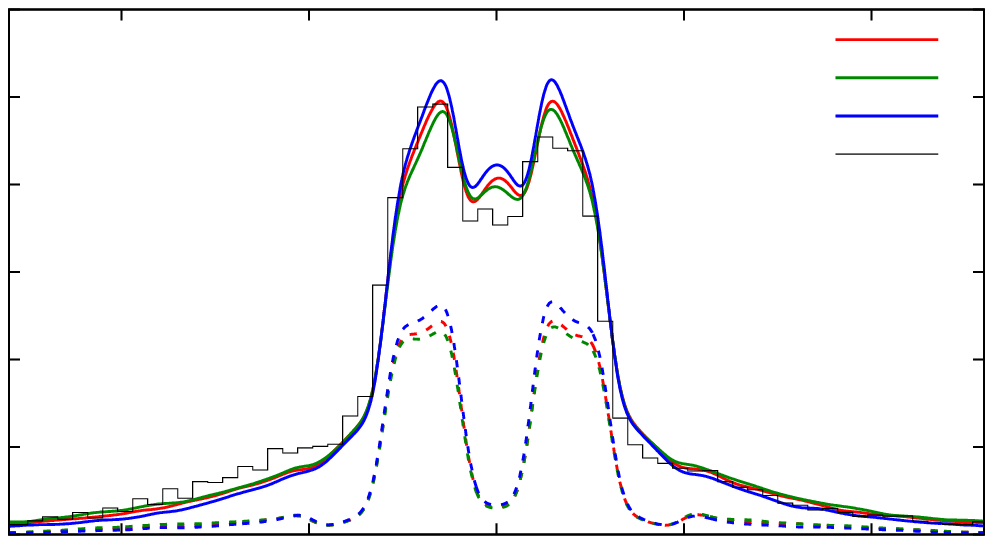}\label{fig:octSim_varyT}}\\ \vspace{-1cm}
\subfloat[]{\input{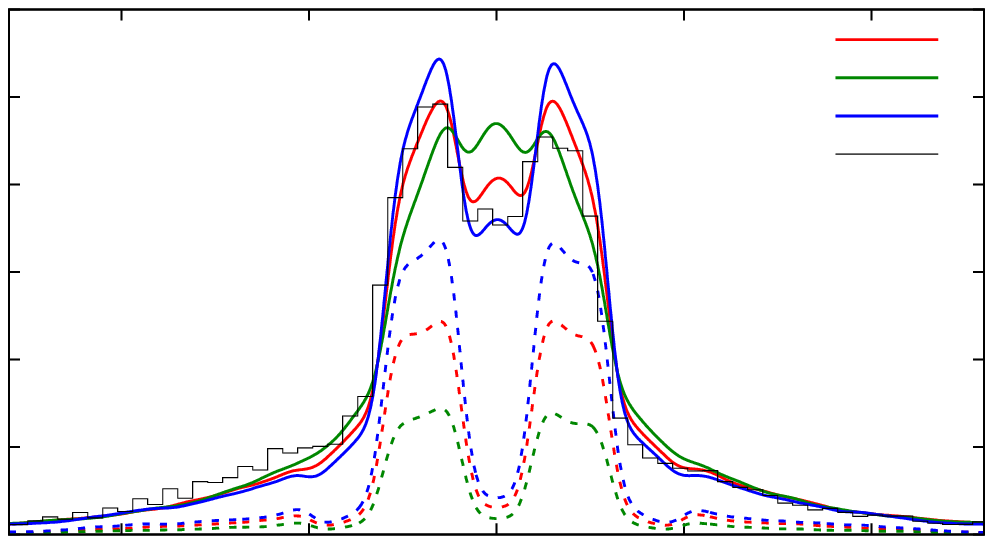}\label{fig:octSim_varyA}}\\ \vspace{-1cm}
\subfloat[]{\input{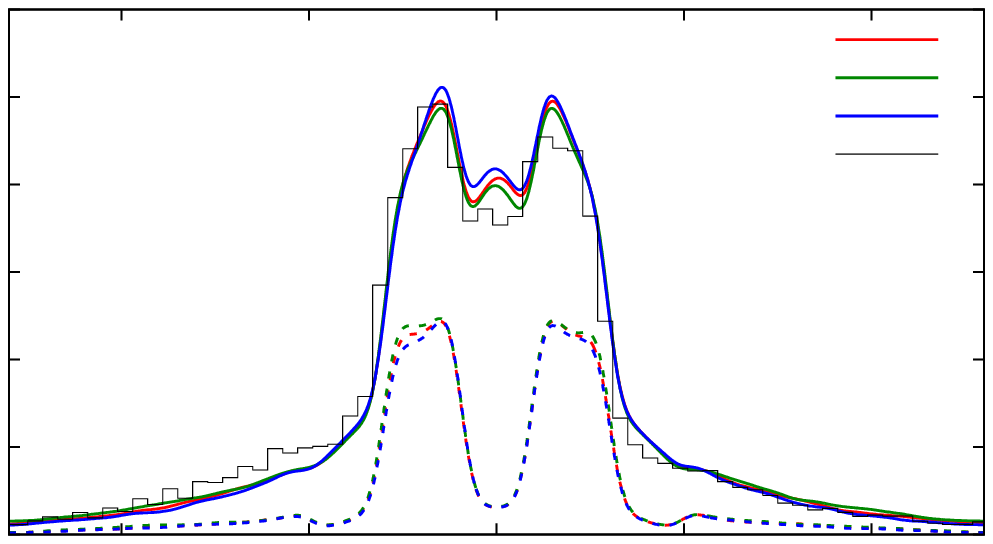}\label{fig:octSim_varyDens}}
\caption[The annihilation distributions obtained when the input parameters to the simulations are varied.]{The annihilation distributions obtained when the input parameters to the simulations are varied. In (a) the temperature is varied, in (b) the exponent of the binding energy distribution `$a$', and in (c) the positron density is varied. Unless otherwise stated, the temperature of the antihydrogen atoms is 140~K, $a$ is 2.0, and the positron density is $7 \times 10^{7}~\mathrm{cm^{-3}}$ (the 2009-2 parameters).}
\label{fig:octSim}
\end{figure}

It is also interesting to compare how the distributions change as the input parameters to the simulation are varied.
Figure \ref{fig:octSim}(a) shows the distributions produced for different antihydrogen temperatures, keeping the state distribution fixed.
The fraction of annihilations contained within the central region is larger for colder temperatures, as expected from the discussion in section \ref{sec:VertexTemperature}, but otherwise the distribution remains fairly static.

If we vary the value of $a$, keeping $T$ constant, the relative fraction of annihilations due to bare antiprotons changes.
For lower values of $a$, proportionally fewer antiprotons are in states with low binding energies, so fewer are field ionised by the fields in the Penning trap.
This behaviour is clearly seen in figure \ref{fig:octSim}(b).

We can also imagine a hypothetical experiment in which the positron density is changed, but all the other parameters remain fixed.
Changing the number of positrons requires re-solving the Poisson and Boltzmann equations to obtain the self-consistent electric potential and positron density.
Figure \ref{fig:octSim}(c) shows the standard 2009-2 data set (using $3 \times 10^6$ positrons, and a density of $7 \times 10^{7}~\mathrm{cm^{-3}}$) along with otherwise identical simulations, but with $1.5 \times 10^{6}$ positrons, giving a density of $\sim 4 \times 10^7 \mathrm{cm^{-3}}$; and $5 \times 10^{6}$ positrons, giving a density of $9 \times 10^7 \mathrm{cm^{-3}}$.
Altering the positron density essentially amounts to changing the electric field, which influences the $\mathbf{E} \times \mathbf{B}$ velocity and, to some extent, the probability of ionisation.
The assumption that the temperature and transverse density profile of the positrons and the distributions of states would remain unchanged when the number of positrons is changed is not likely to be good in reality, but is a second-order effect.
It is however, clear that changing the positron density in this range does not change the annihilation distribution to an appreciable degree.

In the presence of the transverse multipole field, the number of annihilations does not correspond directly to the number of antihydrogen atoms, as there is a contaminant in the form of field ionised antiprotons.
To correctly estimate the number of antihydrogen atoms produced, therefore, it is necessary to evaluate the contribution to the annihilation distribution from antiprotons compared to antihydrogen atoms.
The empirical analysis, modelling the annihilation distribution as the uniform-field distribution combined with Gaussian horns, typically finds that between 70 and 80\% of the annihilations are antihydrogen atoms striking the electrodes.

The analysis presented here makes it easy to separate the components. For the best fit 2009-1 case, $\sim 60\%$ of the annihilations are neutral atoms; for the 2009-2 case, the same number is $\sim 65\%$.
These numbers are in reasonable agreement with the simplified Gaussian model.
This quantity, while useful when analysing an individual experiment, is likely to be dependent on the particular electric field configuration used rather than rooted in the physics of the formation process.

\section{Comparison with other measurements and simulations}

This analysis of the antiproton annihilation distributions, both with and without a transverse multipole field, produces the result that the antihydrogen binding energy distribution follows a power law $\mathcal{E}^{-a}$, with $a$ close to 2.0.

`Arrested' three-body recombination is expected to result in a $\mathcal{E}^{-5}$ dependence \cite{Pohl_review} at deep binding energies as a result of a two-step formation process.
After initial three-body recombination, when the positron is bound by a few times the average positron kinetic energy, the atom can collide with a positron and becomes much more strongly bound.
The dependence follows from a simple density of states argument and is reproduced by guiding-centre calculations \cite{Pohl_review}.
A second simulation \cite{BassAntihydrogen}, in the strongly-magnetised regime, shows similar results.

The measurement of field-ionisation spectrum by ATRAP \cite{GabrielseReview} using a field-ionisation well (see section \ref{sec:fieldionisation}) is stated as `the number of $\Hbar$ atoms that survive $F$ decreases approximately as $F^{-2}$'.
Using equation \ref{eq:Fc_epsilon_au}, this can be transformed into an expression of the form of equation \ref{eq:dN_epsilon} with $a = 5$, which agrees with \cite{Pohl_review}.
The $\mathcal{E}^{-2}$ dependence found in the present analysis is clearly not compatible with the predicted $\mathcal{E}^{-5}$ dependence, or the weak-field dependence seen in the ATRAP experiment.

At electric fields stronger than $\sim~100~\mathrm{V\;cm^{-1}}$, ATRAP's distribution appears to become significantly shallower (smaller $a$), but does not appear to follow a simple  analytical form.
It was suggested that the atoms observed with these deeper binding energies are characterised by chaotic motion of the positron, instead of the guiding-centre picture more appropriate for weakly-bound atoms \cite{GabrielseReview}.
Simulations allowing for chaotic motion produce a good match with the high-field tail \cite{ATRAP_velocityReinterp}.

This analysis is mostly sensitive to the atoms that ionise at high radius, as these can be guided by the octupole fields into the electrode surfaces.
The electric fields at high radius are of the order of $100~\mathrm{V\,cm^{-1}}$, which is in the region where the ATRAP measurement deviates from $\mathcal{E}^{-5}$, so the two measurements might be reconcilable.
We should note that the field ionisation well is only sensitive to atoms that travel along the axis of the Penning trap, while the annihilation vertex detector primarily sees atoms exiting radially, so if there is an angular dependence of the binding energy distribution, we might expect disagreement between the results.

The binding energy distribution of antihydrogen atoms in thermal equilibrium with the positron plasma is $\mathcal{E}^{-5/2}\,\mathrm{exp}\left( - (\mathcal{E}_0 - \mathcal{E})/k_\mathrm{B}T \right)$ (section \ref{sec:HbarBE}).
The distribution at low binding energies is dominated by the $\mathcal{E}^{-5/2}$ term, which could be consistent with the distributions found here.
However, it is thought that the antihydrogen atoms spend too short a time inside the positron plasma to reach thermal equilibrium \cite{FrancisSimulationsHbarFormation}, so this cannot be the complete explanation.

\section{Conclusions}
The analysis  presented in this chapter supports the premise that antihydrogen is produced at a temperature close to that of the positron plasma.
This is an important step forward from the antihydrogen production schemes used in previous experiments by ATHENA, where the measurements indicated that the kinetic energy of the antihydrogen atoms was significantly higher and not at equilibrium with the positron plasma.

Using the shape of the annihilation distribution in the presence of a transverse multipolar magnetic field, which includes a contribution from antiprotons freed from field-ionised antihydrogen atoms, it is possible to find an expression determining the atoms' binding energy distribution.
Assuming a power law distribution -- i.e. $N(\mathcal{E}) \propto \mathcal{E}^{-a}$, it is found that $a = 2.0 \pm 0.1$ produces annihilation distributions that reproduce the measured distributions quite closely.
It should be noted that the model does not contain any free parameters, except for $a$ in the equation above.
The inputs -- the electric potential, the magnetic field and the positron density are determined by measurements on the system.

Measurements of the binding energy distribution can yield valuable insights into the mechanics of recombination in a nested Penning-trap. 
The spatial distribution of annihilation vertices has never before been used to make a measurement of this kind.
It is notable that we find some disagreements with alternative measurements, perhaps indicating that the process is not fully understood.

This analysis is limited in a number of ways.
The first of these stems from the inherent uncertainties in the measured positron conditions. 
The number of positrons contained in the positron plasma, its temperature and its shape are known only to a level of a few to ten percent, at best.
It is also assumed that the positron parameters remain fixed during antihydrogen production.
Temperature and/or density changes have been experimentally observed, but the sizes of the changes are also at the level of a few percent.
A difference between the values used in the simulation and the parameters in the experiment will introduce differences between the simulated distribution and that measured.

The physics included in modelling the ionisation of the atoms is not complete.
We have used the expression for central atomic states, but have shown that polarised atoms behave in almost the same way.
Also, the change in binding energy due to the atom's motion across the magnetic field has been calculated only approximately.
The model expression used for the distribution of binding energies -- a power-law -- is arbitrary; other forms could be equally, or more appropriate. 

The model neglects interactions between particles -- collisions with positrons, antiprotons or residual gas atoms can change the atom's binding energy after formation, possibly even ionising them.
Spontaneous radiative decay from highly-excited, weakly bound states to more strongly bound, harder-to-ionise states is also ignored.

Fully treating all of these effects would require a far more sophisticated and computationally demanding simulation.
With such codes, it takes a prohibitively long time to simulate the large numbers of annihilations required to make quantitative comparisons with the measurements.

The model shows some sensitivity to the antihydrogen temperature, which is the most important parameter to control when attempting to produce cold antihydrogen.
However, it is not capable of producing a quantitative measure of the temperature, which limits the effectiveness.

\begin{figure}[h]
\centering
\input{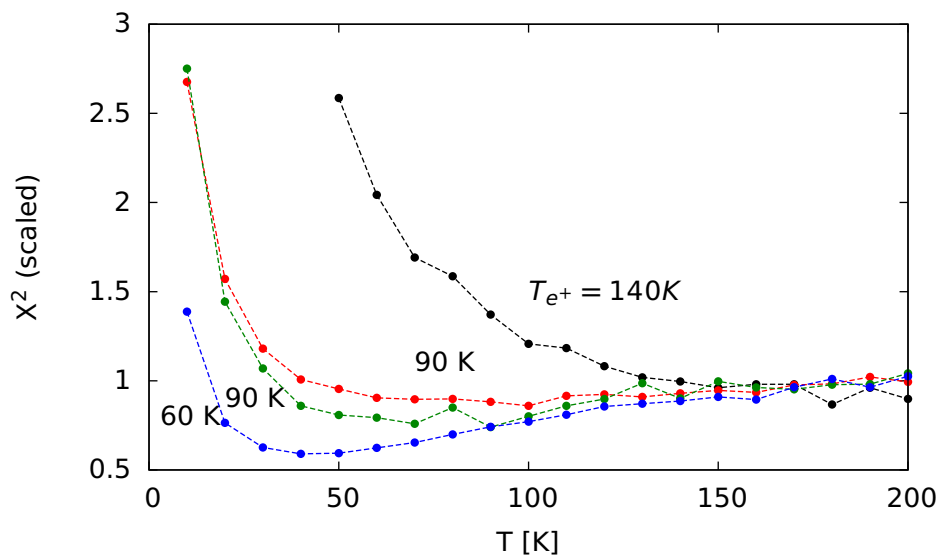}
\caption[A comparison of $\Chi^2$ as a function of $T$ for mixing experiments with different positron temperatures.]{A comparison of $\Chi^2$ as a function of $T$ for mixing experiments with different positron temperatures. $\Chi^2$ has been multiplied by a normalisation factor.}
\label{fig:chi2series}
\end{figure}

Analysis of data sets collected at the ALPHA experiment after most of the work presented in this chapter had been finished shows a distinct trend in the shape of the $\Chi^2$ distribution as a function of the measured positron temperature\footnote{The constraints of time preclude a full discussion of these data sets, which were collected while this thesis was being completed. The positron temperature was reduced by application of the evaporative cooling technique described in section \ref{sec:LeptonEVC}.}.
The dependence of $\Chi^2$ on $T$ is shown in figure \ref{fig:chi2series}.
$\Chi^2$ for each data set has been multiplied by a normalisation factor to correct for the different numbers of vertices in each data set and that forces the plots to tend to the same value at high temperatures.
As the measured positron temperature is reduced, the `corner' feature shifts to the left (lower temperature).
This result shows clearly that the antiproton and antihydrogen temperatures trend in the same direction as the positron plasma temperature.
This supports reduction of the positron temperature as a worthwhile effort to reduce the antihydrogen kinetic energy and increase the number of trapped antihydrogen atoms.
Further analysis of these data sets could yield more information about the antihydrogen temperature (for instance, its time dependence) in the experiments.

    \chapter{Trapped Antihydrogen}
\label{chp:trapping}

\epigraph{Physics is like sex:\\Sure, it may give some practical results, but that's not why we do it.}{Richard Feynman (attrib)}

The previous chapters have described how antihydrogen is produced, in this chapter, we will describe the important parameters that influence the trapping rate, and the experimental steps required to demonstrate trapping of atoms.
The latter includes the detection scheme, the important background contributions and how antihydrogen is positively identified.
These techniques are applied to a data-taking runs, and the first definitive signals of trapped antihydrogen are presented.

\section{Parameter considerations}
\label{sec:TrappingConsiderations}
The two factors that principally determine the number of trapped antihydrogen atoms are the number of atoms produced, and their kinetic energy.
The number of antihydrogen atoms will depend on the parameters entering into the rate equations in section \ref{sec:HbarFormationTheory}, and the number of antiprotons, but producing a quantitative estimate of the number of atoms that will be produced in any experiment is practically impossible.
The number of antihydrogen atoms is fortunately easy to measure, so instead, we focus on the \textit{fraction} of these atoms that would be trapped.

Antihydrogen experiments to date have produced atoms with energy distributions with an average value many times in excess of the atom trap depth.
Thus, only the extreme low-energy tail of the distribution can be trapped.
A model for the velocity distribution of antihydrogen atoms in a nested-well production scheme has already been described in chapter \ref{chp:vertex} -- a Maxwell-Boltzmann distribution in a frame of reference rotating at the $\mathbf{E}\times\mathbf{B}$ frequency.
It is easy to numerically evaluate the fraction of such a velocity distribution that falls below a certain kinetic energy threshold.
We denote this fraction, $f(E)$, where
\begin{equation}
	f(E) = \int_0^E  \frac{\mathrm{d}N}{\mathrm{d}E'}\; \mathrm{d}E,
\end{equation}
where $N(E')$ is the number of atoms with energy in the interval ($E'$, $E'+\mathrm{d}E'$).
With knowledge of the antiproton temperature and the electric and magnetic fields, the number of antihydrogen atoms with energies low enough to be trapped can be calculated.

\begin{figure}[hbt]
	\centering
	\input{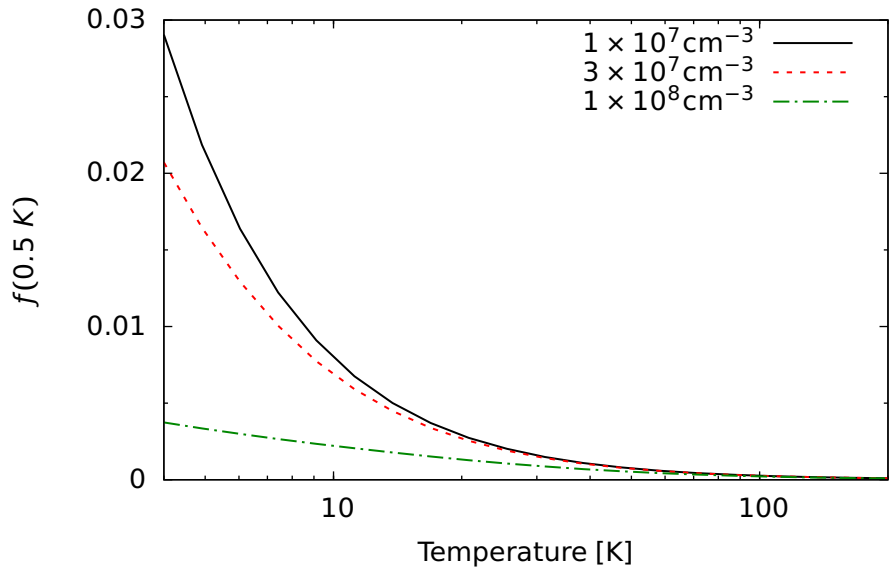}
	\caption[The fraction of atoms with trappable kinetic energies as a function of temperature.]{The fraction of atoms with kinetic energies less than the trap depth (0.5~K) as a function of their temperature for several densities of positron plasmas. The atoms are assumed to form evenly in the volume less than 1~mm from the trap axis.}
	\label{fig:fracBound_temp}
\end{figure}

The dependence of the fraction of atoms with trappable kinetic energy ($f(0.5~K)$) on their temperature in this model is shown for several positron densities in figure \ref{fig:fracBound_temp}.
It is obvious that it is always better in this model to have lower temperature, all else being equal.
The steep rise in $f$ as the temperature is reduced indicates that reducing the positron (and thus, the antihydrogen) temperature is likely to be the most fruitful way to increase the number of trapped atoms.

Also recall that the three-body antihydrogen production rate (equation \ref{eq:GammaTBR}) increases with reducing positron temperature.
It is reasonable to infer that the number of antihydrogen atoms produced in a given experiment will thus also be higher.

\begin{figure}[hbt]
	\centering
	\input{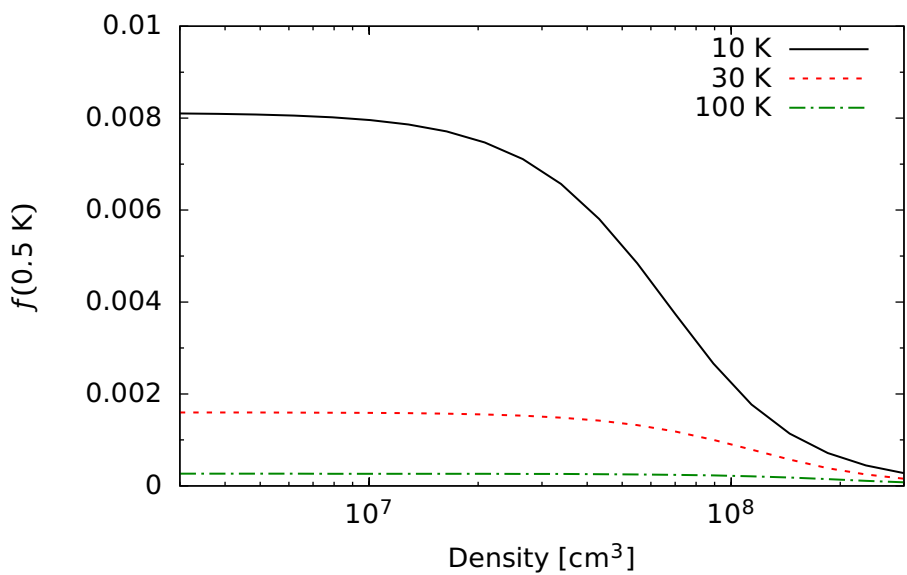}
	\caption[The fraction of atoms with trappable kinetic energies as a function of the positron density.]{The fraction of atoms with kinetic energies less than the trap depth (0.5~K) as a function of the positron plasma density for several temperatures. The atoms are assumed to form evenly in the volume less than 1~mm from the trap axis.}
	\label{fig:fracBound_density}
\end{figure}

The dependence of $f(0.5~\mathrm{K})$ on the positron density, for a constant temperature is shown in figure \ref{fig:fracBound_density}.
Lower density reduces the magnitude of the $\mathbf{E} \times \mathbf{B}$ rotation, increasing the fraction with trappable kinetic energies.
However, this is only true up to a point, as indicated by the plateau behaviour seen in the figure.
This occurs when the major component of the kinetic energy is the thermal motion, not the $\mathbf{E} \times \mathbf{B}$ rotation.
For the temperatures of order 50-100~K seen in ALPHA, it is not useful to reduce the density below between $10^7$ and $10^8~\mathrm{cm^{-3}}$.
However, as has already been discussed (section \ref{sec:plasmaTemperature}), the temperature of the plasma and its density have been seen to be correlated.

The positron density also enters the three-body production rate (equation \ref{eq:GammaTBR}). 
Reducing the density, all else being equal, will reduce the production rate, so reducing the density can be detrimental.
It is not clear how this would affect the number of trapped antihydrogen atoms, as a lower rate can, in principle,  be compensated for by allowing the antiprotons and positrons to interact for a longer time.

The optimum positron density to use for antihydrogen production is still very much an open question.

The antiprotons' magnetron rotation also implies that it is more favourable if the antiprotons form antihydrogen close to the axis of the Penning trap, when the rotational velocity is lower (see figure \ref{fig:isotrop_vr}).
Smaller-radius plasmas are also desirable to minimise the disturbing influence of the octupole's magnetic field on the stored plasmas (section \ref{sec:OctupolePhysics}).
All else being equal, therefore, it is always better to have transversely smaller plasmas.

\section{Detection of trapped antihydrogen}

ALPHA aims to identify trapped antihydrogen atoms by releasing them from the magnetic minimum trap and identifying the annihilation as they strike the surrounding apparatus.
The method of shutting down the magnetic trap has been described in section \ref{sec:magneticTrap}.
The silicon vertex detector (section \ref{sec:siliconDetector}) is used to identify the antiproton annihilations.

In 2009, ALPHA conducted extensive trapping experiments using a fully assembled and commissioned apparatus.
The analysis of the data is made up of two parts - the identification of antiproton annihilations and evidence that the antiproton is bound in an antihydrogen atom.

\section{Identification of annihilations}
\label{sec:cosmicAnalysis}
The depth of the trap falls to less than 0.1\% of its initial value within 30~ms after the start of the shutdown, and any trapped atoms will be expected to escape in this time window.
The short length of this window reduces the number of cosmic rays that pass through the detector and that could be mistaken for annihilations.

In `quiet' operation, the detector trigger rate is approximately 4.31~Hz, and thus an average of 0.13 counts would be expected in each experiment, or approximately one count in eight experiments.
This gives an approximate measure of the minimum amount of antihydrogen that must be produced for detection at a statistically significant level to be feasible.

The rate of background events can be reduced by using the reconstruction capabilities of the silicon detector to distinguish between cosmic events and annihilations.
Cosmic events predominantly consist of one particle passing through the apparatus.
The reconstruction algorithm interprets events such as these to be made up of two particle tracks -- one as the cosmic particle passes through the detector as it enters the apparatus, and another as it leaves.

At the Earth's surface, muons ($\mu^{\pm}$), formed in collisions of heavier particles and nuclei of atmospheric gases, are the most numerous component \cite{pdg}.
The muon flux and energy spectrum have been precisely measured, as understanding of the response of detectors to cosmic rays is important, notably in high-energy physics.
The intensity of cosmic rays is approximately $0.017~\mathrm{Hz\,cm^{-2}}$ \cite{pdg}, which, combined with the horizontal cross-section of the detector ($\sim 550~\mathrm{cm^{2}}$) predicts a rate of $\sim$ 8.5~Hz.
This is within a factor of two of the observed rate -- some disagreement is to be expected because some efficiency of identifying hits is sacrificed to suppress electronic noise.

The mean energy of cosmic muon particles is approximately 4~GeV \cite{pdg}.
Charged particles passing through a magnetic field will curve in the plane perpendicular to the magnetic field due to the Lorentz force.
The bending radius is given by
\begin{equation}
	r = \frac{p^2}{q \mathbf{B} \times \mathbf{p}},
\end{equation}
where $q$ is the charge of the particle.
In the 1~T solenoidal magnetic field of ALPHA, the bending radius of a 4~GeV muon is of order 13~m $/\mathrm{sin}(\theta)$, where $\theta$ is the angle between the muon trajectory and the magnetic field.
Most muons follow a vertical trajectory, and have values of $\theta$ close to $90^\circ$.
A radius of curvature of 13~m is much larger than the $\sim$ 20~cm size of the detector.
Thus, if we ignore the possibility that a muon can undergo scattering with a nucleus of the material making up the apparatus and be deflected, most cosmic events will appear as a pair of straight, co-linear and approximately vertical tracks.

Antiprotons annihilating on matter nucleons produce principally pions ($\pi^{\pm,0})$ \cite{AnnihilationReview}.
Many different combinations of particles are possible, with an average of $\sim 5.0 \pm 0.3$ pions produced in each annihilation, $3.1 \pm 0.3$ of which are charged and $1.8 \pm 0.2$ neutral in proton-antiproton annihilations.
Neutron-antiproton annihilations are broadly similar.
Note that the total electric charge must be conserved, so proton-antiproton annihilations produce an even number of charged pions, while neutron-antiproton annihilations produce an odd number.
Charged pions have a lifetime of 26~ns (in their rest frame) \cite{pdg}, and travel several metres before decaying (usually into muons and muon neutrinos).
Pions from an annihilation inside the apparatus can thus pass through the silicon detector and are readily detected.
This implies that detector events containing at least three charged particle tracks are more likely to be due to antiproton annihilations than cosmic rays.

While powerful, this simple criterion does not go far enough. 
Minor components of cosmic rays include `showers' of electrons, positrons and high-energy photons from the decay of mesons in the atmosphere, and protons and neutrons.
These showers can be quite complex, with the number of particles, their energies and trajectories depending on the energy and species of the original particle.
This can give rise to detector events with more than three tracks, increasing the background.

The signal will also be suppressed - $\sim 47\%$ of proton-antiproton annihilations have only two tracks \cite{AnnihilationReview}, and using the three-track criterion would reject them as being cosmic rays.
The detector does not completely surround the trapping region of the apparatus, and it is possible for one or more tracks to exit in a `blind-spot', further reducing the efficiency for identifying an annihilation.
Thus, we use further characteristics of annihilations to distinguish them from cosmic rays.

In an annihilation resulting in two daughter particles, conservation of momentum would require that the daughter particles have oppositely directed momenta, and are emitted `back-to-back', mimicking a cosmic particle (ignoring the curvature).
However, the final state with two daughters is unlikely compared to the other states -- for example, a proton-antiproton annihilation results in two charged pions (and no other particles) only $\sim 0.4\%$ of the time.
In the specific case relevant to ALPHA -- annihilation of an antiproton in a nucleus, there is always an additional reaction product -- the remainder of the nucleus, so there is only a small chance that the final state includes two back-to-back charged particles.

It has already been calculated that the typical bending radius for cosmic rays in the 1~T field is of order 13~m.
The typical energy of a pion from a proton-antiproton annihilation is $\sim$ 120~MeV \cite{AnnihilationReview} , corresponding to a bending radius of $\sim 0.7~\mathrm{m} / \mathrm{sin}(\theta)$ in a 1~T magnetic field.
This curvature is sufficiently pronounced as to be easily detectable.
Thus, curved tracks are expected of antiproton annihilations, but not cosmic rays.

\begin{figure}
	\centering
	\captionsetup[subfloat]{position=top,captionskip=-20pt, justification=raggedright, font=normalsize, singlelinecheck=false, margin=10pt}
	\captionsetup[subfloat]{captionskip=0pt}
	\subfloat[]{\includegraphics[width=0.4 \textwidth, clip, trim = 4.5cm 4.5cm 4.5cm 4.5cm]{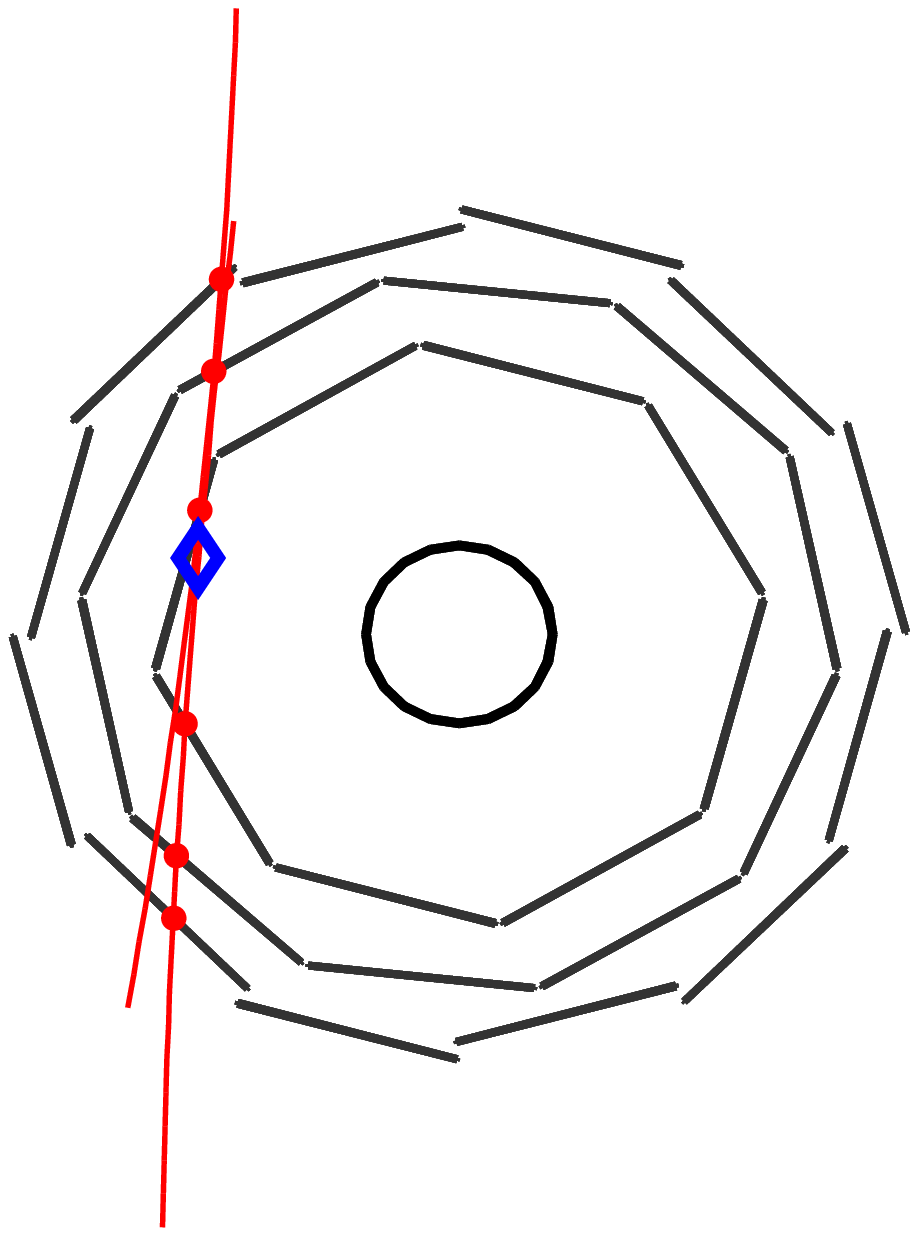}}
	\subfloat[]{\includegraphics[width=0.4 \textwidth, clip, trim = 4.5cm 4.5cm 4.5cm 4.5cm]{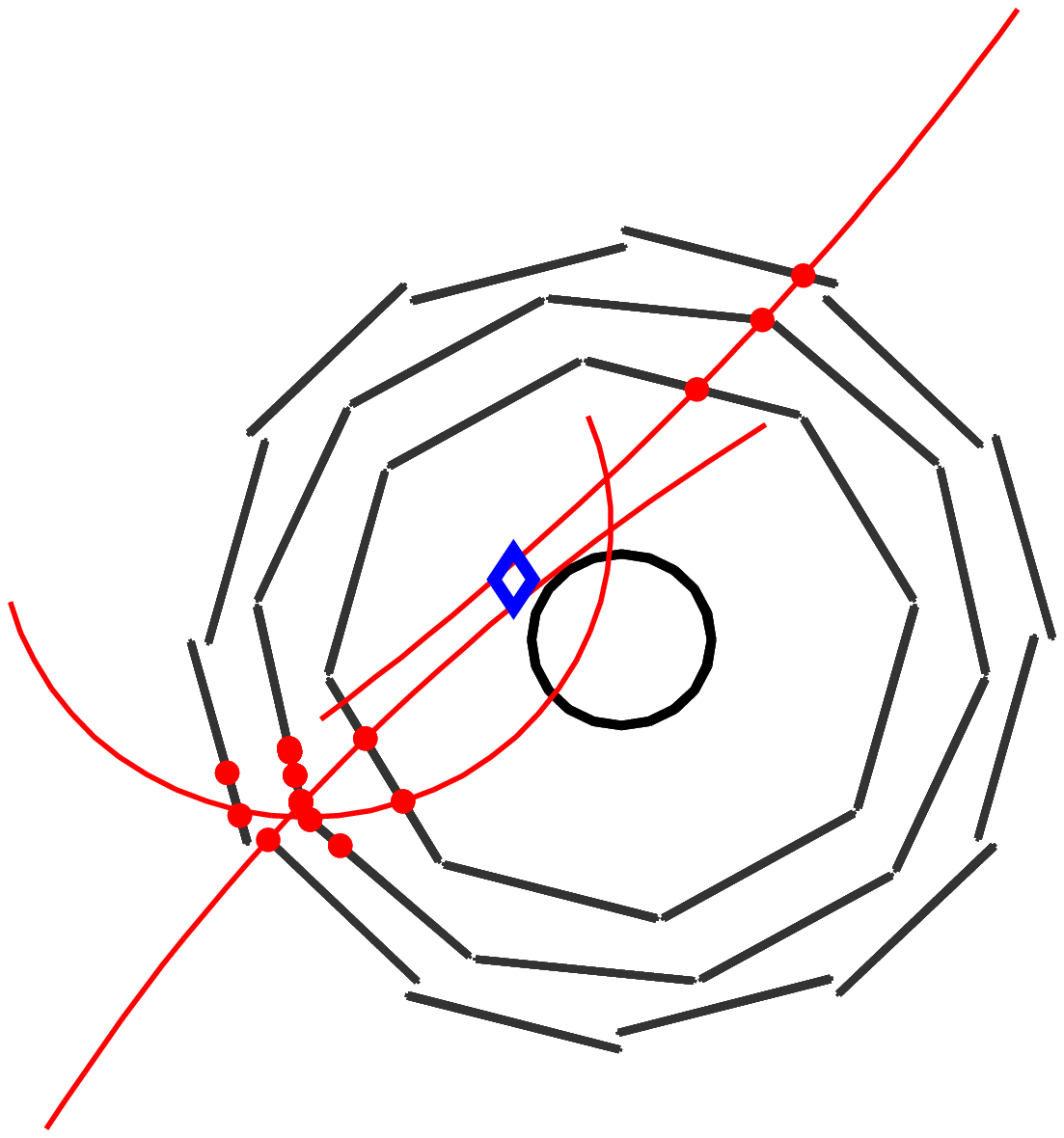}}
	\\
	\subfloat[]{\includegraphics[width=0.4 \textwidth, clip, trim = 4.5cm 4.5cm 4.5cm 4.5cm]{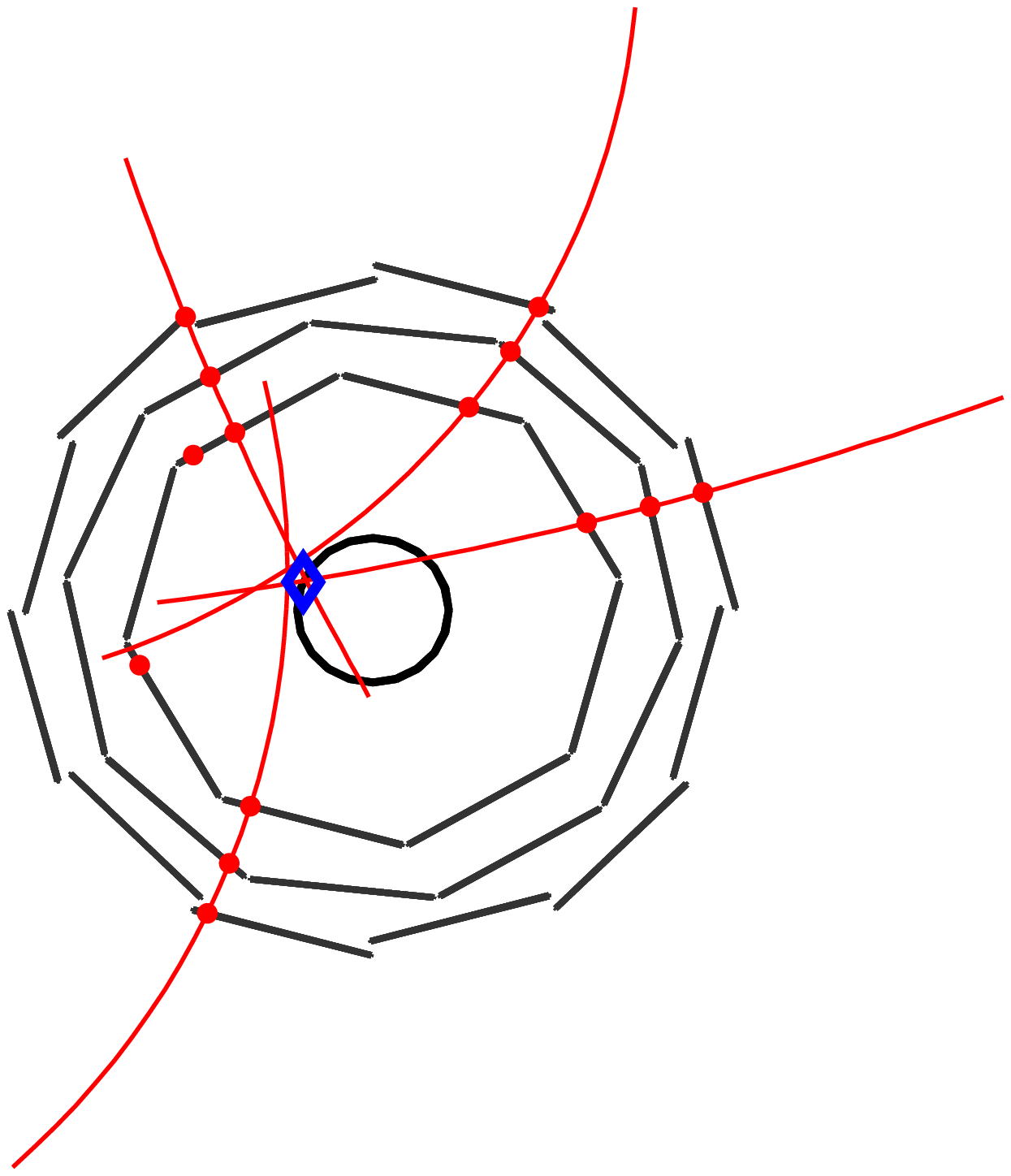}}
	\subfloat[]{\includegraphics[width=0.4 \textwidth, clip, trim = 4.5cm 4.5cm 4.5cm 4.5cm]{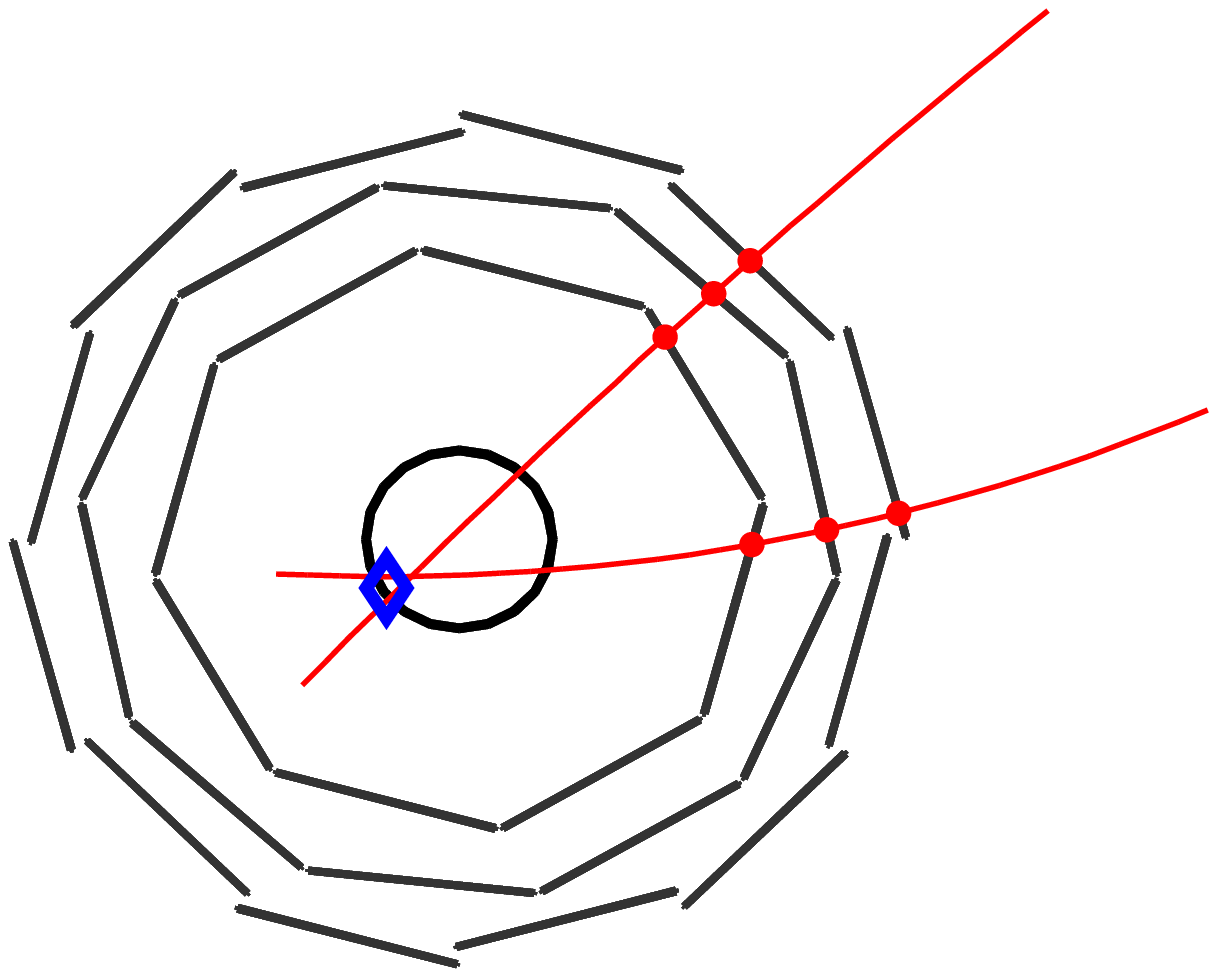}}
	\caption[Examples of annihilation and cosmic events.]{(a) a `classic' recorded cosmic event, (b) a rarer, more complex cosmic event, (c) a multi-track annihilation event, (d) a two-track annihilation event}
	\label{fig:exampleDetectorEvents}
\end{figure}

Several examples of antiproton annihilation and cosmic-ray events, illustrating their different characteristics, are shown in figure \ref{fig:exampleDetectorEvents}.
These events were recorded during periods of high-rate of annihilations and where no antiprotons were present near the apparatus, as appropriate.

\captionsetup[subfloat]{position=top,captionskip=-20pt, justification=raggedright, font=normalsize, singlelinecheck=false, margin=10pt}

While understanding the behaviour of antiproton annihilations is important, to quantitatively evaluate the likelihood of distinguishing an antiproton from a cosmic ray, and to accurately take into account the behaviour of the detector and the influence of the rest of the apparatus, it is necessary to analyse samples of events resulting purely from cosmic rays and purely from antiproton annihilations.

Large samples of cosmic-ray events can be gathered in periods where beam is not delivered to the apparatus.
Annihilation events are collected by producing a high rate of annihilations, so that cosmic rays are a negligible contaminant. 
At a rate of 1~kHz, the cosmic contamination is approximately 0.4\%, which is low enough to be ignored.

\begin{figure}[hbt]
	\centering
	\input{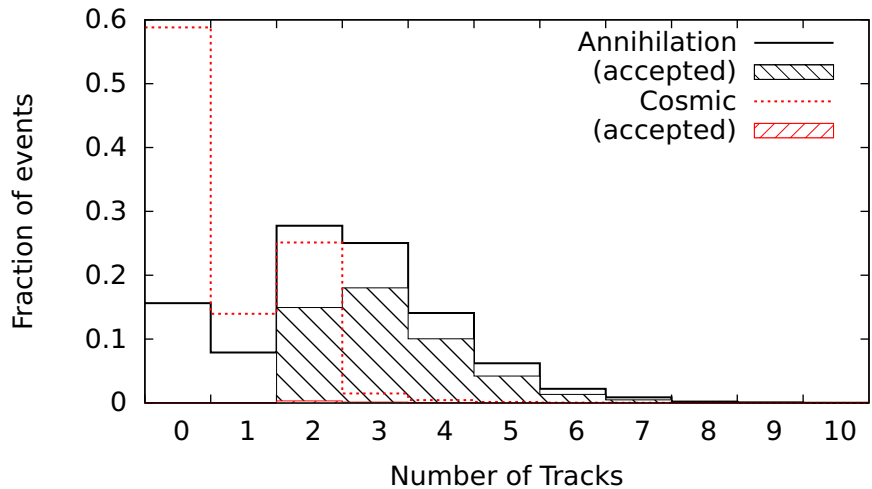}
	\caption[The distributions of number of tracks in samples of antiproton annihilation events and cosmic-ray events.]{The distributions of number of tracks in samples of antiproton annihilation events and cosmic-ray events. The lines are the raw distributions and the hatched regions are the portions remaining when the selection criteria described in the text are applied. The fraction of accepted cosmics is so small as to be hard to distinguish from the x-axis.}
	\label{fig:tracks}
\end{figure}

Figure \ref{fig:tracks} shows the number of tracks per event observed in these samples.
The difference in the number of tracks is clearly visible - the cosmic events mostly have two tracks (ignoring the events with no tracks), while the annihilation events have a wider distribution, extending to large numbers of tracks.
Events with fewer than two tracks can be events in which readout was triggered by electronic noise, where particles exited (or entered) in a region not covered by the detector, or where hits or tracks were not identified by the reconstruction algorithm.
Cosmic events with more than two tracks can be due to showers (discussed above), or a spurious track introduced by electronic noise.

Measurements of the track curvature to a precision high enough to reliably distinguish cosmic rays and annihilation products is not feasible with the ALPHA detector.
Instead, the fact that cosmic events produce straight-line tracks was used indirectly.
A straight line was fit to the hits making up each pair of tracks, and the squared distances from the hits to the straight line were summed to give a single squared residual value.
If there was more than one pair of tracks in an event, the squared residual was calculated for each pair, and the smallest value was taken to parametrise that event.

\begin{figure}[h!]
	\centering
	\subfloat[]{\input{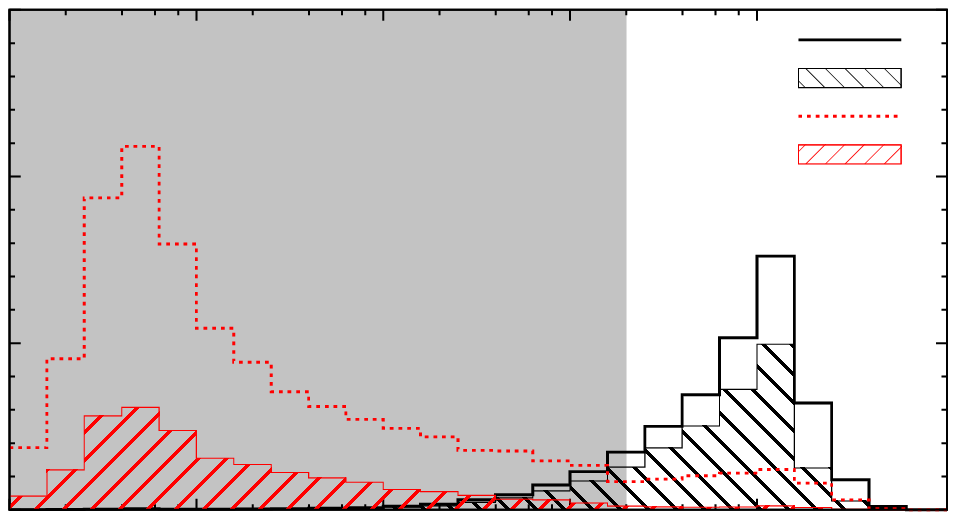}}
	\\
	\subfloat[]{\input{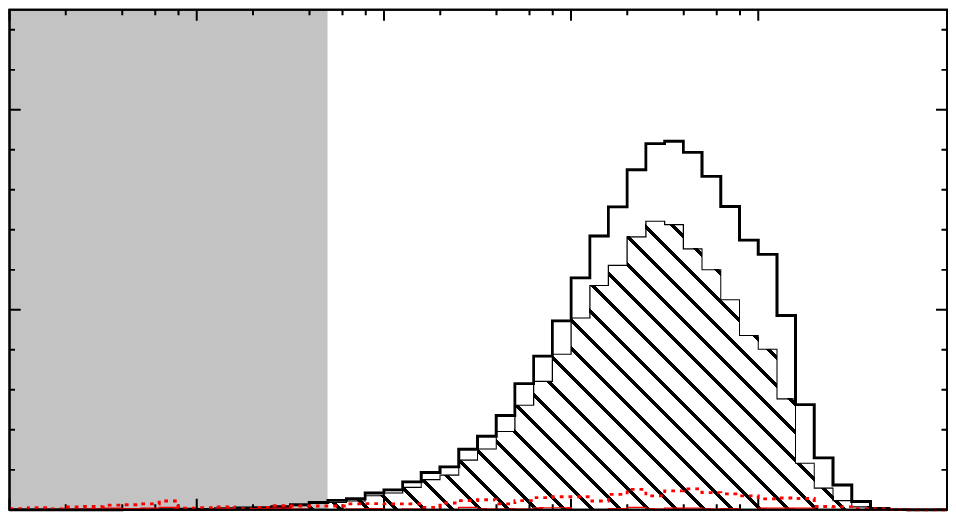}}
	\caption[The distributions of the squared residual quantity for samples of annihilations and cosmics.]{The distributions of the squared residual quantity for samples of annihilations and cosmics. The distributions are shown separately for events with two tracks (a) and three or more tracks (b). The lines are the raw distributions and the hatched regions are the portions accepted when the radial position cut is applied. Events with a squared residual value in the shaded region are rejected as annihilations. In this figure the height, not the area, is proportional to the number of events in each bin.}
	\label{fig:residual}
\end{figure}

The distributions of this variable for the samples of annihilation and cosmic events are shown in figure \ref{fig:residual}.
The distribution is strongly concentrated close to zero for cosmic rays and is much broader for annihilations.
Cosmic events with large values of the squared residual can be due to scattering in the material of the detector, or where the reconstruction algorithm has incorrectly identified the tracks.

\begin{figure}
	\centering
	\input{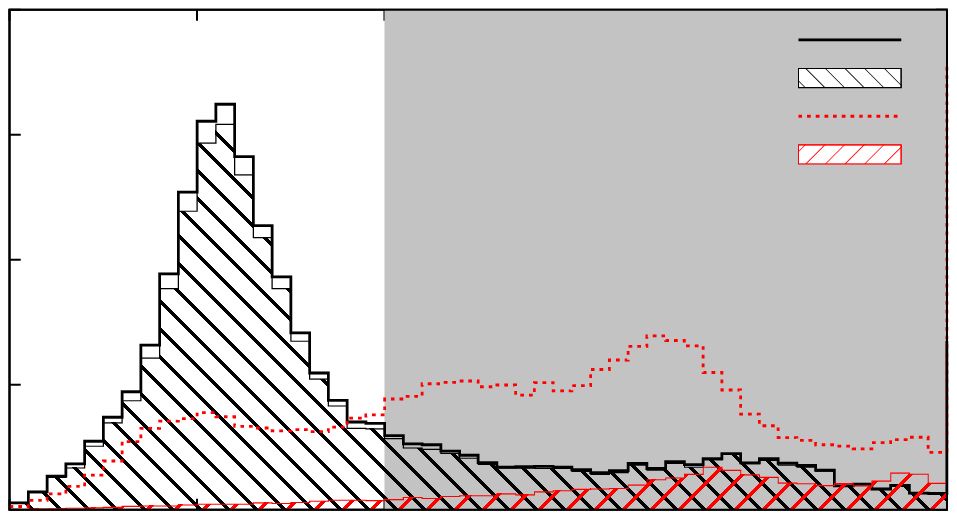}
	\caption[The distributions of the radial coordinate of the reconstructed vertices for samples of annihilations and cosmics.]{The distributions of the radial coordinate of the reconstructed vertices for samples of annihilations and cosmics. The lines are the raw distributions and the hatched regions are the portions accepted when the squared residual cut is applied. Events with radius in the shaded region are rejected as annihilations.}
	\label{fig:vertexRadius}
\end{figure}

A final parameter was used to distinguish between antiproton annihilations and cosmic events.
Antiproton annihilations occur at the inner surface of the Penning-trap electrodes ($r=2.2~\mathrm{cm}$), and the reconstructed annihilation vertex should be close to this radius.
On the other hand, if the algorithm identifies a vertex between two almost-colinear tracks in a cosmic event, there is no reason for a preference for $r=2.2~\mathrm{cm}$, and the distribution will be different.
The distributions of vertex radii are shown in figure \ref{fig:vertexRadius}. 
It is clear that the expected behaviour described above is borne out in reality.

To use the distributions to reject cosmic events, a series of thresholds, or `cuts' were used.
For each parameter, an event either `passes' or `fails' the appropriate cut depending on the value of the parameter relative to the corresponding threshold.
For example, we can place a threshold on the squared residual value at, say, $0.1~\mathrm{cm}^2$, and any event with a squared residual value less than this would be rejected.

The placement of the thresholds is a trade-off between the number of cosmic events that are falsely accepted and the fraction of antiproton events that are falsely rejected.
The figure-of-merit is the probability that random fluctuations in the number of accepted cosmic events could produce a false signal at least as large as that observed.
Cosmic events are independent of each other; thus the number of cosmic rays in any interval follows the Poisson distribution
\begin{equation}
f(N;\lambda) = \frac{\lambda^N \mathrm{e}^{-N}}{N!},
\label{eqn:Poisson}
\end{equation}
where $f(N;\lambda)$ is the probability of observing $N$ events from random fluctuations in an expected background of $\lambda$ events.
The probability of observing $N$ events or more is thus given by
\begin{equation}
p(N;\lambda) = \sum_{i=N}^\infty \, f(i;\lambda),
\label{eqn:poissonPvalue}
\end{equation}
and is termed the p-value \cite{pdg}.

The p-value is usually thought of in terms of the equivalent number of standard deviations (`sigma') one must move from the mean of the normal distribution to have an area equal to $p$ outside this region.
Technically, we take the `one-sided' case of this problem.
This means that we are interested only in the possibility that the observed signal is higher than the background, not that it is lower.

The number of sigma, $S$ is then defined by the relationship
\begin{equation}
	p = \int_{S}^{\infty} \, \frac{1}{\sqrt{2\pi}}\,\mathrm{e}^{-\frac{x^2}{2}}\,\mathrm{d}x,
\end{equation}
or equivalently,
\begin{equation}
	S = \sqrt{2} \, \mathrm{Erfc}^{-1}(2p),
	\label{eqn:significance}
\end{equation}
where $\mathrm{Erfc}^{-1}$ is the inverse of the complementary error function.
Conventionally, $S$> 5 or `$5\sigma$' significance is required to claim discovery or unambiguous observation of an effect.
This is equivalent to the statement that it is 99.99997\% certain that the observation is not due to statistical fluctuations in the background.
The $5\sigma$ criterion is  extremely conservative, more applicable to experiments where the signal must be extracted from a large background, such as in high-energy particle physics experiments, than ALPHA's trapping experiments where the background is small compared to the signal.

The optimum values of the thresholds were chosen by applying the cuts to the samples of annihilation and cosmic events, and calculating the p-value at different thresholds.
A detailed discussion of this procedure can be found in reference \cite{RichardFutureThesis}.
The analysis was performed under `blind' conditions.
That is, the data sample collected in the trapping experiment was not used in any way to determine the analysis procedure or the values of the cut thresholds.
Blind analyses are useful as they eliminate the possibility of bias towards a signal influencing the results.

The results of this analysis found that the p-value would be optimised if the conditions for acceptance were:
\begin{enumerate}
	\renewcommand{\theenumi}{(\alph{enumi})}
	\renewcommand{\labelenumi}{\theenumi}
	\item a reconstructed vertex (this automatically requires at least two tracks).
	\item the vertex lies within 4~cm of the trap axis.
	\item
	\begin{enumerate}
	\renewcommand{\theenumii}{(\roman{enumii})}
	\renewcommand{\labelenumii}{\theenumii}
		\item For events with exactly two tracks - the squared residual exceeds $2~\mathrm{cm}^2$.
		\item For events with three or more tracks - the squared residual exceeds $0.05~\mathrm{cm}^2$.
	\end{enumerate}
\end{enumerate}

These cuts accept an absolute rate of $(2.2 \pm 0.1) \times 10^{-2}~\mathrm{Hz}$, or $(0.51 \pm 0.01)\%$ of the cosmic events and accept $(49.0\pm0.6)\%$ of annihilation events.
This must be combined with the efficiency with which annihilations prompt a detector read-out $(85 \pm 15)\%$ to estimate the efficiency of detecting an annihilation as $(42 \pm 7)\%$.
The total trapping experiment consisted of 212 repetitions of the single experiment, for a total observation window of 6.36~s.
The expected number of cosmic events in this window is $0.14 \pm 0.01$.

When the cuts were applied to the trapping data six events were found to pass the cuts.
The corresponding p-value and significance (equations \ref{eqn:poissonPvalue} and \ref{eqn:significance}) are $9.2 \times 10^{-9}$ and $5.6\sigma$ respectively.
Many consistency checks established that the signals were representative of antiproton annihilations, and that the analysis procedure did not suffer from systematic defects.
A more detailed discussion can be found in \cite{SearchPaper}.

This is extremely strong evidence for observation of antiproton annihilations during shutdown of the magnetic trap.
However, observation of antiprotons does not immediately imply the presence of antihydrogen atoms.
The next section will discuss the possible sources of antiproton annihilations other than antihydrogen.

\section{Antiproton background}

\subsection{Mirror-trapped antiprotons}

The magnetic minimum trap used in ALPHA to form a trap for antihydrogen atoms can also act as a trap for charged particles.
This is most relevant to `bare' antiprotons - i.e., those not bound to a positron in an antihydrogen atom.
This is undesirable, as, when the trap is removed to release the antihydrogen atoms for detection, the bare antiprotons will also be released, and can be misidentified as antihydrogen atoms.

The trapping mechanism arises from the adiabatic invariance of the first magnetic moment (equation \ref{eq:adiabatic1}):
\begin{equation}
	\mu = \frac{\frac{1}{2} m v_\perp^2}{B} = \frac{E_\perp}{B},
	\tag{\ref{eq:adiabatic1}}
\end{equation}
where $E_\perp$ is the kinetic energy of the particle in the plane perpendicular to the magnetic field.
As a particle moves from a region of low magnetic field to higher magnetic field, $E_\perp$ must increase to conserve $\mu$.
Energy is transferred from the degree of freedom parallel to the magnetic field to the perpendicular degree of freedom to achieve this.
This continues until the parallel energy reaches zero, whereupon the particle changes direction.
The particle is said to be `reflected' from the region of high magnetic field, which gives this phenomenon its name of `mirror-trapping'.

In a magnetic field that varies along the trajectory of a particle, this yields an equation for the parallel kinetic energy
\begin{equation}
	E_\parallel = E_0 \left( 1 - \frac{E_{\perp,0}}{E_0}\frac{B}{B_0}  \right),
\end{equation}
where $E_0 = \sqrt{E_{\parallel,0}^2 + E_{\perp,0}^2}$ is the kinetic energy of the particle at a point where the magnetic field is $B_0$.
It can be seen that the trapping condition, $E_\parallel \leq 0$, is met for sufficiently high $B/B_0$ or $E_{\perp,0}/E_{0}$.
This is typically written as the mirror ratio
\begin{equation}
	\alpha = \left( \frac{v_\parallel}{v_\perp}\right)_{\mathrm{crit}} < \sqrt{\frac{B_{max}}{B_{min}} - 1}.
	\label{eq:mirrorRatio}
\end{equation}

\subsection{Energy limits on mirror-trapped antiprotons}
\label{sec:MirrorTrapEnergy}

\begin{figure}[b!]
\centering
\subfloat[]{\input{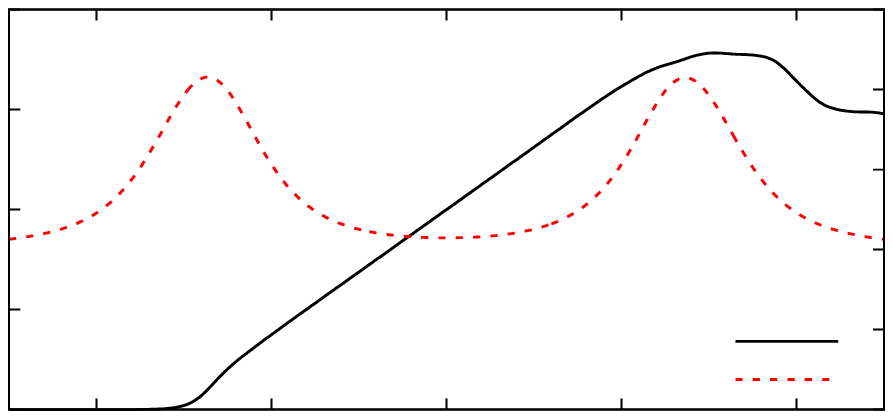}}\\
\subfloat[]{\input{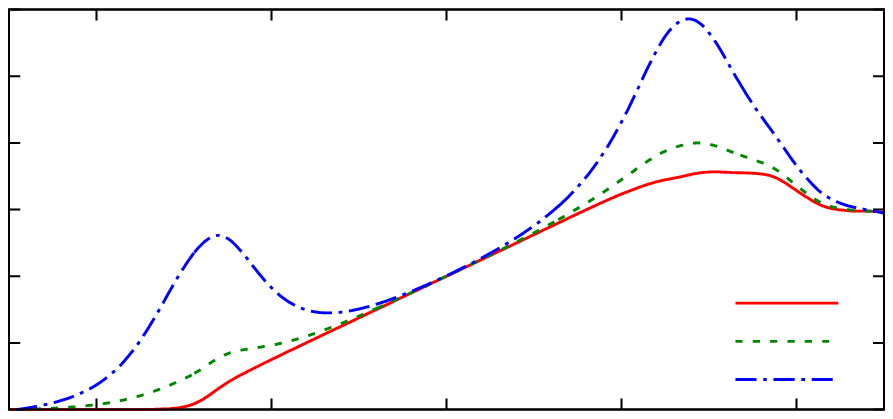}}
\caption[The electric potential, magnetic field and pseudo-potential when removing bare antiprotons.]{(a) the electric potential ($\Phi$) and magnetic field magnitude ($|\mathbf{B}|$) along the axis of the Penning trap used to remove bare antiprotons from the magnetic trap. (b) the pseudo-potential (eq \ref{eq:pseudopotential}) for different values of $E_{\perp}$.}
\label{fig:pseudo-potential}
\end{figure}

In the ALPHA magnetic trap, the magnetic field along the cylindrical axis remains parallel to the axis, even when the magnetic trap is energised.
The magnetic field along the axis is shown in figure \ref{fig:pseudo-potential}(a); clearly there is a minimum of magnetic field around $z=0$, where particles can be mirror-trapped.

The motion can be analysed in the form of a `pseudo-potential', which combines the electrostatic potential energy $\left(-e\right)\Phi$ and the energy of interaction of the magnetic dipole moment (equation \ref{eq:adiabatic1}) with the magnetic field,
\begin{equation}
	U = E_{\perp,0} \left( \frac{B-B_0}{B_0} \right) + (-e) \Phi.
	\label{eq:pseudopotential}
\end{equation}

The presence of $\Phi$ in this equation presents the opportunity to manipulate the shape of the potential so that antiprotons cannot be trapped.
This can be achieved by placing a sloping electric potential across the trap, or equivalently, applying an electric field in one direction.
However, only antiprotons with values of $E_\perp$ below a limited value are untrapped.
This is seen in the examples of pseudo-potentials in figure \ref{fig:pseudo-potential}(b).
At low $E_\perp$, the electric field overcomes the force due to the inhomogeneous magnetic field, and particles are ejected from the trap.
Above 10~eV, a local minimum begins to develop in which particles may be trapped.

Manipulating the electric potentials can place a stricter limit on the minimum perpendicular energy of mirror-trapped particles.
As the potentials are changed, the particles' parallel energies are changed non-adiabatically, and the minimum value of $E_\perp$ required to fulfil the trapping condition is increased.
Following the evolution of the pseudo-potential through the manipulations of this `clearing' procedure leads to the conclusion that a particle must have a value of $E_\perp$ greater than $\simeq$ 25~eV to remain trapped.

Applying the pseudo-potential analysis off the cylindrical axis is much more difficult, as the octupole contributes to the magnetic field and the resulting particle trajectories become very complex.
The only feasible way to approach the problem is from a numerical perspective.
To do this, a series of Monte-Carlo-like simulations were performed.
Single antiprotons with positions and velocities drawn from distributions were propagated through the simulated clearing process with realistic models of the electric potential and magnetic field.
It was found that no particles with $E_\perp$ less than 25~eV remained within the trap after the clearing procedure.
This is the same limit found from the pseudo-potential analysis, but now applies to the entire volume of the trap, not just on the axis.

\subsection{Possible sources of 25~eV, mirror-trapped antiprotons}

With this conclusion, one must therefore address the question of the probability of producing antiprotons with perpendicular kinetic energy greater than 25~eV  in the experiment.
The bulk of antiprotons in the mixing experiment are in a thermal distribution. 
We can thus evaluate the number of mirror-trapped antiprotons by evaluating the Maxwell-Boltzmann distribution within the appropriate limits.
An upper limit on the temperature of the distribution can be obtained from the depth of the side-wells from which the antiprotons are injected into the positron plasma.
The depth of the side-wells is approximately 0.6~eV, and the fraction of a 0.6~eV thermal distribution that lies above 25~eV is less than $10^{-11}$.
Any antiprotons with perpendicular energies greater than 25~eV must therefore come from non-thermal sources.

Spatial redistribution of antiprotons through the formation of antihydrogen atoms and subsequent re-ionisation has already been described in chapter \ref{chp:hbar}.
Re-ionisation, depending on the value of the electric potential at the position of ionisation can be accompanied by large changes in energy.
The maximum energy gain can be estimated from the differences in the voltages applied to the electrodes. 
The largest difference happens to be 25~V, so it is conceivable that 25~eV antiprotons can be produced by this method.

However, this energy gain is purely in the parallel direction.
To become mirror-trapped, the antiproton must undergo a collision with another particle -- the candidates are another antiproton or a gas atom, and convert the parallel energy to energy in the perpendicular direction.
The rate of collisions can evaluated, and thus an upper limit on the number of antiprotons with trappable energies produced.

The antiproton-antiproton collision rate can be approximated using the expression for a thermalised distribution (equation \ref{eq:collisionRate}).
This expression will tend to overestimate the collision rate in a non-thermalised distribution.
The density of antiprotons at high radii, where the large energy gains are possible, can be evaluated from measurements similar to those discussed in section \ref{sec:ballisticLoss}.
The simulations in chapter \ref{chp:vertex} allow us to evaluate the number of antiprotons that remain trapped at high radius relative to the number that annihilate on the wall.
The octupole field effectively removes most of the ionised particles and the ratio of antiproton annihilations to trapped antiprotons is approximately 10:1.
An estimate of the maximum number of trapped antiprotons at any one time would be approximately 300.

Making a further conservative assumption -- that these antiprotons are confined in a shell 1~mm thick shell at $r=1$~cm -- we estimate the density of antiprotons as $\sim 10^2~\mathrm{cm^{-3}}$.
Equation \ref{eq:collisionRate} evaluates to a collision rate per antiproton of $\sim 10^{-6}$~Hz, or a total rate of $\sim 10^{-4}$~Hz (order of magnitude only).
Thus, in 212 runs, with a holding time of 1.2~s, there would be an expected $\sim 0.1$ collisions in total.
This is incompatible with the observed number of annihilation events.

Gas atoms or molecules in the Penning trap are close to temperature of the cryogenic surroundings, and hence are essentially at rest compared to electronvolt-range antiprotons.
The gas density is known from measurements of the antiproton lifetime (section \ref{sec:vacuum}).
Conservation of momentum requires an antiproton impinging on the gas atom to have energy higher than 25~eV so that an $E_\perp = 25~\mathrm{eV}$ final state is possible.
For instance, if the gas atom is atomic hydrogen, the antiproton must have an initial energy of at least 50~eV if a final state with $E_\perp = 25~\mathrm{eV}$ is allowed.
The cross-sections are energy dependent and must be calculated numerically \cite{Svante_atomCollisions}.
However, the cross-sections of interest are vanishingly small, and only upper limits can be placed on the rate.
For 100~eV antiprotons, the lowest energy easily calculable, the rate of collisions resulting in an antiproton with more than 25~eV of perpendicular energy is less than $\sim 10^{-8}~\mathrm{Hz}$ per antiproton.
Note that not all of these antiprotons are mirror trapped, as the trapping criterion is stated as the ratio of perpendicular to parallel energy (equation \ref{eq:mirrorRatio}).

The collision rates calculated in the previous paragraphs show that collisions between antiprotons and gas atoms or other antiprotons are not mechanisms through which antiprotons capable of being mirror-trapped can be produced at a rate approaching the observed rate of annihilations.

While the results of these estimates are convincing evidence that mirror-trapped antiprotons that can remain trapped through the clearing process cannot be produced at a significant rate in the trapping experiment, in practice, theory-based estimates are only as good as the input parameters, which are not very well known.
Thus, it would be better to demonstrate experimentally that mirror trapped antiprotons can (or can not) be trapped in our trap at the rate we observe annihilations.

One possible approach is to perform `null experiments'.
That is, experiments in which the other postulated source of annihilations -- trapped antihydrogen atoms -- is suppressed, but not the mirror trapped antiprotons.
An example of the type of experiment one might consider is to conduct the same experiment, but without energising the octupole magnet. 
In this case, bare antiprotons are still confined transversely by the solenoidal field and axially by the magnetic fields of the mirror coils.
Antihydrogen atoms, however, have no transverse confinement.
If the annihilation signals are due to mirror-trapped antiprotons, they should be observed in the null experiment, and if they are due to antihydrogen, they would not.

Unfortunately, the situation is not as simple as this.
Firstly, the influence of the octupole on the dynamics of the antihydrogen production -- the number of atoms, their temperature and their internal states is not well-known.
It is not possible to guarantee that the redistributed antiproton populations have the same energy and spatial distributions in the null experiment as the trapping experiment.
A further complication is that, in some cases, the octupole field can enhance mirror trapping.
Recall that the magnetic field lines, upon which the particles move, bend radially in the presence of an octupole and a solenoid.
Also recall that the magnetic field strength of the octupole increases as $r^3$ (equation \ref{eq:octupoleField}).
Thus, the magnetic field along the trajectory of the particle increases, even though the $B_z$ component remains fixed.
While the observation of annihilation would be a clear indication that mirror-trapped antiprotons are present in higher quantities than thought, a negative result would not be strong evidence that the observed annihilations were due to antihydrogen atoms, given that possible enhancements to mirror-trapping have been removed in the null experiments.
Thus, we call experiments such as this `partial' null experiments.
A limited number of these experiments were performed, without observing an annihilation at the shutdown of the magnetic trap, but the statistical significance of this measurement is not high enough to allow a conclusion to be drawn.

It is also possible to increase the temperature of the positron plasma by applying radio-frequency power at a suitable frequency \cite{ATHENA_Nature}.
This has the double effect of reducing the rate of antihydrogen production, and increasing the temperature of the antihydrogen atoms produced (see section \ref{sec:HbarFormationTheory}), so that they are less likely to be trapped.
While the magnetic field configuration is unchanged from the trapping experiment, this null experiment suffers from the change in the number of antihydrogen atoms and their internal states, making it also, at best, only a partial null experiment.

Several of the variations of the partial null experiments described above were performed as cross-checks -- 161 trials in all. 
No events satisfying the selection criteria were observed.
If the process that produced the candidate events was still present at the same rate in these experiments, there would be only a $\sim 1\%$ chance of observing zero events.
This is an encouraging result, but not sufficient to definitely identify trapped antihydrogen.

\section{Release signatures of antiprotons and antihydrogen}

Even though the presence of mirror-trapped antiprotons is not experimentally ruled out, further information is available that can help determine if the observed annihilations originate from antihydrogen atoms or bare antiprotons.

The dynamics of trapped antihydrogen atoms in the magnetic minimum differ greatly from those of mirror-trapped antiprotons.
It might therefore be expected that the times and positions at which they escape the trap and annihilate would be different.
One of the differences stems from the differing scale of energies that trapped atoms and mirror-trapped antiprotons would have. 
Trapped ground-state antihydrogen atoms can have energies no more than $0.5~\mathrm{K}$ ($\sim 10^{-4}$~eV), while the energies of mirror-trapped antiprotons are in the electron-volt range.
The speed of a trapped atom is thus at most 100~$\mathrm{m\,s^{-1}}$, and considering that the trap size is of the order of few centimetres, and the time taken by the magnetic field to decay is of order 10~ms, it is clear that they will transit the trap only a small number of times before escaping.
This means that the atom will not, in general, escape the trap at the lowest saddle point, but will have a wide spatial distribution.

On the other hand, the high-energy mirror-trapped antiprotons transit the trap rapidly and have enough time to `explore' the trap edges.
They will tend to escape at the point where the trap depth is minimum.
Thus, their annihilations will tend to be localised around this point.
In the ALPHA trap, the lowest saddle point, where the trap depth is minimised, is at the axial centre of the trap.

The escape signatures of antihydrogen atoms and bare antiprotons were investigated using an extension of the computer simulations discussed above.
In this study, particles that survived the clearing procedure were allowed to propagate while the magnetic field was reduced in the same way as measured in the experiment.
Additional care must be taken when simulating changing magnetic fields, as induced eddy currents and electric fields can have a large effect on the particle trajectories.

\begin{figure}[t]
	\centering
	\subfloat[\label{fig:ztSim_Hbar}]{\input{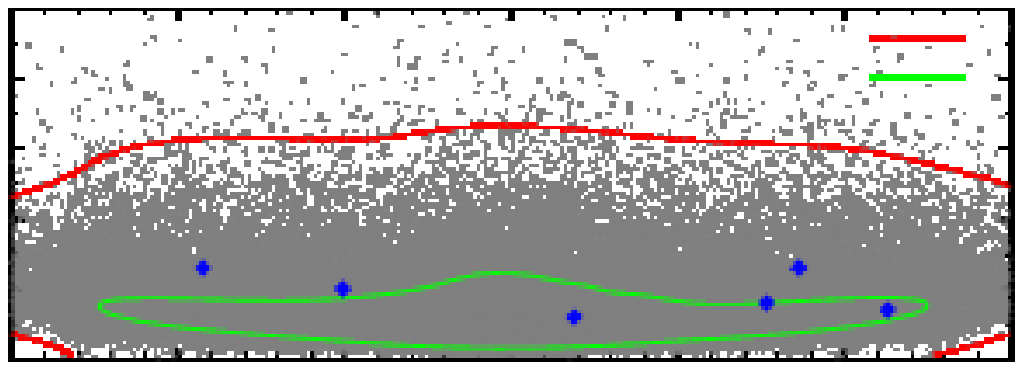}}\\
	\vspace{-1.2cm}
	\subfloat[\label{fig:ztSim_Pbar}]{\input{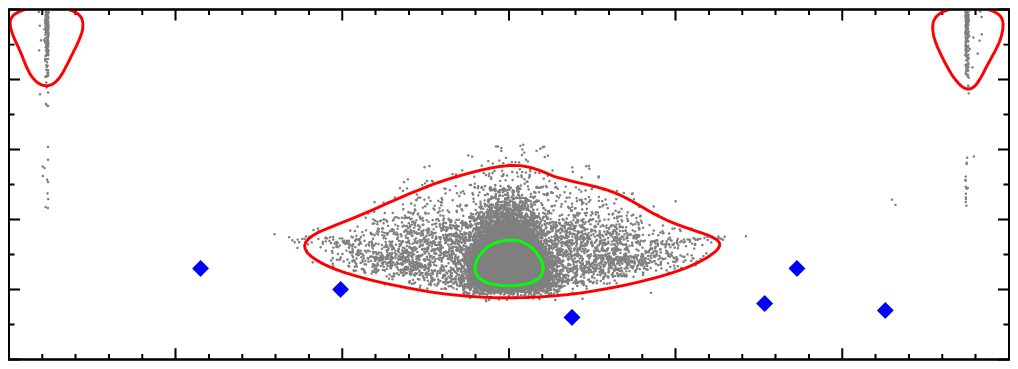}}
	\caption[A scatter plot of the escape time and $z$-position for simulated antihydrogen atoms  and mirror-trapped antiprotons.]{A scatter plot of the escape time and $z$-position for simulated antihydrogen atoms (a) and mirror-trapped antiprotons (b) (grey dots). The contours drawn enclose 50\% (green) and 99\% (red) of the probability when convolved with the detector response function. The blue points are the six candidate events measured in the trapping experiment.   }
	\label{fig:ztSim}
\end{figure}

The results of these simulations are shown in figures \ref{fig:ztSim}(a) and \ref{fig:ztSim}(b).
It is evident that the behaviour qualitatively described above is reproduced by the simulation.
To compare the positions of the simulated points to the reconstructions of the observed annihilation vertices, one must take account of the resolution of the detector and reconstruction algorithm (section \ref{sec:siliconDetector}).
The uncertainty of the time of the event is very small, and is not considered.
If the simulated points are convolved with the detector resolution function, one obtains the probability distribution for the reconstructed vertex coordinates.

By multiplying the probability density at each of the observed positions, it is possible to evaluate the relative probability that the observations are drawn from one distribution over the other.
This is done by performing a Monte-Carlo-like calculation, in which sets of (in this case, six) events are drawn randomly from the probability distribution, scattered according to the resolution function, and the product of the probability densities at their coordinates multiplied to obtain a total probability for this combination of events.
This procedure is repeated a large number of times to obtain a distribution of probabilities.
One then compares the combined probability of the observed set of events to this distribution and determines the fraction of random sets that are less likely than the observation.
This gives the probability that such a set of events at least as improbable as these would have been observed, given the initial distribution.

In the case of the simulated antiproton distribution (figure \ref{fig:ztSim}(b)), several of the observed events lie several centimetres from any of the simulated annihilations, where there is a negligible probability that antiproton events can be observed.
Thus, it is only possible to place an approximate upper bound on the probability.
This is done in figures \ref{fig:ztSim}(a) and (b) by drawing the probability contours.
These are the contours of constant probability density that enclose a given fraction of the probability..
For a contour enclosing $x\%$ of the probability density, it would be expected that, on average, $x\%$ of the observed points would lie inside the contour.
All six of the events lie outside the $99\%$ contour for the antiproton distribution.
The probability of this occurring is $(1\%)^6$, or $10^{-12}$. 
It is thus very unlikely that the observed events are drawn from this distribution.

On the other hand, the events appear compatible with the antihydrogen distribution -- three of the six events lie inside the 50\% contour, and following the procedure outlined above to evaluate the probability gives the result that randomly picked events have less likely configurations around 60\% of the time.
This is close to what would be expected just from chance.

\begin{figure}[hbt]
	\centering
	\subfloat[\label{fig:zphiSim_Hbar}]{\input{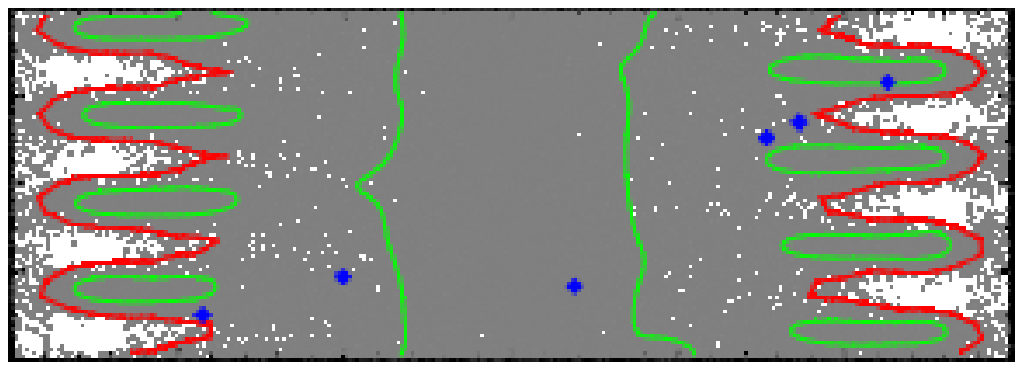}}\\
	\vspace{-1.2cm}
	\subfloat[\label{fig:zphiSim_Pbar}]{\input{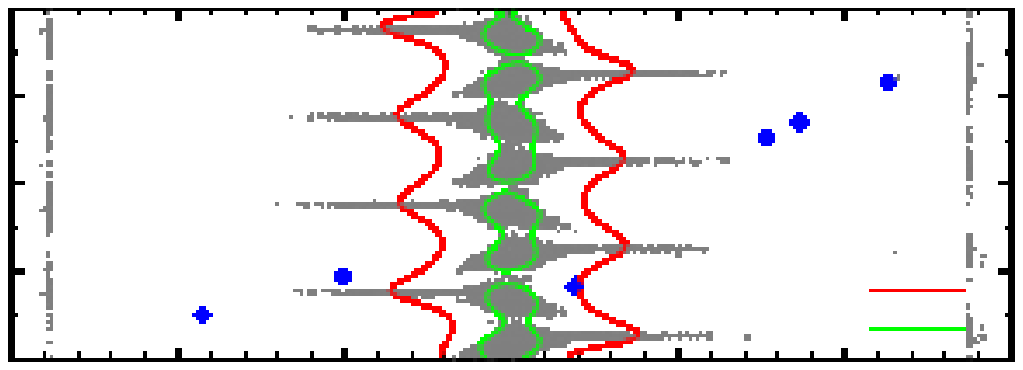}}
	\caption[A scatter plot of the $z$ and $\phi$ coordinates of the escape position for simulated antihydrogen atoms and mirror-trapped antiprotons.]{A scatter plot of the $z$ and $\phi$ coordinates of the escape position for simulated antihydrogen atoms (a) and mirror-trapped antiprotons (b) (grey dots). The contours drawn enclose 50\% (green) and 99\% (red) of the probability density when convolved with the detector response function. The blue points are the six candidate events measured in the trapping experiment.}
	\label{fig:zphiSim}
\end{figure}

The transverse coordinates can also be examined and compared to the observed events.
The two-dimensional $z-\phi$ distributions, shown in figure \ref{fig:zphiSim} are reminiscent of the distributions seen during ballistic antiproton loss in the octupole (section \ref{sec:ballisticLoss}). 
The transverse information is of limited usefulness, for two reasons.
The $\sim$ 1~cm resolution of the detector translates into approximately $30^\circ$ $\phi$ resolution at the radius of the electrodes. Since the structure occurs with a period of $90^\circ$, regions of low and high probability are blended together, and there is little to distinguish one set of events from another.
Also, the peaked features occur at approximately the same $\phi$ coordinates for the antiproton distribution as for the antihydrogen distribution.
Thus, excess numbers of events would be expected at roughly the same positions, regardless of whether the annihilations were due to bare antiprotons or antihydrogen atoms.

While, taken on face value, these results are quite convincing, the analysis relies heavily on numerical simulations of the particles' motion in complex electric and magnetic fields.
In particular, simulations are only as good as their inputs, and in a complex device such as ALPHA, it is practically impossible to include every possible effect in a simulation.
Without supporting experimental evidence or validation that the simulations reflect the true situation in the experiment, it is difficult to argue that the observed events are antihydrogen atoms to a high degree of certainty.
ALPHA decided to pursue further studies before definitively identifying the signals as trapped antihydrogen atoms.

\subsection{Escape signatures in the presence of electric fields}

An experimental demonstration that the observed annihilations do not come from charged particles (i.e are not mirror-trapped antiprotons) can be made by applying an axial electric field during the shutdown of the magnetic trap.
The effect is two-fold -- the electric field will modify the trajectories of antiprotons; this causes their annihilation positions to be shifted longitudinally.
The electric field also reduces the depth of the pseudo-potential (section \ref{sec:MirrorTrapEnergy}), causing the antiprotons to escape at an earlier time.
Antihydrogen atoms, being uncharged, should not be affected in any way \footnote{The electric field is too weak to ionise the atoms, which have by this time been confined for more than 100~ms, and should be deeply bound. The electric dipole moment of the antihydrogen atom is too small to play a role}.

\begin{figure}[hbt]
\centering
\input{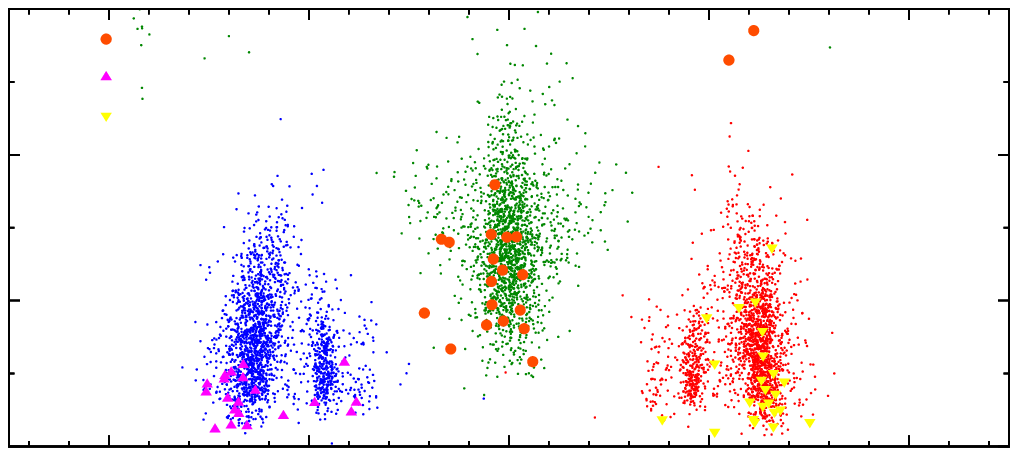}
\caption[A scatter plot of the $z$ and $t$ coordinates of the escape position for simulated mirror-trapped antiprotons, with three bias-field configurations, compared to measurements.]{A scatter plot of the $z$ and $t$ coordinates of the escape position for simulated mirror-trapped antiprotons (small dots) with three bias-field configurations, compared to measured events. The simulated and measured annihilation events with no bias field are shown as green dots and orange circles respectively, those with a left bias field with blue dots or magenta up-pointing triangles, and those with a right bias field with red dots or yellow down-pointing triangles.}
\label{fig:mirrorTrappedData}
\end{figure}

This effect can be demonstrated experimentally by preparing a sample of mirror-trapped antiprotons.
A cloud of $\sim 10^5$ antiprotons was given $\sim 40~\mathrm{eV}$ of kinetic energy in the axial direction, and allowed to collisionally thermalise for 420~s in the magnetic trap, after which a small fraction of the antiprotons can be expected to have a value of $E_\perp/E_\parallel$ that allows them to be mirror-trapped.
The clearing procedure described above was carried out, removing antiprotons with low $E_\perp$.
Before and during the fast shutdown of the magnetic trap, a $5~\mathrm{V\,cm^{-1}}$ electric `bias field' was applied across the trapping region using the Penning trap electrodes.
The $z-t$ distribution of the annihilations recorded in this experiment are shown in figure \ref{fig:mirrorTrappedData}. 
Three configurations of the bias field were used: a negative electric field, which moved the antiprotons to positive $z$; a positive electric field, which moves the particles in the opposite direction, and a case in which no electric field is applied (similar to the data in figure \ref{fig:ztSim}(b)).
The annihilation positions can be simulated using the same procedure already discussed, and these are also shown in figure \ref{fig:mirrorTrappedData}.
The shape of the annihilation distribution without a bias field is not the same as that in figure \ref{fig:ztSim}(b), as the preparation of mirror-trapped antiprotons described above gives very different spatial and energy distributions before clearing and the trap shutdown compared to those expected to be obtained from antiproton-positron mixing.

The deflection of the annihilation positions, both measured and simulated, is clearly visible. 
All of the measured points fall within or close to the simulated distributions -- this validates the ability of the simulations to accurately predict the trajectories of the antiprotons \footnote{The two events in the no-bias field that fall away from the bulk of the annihilations are near a radial step in the electrodes. Some of the simulated antiprotons annihilate impact and annihilate here, but in smaller numbers than the fraction seen here would imply. It is likely that antiprotons on these trajectories are sensitive to small trap imperfections, which cannot be included in the simulation. }

This procedure was implemented in a series of antihydrogen trapping experiments in 2010, reported in \cite{ALPHA_Nature}.
The signal rate increased relative to the experiment described above, likely due to a reduction of the positron temperature to approximately $90~\mathrm{K}$ using evaporative cooling (section \ref{sec:LeptonEVC}).
The number of runs with each bias-field configuration, and the number of events passing the annihilation selection criteria are shown in table \ref{table:NatureTable}.

\begin{table}[hbt]
\centering
\begin{tabular}{|l|r|r|}
	\hline
	\textbf{Bias Field} & \textbf{Number of runs} & \textbf{Events} \\ \hline
	None & 137 & 15 \\ \hline
	Left & 101 & 11 \\ \hline
	Right & 97 & 12 \\ \hline
\end{tabular}
\caption{The number of runs and events satisfying the selection criteria for the three bias-field configurations.}
\label{table:NatureTable}
\end{table}

The expected background from cosmic rays in this experiment was $4.4\times 10^{-2}~\mathrm{Hz}$, different to that found in section \ref{sec:cosmicAnalysis} because of a change in the detector operating conditions leading to an increased acceptance of all types of events.
This part of the analysis was otherwise identical.
Over the total 335 runs, there is an expected contribution of 0.4 events from cosmic rays.
This is incompatible with the observed 38 events at a level $\gtrsim 16\sigma$.

\begin{figure}[hbt]
	\centering
	\subfloat[]{\input{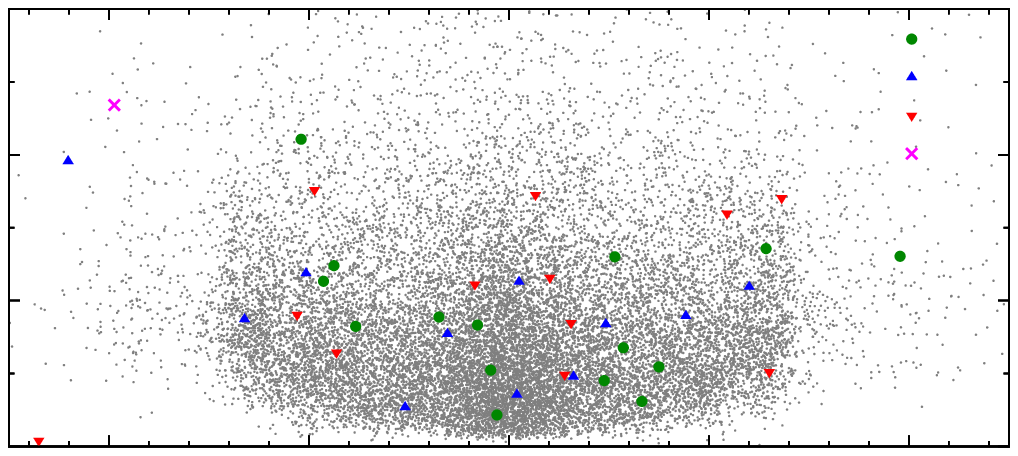}}\\
	\vspace{-1.2cm}
	\subfloat[]{\input{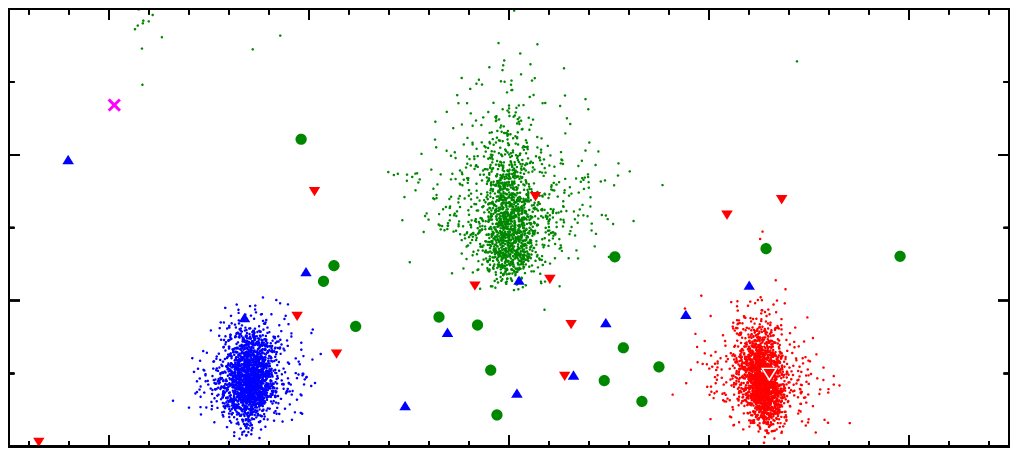}}
\caption[A scatter plot of the $z$ and $t$ coordinates of the escape position for simulated antihydrogen atoms and mirror-trapped antiprotons, with three bias-field configurations.]{A scatter plot of the $z$ and $t$ coordinates of the escape position for simulated antihydrogen atoms (a) and mirror-trapped antiprotons (b) (small dots) with three bias-field configurations. The simulated and measured annihilation events with no bias field are shown as green dots or circles respectively, those with a left bias field with blue dots or up-pointing triangles, and those with a right bias field with red dots or down-pointing triangles. The event observed with heated positrons is marked with a magenta cross.}
\label{fig:NatureFigure}
\end{figure}

The $z-t$ distributions for this data, compared with simulations of the antihydrogen and mirror-trapped antiproton distributions are shown in figure \ref{fig:NatureFigure}.
As before, the measured annihilations (with a few exceptions) are compatible with the simulated antihydrogen distribution.
The data sets from each bias-field configuration are also compatible with each other, indicating that the signals are unaffected by an electric field.
Importantly, the event distributions are clearly incompatible with the mirror trapped distributions.

As stated before, the simulations of the annihilation positions are only as good as the input parameters.
It was a concern that antiprotons with very high transverse energies (for example, captured from the AD beam) could become mirror-trapped in the apparatus.
Such antiprotons are not accounted for in the simulations and could escape the magnetic trap in positions outside the simulated annihilation distributions.
In all cases, these antiprotons would not be associated with antihydrogen production and their presence or absence can be determined by suppressing antihydrogen formation by heating the positron plasma (see section \ref{sec:HbarTemperature}).
Heating the positrons in this experiment has a double effect.
Antihydrogen formation is suppressed (by approximately two orders of magnitude) and in addition, the kinetic energy of the antihydrogen atoms that are produced is increased, as the antiprotons are assumed to be in thermal equilibrium with the positron plasma before formation.
Any annihilation events observed cannot therefore be trapped antihydrogen atoms and can be considered as a background.
One should also realise that the background measured using heated positrons still includes a contribution from cosmic events.

246 runs with the positron plasma heated to 1,100~K were carried out, in which one event passing the annihilation selection criteria was identified -- this event is shown in figure \ref{fig:NatureFigure} as the magenta cross.
When scaled to 335 runs, this event is equivalent to a background $1.4 \pm 1.4$ events.
This number of events cannot account for the observation of 38 annihilation events from mixing with cold antiprotons.
It is notable that this event lies outside both the antiproton and antihydrogen annihilation distributions.
It is hypothesised that the same (unidentified) background process that is responsible for this observation is also responsible for the events seen in the cold-positron experiments that lie outside the distributions.

With two additional pieces of evidence, there is now definitive proof that the signals seen in ALPHA are trapped antihydrogen atoms.
This was announced by the ALPHA Collaboration in November 2010 in \textit{Nature} \cite{ALPHA_Nature}.

\section{Expected trapping rate}

Comparing the number of trapped antihydrogen events to theoretical estimates of same number the can help to further understanding of the processes that determine the number of antihydrogen atoms trapped.
Agreement between the estimates and the observations also helps to support the claim that the events are due to trapped antihydrogen atoms.

The estimate is based on a simplified model of the relevant physics, and requires knowledge of experimental parameters that are not known precisely.
Thus, the result is expected to be accurate only to within an order-of-magnitude.
The estimate uses the number of antihydrogen atoms observed to have been produced and the fraction of them with a trappable energy.
The later quantity has been discussed in section \ref{sec:TrappingConsiderations}.

A complication arises from the fact that the atomic magnetic moment, which depends on the internal atomic state, is not known.
Based on the sign of the magnetic moment, the atoms can be divided into high-field-seeking (positive $\mathbf{\mu})$ and low-field-seeking (negative $\mathbf{\mu}$).
In three-body recombination, and a semi-classical view of the atomic motion, the relative sizes of these populations depends on the position of the antiproton relative to the trajectory of the positron's cyclotron orbit.
The probability that an atom formed in a particular state will be low-field seeking will be approximately the ratio of the areas of the cyclotron orbit and the area enclosed by the positron's orbit \cite{FrancisMagneticMoment}.
As an example, an atom in the n=70 state (a binding energy of $\sim$ 32~K and a radius of $\sim$ 260~nm) formed from a 4~K positron in a 1~T magnetic field, which has a cyclotron radius of $\sim 60~\mathrm{nm}$, will be in a low-field seeking state with probability $(60/260)^2$, or 5\% of the time.
At higher temperatures, the fraction of low-field seekers increases, and the ratio is expected to equalise as the cyclotron radius becomes comparable to the atomic radius.

The magnitude of the atomic moment enters into the expression of the well depth (equation \ref{eq:trapDepth}).
The distribution of states produced from recombination in nested-trap recombination, and crucially, how it evolves over time as the atom moves in the atom trap, is not known.
If an atom is trapped because it possesses a high value of $\mathbf{\mu}$, and undergoes a transition to an atomic state with a smaller value, it could become untrapped.
However, depending on the magnetic field strength at the position where the atom undergoes its transition, the atom can also lose a fraction of its energy \cite{FrancisRadiativeCooling}. 
This can provide a cooling effect on the atom as it de-excites.

The best that can be done is to make an educated guess at the state distributions.
The most conservative approach is to say that only atoms with kinetic energy lower than depth of the trap for a ground state atom (i.e. the minimum possible trap depth) can be trapped.
In this assumption, an atom will remain trapped in any atomic state, as long as the sign of magnetic moment does not change.

The model can be expressed simply as

\begin{equation}
\begin{tabular}{l l l}
$N_{\mathrm{detected}}$ & = & $N_{\mathrm{trapped}} \times f_{\mathrm{detection}}$\\ 
& = & $N_{\mathrm{produced}} \times f_{\mathrm{0.5K}} \times f_{\mathrm{LFS}} \times f_\mathrm{detection}$,
\end{tabular}
\end{equation}

where $f_\mathrm{detection}$ is the efficiency to detect a trapped antihydrogen atom (see section \ref{sec:cosmicAnalysis}), $f_{\mathrm{0.5K}}$ is the fraction of atoms with trappable energies and $f_\mathrm{LFS}$ is the fraction of atoms in low-field seeking states.
$N_\mathrm{detected}$, $N_\mathrm{trapped}$, and $N_\mathrm{produced}$ are the number of antihydrogen atoms detected, trapped and produced respectively.

Entering the measured experimental conditions into the model produces the estimate that approximately 26 atoms should have been trapped over the 212 runs presented in the 2009 set.
Taking into account the detection efficiency (section \ref{sec:cosmicAnalysis}), this means that approximately 11 atoms would have been identified, given the assumptions.
This agrees well with the six observed annihilation events.

When the effect of states with higher magnetic moments and the cooling scheme discussed in reference \cite{FrancisRadiativeCooling} is taken into account, the estimate increases to $\sim 60$ atoms trapped, or $\sim 25$ detected.
This is still agrees with the number of observed events to within the expected order-of-magnitude accuracy.

For the colder positron plasma conditions used in the 2010 data set \cite{ALPHA_Nature}, the simple estimate is approximately 0.35 events detected per run, or 117 for the complete data set, which is again within an order-of-magnitude of the observed number. 

\section{Conclusions}

The work presented in this chapter is the first demonstration of the stable trapping of a pure antimatter atomic system.
The first measurements held only a few atoms for at least 172~ms, which is very long compared to the sub-millisecond time during which an untrapped atom can cross the apparatus and annihilate on the walls.
The importance of this result and the directions for future work will be discussed in the next chapter.

    \chapter{Conclusions and Outlook}

\epigraph{The most incomprehensible thing about the world is that it is at all comprehensible.}{Albert Einstein}

At the time this thesis was written, ALPHA had just demonstrated the confinement of 38 antihydrogen atoms for 172~ms \cite{ALPHA_Nature}.
While this is a ground-breaking result and a major step forward, precision tests of fundamental symmetries are likely to be difficult, though not impossible, with such small numbers of atoms and short confinement times.
Performing spectroscopic experiments becomes much easier if more atoms can be confined for a longer period of time, and the next steps will be to prepare a sample of atoms for spectroscopy.
Since the first experiments, significant progress has been made in this direction -- over the entire 2010 ALPHA run, which had finished shortly before the submission of this work, more than 200 trapped antihydrogen annihilation events have been identified, some after confinement for several minutes.

The parameters that determine the kinetic energies of the antihydrogen atoms, discussed in chapter \ref{chp:trapping} have only begun to be explored.
Calculations of the antiproton-positron thermalisation rate imply that access to the parameter space is limited by the temperature of the positron plasma.
For the first time, the analysis in chapter \ref{chp:vertex}, supports this theoretical prediction with measurements of the antihydrogen annihilation distribution.

Direct measurements of the particles' temperature and density has been key to developing a method for producing antihydrogen atoms with trappable energies and will remain so as the effort for higher trapping rates continues.
Evaporative cooling (section \ref{sec:EVC}) is a powerful technique, newly applied to charged particles in a Penning trap, that can help to realise colder positron plasmas along this path.
The full potential of this technique has still not been exploited.

The internal state of the antihydrogen atom will also influence the number of atoms that can be trapped, and can give insight into the fundamental recombination process, still not completely understood.
The study of how field-ionisation affects the annihilation spatial distribution in chapter \ref{chp:vertex} shows that we can probe this quantity at deeper binding energies than before.

Remembering that the motivation to produced trapped antihydrogen is to make comparisons with matter and tests of fundamental symmetries, we look to the first atomic physics measurements on trapped antihydrogen.
The ALPHA apparatus must be modified to admit laser beams for spectroscopy of the 1S-2S antihydrogen transitions, so the first spectroscopic measurements will be of the hyperfine structure of the atom, where the transition frequencies are in the microwave range.
Introducing microwave radiation into the mixing region is far easier than introducing laser light, and could be implemented within the next year or two.

The first experiments will attempt to induce an atomic transition from a trapped to an untrapped state. 
The transition will be signalled by the annihilation of the atom as it is ejected from the magnetic trap, for which the annihilation identification techniques discussed in section \ref{sec:cosmicAnalysis} will be important.
Identification of the antiproton annihilations is one of the places where antihydrogen has an advantage over the same measurements on hydrogen -- clearer signals can be obtained for much smaller numbers of atoms.
The first limits on the precision will come as the magnetic field homogeneity (presently $\sim 10^{-3}$ over the central 1~cm axial length) and the Doppler shift of the atoms become important.
More advanced trap geometries, with more homogeneous fields in the central region, or colder atoms, possibly employing direct laser cooling of the atoms, will be necessary to progress past this point.

If signals of CPT violation are to be found in the antihydrogen spectrum, it is likely that they are present to only a very small level, as large deviations from CPT symmetry are excluded by measurements on other systems.
Hydrogen spectroscopy has been performed with a precision of $\sim 10^{-14}$, and there is no reason in principle why the same cannot be achieved for antihydrogen.
The task of achieving precision antihydrogen spectroscopy is just as, if not more, challenging than the work to produce trapped atoms, and is likely to take many years to reach.

Trapped antihydrogen atoms open several other avenues to explore the behaviour of neutral antimatter, one of the most obvious being the gravitational interaction of antimatter.
Some kinds of measurements have been proposed (see a fuller discussion in \cite{RouteUltraLowEnergyHBar}), and become feasible as larger numbers of cold, trapped atoms are realised.
The first measurements on antimatter and gravity are still some years away.

Cold-antimatter physics is now at a turning point, where recent discoveries have cleared the path to work not before possible.
The next few years will be an exciting time as these new experiments are carried out.

    \bibliographystyle{bibstyle_eoin_aip}
        \phantomsection
        \addcontentsline{toc}{chapter}{Bibliography}
    \begin{flushleft}
        {\renewcommand\newpage{}\bibliography{bib/alphaPapers,bib/papers,bib/reviews,bib/websites,bib/misc,bib/books,bib/ALPHAInternal}}
    \end{flushleft}

\end{document}